\documentclass[aps,prb,onecolumn,nofootinbib,citeautoscript,10pt]{revtex4-2} 

\usepackage{amsmath,amssymb,bm} 
\usepackage{graphicx,comment}

\usepackage[tight]{subfigure} 

\usepackage[dvipsnames]{xcolor} 
\usepackage[papersize={8.5in,11in}]{geometry}
\usepackage[colorlinks=true]{hyperref}
\hypersetup{
    bookmarks=true,         % show bookmarks bar?
    unicode=false,          % non-Latin characters 
    pdftoolbar=true,        % show Acrobat
    pdfmenubar=true,        % show Acrobat 
    pdffitwindow=false,     % window fit to page when opened
    pdfstartview={FitH},    % fits the width of the page to the window
    pdfkeywords={keyword1} {key2} {key3}, % list of keywords
    pdfnewwindow=true,      % links in new window
    colorlinks=true,       % false: boxed links; true: colored links
    linkcolor=magenta, %red,          % color of internal links (change box color with linkbordercolor)
    citecolor=blue,        % color of links to bibliography
    filecolor=magenta,      % color of file links
    urlcolor=blue           % color of external links
} 

\usepackage{dcolumn}
\usepackage{color}
\usepackage{amssymb,amsmath}
\usepackage{tabularx,graphicx,makecell}
\usepackage{epstopdf}
\usepackage{latexsym}
\usepackage{colortbl}
\usepackage{psfrag}
\usepackage{bbm,bm,array,physics}
\usepackage{dsfont}
\usepackage{float, mathrsfs}

\newcommand{\bk}{\boldsymbol{k}}
\newcommand{\bB}{\boldsymbol{B}}
\newcommand{\bE}{\boldsymbol{E}}
\newcommand{\bw}{\boldsymbol{w}}
\newcommand{\bOm}{\boldsymbol{\Omega}}
\newcommand{\Lhat}{\hat{L}}

\def \nn{\nonumber \\}
\providecommand{\eqrf}[1]{\eqref{#1}}
\def\*#1{\boldsymbol{#1}} %\bold font
\DeclareRobustCommand{\bs}[1]{\bm{#1}}

\begin{document}

\title{Beyond-RTA magnetoconductivity of multifold semimetals under generic orientations of electromagnetic fields}

\author{Ipsita Mandal}
\email{ipsita.mandal@snu.edu.in}

\affiliation{Department of Physics, Shiv Nadar Institution of Eminence (SNIoE), Gautam Buddha Nagar, Uttar Pradesh 201314, India}

\begin{abstract}
%%%%%%%%%%
We present an exact solution of the linearised semiclassical Boltzmann equation (SBE) beyond the relaxation-time approximation for multifold semimetals under arbitrarily oriented electric and magnetic fields. The framework treats the Berry curvature, orbital magnetic moment, Lorentz-force operator, and scattering terms on equal footing. Momentum-independent scattering-amplitudes couple the Fermi-surface sheets through intranode, interband, and internode processes. The scattering kernel has a finite rank and the SBE reduces to a linear system for a few moments, which is closed by the electron-number constraint. We apply this approach to four systems involving quadratic-corrected triple-point, Weyl, and Rarita-Schwinger-Weyl nodes, including tilted triple-point nodes. We obtain analytical results at zero field and to leading orders in the magnetic field, alongside numerical solutions for the full field dependence. Our results elucidate the roles of the topological properties, Lorentz-force operator, tilt, and intra- and internode scatterings in shaping the magnetoelectric conductivity. It cures severe shortcomings of previous treatments like incomplete ansätze and incomplete treatment of the Lorentz-force operator. 
%%%%%%%%%%%%%%%%%
\end{abstract}

\maketitle
\tableofcontents

%%%%%%%%%%%%%%%
\section{Introduction}
\label{secintro}

Three-dimensional (3d) semimetals host symmetry-protected band crossings near the Fermi level, where the density-of-states (DOS) vanishes \cite{burkov11_Weyl, yan17_topological, ips-kush-review, bernevig, grushin-multifold, ips-hermann-review, claudia-multifold}. Chiral crystals admit nodes beyond the twofold nodes of the Weyl semimetals (WSMs) \cite{burkov11_Weyl, armitage_review, yan17_topological}, such as the threefold nodes of pseudospin one that define the triple-point semimetals (TSMs) \cite{bernevig, ady-spin1, grushin-multifold, claudia-multifold}, and nodes of higher pseudospin. Each node is a source or a sink of the Berry curvature (BC) in momentum space, and the sign of its monopole charge defines the chirality $\chi=\pm$. The charges of all the nodes in the Brillouin zone (BZ) add up to zero \cite{nielsen81_no}, and the node that compensates a given node need not have the same pseudospin \cite{claudia-multifold, grushin-multifold}. An example is a TSM node of the charge $+2$ at the $\Gamma$ point and a fourfold node at the $R$ point, which consists of two Weyl nodes of the same chirality and carries the charge $-2$ \cite{claudia-multifold, ips-internode}. A candidate material is CoSi \cite{claudia-multifold}.

At the linear order in the momentum, the TSM node has a flat-band, which is an artefact of the truncation. The symmetries allow a quadratic correction proportional to the identity matrix \cite{ni2021_giant}, which makes this band dispersive. We call the resulting node the quadratic-corrected TSM (QTSM). The correction leaves the spinors unchanged, and the BC and the orbital magnetic moment (OMM) remain those of the TSM. The band that is flat at the linear order has a vanishing BC and a nonvanishing OMM. For $c>0$, with $c$ the coefficient of the quadratic term, and a positive chemical potential, this band contributes to the transport together with the upper band, and the QTSM node has two Fermi-surface (FS) sheets \cite{ni2021_giant, claudia-multifold, ips-exact-spin1}.

Multifold band crossings occur in two distinct situations. In the first, the two conjugate nodes belong to the same pseudospin variety, and the BC of their bands differ by an overall factor of $\chi$. A pair of TSMs is an example of this situation \cite{bernevig, ady-spin1}, and the systems of Secs.~\ref{secsys3} and \ref{secsys4} belong to it. In the second, the two nodes carry different pseudospin quantum numbers and therefore have distinct BC fields. In chiral cubic crystals, a TSM node at the $\Gamma$ point can be compensated by a fourfold-degenerate node at the $R$ point, which consists of two copies of a Weyl node of the same chirality, with equal and opposite net monopole charges of magnitude $2$ \cite{claudia-multifold}. Similarly, an RSW node at the $\Gamma$ point with the monopole charge of magnitude $4$ can be compensated by a sixfold-degenerate node at the $R$ point, which originates from the doubling of the pseudospin-1 excitations \cite{tang2017_multiple, prb108035428, prl119206401, yamakage}. These are the systems of Secs.~\ref{secsys} and \ref{secsys2}. In realistic materials, multifold band crossings therefore do not in general arise in conjugate pairs of the same pseudospin representation. The Nielsen-Ninomiya theorem \cite{nielsen81_no} is satisfied by the vanishing of the net monopole charge, which is summed over all the nodes of the BZ.

The linear response of such systems is usually obtained from the linearised semiclassical Boltzmann equation (SBE) in the relaxation-time approximation (RTA), which discards the in-scattering term. For the chiral anomaly, the RTA treatment of the internode scatterings requires a phenomenological internode relaxation time, as in Ref.~\cite{ips-internode}. Knoll et al.~\cite{timm} provided the pioneering methodology to solve the SBE beyond the RTA under collinear electromagnetic fields, considering a pair of conjugate Weyl nodes with intranode and internode scatterings. Refs.~\cite{ips-exact-spin1, ips-exact-kwn} extend this treatment to multifold nodes. Refs.~\cite{ahmad-sharma-2025, ahmad-phc-2026} also consider multifold nodes, and the latter considers noncollinear fields and tilts. These works have several shortcomings, which we discuss in Sec.~\ref{seclit}. In particular, both of them use an OMM for the pseudospin-1 node that differs by a factor of two from Eq.~\eqref{eqbcomm}. For noncollinear fields, the Lorentz-force operator (LFO) couples the components of the mean free-path vector. Earlier treatments of noncollinear fields are incomplete because they use a truncated ansatz \cite{PhysRevB.103.115146, sharma-nandy-tewari-2020}, or they expand in the LFO \cite{suh-min-2023, ips-rsw-ph, ips-spin1-ph}. Ref.~\cite{ips-exact-noncollinear} has built the formalism to account for the LFO and the in-scattering term together and treated Weyl/multi-Weyl nodes. Here, we apply its framework to multifold nodes. The first system has the QTSM node at $\Gamma$ and two Weyl nodes at $R$. The second has the Rarita-Schwinger-Weyl (RSW) node of spin 3/2 at $\Gamma$ and two TSM nodes at $R$, which first-principles calculations find in chiral cubic crystals \cite{gao2022, prl119206401}. The solution of Ref.~\cite{ips-exact-spin1} for two conjugate QTSM nodes and collinear fields is the simplest scenario of a pair of multifold nodes. We generalise it to arbitrary angles between $\bE$ and $\bB$, and we also add a type-I axial tilt.

The nodes of the first two systems are untilted, as the crystal symmetry requires. All the nodes sit at time-reversal-invariant momenta of a chiral cubic space group. A tilt is a term proportional to the identity matrix and odd in the momentum. Time reversal maps $\bk$ to $-\bk$ and leaves the identity matrix unchanged, and it therefore forbids the tilt at these momenta. The little group of each point, which is the tetrahedral group in the space group 198 and the octahedral group in the space group 212, has no vector that is invariant under all its rotations, and it would forbid the tilt on its own \cite{prl119206401, grushin-multifold}. The quadratic term proportional to the identity matrix is even in the momentum and remains allowed \cite{grushin-multifold}. An axial tilt is allowed at momenta that are not time-reversal invariant and whose little group contains a rotation axis, which is the direction of the tilt. We include it for the pair of conjugate QTSM nodes, restrict ourselves to the type-I regime, and do not specify a space group. The tilt reverses its sign under a rotation by $\pi$ about an axis perpendicular to it. The Onsager relation then connects the conductivity at the tilt $t$ with that at the tilt $-t$ [cf. Eq.~\eqref{eqons4}]. As a consequence, the diagonal elements acquire terms that are odd in $\bB$, and the Hall element has a part that is even in $\bB$ and does not vanish at zero field.

The scattering amplitudes are momentum independent. Compared with a pair of Weyl nodes, the QTSM and Weyl system has a larger span of the finite-rank kernel, because the squared spinor overlap of the QTSM node is of second degree in the pseudospin direction, and it has four FS sheets coupled by interband, intranode, and internode amplitudes. In the untilted systems, the FS of every sheet is axisymmetric about $\bB$. For $\bB$ along the polar axis, the LFO is then diagonal in the azimuthal harmonic, and its resolvent is algebraic. For the tilted nodes, the FS is axisymmetric about $\bB$ only for $\bB$ along the polar axis. For all the other orientations of $\bB$, we evaluate the resolvent by a Galerkin expansion in spherical harmonics. We also give the RTA limit of every system, including the internode contribution that requires a phenomenological internode relaxation time.

The paper is organised as follows. Sec.~\ref{secboltz} sets up the SBE and its exact solution for a finite-rank scattering kernel. The QTSM and Weyl nodes and the RSW and TSM nodes are then treated in Secs.~\ref{secsys} and \ref{secsys2}, followed by two conjugate QTSM nodes without and with an axial tilt in Secs.~\ref{secsys3} and \ref{secsys4}. A comparison with earlier works, the summary, and the outlook conclude the main text. The appendices contain various intermediate details of the results presented in the main text.

We use natural units with $\hbar=k_B=1$ and retain the electronic charge $e$ in the analytical expressions. The BC and OMM contributions are switched off by the factors $\eta_B,\eta_O\in\{0,1\}$ attached to $\bOm$ and $\boldsymbol m$, which are displayed explicitly. The temperature is zero throughout.

%%%%%%%%%%%%%%%
\section{Semiclassical Boltzmann equation}
\label{secboltz}

We follow Refs.~\cite{mermin, sundaram99_wavepacket, xiao_review, timm, ips-exact-spin1, ips-exact-kwn, ips-exact-rsw} for the linearised SBE of the FS sheets. Let $a=(\nu,s)$ label the FS sheets, with $\nu$ the node and $s$ the band index, and let $\chi_\nu$ be the chirality of the node to which the sheet $a$ belongs. The carriers have the charge $-e$ and occupy a band with the BC $\bOm^a$ and the OMM $\boldsymbol m^a$. The OMM shifts the band energy $\epsilon_s(\bk)$ to $\xi_s=\epsilon_s-\eta_O\,\bB\cdot\boldsymbol m^a$, and the group velocity is $\bw_a=\nabla_{\bk}\xi_s$. The semiclassical equations of motion read \cite{sundaram99_wavepacket, xiao_review}
\begin{align}
\dot{\boldsymbol r}=\bw_a-\eta_B\,\dot{\bk}\times\bOm^a\,,\quad
\dot{\bk}=-\,e\left(\bE+\dot{\boldsymbol r}\times\bB\right).
\label{eqeom}
\end{align}
The second term in the equation for $\dot{\boldsymbol r}$ is the anomalous velocity. Solving Eq.~\eqref{eqeom} for $\dot{\boldsymbol r}$ and $\dot{\bk}$ gives
\begin{align}
& \dot{\boldsymbol r}=D_a^{-1}\left[\bw_a+\eta_B\,e\,\bE\times\bOm^a+\eta_B\,e\left(\bOm^a\cdot\bw_a\right)\bB\right],
\quad \dot{\bk}=-\,e\,D_a^{-1}\left[\bE+\bw_a\times\bB+\eta_B\,e\left(\bE\cdot\bB\right)\bOm^a\right],
\nn & D_a=1+\eta_B\,e\,\bB\cdot\bOm^a\,.
\label{eqeomsol}
\end{align}
The factor $D_a$ is the phase-space factor due to the BC, and the invariant momentum-space measure is $D_a\,d^3k/(2\,\pi)^3$.

For time-independent and spatially uniform fields, the distribution function $f_a(\bk)$ of the sheet $a$ obeys $\dot{\bk}\cdot\nabla_{\bk}f_a=I_{\rm coll}$, with $I_{\rm coll}$ the collision integral. To the linear order in $\bE$, $f_a=n_F(\xi_s)+g_{\bk}$ with $n_F$ the Fermi-Dirac distribution, and the deviation is parametrised as
\begin{align}
g_{\bk}=e\,\partial_{\xi} n_F \,\bE\cdot\boldsymbol{\Lambda}_{a}(\bk)\,,
\label{eqansatzg}
\end{align}
where $\boldsymbol{\Lambda}_a(\bk)$ is the vector mean free-path of the sheet $a$. The LFO is
\begin{align}
\Lhat=(\bw\times\bB)\cdot\nabla_{\bk} \,,
\label{eqlfo}
\end{align}
the derivative along the cyclotron orbit. Since $(\bw\times\bB)\cdot\bw=0$, it annihilates $\partial_\xi n_F$ and acts only on $\boldsymbol\Lambda_a$. The kinetic equation to the linear order in $\bE$ is
\begin{align}
-\,e\,D_a^{-1}\,\partial_\xi n_F\,\bE\cdot\left[\tilde{\boldsymbol h}_a+e\,\Lhat\,\boldsymbol\Lambda_a\right]=I_{\rm coll}\,,\quad
\tilde h_a^j=w_a^j+\eta_B\,e\,B^j\left(\bOm^a\cdot\bw_a\right).
\label{eqkinlin}
\end{align}

We consider elastic and nonmagnetic point scatterers with momentum-independent amplitudes $V^{aa'}$. The collision integral is
\begin{align}
& I_{\rm coll}=\sum_{a'}\int\frac{d^3k'}{(2\,\pi)^3}\,D_{a'}(\bk')\,W^{aa'}_{\bk\bk'}\left(g_{\bk'}-g_{\bk}\right),\quad
 W^{aa'}_{\bk\bk'}= 2\,\pi\,n\,|V^{aa'}|^2\;\mathcal O_{aa'}(\hat{\bk},\hat{\bk}')\;\delta\left(\xi_{s'}(\bk')-\xi_s(\bk)\right),
\label{eqicoll}
\end{align}
where $n$ is the density of the scatterers and $\mathcal O_{aa'}$ is the squared overlap of the spinors of the sheets $a$ and $a'$ [cf. Eq.~\eqref{eqoverlap}]. For two sheets of the same node, $\mathcal O_{aa'}$ is the squared overlap of the eigenvectors of the two bands, and it does not vanish for $s\neq s'$. The rate follows from Fermi's golden rule, and the energy-conserving delta function removes the equilibrium part of $f_a$. We write $\sum_{\bk'}\equiv\sum_{a'}\int d^3k'\,D_{a'}(\bk')/(2\,\pi)^3$.

Inserting Eq.~\eqref{eqicoll} in Eq.~\eqref{eqkinlin} and dividing by $-\,e\,\partial_\xi n_F$ yields the SBE for each Cartesian component $j$ of the vector $\boldsymbol\Lambda_a$,
\begin{align}
D_a^{-1}\left[\tilde h^j_a+e\,\Lhat\,\Lambda^j_a\right]=-\sum_{\bk'}W^{aa'}_{\bk\bk'}\left(\Lambda^j_{a'}(\bk')-\Lambda^j_a(\bk)\right).
\label{eqboltz}
\end{align}
Since the scattering is elastic, the sum over $\bk'$ is restricted to the FS. Let $d^2\hat{\bk}'$ be the solid-angle element and $\rho_a=k^2\,D_a/(\bw_a\cdot\hat{\bk})$ the FS weight, which contains the phase-space factor. We assume that the FS of every sheet is a graph over the unit sphere with a single-valued radius $k_a(\hat{\bk})$, which holds when $\xi_s$ increases monotonically along every ray, i.e., for $\bw_a\cdot\hat{\bk}>0$. Functions on the FS, such as $\Lambda^j_a$, can then be expanded in spherical harmonics of $\hat{\bk}$. The collision term of Eq.~\eqref{eqboltz} becomes
\begin{align}
& \sum_{\bk'}W^{aa'}_{\bk\bk'}\left(\Lambda^j_a-\Lambda^{j}_{a'}(\bk')\right)=\left(\hat{\mathcal C}\,\Lambda^j\right)_a\equiv\frac{\Lambda^j_{a}}{\tau_a}-\sum_{a'}\int d^2\hat{\bk}'\,\rho_{a'}(\hat{\bk}')\,\mathcal W_{aa'}(\hat{\bk},\hat{\bk}')\,\Lambda^j_{a'}(\hat{\bk}')\,,\quad
\frac{1}{\tau_a(\hat{\bk})}=\sum_{a'}\int d^2\hat{\bk}'\,\rho_{a'}(\hat{\bk}')\,\mathcal W_{aa'}(\hat{\bk},\hat{\bk}')\,.
\label{eqcoll}
\end{align}
Here $\mathcal W_{aa'}(\hat{\bk},\hat{\bk}')=\mathcal W_{a'a}(\hat{\bk}',\hat{\bk})\geq0$ is proportional to the spinor overlap and absorbs all constant factors. With the source $h^j_a=D_a^{-1}\,\tilde h^j_a$, Eq.~\eqref{eqboltz} becomes
\begin{align}
\left( \hat{\mathcal C}-e\,D_a^{-1}\Lhat\right ) \Lambda^j=h^j_a\,.
\label{eqboltzop}
\end{align}
The operator on the left-hand side has a constant zero mode.

\subsection{Zero mode, number constraint, and pumping}
\label{seczero}

For $\Lambda^j_a=1$ on every sheet, the LFO term vanishes because $\Lhat$ is a derivative, and the collision term vanishes because the in-scattering integral reproduces $1/\tau_a$ of Eq.~\eqref{eqcoll}. Elastic scattering conserves the electron number, and a uniform shift of $\boldsymbol\Lambda$ enters $g_{\bk}\propto\partial_\xi n_F$ as a shift of the chemical potential, which cannot relax. Two properties of the operator follow. First, $\hat{\mathcal C}$ is symmetric with respect to the measure $\rho_a\,d^2\hat{\bk}$, since $\rho_a\,\rho_{a'}'\,\mathcal W_{aa'}$ is symmetric under the exchange of $(a,\hat{\bk})$ and $(a',\hat{\bk}')$. Second, $D_a^{-1}\Lhat$ is antisymmetric with respect to the same measure, because the flow $\bw_a\times\bB$ is tangent to the FS and divergence-free, as $\bw_a=\nabla_{\bk}\xi_s$ is curl-free. Multiplying Eq.~\eqref{eqboltzop} with $h^j_a=0$ by $\rho_a\,\Lambda^j$ and integrating, the antisymmetric part drops out, and
\begin{align}
\sum_a\int d^2\hat{\bk}\,\rho_a\,\Lambda^j_a\left(\hat{\mathcal C}\,\Lambda^j\right)_a=\frac12\sum_{a,a'}\int d^2\hat{\bk}\,d^2\hat{\bk}'\,\rho_a\,\rho_{a'}'\,\mathcal W_{aa'}\left(\Lambda^j_{a}-\Lambda^{j}_{a'}(\hat{\bk}')\right)^2=0\,.
\end{align}
The kernel $\mathcal W_{aa'}$ is nonnegative and vanishes only at isolated directions. The homogeneous solution is therefore constant on every set of sheets connected by the scattering, with one zero mode for each such set. Here the scattering connects all four sheets, and there is a single zero mode. In the finite-rank representation of Eq.~\eqref{eqkernel}, it is the vector of moments of the weight $\rho_a$ itself. Its amplitude is fixed by the conservation of the electron number in the linear response, which gives the number constraint
\begin{align}
\sum_a\int d^2\hat{\bk}\,\rho_a\,\Lambda^j_a=0
\label{eqnumber}
\end{align}
for every $j$. Since constants are also left zero modes of the operator, integrating Eq.~\eqref{eqboltzop} against $\rho_a\,d^2\hat{\bk}$ over the connected set gives the solvability condition
\begin{align}
\sum_a\int d^2\hat{\bk}\,\rho_a\,h^j_a=0\,.
\label{eqsolv}
\end{align}
This is the continuity equation.

The zero mode exists for every orientation of $\bE$ and $\bB$ and is lost only if the scattering does not conserve the electron number.

The pumping rate follows from the geometry of the FS. With $d\mathcal S$ and $\hat n$ the area element and the normal of the FS, $\rho_a\,h^j_a\,d^2\hat{\bk}=d\mathcal S\left(\hat n^j+\eta_B\,e\,B^j\,\bOm^a\cdot\hat n\right)$. The first term integrates to zero over a closed surface, and the second gives the Berry flux,
\begin{align}
\int d^2\hat{\bk}\,\rho_a\,h^j_a=2\,\pi\,\eta_B\,e\,B^j\,\mathcal C_a\,,
\end{align}
where $\mathcal C_a$ is the Chern number of the FS sheet $a$. This is the rate $e^2\,\mathcal C_a\,\bE\cdot\bB/(4\,\pi^2)$ of electron-number change, i.e., the chiral-anomaly term \cite{ips-internode}. The band-velocity term $w^j_a$, which includes the OMM velocity, integrates to zero and carries no anomaly. The rate vanishes for every $\bE\perp\bB$. For $\bE\parallel\bB$, Eq.~\eqref{eqsolv} reduces to $\sum_a\mathcal C_a=0$ over the connected set, which holds for the Chern numbers of the sheets of each system.

\subsection{Exact solution for a finite-rank kernel}
\label{secfinrank}

The scattering kernels of all the systems considered here have a finite rank. For each sheet $a$, the squared spinor overlaps fix a finite set of functions $f^a_p(\hat{\bk})$, which includes the constant function $f^a_0=1$ and spans the range of the in-scattering term of Eq.~\eqref{eqcoll}. The kernel then takes the form $\mathcal W_{aa'}(\hat{\bk},\hat{\bk}')=\sum_{p,p'}f^a_p(\hat{\bk})\,Q^{aa'}_{pp'}\,f^{a'}_{p'}(\hat{\bk}')$, where $p$ and $p'$ label the functions and the constant matrices $Q^{aa'}$ follow from the overlaps. The unknowns of the in-scattering term are the moments $\mathcal M^j_{a,p}=\int d^2\hat{\bk}\,\rho_a\,f^a_p\,\Lambda^j_a$ for each Cartesian component $j$ of $\bE$. The equation for $\Lambda^j_a$ is
\begin{align}
\hat{\mathcal R}_a^{-1}\,\Lambda^j_a=h^j_a+\sum_{a'}\sum_{p,p'}f^a_p\,Q^{aa'}_{pp'}\,\mathcal M^j_{a',p'}\,,\quad
\hat{\mathcal R}_a=\left(\tau_a^{-1}-e\,D_a^{-1}\Lhat\right )^{-1}.
\label{eqtilteq}
\end{align}
Here $\hat{\mathcal R}_a$ is the resolvent, which exists because $D_a^{-1}\Lhat$ is antisymmetric with respect to $\rho_a\,d^2\hat{\bk}$ and $\tau_a^{-1}>0$. The exact solution is
\begin{align}
\Lambda^j_a=\hat{\mathcal R}_a\left[h^j_a\right]+\sum_{a'}\sum_{p,p'}Q^{aa'}_{pp'}\,\mathcal M^j_{a',p'}\,\hat{\mathcal R}_a\left[f^a_p\right].
\label{eqtiltsol}
\end{align}
For $\Lhat\to0$, $\hat{\mathcal R}_a\to\tau_a$ and Eq.~\eqref{eqtiltsol} reduces to the truncated form of the earlier works. Multiplying Eq.~\eqref{eqtiltsol} by $\rho_a\,f^a_p$ and integrating over the sphere gives the linear system for the moments,
\begin{align}
&\sum_{a',p'}\Big( \delta_{aa'}\,\delta_{pp'}-\sum_{p''}K^a_{pp''}\,Q^{aa'}_{p''p'}\Big )
\mathcal M^j_{a',p'}=S^j_{a,p}\,,\quad
 K^a_{pp'}=\int d^2\hat{\bk}\,\rho_a\,f^a_p\,\hat{\mathcal R}_a
\left[f^a_{p'}\right],\quad
S^j_{a,p}=\int d^2\hat{\bk}\,\rho_a\,f^a_p\,\hat{\mathcal R}_a\left[h^j_a\right].
\label{eqmomsys}
\end{align}
The matrix of Eq.~\eqref{eqmomsys} is the same for the three components $j$, which differ only through the sources. Its only null vector is the zero mode of Eq.~\eqref{eqboltzop}, and the number constraint of Eq.~\eqref{eqnumber}, $\sum_a\mathcal M^j_{a,0}=0$, closes the system. The solvability condition of Eq.~\eqref{eqsolv} holds because the Chern numbers of the sheets add up to zero. The equations of the system are linearly dependent, since their combination that corresponds to the integration against $\rho_a\,d^2\hat{\bk}$ is satisfied identically once Eq.~\eqref{eqsolv} holds, and the right-hand side lies in the range of the rank-deficient matrix only under this condition. The number constraint is therefore not an additional condition on an overdetermined system. It supplies the independent equation that fixes the amplitude of the zero mode.

For a general orientation of $\bB$, the resolvent is evaluated by a Galerkin expansion. The ansatz for each sheet $a$ and component $j$ is the expansion in spherical harmonics,
\begin{align}
\Lambda^j_a(\hat{\bk})=\sum_{l=0}^{l_{\max}}\sum_{m=-l}^{l}c^{\,j}_{a,lm}\,Y_{lm}(\hat{\bk})\,,\quad
c^{\,j}_{a,l,-m}=(-1)^m\left(c^{\,j}_{a,lm}\right)^*,
\label{eqansatz}
\end{align}
where the second relation expresses the reality of $\Lambda^j_a$. The coefficient vector $\bs c^{\,j}_a$ is fixed by the equation of motion as
\begin{align}
\bs c^{\,j}_a=\mathsf F_a^{-1}\Big\{\mathsf U\!\left[h^j_a\right]+\sum_{a'}\sum_{p,p'}Q^{aa'}_{pp'}\,\mathcal M^{j}_{a',p'}\,\mathsf U\!\left[f^a_p\right]\Big\},
\label{eqansatzc}
\end{align}
with $\left(\mathsf F_a\right)_{lm,\,l'm'}=\int d^2\hat{\bk}\,Y^*_{lm}\left(\tau_a^{-1}-e\,D_a^{-1}\Lhat\right)Y_{l'm'}$ and $\mathsf U_{lm}=\int d^2\hat{\bk}\,Y^*_{lm}\,u$. The only free parameters are the moments $\mathcal M^j_{a,p}$, which solve Eq.~\eqref{eqmomsys}.

\subsection{Cyclotron flow and resolvent for axisymmetric Fermi surfaces}
\label{secaxires}

The LFO acts on the FS as the cyclotron flow on the unit sphere,
\begin{align}
\Lhat\,u=\bs V\cdot\nabla_{\hat{\bk}}\,u=V_\theta\,\partial_\theta u
+\frac{V_\varphi}{\sin\theta}\,\partial_\varphi u\,,\quad
\bs V=\left[(\bw_a\times\bB)-\hat{\bk}\,\hat{\bk}\cdot(\bw_a\times\bB)\right]/k\,,
\label{eqlfomw}
\end{align}
where $V_\theta=\bs V\cdot\hat{\boldsymbol\theta}$ and $V_\varphi=\bs V\cdot\hat{\boldsymbol\varphi}$ are the components of $\bs V$ along $\hat{\boldsymbol\theta}=\left(\cos\theta\,\cos\varphi,\cos\theta\,\sin\varphi,-\sin\theta\right)$ and $\hat{\boldsymbol\varphi}=\left(-\sin\varphi,\cos\varphi,0\right)$.

When the FS of every sheet is axisymmetric about $\bB$, as for the untilted spherical nodes, we take $\bB=B\,\boldsymbol{\hat z}$ without loss of generality, and the LFO becomes
\begin{align}
\Lhat=-B\,\frac{w_\perp}{k_\perp}\,\partial_\varphi\,,
\label{eqlfoaxi}
\end{align}
with $k_\perp=k\sin\theta$ and $w_\perp$ the cylindrical radial component of $\bw_a$. It is diagonal in the azimuthal harmonic $m$ of $\Lambda\propto e^{i\,m\,\varphi}$ and annihilates the sector $m=0$. The resolvent is then algebraic in $\theta$,
\begin{align}
\hat{\mathcal R}_a\left[u(\theta)\,e^{i\,m\,\varphi}\right]=\frac{u(\theta)\,e^{i\,m\,\varphi}}{\tau_a^{-1}(\theta)+i\,m\,\omega_a(\theta)}\,,\quad
\omega_a(\theta)=\frac{e\,B\,q_a}{k\,D_a}\,,\quad q_a=\frac{w_\perp}{\sin\theta}\,,
\label{eqresalg}
\end{align}
with the explicit forms of $q_a$ and $D_a$ given in Appendix~\ref{secappingr}. The general Galerkin construction of Ref.~\cite{ips-exact-noncollinear} provides an independent evaluation of the resolvent.

The zero mode and the number constraint are confined to the sector $m=0$, which is the component $\Lambda^z$ for $\bE\parallel\bB$. The sectors with $m\neq0$ have no zero mode, number constraint, or pumping. The sources of $\Lambda^z$ and of $\Lambda^x\pm i\,\Lambda^y$ have $m=0$ and $m=\pm1$, respectively, and the sectors with $|m|\geq2$ have no source. The solution for $m=-1$ is the complex conjugate of that for $m=1$. For $\bE=E\left(\sin\vartheta,0,\cos\vartheta\right)$, the deviation from equilibrium is proportional to $\sin\vartheta\,\Lambda^x_a+\cos\vartheta\,\Lambda^z_a$, and the two sectors are solved once at the full $B$. When the kernel or the FS is not invariant under the rotations about $\bB$, the LFO couples the harmonics $m$ and $m\pm1$, and the resolvent is evaluated by the Galerkin matrix of Eq.~\eqref{eqansatzc}.

\subsection{Conductivity tensor}
\label{seccond}

The conductivity tensor is the sum of the Boltzmann part and the Fermi-sea anomalous-Hall (AH) part,
\begin{align}
\sigma_{ij}=\frac{e^2}{8\,\pi^3}\sum_a
\int d^2\hat{\bk}\,\rho_a\,h^i_a\,\Lambda^j_a-\frac{\eta_B\,e^2}{8\,\pi^3}\,\epsilon_{ijl}
\sum_a\int d^2\hat{\bk}\int_0^{k_a(\hat{\bk})}dk\;k^2\,\Omega^a_l\,.
\label{eqsigmw}
\end{align}
The direction of $\bE$ enters only through the sources. The mean free-path is therefore needed for the three Cartesian components $j=x,y,z$, and the deviation from equilibrium for $\bE=E\,\hat{\bE}$ is proportional to $\hat{\bE}\cdot\bs\Lambda_a$. For $\bE=E\,\hat{\bE}$ with $\cos\vartheta=\hat{\bE}\cdot\hat{\bB}$ and $\sin\vartheta\neq0$, the unit vectors along the in-plane direction perpendicular to $\hat{\bE}$ and the out-of-plane direction are
\begin{align}
\hat{\bs\nu}_{\rm PH}=\frac{\cos\vartheta\,\hat{\bE}-\hat{\bB}}{\sin\vartheta}\,,\quad
\hat{\bs\nu}_{\rm H}=\frac{\hat{\bB}\times\hat{\bE}}{\sin\vartheta}\,.
\end{align}
The current components along $\hat{\bE}$, $\hat{\bs\nu}_{\rm PH}$, and $\hat{\bs\nu}_{\rm H}$ are
\begin{align}
j_{\bE}=E\,\hat{\bE}\cdot\bs\sigma\,\hat{\bE}\,,\quad
j_{\rm PH}=E\,\hat{\bs\nu}_{\rm PH}\cdot\bs\sigma\,\hat{\bE}\,,\quad
j_{\rm H}=E\,\hat{\bs\nu}_{\rm H}\cdot\bs\sigma\,\hat{\bE}\,.
\label{eqcurrtilt}
\end{align}

In natural units, the speed of light is unity and $4\,\pi\,\epsilon_0=137$. We take the Fermi velocity $v_F=15\times10^{5}$ m~s$^{-1}$, which equals $0.005$ in these units, from Ref.~\cite{ips-ruiz}. The transport time is of the order of $\tau=10^{-13}$ s, which equals $152$ eV$^{-1}$, following Refs.~\cite{watzman18_dirac, rahul-jpcm}. The magnetic field ranges from $0$ to $10$ Tesla, which equals $0$ to $2000$ eV$^{2}$ \cite{ghosh20_chirality}. The chemical potential ranges from $0.01$ to $0.1$ eV, which equals $1.6\times10^{-21}$ to $1.6\times10^{-20}$ J \cite{ips-ruiz}. The remaining parameters of each system are specified where they are used. The magnetic field enters through the dimensionless parameters $\alpha_a=e\,B/(2\,k_a^2)$ with $k_a$ the FS radius of the sheet $a$ at $B=0$, and through the products $\omega_a\,\tau_a$ of the angle-dependent cyclotron frequency of Eq.~\eqref{eqresalg} and the transport time. The solution depends in addition on the ratios of the scattering amplitudes entering Eq.~\eqref{eqkernel}.
The range of $B$ is restricted by the semiclassical limit, in which the cyclotron energy is small compared with $\mu$, as shown in the Supplemental Material of Ref.~\cite{timm}. The field range quoted above has to be cut accordingly for a given $\mu$. The numerical implementation is described in Appendix~\ref{secnum}.

%%%%%%%%%%%%%%%

%%%%%%%%%%%%%%%%%%%%%%%%%%%%%%%%%%%%%%%%%%%%%%%%%%%%%%%%%%%%
\section{QTSM and Weyl nodes}
\label{secsys}

The system comprises the QTSM node at the $\Gamma$ point and the two Weyl nodes at the $R$ point. It is realised without SOC in chiral cubic crystals of the space group 198, such as CoSi, in which four bands near the Fermi level originate from four orbitals per unit cell \cite{xu2020_optical, ni2021_giant, grushin-multifold, prl119206401, tang2017_multiple}. At $\Gamma$, these four bands form the threefold node of the QTSM and a singlet band. At $R$, all four bands meet in the fourfold node, which consists of two Weyl nodes of the same chirality \cite{grushin-multifold, ni2021_giant}. The Hamiltonians are
\begin{align}
& \mathcal H_\Gamma(\bk)=v_F\left(k_x\,S_x+k_y\,S_y+\chi_\Gamma\,k_z\,S_z\right)+c\,v_F\,k^2\,\mathbb I_{3\times3}\,,\nn
& \mathcal H_{R_i}(\bk)=\tilde v_F\left(k_x\,\sigma_x+k_y\,\sigma_y+\chi_{R}\,k_z\,\sigma_z\right)-\Delta\,\mathbb I_{2\times2}\,,\quad i=1,2\,.
\label{eqham}
\end{align}
Here $\bs S=\{S_x,S_y,S_z\}$ is the vector of the spin-1 matrices of Appendix~\ref{secappingr}, $\bs{\tilde\sigma}$ is the vector of the Pauli matrices, $k=|\bk|$, $v_F$ and $\tilde v_F$ are the isotropic velocities of the QTSM and Weyl nodes, and $c$ is the coefficient of the quadratic correction. The chiralities are $\chi_\Gamma=+1$ and $\chi_R=-1$, and $\Delta$ is the energy offset of the nodal points at $R$ with respect to that at $\Gamma$. The eigenvalues are
\begin{align}
\epsilon_{\Gamma,s}(k)=v_F\left(s\,k+c\,k^2\right),\quad s\in\{1,0,-1\}\,,\qquad
\epsilon_{R}(k)=\tilde v_F\,k-\Delta\,,
\label{eqeval}
\end{align}
where the second expression is the upper band of a Weyl node. We consider $c>0$ and $\mu>0$ and restrict to the momenta with $c\,k<1$, where the band $s=-1$ stays below the nodal point and does not take part in the transport. The chemical potential cuts the bands $s=1$ and $s=0$ at $\Gamma$ and the upper band at each node at $R$, which gives the four FS sheets $a\in\{(\Gamma,1),(\Gamma,0),(R_1,+),(R_2,+)\}$.

The BC and the OMM of the four sheets have the common form
\begin{align}
\bOm^a=-\frac{\chi_\nu\,N_a\,\hat{\bk}}{2\,k^2}\,,\quad
\boldsymbol m^a=-\frac{\chi_\nu\,e\,v_\nu\,G_a\,\hat{\bk}}{2\,k}\,,
\label{eqbcomm}
\end{align}
where $v_\Gamma=v_F$, $v_{R_1}=v_{R_2}=\tilde v_F$, and $\hat{\bk}=\bk/k$. The numbers $\{N_a,G_a\}$ are $\{2,1\}$ for the sheet $(\Gamma,1)$, $\{0,2\}$ for the sheet $(\Gamma,0)$, and $\{1,1\}$ for each sheet at $R$. The Chern number of the sheet $a$ is $\mathcal C_a=-\chi_\nu\,N_a$, which equals $-2$, $0$, $+1$, and $+1$ for $(\Gamma,1)$, $(\Gamma,0)$, $(R_1,+)$, and $(R_2,+)$, and these values add up to zero. The sheet $(\Gamma,0)$ has a vanishing BC and no anomalous pumping, and it enters through the number constraint and the scattering.

The energy shifted by the OMM, the phase-space factor due to the BC, and the group velocity are
\begin{align}
\xi_s(\bk)=\epsilon_s(k)+\eta_O\,\frac{\chi_\nu\,e\,v_\nu\,G_a\,(\bB\cdot\hat{\bk})}{2\,k}\,,\quad
D_a=1-\eta_B\,\frac{\chi_\nu\,N_a\,e\,(\bB\cdot\hat{\bk})}{2\,k^2}\,,\quad
\bw_a=\nabla_{\bk}\,\xi_s\,.
\label{eqxiw}
\end{align}

At $B=0$, the FS radii and the FS weights $\rho_a=k^2/(\partial_k\epsilon_s)$ follow from Eq.~\eqref{eqeval} as
\begin{align}
& k_{\Gamma,1}=\frac{-1+\sqrt{1+4\,c\,\mu/v_F}}{2\,c}\,,\quad
k_{\Gamma,0}=\sqrt{\frac{\mu}{c\,v_F}}\,,\quad
k_{R}=\frac{\mu+\Delta}{\tilde v_F}\,,\quad
 \rho_{\Gamma,1}=\frac{k_{\Gamma,1}^2}{v_F\left(1+2\,c\,k_{\Gamma,1}\right)}\,,\quad
\rho_{\Gamma,0}=\frac{k_{\Gamma,0}}{2\,c\,v_F}\,,\quad
\rho_{R}=\frac{k_{R}^2}{\tilde v_F}\,.
\label{eqfsb0}
\end{align}
For $B\neq0$, the radius $k_a(\hat{\bk})$ is the root of $\xi_s(k\,\hat{\bk})=\mu$ along each ray, which is given in Appendix~\ref{secappingr}.

The scattering amplitudes are $\beta_{aa'}=\sqrt{n/(4\,\pi)}\,|V^{aa'}|=\beta_{a'a}$, and the kernel strengths are $\beta_{aa'}^2$. The two nodes at $R$ are taken to be equivalent. The independent amplitudes are then $\beta^{\Gamma}_{ss'}$ for the scattering between $(\Gamma,s)$ and $(\Gamma,s')$, with $\beta^{\Gamma}_{10}=\beta^{\Gamma}_{01}$, $\beta^{R}_{\rm intra}=\beta_{R_iR_i}$, and $\beta^{\Gamma R}_{s}$ for the scattering between $(\Gamma,s)$ and either node at $R$.

Let $\hat{\bs n}^a_{\bk}=\left(\hat k_x,\hat k_y,\chi_\nu\,\hat k_z\right)$ be the unit vector along the vector $\bs d$ of the node of the sheet $a$. The squared spinor overlaps depend only on $\hat{\bs n}^a_{\bk}\cdot\hat{\bs n}^{a'}_{\bk'}$ and are
\begin{align}
& \mathcal O_{(\Gamma,1)(\Gamma,1)}=\frac14\left(1+\hat{\bs n}\cdot\hat{\bs n}'\right)^2,\quad
\mathcal O_{(\Gamma,0)(\Gamma,0)}=\left(\hat{\bs n}\cdot\hat{\bs n}'\right)^2,\quad
\mathcal O_{(\Gamma,1)(\Gamma,0)}=\frac12\left[1-\left(\hat{\bs n}\cdot\hat{\bs n}'\right)^2\right],\quad \mathcal O_{R_iR_i}=\frac12\left(1+\hat{\bs n}\cdot\hat{\bs n}'\right),
\label{eqoverlap}
\end{align}
with $\hat{\bs n}=\hat{\bs n}^a_{\bk}$ and $\hat{\bs n}'=\hat{\bs n}^{a'}_{\bk'}$, as derived in Appendix~\ref{secappingr}. The spinors of the QTSM node and of a Weyl node have different dimensions, and the overlap between a sheet at $\Gamma$ and a sheet at $R$ is therefore not fixed by the spinors alone. Its form is fixed by the orbital content of the model, as discussed below [cf. Eqs.~\eqref{eqovgr4} and \eqref{eqovgrx}]. The kernel of Eq.~\eqref{eqcoll} is $\mathcal W_{aa'}(\hat{\bk},\hat{\bk}')=\left(\beta_{aa'}^2/\pi\right)\mathcal O_{aa'}(\hat{\bk},\hat{\bk}')$, which reduces to $\left(\beta^2/2\,\pi\right)\left(1+\hat{\bs n}\cdot\hat{\bs n}'\right)$ for two Weyl nodes. The dependence of $\mathcal O_{aa'}$ on $\hat{\bk}$ is a polynomial of second degree in the components of $\hat{\bs n}$ for the sheets at $\Gamma$ and of first degree for the sheets at $R$. The range of the in-scattering term is therefore spanned by the functions
\begin{align}
& f^\nu_0=1\,,\quad f^\nu_i=\hat n_i\ \ (i=x,y,z)\,,\quad
f^\nu_{ij}=\hat n_i\,\hat n_j-\frac{\delta_{ij}}{3}\ \ (\text{five independent})\,,
\label{eqfp}
\end{align}
where the last set is absent for the Weyl nodes. The overlaps of the sheet $(\Gamma,0)$ with every sheet at $\Gamma$ are even under $\hat{\bs n}\to-\hat{\bs n}$, and the range of its in-scattering term is spanned by $f^\nu_0$ and the five functions $f^\nu_{ij}$. The range of the sheet $(\Gamma,1)$ contains all nine functions, and the range of each sheet at $R$ contains four. The moments of a sheet involve the same functions as its range, since the kernel is symmetric. The spans do not depend on the direction of $\bB$. They define the finite-rank form of the kernel of Sec.~\ref{secfinrank}, with the matrices $Q^{aa'}$ of Appendix~\ref{secappingr}.

The two Weyl nodes at $R$ have the same dispersion, and the bands of $R_1$ and $R_2$ are therefore degenerate at every $\bk$. The blocks $\mathcal H_{R_1}$ and $\mathcal H_{R_2}$ are independent of $\bk$ in the copy basis, and the BC, the OMM, and the phase-space factor of the degenerate band are those of each copy times the identity in the copy space.

The two copies at $R$ are nodes of the same nature, namely two Weyl nodes in this section and two TSM nodes in Sec.~\ref{secsys2}. Their Hamiltonians are block diagonal in the copy basis, and the overlap between the two copies therefore vanishes. The amplitude between $R_1$ and $R_2$ is accordingly absent, and Eq.~\eqref{eqoverlap} lists only the overlap within one copy. The copies remain coupled through the collision integral, since both overlap with the sheets at $\Gamma$. The calculation of this section treats $R_1$ and $R_2$ as independent sheets with this coupling. A nonvanishing overlap between the copies requires a $2\times2$ distribution matrix in the copy space, and both cases are outlined in Appendix~\ref{secappdeg}. The systems of Secs.~\ref{secsys3} and \ref{secsys4} have no such copies, and their internode amplitudes refer to the two conjugate nodes.

The nodes at $R$ are equivalent if $\beta^{\Gamma R}_s$ is the same for both and $\beta_{R_1R_1}=\beta_{R_2R_2}$. The exchange of $R_1$ and $R_2$ then maps the configuration onto itself and leaves the number constraint invariant. The solution of Eq.~\eqref{eqboltzop} is unique once the number constraint is imposed, and hence $\Lambda^j_{R_1}=\Lambda^j_{R_2}\equiv\Lambda^j_R$, and only one sheet at $R$ has to be expanded. The two nodes at $R$ enter the equations of the other sheets through $\beta^2_{\Gamma R}$ with the factor two. The number constraint reads $\mathcal M^j_{(\Gamma,1),0}+\mathcal M^j_{(\Gamma,0),0}+2\,\mathcal M^j_{R,0}=0$. Both points act on the same four-dimensional orbital space. At $\Gamma$, the Hamiltonian is $\mathcal H_\Gamma$ in the triplet block, accompanied by the decoupled singlet of the energy $E_A$, and at $R$, it is the block-diagonal matrix of $\mathcal H_{R_1}$ and $\mathcal H_{R_2}$: 
%%%%%%%%%%%%%%
\begin{align}
\mathcal H^{(4)}_\Gamma(\bk)=\begin{bmatrix}
\mathcal H_\Gamma(\bk) & 0\\
0 & E_A
\end{bmatrix},\quad
\mathcal H^{(4)}_R(\bk)=\begin{bmatrix}
\mathcal H_{R_1}(\bk) & 0\\
0 & \mathcal H_{R_2}(\bk)
\end{bmatrix}.
\label{eqhamqw}
\end{align}
%%%%%%%%%%%%%%%%%%%%%%
The energy $E_A$ is set so far from $\mu$ that the singlet provides no FS sheet. The squared overlap between a sheet at $\Gamma$ and a sheet at $R$ is then $|\psi_{R}^\dagger(\bk')\,\psi_{\Gamma}(\bk)|^2$ for the four-component spinors.

The unitary transformations that relate the eigenbasis of the Hamiltonians of Eq.~\eqref{eqham} to the orbital basis are fixed by the spinless four-orbital tight-binding model of the space group 198 of Ref.~\cite{prl119206401}. Let $\tilde\tau_a$ and $\tilde\mu_a$ be two sets of Pauli matrices that act on the four sites of the unit cell, and let $\tilde\tau_0$ and $\tilde\mu_0$ be the corresponding identity matrices, which are not written out. The Hamiltonian of the model is
\begin{align}
\tilde{\mathcal H}_{198}(\bk)=\;& v_1\Big[\tilde\tau_x\cos\!\left(\frac{k_x}{2}\right)\cos\!\left(\frac{k_y}{2}\right)+\tilde\tau_x\tilde\mu_x\cos\!\left(\frac{k_y}{2}\right)\cos\!\left(\frac{k_z}{2}\right)+\tilde\mu_x\cos\!\left(\frac{k_z}{2}\right)\cos\!\left(\frac{k_x}{2}\right)\Big]\nn
& +v_p\Big[\tilde\tau_y\tilde\mu_z\cos\!\left(\frac{k_x}{2}\right)\sin\!\left(\frac{k_y}{2}\right)+\tilde\tau_y\tilde\mu_x\cos\!\left(\frac{k_y}{2}\right)\sin\!\left(\frac{k_z}{2}\right)+\tilde\mu_y\cos\!\left(\frac{k_z}{2}\right)\sin\!\left(\frac{k_x}{2}\right)\Big],
\label{eqtb198}
\end{align}
where $v_1$ and $v_p$ are real parameters. Let $\bk$ be measured from $\Gamma$ and $\bs q$ from $R$. The expansions to the linear order are
\begin{align}
& \tilde{\mathcal H}_{198}(\bk)\simeq v_1\left(\tilde\tau_x+\tilde\tau_x\tilde\mu_x+\tilde\mu_x\right)+\frac{v_p}{2}\left(\tilde\mu_y\,k_x+\tilde\tau_y\tilde\mu_z\,k_y+\tilde\tau_y\tilde\mu_x\,k_z\right),\quad
 \tilde{\mathcal H}_{198}(\bs\pi+\bs q)\simeq-\frac{v_p}{2}\left(\tilde\tau_y\tilde\mu_z\,q_x+\tilde\tau_y\tilde\mu_x\,q_y+\tilde\mu_y\,q_z\right),
\label{eqtbexp}
\end{align}
with $\bs\pi=(\pi,\pi,\pi)$. These agree with the expansions of Ref.~\cite{prl119206401} up to the signs that are suppressed there. The three matrices in each expansion anticommute. At $\Gamma$, the first term separates the singlet, whose orbital-space vector is $(1,1,1,1)/2$, from the triplet, and the second term acts on the triplet as the spin-1 matrices with the velocity $|v_p|/2$. At $R$, each of the two degenerate pairs of bands has the velocity $|v_p|/2$. The matrices at $R$ are those at $\Gamma$ with the cyclic relabelling $x\to y\to z\to x$ of the Cartesian indices. Let $\hat k_a$ and $\hat q_a$ be the Cartesian components of the unit vectors along $\bk$ and $\bs q$. The overlaps depend on the positions of the four orbitals in the unit cell, and we distinguish the scenario of Ref.~\cite{prl119206401} from that of Ref.~\cite{grushin-multifold}.

In the first scenario, which we call the Chang scenario, the orbitals sit at the sites $\bs q_A=(0,0,0)$, $\bs q_B=(1/2,1/2,0)$, $\bs q_C=(1/2,0,1/2)$, and $\bs q_D=(0,1/2,1/2)$ in units of the lattice constant, and Eq.~\eqref{eqtb198} is used as it stands. The overlap of the upper and the middle band of the triplet at $\Gamma$ with the upper pair at $R$, which is summed over the two copies at $R$, is
\begin{align}
\mathcal O_{(\Gamma,1)(R,+)}=\frac12\left(1-\hat q_x\,\hat k_y-\hat q_y\,\hat k_z-\hat q_z\,\hat k_x\right),\quad
\mathcal O_{(\Gamma,0)(R,+)}=\frac12\,.
\label{eqovgr4}
\end{align}

In the second scenario, which we call the Flicker scenario, the orbitals sit at the Wyckoff positions $\bs q_A=(\upsilon,\upsilon,\upsilon)$, $\bs q_B=(1/2+\upsilon,1/2-\upsilon,-\upsilon)$, $\bs q_C=(1/2-\upsilon,-\upsilon,1/2+\upsilon)$, and $\bs q_D=(-\upsilon,1/2+\upsilon,1/2-\upsilon)$. Here, $\upsilon$ is a dimensionless parameter that is fixed by the crystal structure and for $\upsilon=0$ the positions are those of the first scenario. The Hamiltonian is $\tilde{\mathcal H}_\upsilon(\bk)=U_\upsilon^\dagger(\bk)\,\tilde{\mathcal H}_{198}(\bk)\,U_\upsilon(\bk)$ with the diagonal matrix
\begin{align}
U_\upsilon(\bk)={\rm diag}\left(e^{i\,\upsilon\,(k_x+k_y+k_z)},\,e^{i\,\upsilon\,(k_x-k_y-k_z)},\,e^{i\,\upsilon\,(k_z-k_y-k_x)},\,e^{i\,\upsilon\,(k_y-k_x-k_z)}\right)
\label{equx}
\end{align}
for the orbitals ordered as $A$, $B$, $C$, and $D$ \cite{grushin-multifold}. This transformation leaves the bandstructure unchanged and modifies the eigenvectors. At $\Gamma$, the matrix $U_\upsilon$ is the identity. At $R$, the orbitals $B$, $C$, and $D$ acquire the same phase, and the orbital $A$ acquires a different one, so that $U_\upsilon(\bs\pi)=e^{-i\pi \upsilon}\,{\rm diag}\left(e^{i\phi},1,1,1\right)$ with $\phi=4\,\pi\,\upsilon$. Only the relative phase $\phi$ enters the overlaps. Let $s_p={\rm sgn}(v_p)$, and let the indices $a$, $a+1$, and $a+2$ run cyclically over $x$, $y$, and $z$. The overlaps are then
\begin{align}
\mathcal O_{(\Gamma,1)(R,+)}&=\frac12\bigg\{1-\sum_a\hat q_a\Big[\hat k_{a+1}+\frac{\cos\phi-1}{2}\left(\hat k_{a+1}-\hat k_{a+2}\right)+\frac{s_p\sin\phi}{2}\left(\hat k_a^2-\hat k_{a+1}\,\hat k_{a+2}\right)\Big]\bigg\},\nn
\mathcal O_{(\Gamma,0)(R,+)}&=\frac12\bigg\{1+\frac{s_p\sin\phi}{2}\sum_a\hat q_a\Big[2\left(\hat k_a^2-\hat k_{a+1}\,\hat k_{a+2}\right)-1\Big]\bigg\}.
\label{eqovgrx}
\end{align}

For $\phi=0$, Eq.~\eqref{eqovgrx} reduces to Eq.~\eqref{eqovgr4}, which is the case of the first scenario. For $\phi=\pi$, the overlap of the upper band is that of Eq.~\eqref{eqovgr4} with $\hat k_{a+2}$ in place of $\hat k_{a+1}$, which reverses the cyclic order, and the overlap of the middle band is $1/2$. For the other values of $\upsilon$, the terms proportional to $\sin\phi$ appear, and they change sign with $v_p$. The two scenarios thus give different overlaps between $\Gamma$ and $R$, although they describe the same bandstructure. In both scenarios, the overlap depends on the azimuths of $\bk$ and $\bs q$ separately and not only on their difference. The scattering between $\Gamma$ and $R$ therefore couples different azimuthal harmonics, and the sector equations of Eq.~\eqref{eqsector} hold only without this scattering. This is the second of the two cases distinguished in Sec.~\ref{sectilt}, where its set-ups are classified.

The cancellation of the Chern numbers of the FS sheets, which Eq.~\eqref{eqsolv} requires, follows from the Nielsen-Ninomiya theorem \cite{nielsen81_no} and can be read off from Eq.~\eqref{eqtbexp}. Let $\Sigma^\Gamma_a=\left(\tilde\mu_y,\,\tilde\tau_y\tilde\mu_z,\,\tilde\tau_y\tilde\mu_x\right)$ and $\Sigma^R_a=\left(\tilde\tau_y\tilde\mu_z,\,\tilde\tau_y\tilde\mu_x,\,\tilde\mu_y\right)$ with $a=x,y,z$ be the three matrices in the expansions at $\Gamma$ and at $R$, respectively. Both sets satisfy
\begin{align}
\left\{\Sigma_a,\Sigma_b\right\}=2\,\delta_{ab}\,,\quad
\left[\Sigma_a,\Sigma_b\right]=2\,i\,\epsilon_{abc}\,\Sigma_c\,,
\label{eqsigalg}
\end{align}
and hence $\Sigma_x\Sigma_y\Sigma_z=i$ in the whole four-dimensional space. This product fixes the handedness of every Weyl block into which the representation splits. The two copies at $R$ therefore have the same chirality, which is the property of the fourfold node that Eq.~\eqref{eqham} assumes. At $\Gamma$, the matrices $\Sigma^\Gamma_a$ projected onto the triplet are spin-1 matrices $S_a$ with $\left[S_a,S_b\right]=i\,\epsilon_{abc}\,S_c$. The Hamiltonian of the triplet is then $\mathcal H=\left(v_p/2\right)\bk\cdot\bs S$, and by Eq.~\eqref{eqbcomm}, its upper band has the Chern number $-2\,{\rm sgn}(v_p)$. At $R$, the Hamiltonian is $\mathcal H=-\,v_p\,\bs q\cdot\bs s$ with $s_a=\Sigma^R_a/2$, which satisfy the spin-1/2 algebra, and each of the two copies contributes $+{\rm sgn}(v_p)$ to the Chern number of the upper pair of bands. The total of the pair is $+2\,{\rm sgn}(v_p)$, and the Chern numbers of the two pockets add up to zero for either sign of $v_p$. The quadratic terms proportional to the identity matrix do not change these values, since they leave the spinors unchanged. No further pocket with a net Chern number is present near the Fermi level in the DFT bandstructures of RhSi and CoSi, whose Fermi surfaces consist of the pockets around $\Gamma$ and $R$ only \cite{prl119206401, xu2020_optical}. The sheets $(\Gamma,1)$, $(R_1,+)$, and $(R_2,+)$ therefore realise the cancellation required by Eq.~\eqref{eqsolv}, and the sheet $(\Gamma,0)$ does not carry a Chern number.

%%%%%%%%%%%%%%%
\subsection{Solution for the magnetic field along the polar axis}
\label{secuntiltsol}

The equation for each azimuthal sector follows from Eq.~\eqref{eqboltzop}. Let $\Lambda^{(m)}_a(\theta)$ be the coefficient of $e^{i\,m\,\varphi}$ in $\Lambda_a$, and let $h^{(m)}_a(\theta)$ be the corresponding harmonic of the source $h_a$. By Eq.~\eqref{eqlfoaxi}, the LFO term acts on $e^{i\,m\,\varphi}$ as the multiplication by $i\,m\,\omega_a$. Together with the collision term of Eq.~\eqref{eqcoll}, this gives
\begin{align}
\left[\tau_a^{-1}(\theta)+i\,m\,\omega_a(\theta)\right]\Lambda^{(m)}_a(\theta)=h^{(m)}_a(\theta)+\sum_{a'}\int_0^\pi d\theta'\,\sin\theta'\,\rho_{a'}(\theta')\,\mathcal K^{(m)}_{aa'}(\theta,\theta')\,\Lambda^{(m)}_{a'}(\theta')\,,
\label{eqsector}
\end{align}
where $\mathcal K^{(m)}_{aa'}(\theta,\theta')$ denotes the kernel of Eq.~\eqref{eqkernel} for the harmonic $m$. The sectors $m=\pm2$ are reached by the kernel of the sheets at $\Gamma$, but they have no source and give no contribution. The kernel vanishes for $|m|\geq3$, and the out-scattering rate follows from Eq.~\eqref{eqtau} with the FS weights of Appendix~\ref{secappingr} and the amplitudes of Eq.~\eqref{eqkernel}.

We first consider $\bE\parallel\bB$, which is the sector $m=0$ of $\Lambda_z$ and is annihilated by the LFO. The source is
\begin{align}
h_a=D_a^{-1}\left[w_{a,z}-\eta_B\,\chi_\nu\,N_a\,e\,B\,\frac{w_{a,k}}{2\,k^2}\right],
\end{align}
where $w_{a,k}$ and $w_{a,z}$ are the radial and the $z$ components of $\bw_a$. Let $M^{(n)}_a=\int_0^\pi d\theta\,\sin\theta\,\cos^n\theta\;\rho_a\,\Lambda_a$ be the moments of the solution, with $n=0,1,2$ for the sheet $(\Gamma,1)$, $n=0,2$ for the sheet $(\Gamma,0)$, and $n=0,1$ for the sheets at $R$. Inserting the kernel of Eq.~\eqref{eqkernel} into Eq.~\eqref{eqsector}, the in-scattering term becomes a polynomial in $\cos\theta$ of the degree given by the span, with the moments as coefficients. We obtain
\begin{align}
\Lambda_a(\theta)=\tau_a(\theta)\Big[h_a(\theta)+\sum_{a'}\sum_{n}\mathcal Q^{(n)}_{aa'}(\theta)\,M^{(n)}_{a'}\Big],
\label{eqlongsol}
\end{align}
where the polynomials $\mathcal Q^{(n)}_{aa'}(\theta)$ follow from Eq.~\eqref{eqkernel}. This form is exact, because the right-hand side of Eq.~\eqref{eqsector} lies in the span of $h_a$ and the functions listed in Eq.~\eqref{eqspans}. Inserting Eq.~\eqref{eqlongsol} into the definition of the moments gives a linear system with one null vector, which is the zero mode of Eq.~\eqref{eqboltzop}. The number constraint of Eq.~\eqref{eqnumber}, which reads $\sum_aM^{(0)}_a=0$, closes the system. The solvability condition of Eq.~\eqref{eqsolv} holds because the pumping rates of the four sheets add up to zero. The conductivity is
\begin{align}
\sigma_\parallel=\frac{e^2}{4\,\pi^2}\sum_a\int_0^\pi d\theta\,\sin\theta\;\rho_a\,h_a\,\Lambda_a\,.
\label{eqsigpar}
\end{align}

Next, we consider $\bE\perp\bB$, which is the sector $m=\pm1$ of $\Lambda_x\pm i\,\Lambda_y$. For $i=x,y$, the component $B_i$ vanishes, and the source is $h^i_a=D_a^{-1}w_{a,i}$. Since $w_x\pm i\,w_y=w_\perp\,e^{\pm i\,\varphi}=q_a\,\sin\theta\,e^{\pm i\,\varphi}$, the source has the harmonic $h^{(1)}_a=(q_a/D_a)\sin\theta$. Writing $\Lambda_x+i\,\Lambda_y=\ell_a(\theta)\,e^{i\,\varphi}$, the kernel of Eq.~\eqref{eqkernel} turns the in-scattering term into $\check\lambda^{(0)}_a\,\sin\theta+\check\lambda^{(1)}_a\,\sin\theta\,\cos\theta$. The coefficients $\check\lambda^{(0)}_a$ and $\check\lambda^{(1)}_a$ are complex numbers, which are linear combinations of the moments $\int_0^\pi d\theta'\,\sin^2\theta'\,\rho_{a'}\,\ell_{a'}$ and $\int_0^\pi d\theta'\,\sin^2\theta'\cos\theta'\,\rho_{a'}\,\ell_{a'}$ with the coefficients of Appendix~\ref{secappingr}. For the sheets at $R$, only $\check\lambda^{(0)}_a$ is present, and for the sheet $(\Gamma,0)$, only $\check\lambda^{(1)}_a$ is present. With $\Lhat\to-i\,B\,w_\perp/k_\perp$, the equation is algebraic in $\theta$ up to these numbers,
\begin{align}
\ell_a(\theta)=\frac{\left(q_a/D_a+\check\lambda^{(0)}_a\right)\sin\theta+\check\lambda^{(1)}_a\,\sin\theta\,\cos\theta}{\tau_a^{-1}(\theta)+i\,\omega_a(\theta)}\,.
\label{eqell}
\end{align}
The LFO enters only as an imaginary, angle-dependent addition $i\,\omega_a$ to the scattering rate. Inserting Eq.~\eqref{eqell} into the definitions of $\check\lambda^{(0)}_a$ and $\check\lambda^{(1)}_a$ gives a complex linear system. The conductivities from the Boltzmann part are
\begin{align}
\sigma_\perp=\frac{e^2}{8\,\pi^2}\sum_a\int_0^\pi d\theta\;\sin^3\theta\,\rho_a\,
\frac{q_a}{D_a}\,\Re\! \left( \frac{\ell_a}{\sin\theta}\right) ,\quad
\sigma_H^{\rm B}=-\frac{e^2}{8\,\pi^2}\sum_a
\int_0^\pi d\theta\;\sin^3\theta\,\rho_a\,\frac{q_a}{D_a}\,
\Im \! \left(\frac{\ell_a}{\sin\theta} \right).
\label{eqsigperp}
\end{align}
Here $\sigma_H\equiv\sigma_{yx}$ for $\bB\parallel\hat{\boldsymbol z}$. The equilibrium AH term follows from the second term of Eq.~\eqref{eqsigmw}.

By axisymmetry, the full tensor is $\sigma_{ij}=\sigma_\perp(\delta_{ij}-b_i\,b_j)+\sigma_\parallel\,b_i\,b_j-\sigma_H\,\epsilon_{ijl}\,b_l$ with $\boldsymbol{b}=\hat{\bB}$, and three scalar functions therefore determine everything. For $\bE=E\,(\sin\vartheta,0,\cos\vartheta)$, the current has the components
\begin{align}
j_{\bE} =\left(\sigma_\perp\,\sin^2\vartheta+\sigma_\parallel\,\cos^2\vartheta\right)E\,,\quad
j_{\rm PH}=\left(\sigma_\perp-\sigma_\parallel\right)\sin\vartheta\,\cos\vartheta\;E\,,\quad
j_{\rm H} =\sigma_H\,\sin\vartheta\;E\,,
\label{eqcurr}
\end{align}
along $\bE$, along the in-plane perpendicular direction $(\cos\vartheta,0,-\sin\vartheta)$ (in-plane transverse), and out of the plane spanned by $\bE$ and $\bB$ (out-of-plane transverse), respectively. The solution is not a superposition of the results for the components of $\bB$ along and perpendicular to $\bE$, because $\xi_s$, $D_a$, $\rho_a$, $\omega_a$, and $\tau_a$ depend on the full magnitude $B$. The two sectors of Eqs.~\eqref{eqlongsol} and \eqref{eqell} are therefore solved once at the full $B$, and the angle enters only through the weights in Eq.~\eqref{eqcurr}.

%%%%%%%%%%%%%%%
\subsection{Symmetries and classification of the set-ups}
\label{sectilt}

With $\Lambda^j_a$ known for $j=x,y,z$, the conductivity tensor $\bs\sigma$ follows from Eq.~\eqref{eqsigmw}. Since $\bE$ enters only through the sources [see Eq.~\eqref{eqcurrtilt}], the tensor $\bs\sigma$ contains the response to every direction of $\bE$, and the geometry is specified by $\bB$ alone. The orientation of $\bE$ does not define a separate case. All the nodes are untilted, and the dispersions are spherical. A rotation $\mathcal R$ of $\bB$ rotates the tensor whenever the kernel is invariant under the rotation of all the momenta,
\begin{align}
\bs\sigma\!\left(\mathcal R\,\bB\right)=\mathcal R\,\bs\sigma\!\left(\bB\right)\,\mathcal R^{\rm T}\,.
\label{eqrot}
\end{align}
Time reversal maps a node onto a node of the same chirality at $-\bk$. The quantities $\xi_s$, $D_a$, and $\rho_a$ are unchanged under $\bk\to-\bk$ and $\bB\to-\bB$, whereas $\bw_a$ and $\Lhat$ change sign, and the in-scattering kernel is symmetric. The operator of Eq.~\eqref{eqtilteq} is thereby mapped onto its adjoint, which gives the Onsager relation
\begin{align}
\sigma_{ij}\!\left(\bB\right)=\sigma_{ji}\!\left(-\bB\right).
\label{eqons}
\end{align}
The intranode overlaps and the overlap between the two Weyl nodes depend on $\hat{\bs n}\cdot\hat{\bs n}'$ with equal chiralities and satisfy this condition. Two cases are distinguished by the overlaps between $\Gamma$ and $R$. In the first case, the impurity matrix is symmetry adapted, so that these overlaps depend on $\varphi-\varphi'$ only and satisfy the condition above. The sectors of different $m$ decouple, the exchange of $R_1$ and $R_2$ is a symmetry, and Eq.~\eqref{eqrot} holds. The tensor $\bs\sigma$ is fixed by $\sigma_\parallel$, $\sigma_\perp$, and $\sigma_H$ at the full magnitude of $\bB$. The overlaps of this case are not fixed by the spinors alone. In the second case, the impurity potential is the identity matrix in the orbital space, and the overlaps of Eqs.~\eqref{eqovgr4} and \eqref{eqovgrx} depend on the azimuthal angles at $\Gamma$ and at $R$ separately. Eq.~\eqref{eqrot} is then not available for a general rotation. A direct evaluation shows that the overlaps of both scenarios are invariant under the rotations of all the momenta about the axis $\hat{\bs c}=\left(1,1,1\right)/\sqrt3$ and under no other rotation. The intranode overlaps and the overlap between the two Weyl nodes are invariant under every rotation. Improper elements are not symmetries, since they map the QTSM node of chirality $+1$ onto a node of the opposite chirality, which is absent. The orientations of $\bB$ with the same polar angle $\theta_B$ to $\hat{\bs c}$ are therefore related by a rotation, and Eq.~\eqref{eqons} relates $\theta_B$ and $\pi-\theta_B$. We use the frame with the polar axis along $\hat{\bs c}$ for the second case, and the classification is collected in Table~\ref{tabgeom}.

In the first case, all the orientations of $\bB$ are equivalent, and the reference set-up is $\bB=B\,\boldsymbol{\hat z}$. In the second case, the same set-up is $\bB\parallel\hat{\bs c}$, for which the overlaps depend on the azimuths about $\hat{\bs c}$ through their difference only. In both cases, the sectors $m=0$ and $m=\pm1$ decouple, the sector $m=\pm2$ has no source, and the tensor $\bs\sigma$ is fixed by $\sigma_\parallel$, $\sigma_\perp$, and $\sigma_H$. The exchange symmetry of the two nodes at $R$, which is discussed after Eq.~\eqref{eqansatzc}, reduces the number of unknowns in both sectors. For $\bB\perp\hat{\bs c}$ in the second case, the rotation $\mathcal R_{\hat c}(\pi)$ maps $\bB$ onto $-\bB$ and leaves the kernel invariant. Together with Eq.~\eqref{eqons}, it gives
\begin{align}
\sigma_{ij}=\eta_i\,\eta_j\,\sigma_{ji}\,,
\label{eqplanar5}
\end{align}
where $i$ and $j$ label the axes of the frame with $\boldsymbol{\hat z}=\hat{\bs c}$, and $\eta_z=+1$ and $\eta_x=\eta_y=-1$. This leaves six independent elements. For the other orientations, no symmetry beyond Eq.~\eqref{eqons} exists, and the nine elements of $\bs\sigma$ are independent.

\begin{table}[t!]
\centering
\small
\renewcommand{\arraystretch}{1.6}
\begin{tabular}{|c|c|c|c|c|}
\hline
Case & Reference set-up & Relations & \makecell{Independent elements \\ of $\bs\sigma$} & Solver \\
\hline
(a) & \makecell{$\bB=B\,\hat{\boldsymbol{z}}$ \\ ($\bB\parallel\hat{\bs c}$ in the second case)} & \makecell{axisymmetry about $\hat{\bB}$, \\ Eq.~\eqref{eqons}} & \makecell{three: \\ $\sigma_\parallel$, $\sigma_\perp$, $\sigma_H$} & \makecell{sectors of Eqs.~\eqref{eqlongsol} \\ and \eqref{eqell}} \\
\hline
(b) & $\bB=B\,\hat{\boldsymbol{x}}$, $\boldsymbol{\hat z}=\hat{\bs c}$ & \makecell{$\mathcal R_{\hat c}(\pi)$ and Eq.~\eqref{eqons}, \\ Eq.~\eqref{eqplanar5}} & \makecell{six: $\sigma_{xx}$, $\sigma_{yy}$, $\sigma_{zz}$, \\ $\sigma_{xy}=\sigma_{yx}$, \\ $\sigma_{xz}=-\sigma_{zx}$, \\ $\sigma_{yz}=-\sigma_{zy}$} & \makecell{general solver, \\ Eq.~\eqref{eqmomsys}} \\
\hline
(c) & \makecell{$\bB=B\left(\sin\theta_B,0,\cos\theta_B\right)$, \\ $0<\theta_B<\pi/2$, $\boldsymbol{\hat z}=\hat{\bs c}$} & Eq.~\eqref{eqons} & nine & \makecell{general solver, \\ Eq.~\eqref{eqmomsys}} \\
\hline
\end{tabular}
\caption{\label{tabgeom} Independent geometries of the QTSM and Weyl nodes. The set-up (a) is the only one of the first case, and the set-ups (b) and (c) arise in the second case. All directions of $\bE$ follow from the reference set-ups.}
\end{table}

%%%%%%%%%%%%%%%
\subsection{Ansatz for the mean free-path}
\label{secansatz}

The exact form of $\Lambda^j_a$ is Eq.~\eqref{eqtiltsol} with the functions $f^a_p$ of Eq.~\eqref{eqfp}. The sheet $(\Gamma,1)$ carries nine functions, the sheet $(\Gamma,0)$ carries six, and each sheet at $R$ carries four. Only one sheet at $R$ is expanded, since $\Lambda^j_{R_1}=\Lambda^j_{R_2}$. This gives nineteen moments $\mathcal M^j_{a,p}$ for each $j$, which are reduced by one through the number constraint. The set-ups are the cases (a) to (c) of Table~\ref{tabgeom}.

For case (a), the ansatz is closed in the azimuthal sectors $m$ about $\hat{\bB}$. In each sector, the in-scattering term of the sheet $a$ is spanned by a set $\mathscr F^m_a$ of functions of $\theta$, which are multiplied by $e^{i\,m\,\varphi}$. For the sheets at $\Gamma$ and at $R_i$, these sets are
\begin{align}
\mathscr F^m_{(\Gamma,1)}&=\begin{cases}
\{1,\ \cos\theta,\ 3\cos^2\theta-1\}\,, & m=0\,,\\
\{\sin\theta,\ \sin\theta\cos\theta\}\,, & m=\pm1\,,\\
\{\sin^2\theta\}\,, & m=\pm2\,,
\end{cases}\notag\\
\mathscr F^m_{(\Gamma,0)}&=\begin{cases}
\{1,\ 3\cos^2\theta-1\}\,, & m=0\,,\\
\{\sin\theta\cos\theta\}\,, & m=\pm1\,,\\
\{\sin^2\theta\}\,, & m=\pm2\,,
\end{cases}\notag\\
\mathscr F^m_{(R_i,+)}&=\begin{cases}
\{1,\ \cos\theta\}\,, & m=0\,,\\
\{\sin\theta\}\,, & m=\pm1\,.
\end{cases}
\label{eqspans}
\end{align}
The sheets $(R_i,+)$ contain no function in the sector $m=\pm2$.

The components of $\bs\Lambda$ are those of Eqs.~\eqref{eqlongsol} and \eqref{eqell}. The unknowns are the seven real moments $M^{(n)}_a$ of the sector $m=0$, which are subject to Eq.~\eqref{eqnumber}, and the four complex coefficients $\check\lambda^{(n)}_a$ of the sector $m=1$.

For the cases (b) and (c), no rotation fixes $\bB$, and the LFO couples the harmonics $m$ and $m\pm1$ about $\hat{\bs c}$. No pointwise relation between the components of $\bs\Lambda$ exists. The relation of Eq.~\eqref{eqplanar5} follows from the Onsager relation, which involves the adjoint operator, and it constrains $\bs\sigma$ but not $\Lambda^j_a$. The ansatz is Eq.~\eqref{eqtiltsol}, where the sheets at $R$ enter the sum over $a'$ with the factor two. The same nineteen moments are unknown for each $j$ in both cases, with the constraint $\mathcal M^j_{(\Gamma,1),0}+\mathcal M^j_{(\Gamma,0),0}+2\,\mathcal M^j_{R,0}=0$.

%%%%%%%%%%%%%%%
\subsection{Limit of the relaxation-time approximation}
\label{secrta}

In the RTA, the in-scattering term is dropped, and every sheet relaxes with the constant time $\tau$. For $\bE\parallel\bB$, Eq.~\eqref{eqsigpar} reduces to $\sigma_{\parallel,a}=\left(e^2\tau/4\pi^2\right)\int_0^\pi d\theta\,\sin\theta\,\rho_a\,h_a^2$. The sheets at $\Gamma$ are those of Eq.~\eqref{eqham}, and the sheets at $R$ have the linear dispersion. We consider a sheet with $\epsilon_a=\mathsf u_a\,k$ measured from its node, the numbers $N_a$ and $G_a$ of Eq.~\eqref{eqbcomm}, and the velocity $v_\nu$ in the OMM. Let $\mu_a$ be the chemical potential measured from the node, so that $k_a=\mu_a/\mathsf u_a$ at $B=0$. To the first order in $B$, the radius is
\begin{align}
k_a(\theta)=\frac{\mu_a}{\mathsf u_a}-\eta_O\,\frac{\chi_\nu\,e\,v_\nu\,G_a\,B\cos\theta}{2\,\mu_a}\,,
\end{align}
and the expansion of $\rho_a\,h_a^2$ in $B$ with the quantities of Eq.~\eqref{eqwqd}, followed by the integration over $\cos\theta$, gives
\begin{align}
\sigma_{\parallel,a}^{\rm RTA}=\frac{e^2\,\tau\,\mu_a^2}{6\,\pi^2\,\mathsf u_a}+\frac{e^4\,\tau\,\mathsf u_a\,B^2}{120\,\pi^2\,\mu_a^2}\left[8\,\eta_B\,N_a^2\,\mathsf u_a^2-9\,\eta_B\,\eta_O\,N_a\,G_a\,\mathsf u_a\,v_\nu+5\,\eta_O\,G_a^2\,v_\nu^2\right]+\mathcal O\!\left(B^4\right).
\label{eqrtagen}
\end{align}
The term linear in $B$ vanishes. For each Weyl node at $R$, we have $\mathsf u_a=v_\nu=\tilde v_F$, $N_a=G_a=1$, and $\mu_a=\mu+\Delta$, which gives
\begin{align}
\sigma_{\parallel,R_i}^{\rm RTA}=\frac{e^2\,\tau\left(\mu+\Delta\right)^2}{6\,\pi^2\,\tilde v_F}+\frac{e^4\,\tau\,\tilde v_F^3\,B^2}{\pi^2\left(\mu+\Delta\right)^2}\left(\frac{\eta_B}{15}-\frac{3\,\eta_B\,\eta_O}{40}+\frac{\eta_O}{24}\right).
\label{eqrtaweyl}
\end{align}
The term proportional to $\eta_B$ is the contribution of the BC, that proportional to $\eta_O$ is the contribution of the OMM, and the product is their cross term. The contribution of the two nodes at $R$ is twice this value. The RTA does not contain the scattering between the sheets, and the interband scattering within the QTSM node is absent. The internode scattering between $\Gamma$ and $R$ is added through the relaxation time $\tau_{\rm inter}$ in Eq.~\eqref{eqrtainter6} below. The derivation of Eqs.~\eqref{eqrtagen} and \eqref{eqrtaweyl}, and of the corresponding expressions for the QTSM node at $\Gamma$, is given in Appendix~\ref{secrtab}.

For noncollinear fields, the RTA drops the in-scattering term and keeps the LFO, as in Sec.~\ref{secrta3}. For $\bE\perp\bB$, the sources carry no pumping term, and every sheet has the mean free-path $\ell_a=\left(q_a/D_a\right)\sin\theta\,\tau/\left(1+i\,\omega_a\,\tau\right)$. Equation~\eqref{eqsigperp} then gives
\begin{align}
\sigma_\perp-i\,\sigma_H^{\rm B}=\frac{e^2\,\tau}{8\,\pi^2}\sum_a\int_0^\pi d\theta\,\sin^3\theta\,\rho_a\,\frac{\left(q_a/D_a\right)^2}{1+i\,\omega_a\,\tau}\,,
\label{eqrtaperp1}
\end{align}
where the sum runs over the four sheets $(\Gamma,1)$, $(\Gamma,0)$, $(R_1,+)$, and $(R_2,+)$. The expansion in $B$ of one sheet is given by Eqs.~\eqref{eqrtaexp3}, \eqref{eqrtaHexp3}, \eqref{eqrtaAH3}, and \eqref{eqrtaph3}, which hold for every sheet. In these equations, $b_a$ is the coefficient of the linear term of the dispersion, $\bar v_a=\partial_k\epsilon_s$ at $k_a$, and $N_a$, $G_a$, and $v_\nu$ are those of Eq.~\eqref{eqbcomm}. For the sheets at $\Gamma$, we have $b_a=v_F$, $\bar v_a=v_F\left(1+2\,c\,k_{\Gamma,1}\right)$, $N_a=2$, and $G_a=1$ for $s=1$, and $b_a=0$, $\bar v_a=2\,c\,v_F\,k_{\Gamma,0}$, $N_a=0$, and $G_a=2$ for $s=0$, with $v_\nu=v_F$ in both cases. For each sheet at $R$, the dispersion is linear, so that $b_a=\bar v_a=v_\nu=\tilde v_F$ and $N_a=G_a=1$. The term of Eq.~\eqref{eqrtaexp3} proportional to $\eta_O$ then vanishes, and with $\mu_R=\mu+\Delta$ we obtain
\begin{align}
& \sigma^{\rm RTA}_{\perp,R_i}=\frac{e^2\,\tau\,\mu_R^2}{6\,\pi^2\,\tilde v_F}+\frac{e^4\,\tau\,\tilde v_F^3\,B^2}{\pi^2\,\mu_R^2}\left(\frac{\eta_B}{120}-\frac{\eta_B\,\eta_O}{40}\right)-\frac{e^4\,\tau^3\,\tilde v_F^3\,B^2}{6\,\pi^2}\,,\nn
& \sigma^{\rm RTA}_{\perp,R_i}-\sigma^{\rm RTA}_{\parallel,R_i}=\frac{e^4\,\tau\,\tilde v_F^3\,B^2}{\pi^2\,\mu_R^2}\left(\frac{\eta_B\,\eta_O}{20}-\frac{7\,\eta_B}{120}-\frac{\eta_O}{24}\right)-\frac{e^4\,\tau^3\,\tilde v_F^3\,B^2}{6\,\pi^2}\,.
\label{eqrtaweylperp}
\end{align}
The leading terms of the Hall conductivity of one Weyl node are $\sigma^{\rm B,RTA}_{H,R_i}=e^3\,\tau^2\,\mu_R\,\tilde v_F\,B/\left(6\,\pi^2\right)$ and $\sigma^{\rm AH}_{H,R_i}=\eta_B\,\eta_O\,e^3\,\tilde v_F\,B/\left(24\,\pi^2\,\mu_R\right)$. The sheet $(\Gamma,0)$ has a vanishing BC and therefore no AH part, and the AH part of the sheet $(\Gamma,1)$ is $\eta_B\,\eta_O\,e^3\,v_F\,B/\left(12\,\pi^2\,k_{\Gamma,1}\,\bar v_a\right)$ to the leading order in $B$. The totals are the sums over the four sheets, with the two nodes at $R$ counted twice. The overlaps between $\Gamma$ and $R$ do not enter these expressions, since the RTA contains no scattering between the sheets.

The internode scattering between $\Gamma$ and $R$ is included with the relaxation time $\tau_{\rm inter}$ following Ref.~\cite{ips-internode}. In this construction, the node at $\Gamma$ and the fourfold node at $R$ are the two nodes, and the two degenerate copies at $R$ form a single node. The nodes have different DOS and different pumping vertices, and the intranode collision term relaxes the distribution of each node towards its own local chemical potential. The internode term relaxes it towards the mean of the two local chemical potentials. The tensor of the internode current is proportional to $B_i\,B_j$ at the order $B^2$. It therefore acts on the component of $\bE$ along $\bB$, and it vanishes for $\bE\perp\bB$ in line with the sector $m=1$, in which $\tau_{\rm inter}$ does not enter. Let $\varrho_\nu$ be the density-of-states of the node $\nu$ at $B=0$ and $\Upsilon_\nu$ its pumping vertex of Eq.~\eqref{eqrtab8}. For $\bE\parallel\bB=B\,\boldsymbol{\hat z}$, the contribution of the node $\nu$ is, in terms of the node $\bar\nu$ that is the other one,
\begin{align}
\sigma^{\rm inter}_{\parallel,\nu}=\frac{e^2\,\Upsilon_\nu^2}{\varrho_\nu}\left[\frac{2\,\tau_{\rm inter}\,\varrho_{\bar\nu}}{\varrho_\Gamma+\varrho_R}-\tau\right].
\label{eqrtainter6}
\end{align}
The pumping vertices follow from the BC and the OMM of Eq.~\eqref{eqbcomm}. With $\bar v_a=\partial_k\epsilon_s$ at $k_a$, the vertex of a sheet is $\Upsilon_a=-\chi_\nu\,e\,B\left[\eta_B\,N_a/\left(4\,\pi^2\right)+\eta_O\,v_\nu\,G_a/\left(12\,\pi^2\,\bar v_a\right)\right]$, and the sums over the sheets of the two nodes are given by
\begin{align}
\Upsilon_\Gamma=-\frac{e\,B}{12\,\pi^2}\,\mathcal E_\Gamma\,,\quad
\Upsilon_R=\frac{e\,B}{12\,\pi^2}\,\mathcal E_R\,,\quad
\mathcal E_\Gamma=6\,\eta_B+\eta_O\,v_F\left(\frac{1}{w_1}+\frac{2}{w_0}\right),\quad
\mathcal E_R=6\,\eta_B+2\,\eta_O\,.
\label{eqrtavert6}
\end{align}
With $\bar\rho_\Gamma=\rho_1+\rho_0$ and $\bar\rho_R=2\,\rho_R$, the densities-of-states are $\varrho_\Gamma=\bar\rho_\Gamma/\left(2\,\pi^2\right)$ and $\varrho_R=\bar\rho_R/\left(2\,\pi^2\right)$, and the sum over the two nodes of Eq.~\eqref{eqrtainter6} is
\begin{align}
\sigma^{\rm inter}_\parallel=\frac{e^4\,B^2}{72\,\pi^2}\sum_{\nu=\Gamma,R}\frac{\mathcal E_\nu^2}{\bar\rho_\nu}\left[\frac{2\,\tau_{\rm inter}\,\bar\rho_{\bar\nu}}{\bar\rho_\Gamma+\bar\rho_R}-\tau\right]+\mathcal O\!\left(B^3\right).
\label{eqrtainter6b}
\end{align}
The derivation is given in Appendix~\ref{secrtab7}. The term is of the order $B^2$ and it adds to the intranode expressions of this subsection. For $\bar\rho_\Gamma=\bar\rho_R$, it is proportional to $\tau_{\rm inter}-\tau$ as for the conjugate nodes of Sec.~\ref{secrta3}. For unequal densities-of-states, it does not vanish at $\tau_{\rm inter}=\tau$. As in Sec.~\ref{secrta3}, the OMM part of the vertices is specific to the RTA, and the exact solution contains the OMM through $\rho_a$, $h_a$, and $\omega_a$.

The RTA treatment of the TSM node with the linear dispersion was carried out in Ref.~\cite{ips-spin1-ph}, and that of the Weyl nodes in Refs.~\cite{rahul-jpcm, ips-ruiz, ips-tilted}, for an arbitrary orientation of $\bB$ relative to $\bE$. The works on the Weyl nodes do not include the LFO or the internode scattering, and Ref.~\cite{rahul-jpcm} also neglects the OMM. The longitudinal RTA conductivity of the QTSM node was given in Ref.~\cite{ips-exact-spin1}, and the RTA contribution of the internode scattering between a TSM node at $\Gamma$ and a double Weyl node at $R$ was reported in Ref.~\cite{ips-internode} for linear dispersions. The noncollinear expressions of the sheets at $\Gamma$ with the quadratic correction $c$ are reported here. The internode term of Eq.~\eqref{eqrtainter6b} extends that of Ref.~\cite{ips-internode} to the QTSM node at $\Gamma$ with the quadratic correction.

%%%%%%%%%%%%%%%
\subsection{Results}
\label{secres}

The analytical results for the QTSM and Weyl nodes are collected here, and the intermediate steps are given in Appendix~\ref{secappres}. They concern the limit $B\to0$ and the terms that are linear in $B$, for which the moment system of Eq.~\eqrf{eqmomsys} closes in a few unknowns. The overlaps between the sheets at $\Gamma$ and at $R$ enter through the isotropic form that is required by Eq.~\eqrf{eqrot}. The span $\{1,\cos\theta\}$ of the sheets at $R$ in Eq.~\eqref{eqspans} makes these overlaps linear in $\hat{\bk}'$, and the absence of $\sin\theta$ in the span of the sheet $(\Gamma,0)$ removes the dependence on $\hat{\bk}$ for this sheet. With $C=\hat{\bk}\cdot\hat{\bk}'$, the overlaps are therefore
\begin{align}
\mathcal O_{(\Gamma,1)(R_i,+)}=\bar o_1+\bar p_1\,C\,,\quad
\mathcal O_{(\Gamma,0)(R_i,+)}=\bar o_0\,,
\label{eqovgriso}
\end{align}
where the real constants $\bar o_1$, $\bar p_1$, and $\bar o_0$ are fixed by the impurity matrix. We abbreviate $k_1=k_{\Gamma,1}$, $k_0=k_{\Gamma,0}$, $k_R$, and $\rho_1=\rho_{\Gamma,1}$, $\rho_0=\rho_{\Gamma,0}$, $\rho_R$ of Eq.~\eqrf{eqfsb0}, as well as $\beta_{ss'}=\beta^{\Gamma}_{ss'}$, $g_s=\beta^{\Gamma R}_s$, and $\gamma_R=\left(\beta^R_{\rm intra}\right)^2+\left(\beta^R_{\rm inter}\right)^2$. The velocities at the FS for $B=0$ are $w_1=v_F\left(1+2\,c\,k_1\right)$, $w_0=2\,c\,v_F\,k_0$, and $w_R=\tilde v_F$.

\subsubsection{Longitudinal magnetoconductivity for parallel fields}
\label{secresll}

At $B=0$, the mean free-path of every sheet is proportional to $\cos\theta$ for $\bE\parallel\hat{\boldsymbol z}$. We write $\Lambda_a=\mathcal A_a\cos\theta$, with $\mathcal A_{R_1}=\mathcal A_{R_2}\equiv\mathcal A_R$. The out-scattering rates of Eq.~\eqrf{eqtau} become
\begin{align}
& \tau_{1}^{-1}=\frac43\left(\beta_{11}^2\,\rho_1+\beta_{10}^2\,\rho_0\right)+8\,\bar o_1\,g_1^2\,\rho_R\,,\quad
\tau_{0}^{-1}=\frac43\left(\beta_{10}^2\,\rho_1+\beta_{00}^2\,\rho_0\right)+8\,\bar o_0\,g_0^2\,\rho_R\,,\nn
& \tau_{R}^{-1}=2\,\gamma_R\,\rho_R+4\left(\bar o_1\,g_1^2\,\rho_1+\bar o_0\,g_0^2\,\rho_0\right).
\label{eqtauB0}
\end{align}
The sheet $(\Gamma,0)$ decouples from the in-scattering term, and the amplitudes of the other two sheets obey a $2\times2$ system with the entries
\begin{align}
& N_{\Gamma\Gamma}=\frac23\,\beta_{11}^2\,\rho_1+\frac43\,\beta_{10}^2\,\rho_0+8\,\bar o_1\,g_1^2\,\rho_R\,,\quad
N_{\Gamma R}=\frac83\,\bar p_1\,g_1^2\,\rho_R\,,\nn
& N_{R\Gamma}=\frac43\,\bar p_1\,g_1^2\,\rho_1\,,\quad
N_{RR}=\frac43\,\gamma_R\,\rho_R+4\left(\bar o_1\,g_1^2\,\rho_1+\bar o_0\,g_0^2\,\rho_0\right).
\label{eqNmat}
\end{align}
With $\mathcal N=N_{\Gamma\Gamma}\,N_{RR}-N_{\Gamma R}\,N_{R\Gamma}$, the amplitudes are
\begin{align}
\mathcal A_1=\frac{N_{RR}\,w_1+N_{\Gamma R}\,w_R}{\mathcal N}\,,\quad
\mathcal A_R=\frac{N_{\Gamma\Gamma}\,w_R+N_{R\Gamma}\,w_1}{\mathcal N}\,,\quad
\mathcal A_0=\tau_0\,w_0\,.
\label{eqAamp}
\end{align}
The zero-field conductivity, which is the reference for the magnetoconductivity, is
\begin{align}
\sigma_0=\frac{e^2}{6\,\pi^2}\left(k_1^2\,\mathcal A_1+k_0^2\,\mathcal A_0+2\,k_R^2\,\mathcal A_R\right).
\label{eqsig0}
\end{align}
The number constraint of Eq.~\eqrf{eqnumber} is satisfied identically, since the weights are constants at $B=0$ and the integral of $\cos\theta$ vanishes. The term with $\bar p_1$ is the vertex correction that couples the sheet $(\Gamma,1)$ to the sheets at $R$. For $\bar p_1=0$, we obtain $\mathcal A_1=w_1/N_{\Gamma\Gamma}$ and $\mathcal A_R=w_R/N_{RR}$.

\subsubsection{Planar-Hall and Hall response for perpendicular fields and a general angle}
\label{secresph}

For $B=0$, the sector $m=1$ obeys the same equations as the sector $m=0$ [cf. Appendix~\ref{secappres}]. Consequently, $\sigma_\perp=\sigma_\parallel=\sigma_0$ and $\sigma_H=0$, and Eq.~\eqrf{eqcurr} gives $j_{\rm PH}=0$ and $j_{\bE}=\sigma_0\,E$ for every angle $\vartheta$ between $\bE$ and $\bB$. A nonvanishing planar-Hall response therefore requires $\sigma_\perp\neq\sigma_\parallel$ at finite $B$.

To the leading order in $B$, the Boltzmann part of the Hall conductivity follows from the imaginary part of the sector $m=1$. With $\mathcal Y_a=w_a\,\mathcal A_a/k_a$ and
\begin{align}
\mathcal Z_1=\frac{N_{RR}\,\mathcal Y_1+N_{\Gamma R}\,\mathcal Y_R}{\mathcal N}\,,\quad
\mathcal Z_R=\frac{N_{\Gamma\Gamma}\,\mathcal Y_R+N_{R\Gamma}\,\mathcal Y_1}{\mathcal N}\,,\quad
\mathcal Z_0=\tau_0\,\mathcal Y_0\,,
\end{align}
it reads
\begin{align}
\sigma_H^{\rm B}\simeq\frac{e^3\,B}{6\,\pi^2}\left(k_1^2\,\mathcal Z_1+k_0^2\,\mathcal Z_0+2\,k_R^2\,\mathcal Z_R\right).
\label{eqsigHB}
\end{align}
The AH part of Eq.~\eqrf{eqsigmw} is linear in $B$ as well. Only the product of the BC and the OMM contributes at this order, and it is
\begin{align}
\sigma_H^{\rm AH}\simeq\frac{\eta_B\,\eta_O\,e^3\,B}{24\,\pi^2}\left[\frac{2}{k_1\left(1+2\,c\,k_1\right)}+\frac{2\,\tilde v_F}{\mu+\Delta}\right].
\label{eqsigHAH}
\end{align}
The sheet $(\Gamma,0)$ does not contribute, since its BC vanishes. The term of the sheet $(\Gamma,1)$ is half of the corresponding expression of Eq.~\eqref{eqsigah3} for the two conjugate QTSM nodes.

\subsubsection{Expansion to the third order in $B$}
\label{secresB3}

The expansion of Sec.~\ref{secresB33} applies to the sheets of this section. Equations~\eqref{eqexp3}, \eqref{eqkser}, \eqref{eqforms}, \eqref{eqlam1}, \eqref{eqSn}, and \eqref{eqsigcoef} hold with the sums over the four sheets $(\Gamma,1)$, $(\Gamma,0)$, $R_1$, and $R_2$, and with the kernel $\mathcal K^{(m)}_{aa'}$ of Eq.~\eqref{eqsector}. Let $\tilde\lambda_a=1$ for the sheets $(\Gamma,1)$ and $(R_i,+)$, and let $\tilde\lambda_a=0$ for the sheet $(\Gamma,0)$. Let $v_a$ and $c_a$ be $v_F$ and $c$ for the sheets at $\Gamma$, and $\tilde v_F$ and $0$ for the sheets at $R$. We define $\tilde\nu_a=\tilde\lambda_a+2\,c_a\,k_a$, such that $w_a=v_a\,\tilde\nu_a$. The numbers $N_a$ and $G_a$ are $2$ and $1$ for $(\Gamma,1)$, $0$ and $2$ for $(\Gamma,0)$, and $1$ and $1$ for each sheet at $R$. The radius of each sheet is the positive root of the cubic equation of Appendix~\ref{secappqtsm}, which holds for the sheets at $R$ with $c_a=0$, and its expansion is Eq.~\eqref{eqkser} with these parameters. The first-order coefficients of the weights, of the sources, and of the frequency are those of Eq.~\eqref{eqing1}. Since $\sigma_\perp^{(0)}=\sigma_\parallel^{(0)}$ in this section, Eq.~\eqref{eqcurr} gives the planar-Hall current $j_{\rm PH}=\left(\sigma_\perp^{(2)}-\sigma_\parallel^{(2)}\right)B^2\sin\vartheta\,\cos\vartheta\,E+\mathcal O(B^4)$.

The sum of Eq.~\eqref{eqsigHAH3} over the four sheets gives the AH part. The sheet $(\Gamma,0)$ does not contribute, since $N_a=0$. With $k_1=k_{\Gamma,1}$ and $k_R=\left(\mu+\Delta\right)/\tilde v_F$, the sums are
\begin{align}
\sigma_H^{\rm AH}\simeq\frac{\eta_B\,\eta_O\,e^3\,B}{24\,\pi^2}\left[\frac{2}{k_1\left(1+2\,c\,k_1\right)}+\frac{2}{k_R}\right]+\frac{\eta_B\,\eta_O^3\,e^5\,B^3}{160\,\pi^2}\left[\frac{2\left(2+11\,c\,k_1+16\,c^2\,k_1^2\right)}{k_1^5\left(1+2\,c\,k_1\right)^5}+\frac{4}{k_R^5}\right].
\end{align}
The first term is Eq.~\eqref{eqsigHAH}. Neither term depends on the scattering amplitudes.

The weights and the rates are constant at $B=0$, and the overlaps of Eq.~\eqref{eqovgriso} are invariant under $\left(\theta,\theta'\right)\to\left(\pi-\theta,\pi-\theta'\right)$. Every sheet then has a definite parity in $\cos\theta$ separately. For $m=0$, the function $\Lambda_0$ is odd, as in Sec.~\ref{secresll3}, and $\Lambda_1$ is even. The even function $\Lambda_1$ contains the zero mode, and the constant drops out of Eq.~\eqref{eqSn}, because $\mathcal G_B[1,f]=0$ follows from $\tau_a^{-1}=\sum_{a'}\int\rho_{a'}\,\mathcal K^{(0)}_{aa'}$ and $\mathfrak H_B[1]=0$ follows from Eq.~\eqref{eqsolv}. For $m=1$, the function $\Lambda_0$ is even, and we write $\Lambda_1=\Lambda_1^{\rm R}+i\,\Lambda_1^{\rm I}$. The imaginary part $\Lambda_1^{\rm I}$ is even, and the real part $\Lambda_1^{\rm R}$ is odd in $\cos\theta$. The coefficients $S_2^{(0)}$ and $S_2^{(1)}$ are given in terms of elementary quantities in Appendix~\ref{secappB3}.

\subsubsection{Dependence on the scattering amplitudes}
\label{secresbeta}

The intranode and internode amplitudes of the Weyl nodes enter Eqs.~\eqrf{eqtauB0} and \eqrf{eqNmat} only through $\gamma_R$. The zero-field conductivity cannot discriminate between $\beta^R_{\rm intra}$ and $\beta^R_{\rm inter}$. For $g_0=g_1=0$, the sheets at $\Gamma$ and at $R$ decouple. The Weyl nodes then give
\begin{align}
\sigma_R=\frac{e^2\,\tilde v_F^2}{4\,\pi^2\,\gamma_R}\,,
\end{align}
which does not depend on $\mu$ and $\Delta$. The sheets at $\Gamma$ give
\begin{align}
\sigma_\Gamma=\frac{e^2}{6\,\pi^2}\left[\frac{k_1^2\,w_1}{\frac23\,\beta_{11}^2\,\rho_1+\frac43\,\beta_{10}^2\,\rho_0}+\frac{k_0^2\,w_0}{\frac43\left(\beta_{10}^2\,\rho_1+\beta_{00}^2\,\rho_0\right)}\right].
\end{align}
The zero-field result is continuous at $g_0=g_1=0$, and its sum is $\sigma_0=\sigma_\Gamma+\sigma_R$ in this limit. The limit does not extend to $B\neq0$, since the Chern numbers of the two decoupled sets of sheets do not add up to zero separately, and Eq.~\eqrf{eqsolv} is violated for $\bE\parallel\bB$.

\subsubsection{Role of the quadratic correction, the velocity ratio, and the energy offset}
\label{secrespar}

The parameters $\tilde v_F$ and $\Delta$ enter the zero-field results through $k_R=(\mu+\Delta)/\tilde v_F$, $\rho_R=k_R^2/\tilde v_F$, and $w_R=\tilde v_F$ only. The part of Eq.~\eqrf{eqsigHAH} due to the Weyl nodes depends on them through $\tilde v_F/(\mu+\Delta)=1/k_R$.

The limit $c\to0$ is singular for the sheet $(\Gamma,0)$, because $k_0=\sqrt{\mu/(c\,v_F)}$ and $\rho_0=k_0/(2\,c\,v_F)$ diverge, whereas $k_1$, $\rho_1$, and $\rho_R$ stay finite. For nonzero $\beta_{10}$, $\beta_{00}$, and $\bar o_0\,g_0$, the in-scattering into the sheet $(\Gamma,0)$ suppresses the contributions of the sheets $(\Gamma,1)$ and $R$. The zero-field conductivity is then dominated by the sheet $(\Gamma,0)$, and
\begin{align}
\sigma_0=\frac{e^2\,c\,v_F\,\mu}{2\,\pi^2\,\beta_{00}^2}+\mathcal O\!\left(c^{3/2}\right).
\label{eqsigc0}
\end{align}
This leading term vanishes linearly with $c$ and is independent of $\tilde v_F$, $\Delta$, and of all the amplitudes except $\beta_{00}$.

\subsubsection{Numerical results}
\label{secresnum}

The figures show the exact solution of the moment system of Eq.~\eqrf{eqmomsys} for $\bB=B\,\boldsymbol{\hat z}$, which is case (a) of Table~\ref{tabgeom}. The field enters through $\alpha=e\,B/(2\,k_1^2)$, where $k_1$ is the radius of the sheet $(\Gamma,1)$ at $B=0$, and both signs of $\alpha$ are shown. The parameters are taken from the tight-binding model that was fitted to the DFT bandstructure of RhSi in the Supplemental Material of Ref.~\cite{prl119206401}. The fitted values in eV, with the lattice constant set to unity, are $v_1=0.55$, $v_p=-0.76$, and $v_2=0.16$, where $v_2$ multiplies the second-neighbour hopping $\left[\cos k_x+\cos k_y+\cos k_z\right]\mathbb I$. The velocities of the QTSM node and of the Weyl nodes are $v_F=\tilde v_F=|v_p|/2$ [cf. Eq.~\eqref{eqtbexp}]. The second-neighbour term contributes $-v_2\,k^2/2$ at $\Gamma$ and $+v_2\,q^2/2$ at $R$, and we set $c\,v_F=-v_2/2$ for the QTSM node, which gives $c=-5.5\times10^{-3}\,v_F/\mu$ for $\mu=0.01$ eV. The corresponding term of the Weyl nodes is not part of Eq.~\eqref{eqham}. The energies of the nodes at $\Gamma$ and at $R$ are $-v_1+3\,v_2$ and $-3\,v_2$, which gives the offset $\Delta=0.41$ eV, in agreement with the Supplemental Material of Ref.~\cite{prl119206401}. The chemical potential is $\mu=0.01$ eV, measured from the node at $\Gamma$, and $\mu\,\tau_0=15$ with $\tau_0$ the out-scattering time of the sheet $(\Gamma,1)$ at $B=0$. Since $c<0$, the band $s=0$ of the QTSM node does not cross $\mu$, and the sheet $(\Gamma,0)$ is absent. The system has the three sheets $(\Gamma,1)$, $R_1$, and $R_2$, and their Chern numbers add up to zero. One amplitude $\beta$ is used for all pairs of sheets within the QTSM node and within each Weyl node, and the amplitude $r_\beta\,\beta$ is used for all pairs of sheets that belong to different nodes, which are $(\Gamma,1)R_i$ and $R_1R_2$. We set $r_\beta=1/2$ unless stated otherwise. The overlaps of Eq.~\eqrf{eqovgriso} are $\bar o_1=1/4$ and $\bar p_1=0$ for each copy at $R$. They are half of the average of Eq.~\eqref{eqovgr4} over the rotations of all the momenta, with equal weights for the two copies. The average of $\hat q_x\,\hat k_y+\hat q_y\,\hat k_z+\hat q_z\,\hat k_x$ vanishes, since the trace of the cyclic permutation matrix is zero. All the quantities are normalised to the corresponding values at $B=0$, and the numerical implementation is described in Appendix~\ref{secnum}.

\begin{figure}[t!]
\centering
\subfigure[\label{figqw}]{\includegraphics[width=0.85\textwidth]{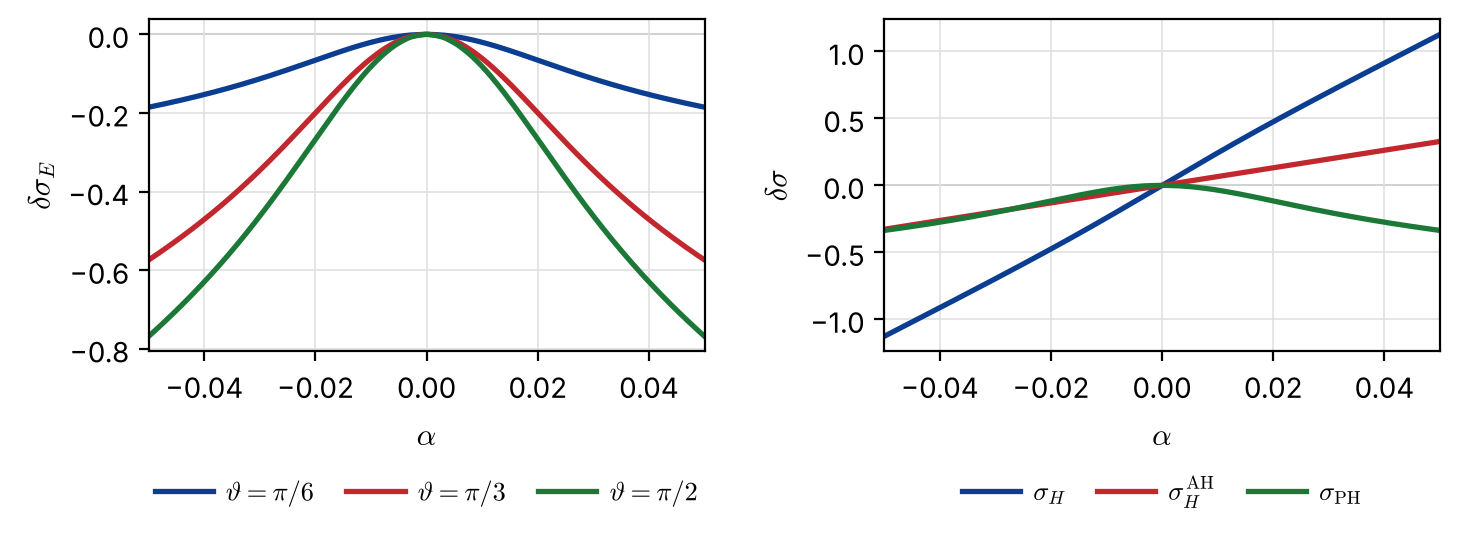}}\\
\subfigure[\label{figqwratio}]{\includegraphics[width=0.85\textwidth]{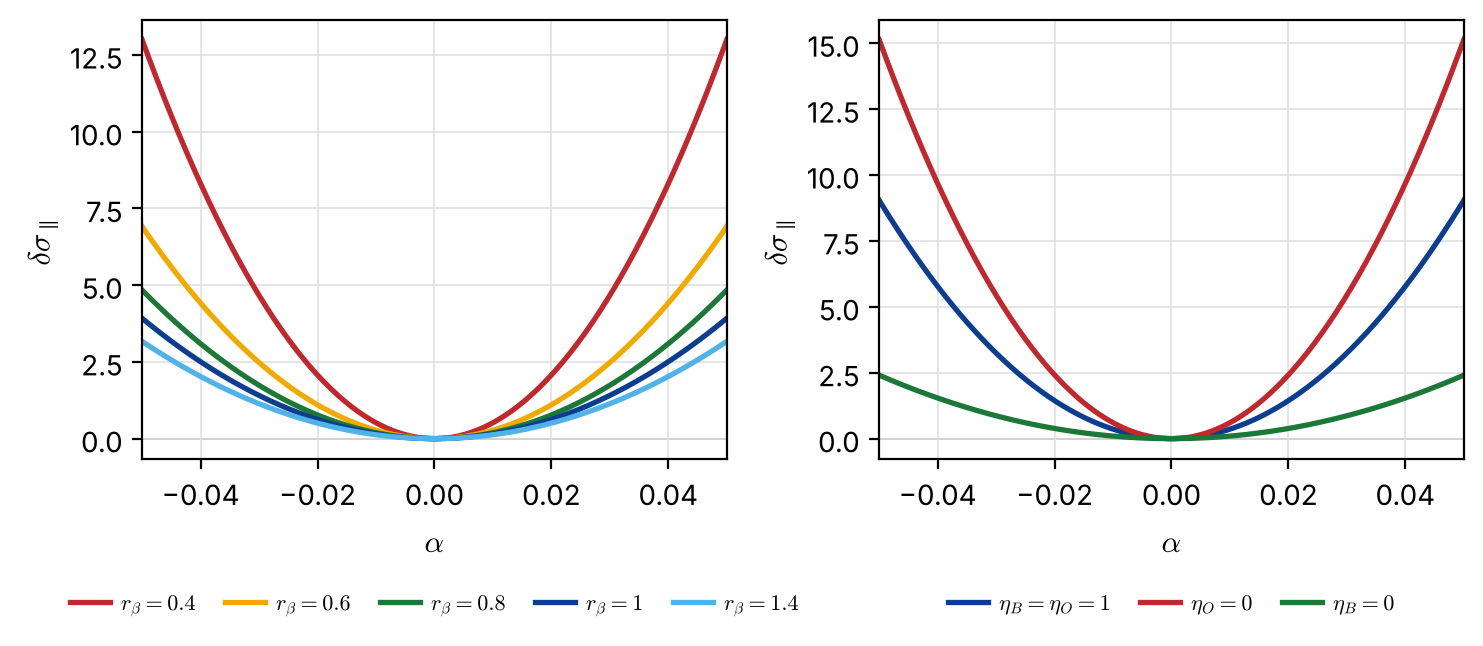}}
\caption{QTSM and Weyl nodes for $\bB\parallel\hat{\boldsymbol z}$. (a) Left: $\delta\sigma_{E}$ multiplied by $10^3$ for $\bE$ at the angle $\vartheta=\pi/6$ (navy), $\pi/3$ (crimson), and $\pi/2$ (green) to $\bB$. Right: $\sigma_H$ multiplied by $10^2$ (navy), $\sigma_H^{\rm AH}$ multiplied by $10^5$ (crimson), and the planar-Hall change for $\vartheta=\pi/3$ multiplied by $10^3$ (green), in units of $\sigma_\perp$ at $B=0$. (b) For $\bE\parallel\bB$, $\delta\sigma_\parallel$ is multiplied by $10^6$. Left: $r_\beta=0.4$ (crimson), $0.6$ (amber), $0.8$ (green), $1$ (navy), and $1.4$ (sky). Right: $r_\beta=1/2$ with $\eta_B=\eta_O=1$ (navy), $\eta_O=0$ (crimson), and $\eta_B=0$ (green).}
\label{figqwbz}
\end{figure}

By Eq.~\eqref{eqons} and the axisymmetric form of the tensor, $\sigma_\parallel$ and $\sigma_\perp$ are even and $\sigma_H$ is odd in $\alpha$, and the curves of Fig.~\ref{figqw} display this symmetry. The diagonal response of this set-up has therefore no term odd in $B$. The radius of the sheets at $R$ exceeds that of the sheet at $\Gamma$ by the factor $\left(\mu+\Delta\right)/\mu$, and the sheets at $R$ carry almost the whole of $\sigma_0$. The changes in units of $\sigma_\perp$ at $B=0$ are therefore small, which is why the factors in the captions are large. For an oblique $\bE$, the change of the current along $\bE$ is negative and grows in magnitude as $\vartheta$ increases. In the right panel, the Hall response is dominated by the Boltzmann part, and the AH part is much smaller. The Hall response increases monotonically with $\alpha$ in the range shown. The planar-Hall change is negative for both signs of $\alpha$, in accordance with the decrease of $\sigma_\perp$ and the increase of $\sigma_\parallel$ in Eq.~\eqref{eqcurr}.

For $\bE\parallel\bB$, the change of $\sigma_\parallel$ is positive and even in $\alpha$ for all the ratios in Fig.~\ref{figqwratio}, and it decreases with increasing $r_\beta$. No change of sign occurs in this range. The right panel separates the contributions of the BC and the OMM to the longitudinal change. Both are positive, and the combined change is smaller than the BC contribution alone, which shows that the cross term is negative. The OMM contribution alone is the smallest of the three.

\begin{figure}[t!]
\centering
\subfigure[\label{figqwscan}]{\includegraphics[width=\textwidth]{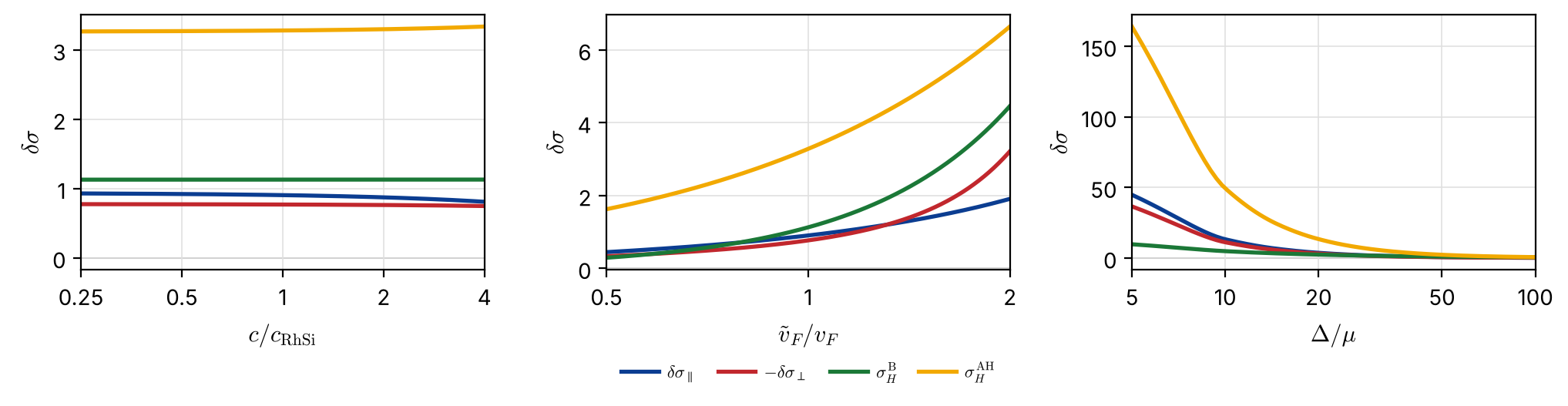}}\\
\subfigure[\label{figqwcosi}]{\includegraphics[width=\textwidth]{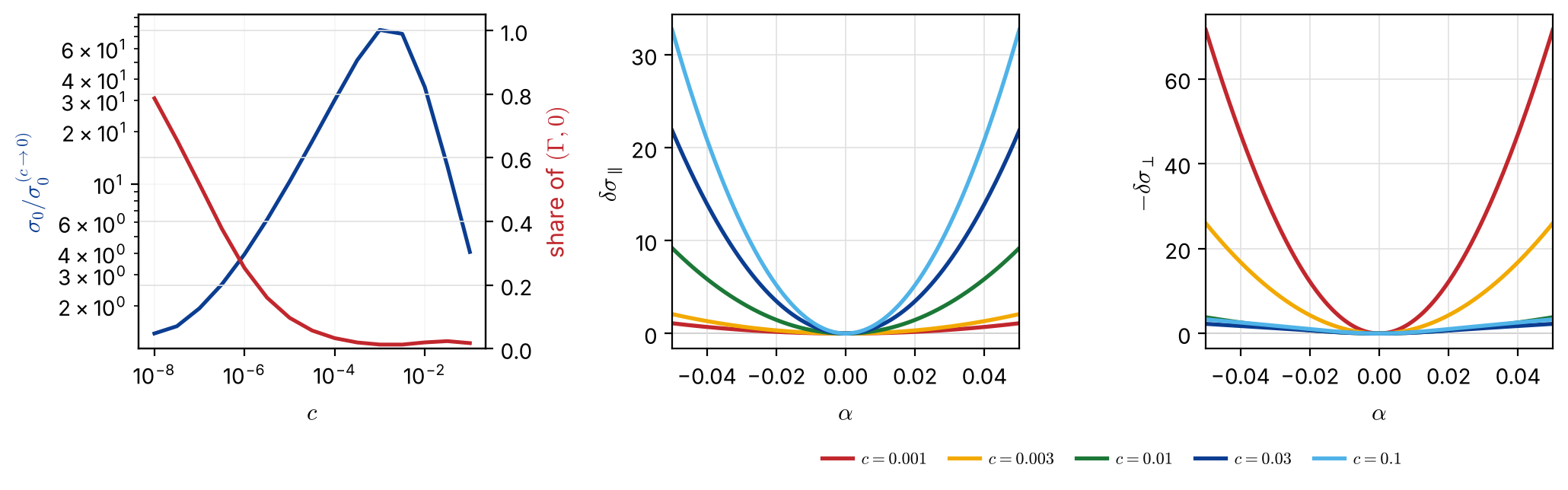}}
\caption{QTSM and Weyl nodes for $\bB\parallel\hat{\boldsymbol z}$. (a) The field is fixed at $\alpha=0.05$. Left, middle, and right: dependence on $c/c_{\rm RhSi}$, $\tilde v_F/v_F$, and $\Delta/\mu$. In every panel, $\delta\sigma_\parallel$ multiplied by $10^5$ (navy), $-\delta\sigma_\perp$ multiplied by $10^3$ (crimson), $\sigma_H^{\rm B}$ multiplied by $10^2$ (green), and $\sigma_H^{\rm AH}$ multiplied by $10^6$ (amber), in units of the values at $B=0$. (b) The parameters are $c>0$ and $\Delta=0.225$ eV. Left: $\sigma_0$ in units of its leading term of Eq.~\eqref{eqsigc0} (navy, left axis) and the contribution of the sheet $(\Gamma,0)$ to $\sigma_0$ in units of $\sigma_0$ (crimson, right axis) versus $c$. Middle and right: $\delta\sigma_\parallel$ multiplied by $10^6$ and $-\delta\sigma_\perp$ multiplied by $10^3$ for $c=10^{-3}$ (crimson), $3\times10^{-3}$ (amber), $10^{-2}$ (green), $3\times10^{-2}$ (navy), and $10^{-1}$ (sky), in units of the values at $B=0$.}
\label{figqwpar}
\end{figure}

Figure~\ref{figqwscan} shows the dependence on the parameters of Sec.~\ref{secrespar}. In each panel, one of $c$, $\tilde v_F$, and $\Delta$ is varied about the RhSi values, the other two are kept fixed, and the field is fixed at the value $\alpha=0.05$ of the RhSi parameters. The changes are in units of the values at $B=0$ of the same parameters. The quadratic coefficient is varied in the range for which the sheet $(\Gamma,0)$ is absent, since the limit $c\to0$ is singular [cf. Eq.~\eqref{eqsigc0}].

The dependence on $c$ is weak. The change $\delta\sigma_\parallel$ and the magnitude of $\delta\sigma_\perp$ decrease slightly with $|c|$, and $\sigma_H^{\rm B}$ and $\sigma_H^{\rm AH}$ increase slightly. This is consistent with the fact that the sheets at $R$, which are independent of $c$, carry almost the whole of $\sigma_0$. The dependence on $\tilde v_F$ and $\Delta$ is strong, since they set the radius $k_R=(\mu+\Delta)/\tilde v_F$ of the sheets at $R$. All four quantities increase monotonically with $\tilde v_F$ and decrease monotonically with $\Delta$. The curves have no sign change in the ranges shown, and the Boltzmann Hall part remains larger than the AH part throughout.

The RhSi parameters give $c<0$, and the role of the sheet $(\Gamma,0)$ is therefore not visible in the figures above. Figure~\ref{figqwcosi} shows the system with $c>0$, in which this sheet is present. The offset is $\Delta=0.225$ eV, which is the value predicted for CoSi in Ref.~\cite{sanchez2019}, where the SOC is weak, and the velocities are those of RhSi. The values of $c$, in units of $v_F/\mu$, are those of a model and do not follow from a bandstructure fit. All the other parameters are as above.

In the left panel of Fig.~\ref{figqwcosi}, the ratio of $\sigma_0$ to the leading term of Eq.~\eqref{eqsigc0} decreases towards unity as $c\to0$, and it is about $1.4$ at the smallest value shown. The amplitude $\beta_{00}$ in Eq.~\eqref{eqsigc0} is the one fixed by $\mu\,\tau_0$ for the given $c$. The contribution of the sheet $(\Gamma,0)$ increases towards unity in the same limit, and it is small for larger $c$. Values of $c$ below the range shown are not included, because the linear system is ill-conditioned there. In the middle and right panels, $\delta\sigma_\parallel$ is positive and increases with $c$, and the change $\delta\sigma_\perp$ is negative. Its magnitude decreases with $c$ up to $c=3\times10^{-2}$ and is larger again at $c=10^{-1}$. The curves are even in $\alpha$ in both panels.

\label{secresnumbc}

The cases (b) and (c) of Table~\ref{tabgeom} are solved with the general solver of Eq.~\eqref{eqmomsys}, which treats the three components $j$ together and does not use the azimuthal sectors. The implementation is described in Appendix~\ref{secnum}. These cases belong to the second case of Sec.~\ref{sectilt}, and the overlap between $(\Gamma,1)$ and one copy at $R$ is therefore taken from Eq.~\eqref{eqovgr4} of the Chang scenario. It reads $\left(1-\hat q_x\,\hat k_y-\hat q_y\,\hat k_z-\hat q_z\,\hat k_x\right)/4$, since the two copies have equal weights. The Flicker scenario is not shown, and the other parameters are those of Sec.~\ref{secresnum}. The tensor is written in the frame with $\boldsymbol{\hat z}=\hat{\bs c}$, and we take $\bB=B\left(\sin\theta_B,0,\cos\theta_B\right)$. The value $\theta_B=\pi/2$ is the case (b), and the value $\theta_B=0$ is the case (a) of the second case, in which the overlap of Eq.~\eqref{eqovgr4} replaces its average over the rotations that was used in Fig.~\ref{figqwbz}.

The kernel is invariant under the rotations about $\hat{\bs c}$ only, and $\bs\sigma$ at $B=0$ is therefore uniaxial about $\hat{\bs c}$. Its anisotropy is small. All the changes are measured from the tensor at $B=0$, and they are given in units of $\sigma_0=\mathrm{tr}\,\bs\sigma(0)/3$. For $\bB\perp\hat{\bs c}$, the rotation $\mathcal R_{\hat c}(\pi)$ maps $\bB$ onto $-\bB$ and leaves the kernel invariant, which gives $\sigma_{ij}(-\bB)=\eta_i\,\eta_j\,\sigma_{ij}(\bB)$ with $\eta_i$ of Eq.~\eqref{eqplanar5}. The diagonal elements and $\sigma_{xy}$ are therefore even in $\alpha$, and $\sigma_{xz}$ and $\sigma_{yz}$ are odd. For a general orientation, Eq.~\eqref{eqons} makes the symmetric part of $\bs\sigma$ even and the antisymmetric part odd in $\alpha$. The diagonal elements are therefore even in every case, and the terms that are odd in $B$ are confined to the antisymmetric part. The numerical tensor has these properties and satisfies the relations of Table~\ref{tabgeom}.

\begin{figure}[t!]
\centering
\subfigure[\label{figqwb}]{\includegraphics[width=0.85\textwidth]{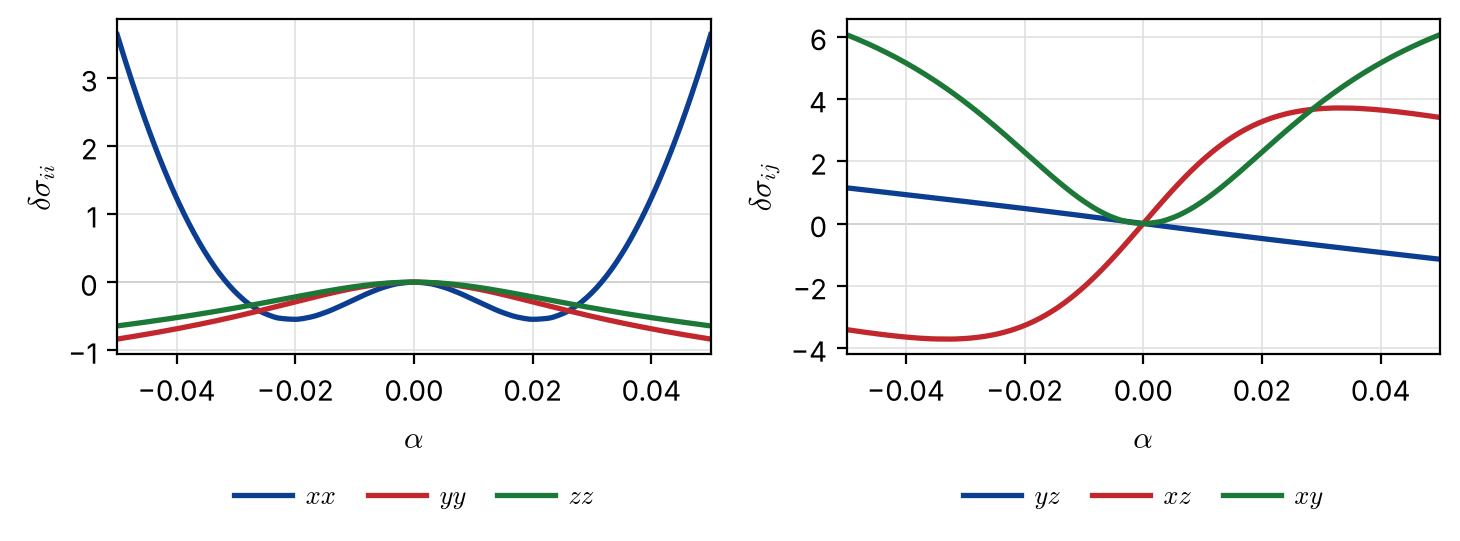}}\\
\subfigure[\label{figqwc}]{\includegraphics[width=\textwidth]{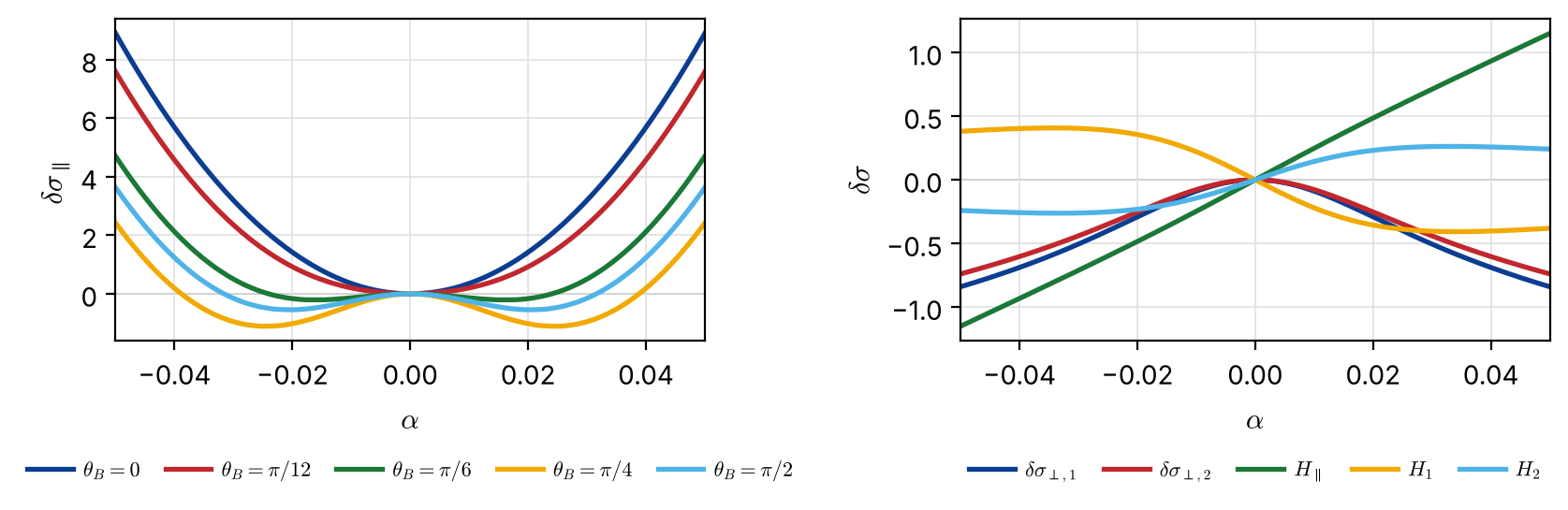}}
\caption{QTSM and Weyl nodes for $\hat{\boldsymbol z}=\hat{\bs c}$. (a) Case (b) with $\bB\parallel\hat{\boldsymbol x}$. Left: $\delta\sigma_{xx}$ multiplied by $10^6$ (navy), $\delta\sigma_{yy}$ (crimson), and $\delta\sigma_{zz}$ (green) multiplied by $10^3$. Right: $\delta\sigma_{yz}$ multiplied by $10^2$ (navy), $\delta\sigma_{xz}$ (crimson), and $\delta\sigma_{xy}$ (green) multiplied by $10^5$, in units of $\sigma_0$. (b) Case (c) with $\bB=B\left(\sin\theta_B,0,\cos\theta_B\right)$. Left: $\delta\sigma_\parallel=\hat{\bB}\cdot\delta\bs\sigma\cdot\hat{\bB}$ multiplied by $10^6$ for $\theta_B=0$ (navy), $\pi/12$ (crimson), $\pi/6$ (green), $\pi/4$ (amber), and $\pi/2$ (sky). Right: for $\theta_B=\pi/4$, the eigenvalues $\delta\sigma_{\perp,1}$ (navy) and $\delta\sigma_{\perp,2}$ (crimson) of the symmetric part of $\delta\bs\sigma$ in the plane perpendicular to $\bB$ multiplied by $10^3$, the component $H_\parallel$ of the Hall vector along $\bB$ (green) multiplied by $10^2$, and its components $H_1$ (amber) and $H_2$ (sky) along $\left(\cos\theta_B,0,-\sin\theta_B\right)$ and along $\boldsymbol{\hat y}$ multiplied by $10^4$, in units of $\sigma_0$.}
\label{figqwbc}
\end{figure}

In Fig.~\ref{figqwb}, the element $\delta\sigma_{xx}$ is the change for $\bE\parallel\bB$. It is positive at the larger values of $|\alpha|$, and it has a shallow negative minimum at smaller values. The elements $\delta\sigma_{yy}$ and $\delta\sigma_{zz}$ are the changes for $\bE\perp\bB$ along the axis perpendicular to $\hat{\bs c}$ and along $\hat{\bs c}$, respectively. Both are negative, and they differ, with the larger magnitude for $\delta\sigma_{yy}$. The transverse response therefore depends on the orientation of $\bE$ relative to the crystal axis $\hat{\bs c}$, which is absent in the case (a) of the first case. The element $\sigma_{yz}$ is the Hall response. It is odd in $\alpha$ and close to linear. The elements $\sigma_{xz}$ and $\sigma_{xy}$ do not exist for an axisymmetric tensor, and they are smaller than $\sigma_{yz}$ by more than two orders of magnitude. The element $\sigma_{xz}$ is odd, and $\sigma_{xy}$ is even, in agreement with the parities above.

The Hall vector $\bs H$ of Fig.~\ref{figqwc} is defined by the antisymmetric part of the tensor, $\sigma^{\rm A}_{ij}=-\epsilon_{ijl}\,H_l$. For $\bE\parallel\bB$, the change $\delta\sigma_\parallel$ is positive at the largest $|\alpha|$ for all the orientations shown. It is largest for $\theta_B=0$, decreases with $\theta_B$ up to $\pi/4$, and is larger again at $\pi/2$. For $\theta_B\geq\pi/6$, it is negative in an interval at smaller $|\alpha|$, and it changes sign. In the right panel, the two eigenvalues of the symmetric transverse block are negative and close to each other, and their difference measures the dependence of the transverse change on the direction of $\bE$ in the plane perpendicular to $\bB$. The Hall vector is dominated by $H_\parallel$, which is close to linear in $\alpha$. The components $H_1$ and $H_2$ perpendicular to $\bB$ are odd and smaller than $H_\parallel$ by more than two orders of magnitude, and they are not monotonic. They vanish for $\theta_B=0$, since $\bs H$ is then parallel to $\bB$ by the axisymmetry.

\begin{figure}[t!]
\centering
\includegraphics[width=\textwidth]{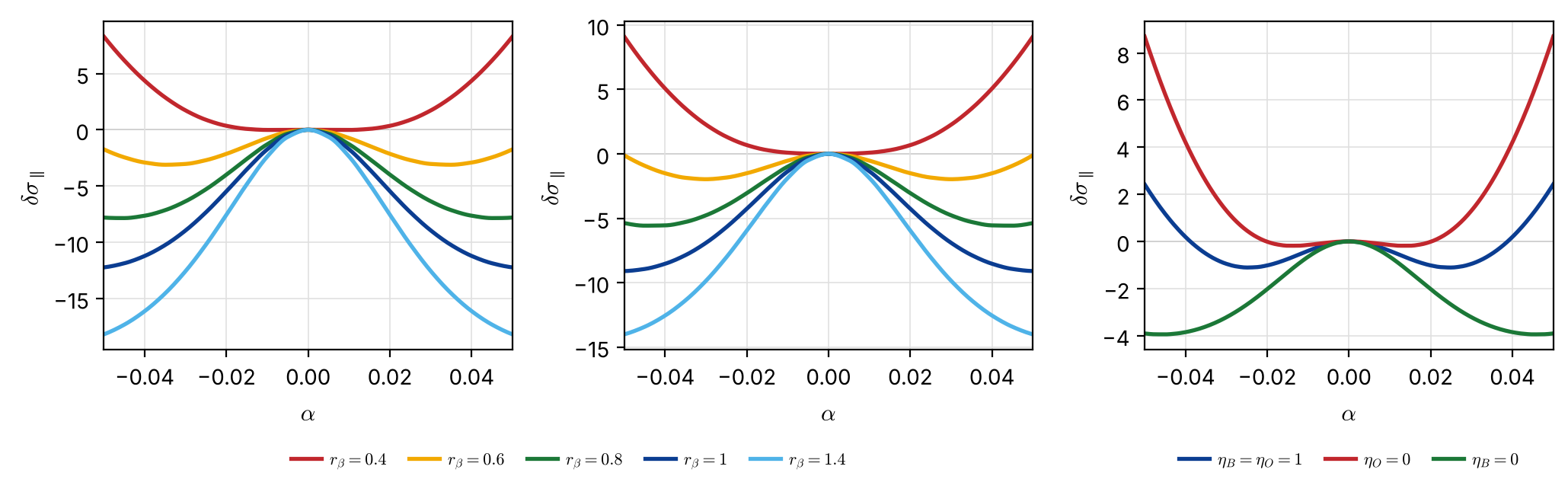}
\caption{QTSM and Weyl nodes for $\bB=B\left(\sin\theta_B,0,\cos\theta_B\right)$ with $\boldsymbol{\hat z}=\hat{\bs c}$. Left and middle: $\delta\sigma_\parallel$ multiplied by $10^6$ for $\theta_B=\pi/4$ [case (c)] and $\theta_B=\pi/2$ [case (b)], respectively, with $r_\beta=0.4$ (crimson), $0.6$ (amber), $0.8$ (green), $1$ (navy), and $1.4$ (sky). Right: $\theta_B=\pi/4$ and $r_\beta=1/2$ with $\eta_B=\eta_O=1$ (navy), $\eta_O=0$ (crimson), and $\eta_B=0$ (green), multiplied by $10^6$. All changes are in units of $\hat{\bB}\cdot\bs\sigma(0)\cdot\hat{\bB}$.}
\label{figqwcratio}
\end{figure}

Figure~\ref{figqwcratio} shows the dependence of the longitudinal change on $r_\beta$ and on the BC and the OMM. In the left and middle panels, $\delta\sigma_\parallel$ is positive at the largest $|\alpha|$ for $r_\beta=0.4$ only. For $r_\beta\geq0.6$, it is negative, and its magnitude increases monotonically with $r_\beta$. The curves for $r_\beta=0.6$ have a minimum at an intermediate $|\alpha|$, and they are flatter at the largest $|\alpha|$. The sign of the longitudinal change therefore depends on $r_\beta$, which differs from the case (a) of Fig.~\ref{figqwratio}, where the overlap of Eq.~\eqref{eqovgriso} was used. The two orientations give the same ordering, and for $r_\beta\geq0.6$ the magnitudes are smaller for $\theta_B=\pi/2$. In the right panel, the change with $\eta_O=0$ is positive, and the change with $\eta_B=0$ is negative. The combined change is smaller than the BC contribution alone, and it has a negative minimum at an intermediate $|\alpha|$ and turns positive at the largest $|\alpha|$. It differs from the sum of the two contributions, and the cross term is negative at the largest $|\alpha|$.

%%%%%%%%%%%%%%%
\section{RSW and TSM nodes}
\label{secsys2}

In the chiral cubic crystals, the spin-orbit coupling (SOC) splits the multifold nodes of the spinless bandstructure. At $\Gamma$, the threefold node of the pseudospin-1 operator $\bs L$ combines with the electron spin $\bs{\tilde\sigma}/2$. The sixfold space splits into the $J=3/2$ and $J=1/2$ multiplets by the SOC term $\tilde\epsilon\,\bs L\cdot\bs{\tilde\sigma}$, whose eigenvalues are $\tilde\epsilon$ and $-2\,\tilde\epsilon$ \cite{xu2020_optical}. At $R$, the double Weyl node splits into a sixfold node, which is a double pseudospin-1 node, and a twofold Kramers pair \cite{xu2020_optical, gao2022}. The Hamiltonians below are written in the isotropic limit, with the quadratic corrections, and with the couplings between the multiplets dropped. The case $c_{3/2}=c_{1/2}=c_R=0$ is the one with linear dispersions, in which the node at $R$ is the TSM node. Both cases are kept below, and the linear one is stated explicitly for the FS radii and for the RTA.

The Hamiltonian at $\Gamma$ is the six-dimensional block-diagonal matrix of the RSW node with the quadratic correction, and of the Weyl node of the $J=1/2$ multiplet. The Hamiltonian at $R$ consists of the two copies $R_1$ and $R_2$ of the pseudospin-1 node, which are its upper and lower blocks. At $\Gamma$, $\bs J$ is the vector of the spin-3/2 matrices of Appendix~\ref{secapprsw}, $\bs{\tilde\sigma}$ is the vector of the Pauli matrices, $E_\Gamma$ and $E_W$ are the energies of the RSW node and of the Weyl node, and $v_{3/2}>0$ and $v_{1/2}>0$ are the corresponding velocities. At $R$, $\bs\Lambda$ is the vector of the spin-1 matrices of Appendix~\ref{secappsp2}, $E_R$ is the energy of the nodes, and $v_R>0$ is their velocity. The coefficients $c_{3/2}$, $c_{1/2}$, and $c_R$ multiply the quadratic terms. The Hamiltonians are
\begin{align}
& \mathcal H_\Gamma(\bk)=\begin{bmatrix}
\left(E_\Gamma+c_{3/2}\,v_{3/2}\,k^2\right)\mathbb I_4+v_{3/2}\,\bk\cdot\bs J & 0\\
0 & \left(E_W+c_{1/2}\,v_{1/2}\,k^2\right)\mathbb I_2+v_{1/2}\,\bk\cdot\bs{\tilde\sigma}
\end{bmatrix},\nn
& \mathcal H_R(\bk)=\left(E_R+c_R\,v_R\,k^2\right)\mathbb I_6+v_R\begin{bmatrix}
\bk\cdot\bs\Lambda & 0\\
0 & \bk\cdot\bs\Lambda
\end{bmatrix}.
\label{eqham2}
\end{align}
The projection of $\bk\cdot\bs L$ on the two multiplets gives $v_{3/2}=v_{1/2}=2\,v/3$ and $E_W-E_\Gamma=-3\,\epsilon$ in the isotropic limit, with the quadratic term $c\,v\,k^2\,\mathbb I_6$ of the spinless model, that is $c_{3/2}\,v_{3/2}=c_{1/2}\,v_{1/2}=c\,v$. The cubic anisotropy makes the two velocities different in general. In the basis of Ref.~\cite{grushin-multifold}, the Hamiltonian at $R$ is $\mathcal H_R=\mathrm{diag}\left[\mathcal H_{3f}\!\left(\pi/6,\bk\right),\mathcal H_{3f}\!\left(5\,\pi/6,\bk\right)\right]$ with
\begin{align}
\mathcal H_{3f}(\phi,\bk)=v_R\begin{pmatrix}
0 & e^{i\,\phi}\,k_x & e^{-i\,\phi}\,k_y\\
e^{-i\,\phi}\,k_x & 0 & e^{i\,\phi}\,k_z\\
e^{i\,\phi}\,k_y & e^{-i\,\phi}\,k_z & 0
\end{pmatrix},
\label{eqflicker}
\end{align}
which has the eigenvalues $0$ and $\pm v_R\,k$ for $\phi=\pi/6,\,\pi/2,\,5\,\pi/6$, and both copies are therefore isotropic. In the model of Ref.~\cite{gao2022}, the two copies are coupled, and this coupling is dropped here. Without SOC, the node at $R$ is the double Weyl node $\mathcal H_R^{(0)}=E_R\,\mathbb I_4+\chi_R\,v_R\,\bk\cdot\bs{\tilde\sigma}\otimes\tilde\tau_0$, in which the two copies have the same chirality $\chi_R$ and $\tilde\tau_0$ is the identity matrix in the copy space. The Weyl nodes of Eq.~\eqref{eqham} correspond to Eq.~\eqref{eqham2} with $\bk\cdot\bs\Lambda$ replaced by $\bk\cdot\bs{\tilde\sigma}$.

The chirality of the RSW node is $\chi_\Gamma=+1$ for $v_{3/2}>0$. The matrices $\Lambda_a$ are left-handed, and each copy at $R$ has the chirality $\chi_R=-1$. The Weyl node at $\Gamma$ has the same chirality as the RSW node. Its energy $E_W$ will be set so far from $\mu$ that it provides no FS sheet. It remains in the matrix of $\Gamma$, so that the spinors at $\Gamma$ and at $R$ have the same dimension. The net topological charge of the nodes at $\Gamma$ and $R$ does not vanish if this Weyl node is included, since the charges of the RSW node and of the Weyl node at $\Gamma$ add up to $4+1$ in magnitude, whereas the two copies at $R$ carry $-2-2$. The Nielsen--Ninomiya theorem \cite{nielsen81_no} is satisfied in the actual bandstructures by additional Weyl nodes at generic points of the BZ between $\Gamma$ and $R$, which are not part of the model \cite{gao2022}. We choose $\mu$ such that neither the Weyl node at $\Gamma$ nor these additional nodes provide a FS sheet. The charge bookkeeping below is then that of the four sheets at the Fermi level, for which it closes within the model.

The overlaps between the sheets at $\Gamma$ and at $R$ require the unitary matrix $\mathcal U$ that relates the eigenbasis of $\mathcal H_\Gamma$ to that of $\mathcal H_R$ through the common orbital basis. The squared overlap of the sheets is $|\psi_R^\dagger(\bk')\,\mathcal U\,\psi_\Gamma(\bk)|^2$. The spinors of the RSW node vanish on the two orbitals of the Weyl node, and only the $6\times4$ block $\mathcal Y$ of $\mathcal U$ between the four orbitals of the RSW node and the six orbitals at $R$ enters. The $6\times2$ block that involves the Weyl node at $\Gamma$ gives no contribution to the overlaps and thus not to the conductivity. Unitarity of $\mathcal U$ requires $\mathcal Y^\dagger\,\mathcal Y=\mathbb I_4$. Let $\mathcal Y_1$ and $\mathcal Y_2$ be the $3\times4$ blocks that belong to the copies $R_1$ and $R_2$.

A rotation by an angle $\alpha$ about the $z$ axis is generated by $J_z$ at $\Gamma$ and by the third spin-1 matrix at $R$ in the bases of Appendix~\ref{secappsp2}. Let $j_\Gamma$ and $j_R$ be the eigenvalues of these generators, which label the components of the spinors. The squared overlap depends on $\varphi$ and $\varphi'$ only through $\varphi-\varphi'$ if every nonzero element of $\mathcal Y_i$ connects the components with $j_R-j_\Gamma=d_i$, where $d_i$ has the same value for all the elements of the block (Appendix~\ref{secappsp2}). We take $d_1=-1/2$ and $d_2=+1/2$ for the two time-reversed copies. With the components of $\psi_\Gamma$ ordered by $j_\Gamma=3/2,1/2,-1/2,-3/2$, and those of each copy ordered by $j_R=+1,0,-1$, unitarity then gives, after the phases are absorbed in the spinors,
\begin{align}
\mathcal Y_1=\begin{bmatrix}
1&0&0&0\\0&\zeta&0&0\\0&0&\eta&0
\end{bmatrix},\quad
\mathcal Y_2=\begin{bmatrix}
0&\eta&0&0\\0&0&\zeta&0\\0&0&0&1
\end{bmatrix},\quad
\zeta^2+\eta^2=1\,.
\label{eqyblocks}
\end{align}
We take the same ratio in the two columns that are shared by the copies. The ratio of $\zeta$ and $\eta$ is not fixed by the symmetry and is a parameter of the microscopic model.

The helicity operator $\hat{\bk}\cdot\bs J$ commutes with the RSW block of $\mathcal H_\Gamma$, and its eigenvalues $\tilde\psi\in\{3/2,1/2,-1/2,-3/2\}$ label the four nondegenerate bands of the RSW node. The bands of the RSW node, of the Weyl node, and of each copy at $R$ are
\begin{align}
& \epsilon_{\Gamma,\tilde\psi}(k)=E_\Gamma+v_{3/2}\left(\tilde\psi\,k+c_{3/2}\,k^2\right),\quad
\epsilon_{W,\pm}(k)=E_W+v_{1/2}\left(\pm k+c_{1/2}\,k^2\right),\nn
& \epsilon_{R,s}(k)=E_R+v_R\left(s\,k+c_R\,k^2\right) \text{ for } s\in\{1,0,-1\}\,.
\label{eqeval2}
\end{align}
For $\mu>E_\Gamma$, the chemical potential cuts the bands $\tilde\psi=3/2$ and $\tilde\psi=1/2$ at $\Gamma$. For $\mu>E_R$, it cuts the band $s=1$ of each copy at $R$. We consider these four sheets, $a\in\{(\Gamma,3/2),(\Gamma,1/2),(R_1,+),(R_2,+)\}$, which are continuously connected to the sheets of the linear model. The two sheets at $\Gamma$ are nested, and the sheet $(\Gamma,3/2)$ has the smaller radius. For $c_{3/2}\leq0$ and $c_R\leq0$, the remaining bands of the RSW node and of the nodes at $R$ are not cut. For a positive coefficient, the bands with $\tilde\psi<0$ at $\Gamma$ and with $s\leq0$ at $R$ are also cut at a large radius, and these sheets are not included. At $B=0$, with $\mu_\Gamma=\mu-E_\Gamma$ and $\mu_R=\mu-E_R$, the FS radii and the FS weights $\rho_a=k^2/(\partial_k\epsilon_s)$ are
\begin{align}
& k_{\Gamma,\tilde\psi}=\frac{-\tilde\psi+\sqrt{\tilde\psi^2+4\,c_{3/2}\,\mu_\Gamma/v_{3/2}}}{2\,c_{3/2}}\,,\quad
k_{R}=\frac{-1+\sqrt{1+4\,c_R\,\mu_R/v_R}}{2\,c_R}\,,\nn
& \rho_{\Gamma,\tilde\psi}=\frac{k_{\Gamma,\tilde\psi}^2}{v_{3/2}\sqrt{\tilde\psi^2+4\,c_{3/2}\,\mu_\Gamma/v_{3/2}}}\,,\quad
\rho_{R}=\frac{k_{R}^2}{v_R\sqrt{1+4\,c_R\,\mu_R/v_R}}\,.
\label{eqfsb02}
\end{align}
For $c_{3/2}\to0$ and $c_R\to0$, the radii become $k_{\Gamma,3/2}=2\,\mu_\Gamma/(3\,v_{3/2})$, $k_{\Gamma,1/2}=2\,\mu_\Gamma/v_{3/2}$, and $k_R=\mu_R/v_R$. The parameters of Eq.~\eqref{eqham} correspond to these by $E_\Gamma=0$ and $E_R=-\Delta$ for the offset of the nodes at $R$.

The BC and the OMM of the four sheets have the common form of Eq.~\eqref{eqbcomm}, with $v_\Gamma=v_{3/2}$ and $v_{R_1}=v_{R_2}=v_R$. For a band of helicity $\tilde\psi$ of a node $E+v\,\bk\cdot\bs J$ with the spin $J$, the numbers in that equation are $N=2\,\tilde\psi$ and $G=J(J+1)-\tilde\psi^2$. They reproduce the values of the pseudospin-1 node of Eq.~\eqref{eqbcomm}. For the RSW node, they agree with the BC and the OMM of Ref.~\cite{ips-exact-rsw}, whose OMM is written without the factor $1/2$ of Eq.~\eqref{eqbcomm}, so that its values $7/4$ and $3/4$ correspond to $G_a=7/2$ and $3/2$ here. The quadratic terms are proportional to the identity matrix, and they leave the spinors, the BC, and the OMM unchanged. The values for the four sheets are collected in Table~\ref{tabnodes2}. The quantities $\xi_s$, $D_a$, and $\bw_a$ follow from Eq.~\eqref{eqxiw} with these numbers. The Chern number of the sheet $a$ is $\mathcal C_a=-\chi_\nu\,N_a$, and the four values of the sheets at the Fermi level add up to zero. The magnitudes $3$ and $1$ of the two sheets at $\Gamma$ add up to $4$, and the magnitude $2$ at each copy at $R$ gives the same total.

\begin{table}[t!]
\centering
\renewcommand{\arraystretch}{1.4}
\begin{tabular}{|c|c|c|c|c|}
\hline
Sheet $a$ & $\chi_\nu$ & $N_a$ & $G_a$ & $\mathcal C_a$ \\
\hline
$(\Gamma,3/2)$ & $+1$ & $3$ & $3/2$ & $-3$ \\
\hline
$(\Gamma,1/2)$ & $+1$ & $1$ & $7/2$ & $-1$ \\
\hline
$(R_i,+)$ & $-1$ & $2$ & $1$ & $+2$ \\
\hline
\end{tabular}
\caption{\label{tabnodes2} The numbers $N_a$ and $G_a$ entering the BC and the OMM of Eq.~\eqref{eqbcomm}, and the resulting Chern numbers, for the four FS sheets of the RSW and TSM nodes.}
\end{table}

%%%%%%%%%%%%%%%
\subsection{Fermi surfaces and scattering kernel}
\label{secfsker2}

The scattering amplitudes are defined as in Eq.~\eqref{eqkernel}. The independent amplitudes are $\beta^{\Gamma}_{\tilde\psi\tilde\psi'}$ between the sheets $(\Gamma,\tilde\psi)$ and $(\Gamma,\tilde\psi')$ with $\beta^{\Gamma}_{3/2,1/2}=\beta^{\Gamma}_{1/2,3/2}$, $\beta^{R}_{\rm intra}$ within one copy at $R$, and $\beta^{\Gamma R_i}_{\tilde\psi}$ between $(\Gamma,\tilde\psi)$ and the copy $R_i$. The squared overlaps follow from the spinors and from the block $\mathcal Y$ of Eq.~\eqref{eqyblocks}, as given in Appendix~\ref{secappsp2}.

The overlaps between two sheets at $\Gamma$ and between two sheets of the same copy at $R$ depend only on $C=\hat{\bk}\cdot\hat{\bk}'$ and are given in Eqs.~\eqref{eqovrsw} and \eqref{eqovtsm}. The overlap between the two copies vanishes, because the spinors of the two copies have no common nonzero component. The copies are therefore coupled only through the sheets at $\Gamma$. The overlaps between $\Gamma$ and $R$ of Eq.~\eqref{eqovgr} depend on $\varphi-\varphi'$ only, and they have the azimuthal decomposition of Eq.~\eqref{eqoverlapm} with the harmonics $0$, $1$, and $2$. The overlaps with $R_1$ and $R_2$ differ [cf. Eq.~\eqref{eqovgr2}], and the exchange of the two copies alone is therefore not a symmetry of the kernel. It is a symmetry only when it is combined with a rotation or with time reversal (Sec.~\ref{secclass2} and Appendix~\ref{secappsym2}).

The two copies at $R$ have the same dispersion, and their bands are therefore degenerate at every $\bk$. The coupling between the copies is dropped, and the copy basis is independent of $\bk$. The BC, the OMM, and the phase-space factor of the degenerate band are those of each copy times the identity in the copy space. The calculation below treats the copies as independent sheets that are coupled through the sheets at $\Gamma$. This neglects the coherences between the copies, which the sheets at $\Gamma$ generate and which are kept in the matrix treatment of Appendix~\ref{secappdeg}.

The kernel between two sheets at $\Gamma$ is a polynomial of degree three in $C$, and the range of its in-scattering term is spanned by the spherical harmonics with $l\leq3$. For the sheets at $R$, the kernel is of degree two in $C$, and the range is spanned by the harmonics with $l\leq2$. The overlaps between $\Gamma$ and $R$ are polynomials of the same degrees in the direction at $\Gamma$ and at $R$ (Appendix~\ref{secappkernel}). The kernel for the harmonic $m$ is given by Eq.~\eqref{eqkernel} with $m=0,\pm1,\pm2$ for the overlaps between $\Gamma$ and $R$, and with $|m|\leq3$ for the sheets at $\Gamma$. The kernel vanishes for $|m|\geq4$. For $\bB$ along the polar axis, the LFO is diagonal in $m$, and the sectors decouple.

The overlaps of Eq.~\eqref{eqovgr} are defined in the bases of Eq.~\eqref{eqham2}, which are not tied to the orbitals of a crystal. A model with the orbital content of the space group 198 exists for the spin-orbit-coupled case as well. It is the eight-band tight-binding model of Ref.~\cite{prl119206401}, in which an $s$ orbital with spin sits on each of the four sites of the unit cell. Let $\tilde\sigma_a$ be the Pauli matrices of the spin, and let $V_{r,i}$ and $V_{s,i}$ be the spin-orbit terms of that reference. The Hamiltonian of the model is $\mathcal H_{198}(\bk)=\tilde{\mathcal H}_{198}(\bk)\otimes\tilde\sigma_0+\sum_{i=1}^3\left[V_{r,i}(\bk)+V_{s,i}(\bk)\right]$, where $\tilde{\mathcal H}_{198}$ is the spinless Hamiltonian of Eq.~\eqref{eqtb198}. Its bands near $\Gamma$ consist of a fourfold node, which is the RSW node, and two twofold nodes. Its bands near $R$ consist of a sixfold node, which is the double spin-1 node, and a twofold band \cite{prl119206401}. The orbital embedding enters as in the spinless case, through the matrix $U_\upsilon(\bk)\otimes\tilde\sigma_0$ with $U_\upsilon$ of Eq.~\eqref{equx} \cite{grushin-multifold}. At $\Gamma$, this matrix is the identity. At $R$, it is $e^{-i\pi \upsilon}\,{\rm diag}\left(e^{i\phi},1,1,1\right)\otimes\tilde\sigma_0$ with $\phi=4\,\pi\,\upsilon$. The Chang scenario with $\upsilon=0$ and the Flicker scenario with $\upsilon\neq0$ of the QTSM and Weyl system therefore apply here with the same phase $\phi$. In both scenarios, the overlap between a state $\psi_\Gamma$ at $\Gamma$ and a set of bands at $R$ with the projector $P_R$ is $\langle\psi_\Gamma|\,U_\upsilon^\dagger(\bs\pi)\,P_R\,U_\upsilon(\bs\pi)\,|\psi_\Gamma\rangle$. The matrix $U_\upsilon(\bs\pi)$ changes the relative phase of the orbital $A$, and the two scenarios thus give different overlaps for the same bandstructure. The eight-band model has only the symmetry of the chiral tetrahedral group, which acts jointly on $\bk$ and $\bs q$. Its overlaps between $\Gamma$ and $R$ are therefore not required to depend on $\varphi-\varphi'$ only, in contrast to Eq.~\eqref{eqovgr}.

To the leading order in the spin-orbit coupling, the overlaps of the eight-band model can be written in closed form. Let $\Sigma^\Gamma_a$ and $\Sigma^R_a$ be the matrices defined above Eq.~\eqref{eqsigalg}, extended to the spin space by the identity matrix, and let $|s\rangle=(1,1,1,1)/2$ be the singlet at $\Gamma$ in the orbital space. The projector onto the triplet is $Q=\left(1-|s\rangle\langle s|\right)\otimes\tilde\sigma_0$, and $S_a=Q\,\Sigma^\Gamma_a\,Q$ are the spin-1 matrices of the triplet. We define $\bs J_\Gamma=\bs S+\bs{\tilde\sigma}/2$ and $\bs J_R=\left(\bs\Sigma^R+\bs{\tilde\sigma}\right)/2$. The mass term of the spin-orbit coupling at $R$ is proportional to $\bs\Sigma^R\cdot\bs{\tilde\sigma}$ \cite{prl119206401}, and the sixfold node is the eigenspace of this operator with the eigenvalue $1$. Its projector is $P_6=\left(3+\bs\Sigma^R\cdot\bs{\tilde\sigma}\right)/4$. At $\Gamma$, the fourfold node is the sector with $J_\Gamma=3/2$ of the triplet and the spin, and its projector is $P_4=\left(\bs S\cdot\bs{\tilde\sigma}+2\,Q\right)/3$. The subspaces of the fourfold node and of the sixfold node coincide with these projectors. This identification is not stated in Ref.~\cite{prl119206401}. We built the screws, the threefold rotation, and time reversal in the eight-band representation of that reference, and we found the Hermitian matrices that they allow at each of the two points. These form spaces of dimensions $4$ at $\Gamma$ and $2$ at $R$. For the fitted parameters of the reference and for random members of these spaces, the fourfold subspace at $\Gamma$ is the image of $P_4$, and the sixfold subspace at $R$ is the image of $P_6$. The identification does not depend on the values of the spin-orbit parameters, as long as the representation is that of the eight-band model. Projecting the linear terms of Eq.~\eqref{eqtbexp} onto them, and neglecting the linear terms of the spin-orbit coupling, we obtain $\mathcal H_\Gamma\simeq\left(v_p/3\right)\bk\cdot\bs J_\Gamma$ and $\mathcal H_R\simeq-\left(v_p/2\right)\bs q\cdot\bs J_R$. The first is the RSW Hamiltonian with the velocity $|v_p|/3$, and the second consists of two copies of the spin-1 Hamiltonian with the velocity $|v_p|/2$.

Let $\tilde\psi$ and $\tilde\psi'$ take the values $\pm3/2$ and $\pm1/2$. The projector onto the helicity $\tilde\psi$ at $\Gamma$ is $P_{\tilde\psi}(\hat{\bk})=P_4\prod_{\tilde\psi'\neq \tilde\psi}\left(\hat{\bk}\cdot\bs J_\Gamma-\tilde\psi'\right)/\left(\tilde\psi-\tilde\psi'\right)$. The projector onto the helicity $m=\pm1$ of the pairs of bands at $R$ is $P^R_m(\hat{\bs q})=\left(\hat{\bs q}\cdot\bs J_R\right)\left(\hat{\bs q}\cdot\bs J_R+m\right)/2$. The upper bands of the two nodes have the helicities $s_p\,\tilde\psi$ and $-s_p$, respectively. The overlaps of the sheets $(\Gamma,\tilde\psi)$ with the upper pair at $R$ are then
\begin{align}
\mathcal O_{(\Gamma,\tilde\psi)(R,+)}={\rm Tr}\left[P_{s_p \tilde\psi}(\hat{\bk})\;U_x^\dagger(\bs\pi)\,P^R_{-s_p}(\hat{\bs q})\,U_x(\bs\pi)\right],\quad \tilde\psi=\frac32,\frac12\,,
\label{eqovgr8}
\end{align}
where the trace runs over the eight-dimensional space and $U_\upsilon(\bs\pi)$ is the matrix given above. For $\upsilon=0$, Eq.~\eqref{eqovgr8} is the Chang scenario, and for $\upsilon\neq0$, it is the Flicker scenario. The overlap is a polynomial of the third degree in the components of $\hat{\bk}$ and of the second degree in the components of $\hat{\bs q}$. It depends on the azimuths of $\hat{\bk}$ and $\hat{\bs q}$ separately, since the matrices $\Sigma^R_a$ are the matrices $\Sigma^\Gamma_{a+1}$ with the cyclically shifted index, whereas the matrices $\tilde\sigma_a$ are not shifted.

The trace of Eq.~\eqref{eqovgr8} reduces to a short closed form. Let $\hat k_a$ and $\hat q_a$ be the Cartesian components of $\hat{\bk}$ and $\hat{\bs q}$, with the indices $a$, $a+1$, and $a+2$ running cyclically over $x$, $y$, and $z$. We define
%%%%%%%%%%%%%%%%%
\begin{align}
& c_{\Gamma R}=\hat{\bk}\cdot\hat{\bs q}\,,\quad
\varpi_\pm=\sum_a\hat q_a\,\hat k_{a\pm1}\,,\quad
\varsigma_k=\sum_a\hat k_a\,,\quad
\varsigma_q=\sum_a\hat q_a\,,\quad
\varsigma_{qq}=\sum_a\hat q_a\,\hat q_{a+1}\,,\quad
\varsigma_{qkk}=\sum_a\hat q_a\left(\hat k_a^2-\hat k_{a+1}\,\hat k_{a+2}\right).
\label{eqovgr8defs}
\end{align}
%%%%%%%%%%%%%%%%%%%%%%%%%%
Let $\mathcal O^{\rm orb}_1$ and $\mathcal O^{\rm orb}_0$ be the overlaps $\mathcal O_{(\Gamma,1)(R,+)}$ and $\mathcal O_{(\Gamma,0)(R,+)}$ of Eq.~\eqref{eqovgrx} of the four-orbital model. In the present notation, they read
%%%%%%%%%%%%%%%%
\begin{align}
\mathcal O^{\rm orb}_1&=\frac12\left(1-\varpi_+\right)-\frac{1-\cos\phi}{4}\left(\varpi_--\varpi_+\right)-\frac{s_p\,\sin\phi}{4}\,\varsigma_{qkk}\,,\quad
\mathcal O^{\rm orb}_0 =
\frac12+\frac{s_p\,\sin\phi}{4}\left(2\,\varsigma_{qkk}-\varsigma_q\right).
\label{eqovgr8orb}
\end{align}
The overlaps of Eq.~\eqref{eqovgr8} are then
%%%%%%%%%%%
\begin{align}
\mathcal O_{(\Gamma,3/2)(R,+)}&=\frac12\left(1-c_{\Gamma R}\right)\mathcal O^{\rm orb}_1\,,\quad
\mathcal O_{(\Gamma,1/2)(R,+)} =
\frac16\left(1+c_{\Gamma R}\right)\mathcal O^{\rm orb}_1+\frac13\left(1-c_{\Gamma R}\right)\mathcal O^{\rm orb}_0+\mathcal I_{\Gamma R}\,,\nn
%%%%%%%%%%%%%%
\mathcal I_{\Gamma R}&=\frac16\left(\varsigma_{qq}-c_{\Gamma R}\,\varpi_+\right)-\frac{1-\cos\phi}{12}\,c_{\Gamma R}\left(\varpi_--\varpi_+\right)
+\frac{s_p\,\sin\phi}{6}\left[c_{\Gamma R}\left(\varsigma_q-\varsigma_{qkk}\right)-\varsigma_k\,\varsigma_{qq}\right].
\label{eqovgr8c}
\end{align}
%%%%%%%%%%%%%%%%%%%%%%%%%%%%%%
For the Chang scenario with $\phi=0$, these overlaps become
\begin{align}
\mathcal O_{(\Gamma,3/2)(R,+)}&=\frac14\left(1-c_{\Gamma R}\right)\left(1-\varpi_+\right),\quad
\mathcal O_{(\Gamma,1/2)(R,+)} =
\frac1{12}\left(1+c_{\Gamma R}\right)\left(1-\varpi_+\right)+\frac16\left(1-c_{\Gamma R}\right)+\frac16\left(\varsigma_{qq}-c_{\Gamma R}\,\varpi_+\right).
\label{eqovgr8ch}
\end{align}
We obtained Eq.~\eqref{eqovgr8c} by fitting the trace of Eq.~\eqref{eqovgr8} to polynomials in the components of $\hat{\bk}$ and $\hat{\bs q}$. The factor $\left(1-c_{\Gamma R}\right)/2$ is the squared overlap of the spin states. The state of the helicity $s_p\,3/2$ at $\Gamma$ is the product of the orbital state of the upper band of the triplet and a spin state, and the overlap factorises accordingly. The state of the helicity $s_p/2$ is a superposition of two such products with the weights $1/3$ and $2/3$, whose spin factors are $\left(1+c_{\Gamma R}\right)/2$ and $\left(1-c_{\Gamma R}\right)/2$. This gives the first two terms of the overlap, and $\mathcal I_{\Gamma R}$ is the interference between the two products. The overlap is of the third degree in the components of $\hat{\bk}$ and of the second degree in those of $\hat{\bs q}$. Its azimuthal harmonics therefore have $|m|\leq3$ in the azimuth of $\hat{\bk}$ and $|m'|\leq2$ in that of $\hat{\bs q}$, for any orientation of the polar axis. They follow from Eq.~\eqref{eqovgr8c} by inserting the polar angles. They do not depend on $\varphi-\varphi'$ only, because the cyclic sums of Eq.~\eqref{eqovgr8defs} contain the components of the two vectors separately.

%%%%%%%%%%%%%%%
\subsection{Solution for the magnetic field along the polar axis}
\label{secsol2}

For $\bB=B\,\boldsymbol{\hat z}$, the equation of each azimuthal sector is Eq.~\eqref{eqsector}, with the four sheets of Eq.~\eqref{eqham2} and the kernel of Eq.~\eqref{eqkernel}. The frequency $\omega_a$ is that of Eq.~\eqref{eqresalg}, evaluated with the quantities of Eq.~\eqref{eqwqd} for the numbers of Table~\ref{tabnodes2}.

Let $\Theta_{lm}(\theta)$ be the normalised associated Legendre functions, $\int_0^\pi d\theta\,\sin\theta\,\Theta_{lm}\,\Theta_{l'm}=\delta_{ll'}$, so that $Y_{lm}=\Theta_{lm}(\theta)\,e^{i\,m\,\varphi}/\sqrt{2\,\pi}$. The in-scattering term of the sheet $a$ in the sector $m$ is then $\sum_{a'}\sum_{l,l'}\Theta_{lm}(\theta)\,K^{(m)}_{aa',ll'}\,M^{(m)}_{a',l'}$ with the moments
\begin{align}
M^{(m)}_{a,l}=\int_0^\pi d\theta\,\sin\theta\;\Theta_{lm}(\theta)\,\rho_a(\theta)\,\Lambda^{(m)}_a(\theta)\,,
\label{eqmom2}
\end{align}
where $l\leq3$ for the sheets at $\Gamma$, $l\leq2$ for the sheets at $R$, and $l\geq|m|$. The matrices $K^{(m)}_{aa'}$ follow from Appendix~\ref{secappkernel}. They are diagonal in $l$ between two sheets of the same node, and they couple $l$ and $l'$ between $\Gamma$ and $R$. The solution of the sector is
\begin{align}
\Lambda^{(m)}_a(\theta)=\frac{h^{(m)}_a(\theta)+\sum_{a'}\sum_{l,l'}\Theta_{lm}(\theta)\,K^{(m)}_{aa',ll'}\,M^{(m)}_{a',l'}}{\tau_a^{-1}(\theta)+i\,m\,\omega_a(\theta)}\,.
\label{eqsol2}
\end{align}
This form is exact, because the in-scattering term lies in the span of $\Theta_{lm}$ with the above values of $l$. Inserting Eq.~\eqref{eqsol2} into Eq.~\eqref{eqmom2} gives the linear system
\begin{align}
\sum_{a'}\sum_{l'}\Big[\delta_{aa'}\,\delta_{pl'}-\sum_{q}G^{(m)}_{a,pq}\,K^{(m)}_{aa',ql'}\Big]M^{(m)}_{a',l'}=s^{(m)}_{a,p}\,,
\label{eqmomsys2}
\end{align}
with
\begin{align}
G^{(m)}_{a,pq}=\int_0^\pi d\theta\,\sin\theta\,\frac{\rho_a\,\Theta_{pm}\,\Theta_{qm}}{\tau_a^{-1}+i\,m\,\omega_a}\,,\quad
s^{(m)}_{a,p}=\int_0^\pi d\theta\,\sin\theta\,\frac{\rho_a\,\Theta_{pm}\,h^{(m)}_a}{\tau_a^{-1}+i\,m\,\omega_a}\,.
\end{align}
For $m=0$ and $\bE\parallel\bB$, the source is $h_a=D_a^{-1}\left[w_{a,z}-\eta_B\,\chi_\nu\,N_a\,e\,B\,w_{a,k}/(2\,k^2)\right]$ as in Eq.~\eqref{eqlongsol}. The system has $2\times4+2\times3=14$ real unknowns, and the number constraint of Eq.~\eqref{eqnumber}, $\sum_aM^{(0)}_{a,0}=0$, removes its null vector. For $m=1$ and $\bE\perp\bB$, the source is $h^{(1)}_a=(q_a/D_a)\sin\theta$ with $q_a$ of Eq.~\eqref{eqwqd}. The system has $2\times3+2\times2=10$ complex unknowns.

The conductivities are given by Eqs.~\eqref{eqsigpar} and \eqref{eqsigperp} with the sums over the four sheets of this section. The Fermi-sea term of Eq.~\eqref{eqsigmw} has only the component along $\bB$, since the FS of every sheet is axisymmetric about $\bB$ [cf. Eq.~\eqref{eqsigah3}]. Its value is
\begin{align}
\sigma_H^{\rm AH}=-\,\frac{\eta_B\,e^2}{8\,\pi^2}\sum_a\chi_\nu\,N_a\int_{-1}^{1}du\,u\,k_a(u)\,,
\label{eqsigah2}
\end{align}
with $u=\cos\theta$ and $k_a(u)$ the radius of the sheet $a$, and the numbers of Table~\ref{tabnodes2}. For $B\neq0$, the radius is the positive root, continued from Eq.~\eqref{eqfsb02}, of the cubic equation
\begin{align}
c\,v\,k^3+\tilde\psi\,v\,k^2-\mu_\Gamma\,k+\frac{\eta_O\,\chi_\nu\,e\,v\,G_a\,B\,u}{2}=0
\end{align}
for the sheets at $\Gamma$ with $c=c_{3/2}$, $v=v_{3/2}$, and the same equation with $\tilde\psi\to1$, $c\to c_R$, $v\to v_R$, $\mu_\Gamma\to\mu_R$ for the sheets at $R$.

%%%%%%%%%%%%%%%
\subsection{Symmetries and classification of the set-ups}
\label{secclass2}

The set-up is classified by symmetry in the same way as in Table~\ref{tabgeom}, except that the nodes are untilted and the geometry is therefore fixed by $\bB$ alone. The intranode overlaps depend on $C$ only, and the overlap between the two copies vanishes. These overlaps are invariant under the rotation of all the momenta. The overlap between the sheets at $\Gamma$ and $R$ is not fixed by the spinors alone, since it depends on the block $\mathcal Y$ of Eq.~\eqref{eqyblocks}. It depends on $\varphi-\varphi'$ only, and it is invariant under (i) the rotations $\mathcal R_z(\delta)$ about the polar axis, (ii) the rotations $C_{2x}$ and $C_{2y}$ combined with the exchange $\mathcal P$ of the copies $R_1$ and $R_2$, and (iii) the inversion of all the momenta combined with $\mathcal P$, which is time reversal. These statements are proved in Appendix~\ref{secappsym2}. The exchange $\mathcal P$ alone, the rotations about axes in the $xy$ plane by generic angles, and the improper elements are not symmetries. Time reversal gives the Onsager relation of Eq.~\eqref{eqons} for every orientation of $\bB$.

By (i), $\bB$ is brought into the $xz$ plane, $\bB=B\,(\sin\beta,0,\cos\beta)$. By (ii), $C_{2x}\,\mathcal P$ maps $\beta\to\pi-\beta$, and $0\leq\beta\leq\pi/2$ is sufficient. The direction of $\bE$ is arbitrary in all the cases. The three distinct set-ups are collected in Table~\ref{tabgeom2}.

\begin{table}[t!]
\centering
\small
\renewcommand{\arraystretch}{1.6}
\begin{tabular}{|c|c|c|c|c|}
\hline
Case & $\bB$ & Relations & \makecell{Independent elements \\ of $\bs\sigma$} & Unknowns \\
\hline
(a) & $B\,\boldsymbol{\hat z}$ & \makecell{$\mathcal R_z(\delta)$, \\ Eq.~\eqref{eqons}} & \makecell{three: \\ $\sigma_\parallel$, $\sigma_\perp$, $\sigma_H$} & \makecell{$14$ and $10$ \\ ($m=0$, $1$)} \\
\hline
(b) & $B\,\boldsymbol{\hat x}$ & \makecell{$C_{2x}\,\mathcal P$, \\ Eq.~\eqref{eqons}} & \makecell{four: $\sigma_{xx}$, $\sigma_{yy}$, $\sigma_{zz}$, \\ $\sigma_{yz}=-\sigma_{zy}$} & \makecell{$25$ for each $j$ \\ ($j=x,y,z$)} \\
\hline
(c) & \makecell{$B\,(\sin\beta,0,\cos\beta)$, \\ $0<\beta<\pi/2$} & \makecell{$C_{2y}\,\mathcal P$, \\ Eq.~\eqref{eqons}} & \makecell{six: $\sigma_{xx}$, $\sigma_{yy}$, $\sigma_{zz}$, \\ $\sigma_{xz}=\sigma_{zx}$, \\ $\sigma_{xy}=-\sigma_{yx}$, $\sigma_{yz}=-\sigma_{zy}$} & \makecell{$50$ for each $j$ \\ ($j=x,y,z$)} \\
\hline
\end{tabular}
\caption{\label{tabgeom2} Independent geometries of the RSW node at $\Gamma$ and the two TSM nodes at $R$. The nodes are untilted, and the geometry is fixed by $\bB$. All directions of $\bE$ follow from the three solutions $j=x,y,z$.}
\end{table}

For $\bB=B\,\boldsymbol{\hat z}$, the rotations $\mathcal R_z(\delta)$ give the form below Eq.~\eqref{eqsigperp}. For case (b), Eq.~\eqref{eqsymg2} with $g=C_{2x}$ removes $\sigma_{xy}$, $\sigma_{xz}$, $\sigma_{yx}$, and $\sigma_{zx}$. The element $C_{2y}$ maps $\bB$ onto $-\bB$, and together with Eq.~\eqref{eqons} it gives $\sigma_{yz}=-\sigma_{zy}$. For case (c), only this last argument applies, and it gives $\sigma_{xy}=-\sigma_{yx}$, $\sigma_{yz}=-\sigma_{zy}$, and $\sigma_{xz}=\sigma_{zx}$ with three independent diagonal elements. The sectors $m=0$ and $m=\pm1$ decouple only in case (a), where the sectors $|m|\geq2$ have no source. The Fermi-sea term of Eq.~\eqref{eqsigah2} has only the component along $\bB$ for every orientation, since the FS of every sheet is axisymmetric about $\bB$ [cf. Eq.~\eqref{eqsigah3}].

%%%%%%%%%%%%%%%
\subsection{Ansatz for the mean free-path}
\label{secansatz2}

The mean free-path of the sheet $a$ is given by Eq.~\eqref{eqsol2}, and no truncation in $l$ is needed. For case (a) of Table~\ref{tabgeom2}, the sectors of Eq.~\eqref{eqsol2} contain $14$ and $10$ unknown moments for $m=0$ and $m=1$, respectively, and the in-scattering term of the sheet $a$ in the azimuthal sector $m$ is spanned by a set $\mathscr F^m_a$ of functions of $\theta$, which are multiplied by $e^{i\,m\,\varphi}$. For the sheets of the RSW and TSM nodes, these sets are
\begin{align}
\mathscr F^m_{(\Gamma,\tilde\psi)}&=\begin{cases}
\{1,\ \cos\theta,\ 3\cos^2\theta-1,\ 5\cos^3\theta-3\cos\theta\}\,, & m=0\,,\\
\{\sin\theta,\ \sin\theta\cos\theta,\ \sin\theta\left(5\cos^2\theta-1\right)\}\,, & m=\pm1\,,\\
\{\sin^2\theta,\ \sin^2\theta\cos\theta\}\,, & m=\pm2\,,\\
\{\sin^3\theta\}\,, & m=\pm3\,,
\end{cases}\notag\\
\mathscr F^m_{(R_i,+)}&=\begin{cases}
\{1,\ \cos\theta,\ 3\cos^2\theta-1\}\,, & m=0\,,\\
\{\sin\theta,\ \sin\theta\cos\theta\}\,, & m=\pm1\,,\\
\{\sin^2\theta\}\,, & m=\pm2\,.
\end{cases}
\label{eqspans2}
\end{align}
The sheets $(R_i,+)$ contain no function in the sector $m=\pm3$.

The numbers of the unknown moments of the sectors are given in Appendix~\ref{secappcases}.

For $\bB$ not along the axis $z$ of the nodes, the sectors do not decouple, and the sources $j=x,y,z$ are solved separately by Eq.~\eqref{eqtiltsol}. The basis of the in-scattering term is the set of real harmonics $f_p(\hat{\bk})$ with $l\leq3$ at $\Gamma$ and with $l\leq2$ at $R$. The matrices $Q^{ab}$ are the matrix elements of the kernel $\mathcal W_{ab}=\left(\beta_{ab}^2/\pi\right)\mathcal O_{ab}$ in the orthonormal real harmonics,
\begin{align}
Q^{ab}_{pp'}=\int d^2\hat{\bk}\int d^2\hat{\bk}'\,f_p(\hat{\bk})\,\mathcal W_{ab}(\hat{\bk},\hat{\bk}')\,f_{p'}(\hat{\bk}')\,.
\end{align}
Between two sheets of the same node, they are diagonal, $Q^{ab}_{lm,l'm'}=\delta_{ll'}\,\delta_{mm'}\,4\,\beta_{ab}^2\,o_l/(2\,l+1)$, with $o_l$ of Eq.~\eqref{eqlegendre}. Between $(\Gamma,\tilde\psi)$ and $R_i$, they follow from the overlap of Eq.~\eqref{eqovgr} and the harmonics $0$, $1$, and $2$ of Eq.~\eqref{eqoverlapm}, and they are nonzero for $|m-m'|\leq2$ with the selection rules of the azimuthal decomposition. The cases of Table~\ref{tabgeom2} differ by the reduction of Eq.~\eqref{eqmomsys} with the relations of Appendix~\ref{secappsym2}.

Case (b), $\bB=B\,\boldsymbol{\hat x}$. The element $C_{2x}$ fixes $\bB$, and Eq.~\eqref{eqmomrel2} holds with $g_{xx}=+1$ and $g_{yy}=g_{zz}=-1$. The ansatz of the sheets at $\Gamma$ then contains only the moments with $g_{jj}\,\varepsilon_p=+1$,
\begin{align}
\Lambda^j_{(\Gamma,\tilde\psi)}=\hat{\mathcal R}_{(\Gamma,\tilde\psi)}\Big[h^j_{(\Gamma,\tilde\psi)}+\sum_{\tilde\psi'}\sum_{p,p'}Q^{(\Gamma,\tilde\psi)(\Gamma,\tilde\psi')}_{pp'}\,M^j_{(\Gamma,\tilde\psi'),p'}\,f_p\Big]
+\hat{\mathcal R}_{(\Gamma,\tilde\psi)}\Big[\sum_{p,p'}\left(Q^{(\Gamma,\tilde\psi)R_1}_{pp'}+g_{jj}\,\varepsilon_{p'}\,Q^{(\Gamma,\tilde\psi)R_2}_{pp'}\right)M^j_{R_1,p'}\,f_p\Big]\,,
\end{align}
and the copy $R_1$ has the same form with the in-scattering from $(\Gamma,\tilde\psi')$ and from itself, whereas $\Lambda^j_{R_2}(\hat{\bk})$ follows from $\Lambda^j_{R_1}$ by Eq.~\eqref{eqsymg2}. The unknowns are the eight moments of each sheet at $\Gamma$ with $g_{jj}\,\varepsilon_p=+1$ and the nine moments of $R_1$, which gives $25$ for each $j$ (Table~\ref{tabunknowns2b}). Only $j=x$ has the null vector.

Case (c), $\bB=B\,(\sin\beta,0,\cos\beta)$ with $0<\beta<\pi/2$. No proper rotation fixes $\bB$, and the ansatz of Eq.~\eqref{eqtiltsol} is used with all the moments of the four sheets, $2\times16+2\times9=50$ for each $j$. The element $C_{2y}$ relates the solution at $\bB$ to that at $-\bB$, and it enters only in the form of the tensor. The null vector is removed by the number constraint for each $j$.

For $\beta=0$, the ansatz of Eq.~\eqref{eqtiltsol} reproduces the sector form of Eq.~\eqref{eqsol2}.

%%%%%%%%%%%%%%%
\subsection{Limit of the relaxation-time approximation}
\label{secrta2}

For $c_{3/2}=c_R=0$, all the sheets have linear dispersions, and the RTA with the constant time $\tau$ gives the contribution of each sheet by Eq.~\eqref{eqrtagen}. With $\mu_\Gamma=\mu-E_\Gamma$ and $\mu_R=\mu-E_R$, the sheets of the RSW node have $\mathsf u_a=\tilde\psi\,v_{3/2}$, $v_\nu=v_{3/2}$, and $\left(N_a,G_a\right)$ of Table~\ref{tabnodes2}, and the contributions are
\begin{align}
& \sigma_{\parallel,(\Gamma,3/2)}^{\rm RTA}=\frac{e^2\,\tau\,\mu_\Gamma^2}{9\,\pi^2\,v_{3/2}}+\frac{e^4\,\tau\,v_{3/2}^3\,B^2}{\pi^2\,\mu_\Gamma^2}\left(\frac{81\,\eta_B}{40}-\frac{243\,\eta_B\,\eta_O}{320}+\frac{9\,\eta_O}{64}\right),\nn
& \sigma_{\parallel,(\Gamma,1/2)}^{\rm RTA}=\frac{e^2\,\tau\,\mu_\Gamma^2}{3\,\pi^2\,v_{3/2}}+\frac{e^4\,\tau\,v_{3/2}^3\,B^2}{\pi^2\,\mu_\Gamma^2}\left(\frac{\eta_B}{120}-\frac{21\,\eta_B\,\eta_O}{320}+\frac{49\,\eta_O}{192}\right),\nn
& \sigma_{\parallel,(R_i,+)}^{\rm RTA}=\frac{e^2\,\tau\,\mu_R^2}{6\,\pi^2\,v_R}+\frac{e^4\,\tau\,v_R^3\,B^2}{\pi^2\,\mu_R^2}\left(\frac{4\,\eta_B}{15}-\frac{3\,\eta_B\,\eta_O}{20}+\frac{\eta_O}{24}\right).
\label{eqrtarsw}
\end{align}
The total is the sum over the two sheets at $\Gamma$ and the two copies at $R$. In the RTA, the overlaps between $\Gamma$ and $R$ do not enter, and the dependence on the orientation of $\bB$ relative to the axes of the spinor basis is absent.

For nonzero $c_{3/2}$ and $c_R$, in the RTA, the in-scattering term is dropped, and every sheet relaxes with the constant time $\tau$. For $\bE\parallel\bB$, Eq.~\eqref{eqsigpar} reduces to $\sigma_{\parallel,a}=\left(e^2\tau/4\pi^2\right)\int_0^\pi d\theta\,\sin\theta\,\rho_a\,h_a^2$. The sheets at $\Gamma$ and at $R$ have the quadratic dispersion $\epsilon_a=\mathsf u_a\,k+\mathsf w_a\,k^2$ measured from the node, with $\mathsf u_a=\tilde\psi\,v_{3/2}$ and $\mathsf w_a=c_{3/2}\,v_{3/2}$ at $\Gamma$, and with $\mathsf u_a=v_R$ and $\mathsf w_a=c_R\,v_R$ at $R$. Let $v_\nu$ be the velocity in the OMM, which is $v_{3/2}$ at $\Gamma$ and $v_R$ at $R$. With $\mu_a$ the chemical potential measured from the node, let $\bar v_a=\sqrt{\mathsf u_a^2+4\,\mathsf w_a\,\mu_a}$ be the group velocity at the FS for $B=0$, so that $k_a=\left(\bar v_a-\mathsf u_a\right)/(2\,\mathsf w_a)$. The expansion of $\rho_a\,h_a^2$ in $B$ with the quantities of Eq.~\eqref{eqwqd}, in which $\partial_k\epsilon_a=\mathsf u_a+2\,\mathsf w_a\,k$, followed by the integration over $\cos\theta$, gives
\begin{align}
\sigma_{\parallel,a}^{\rm RTA}=\frac{e^2\,\tau\,k_a^2\,\bar v_a}{6\,\pi^2}+\frac{e^4\,\tau\,B^2}{120\,\pi^2\,k_a^2}\Big[8\,\eta_B\,N_a^2\,\bar v_a-\eta_B\,\eta_O\,N_a\,G_a\,v_\nu\,\frac{7\,\mathsf u_a+2\,\bar v_a}{\bar v_a}+\eta_O\,G_a^2\,v_\nu^2\,\frac{7\,\bar v_a^2+6\,\mathsf u_a\,\bar v_a-3\,\mathsf u_a^2}{2\,\bar v_a^3}\Big]+\mathcal O\!\left(B^4\right).
\label{eqrtaquad}
\end{align}
The term linear in $B$ vanishes. For $\mathsf w_a\to0$, we have $\bar v_a=\mathsf u_a$ and $k_a=\mu_a/\mathsf u_a$, and Eq.~\eqref{eqrtaquad} reduces to Eq.~\eqref{eqrtagen}. The term proportional to $\eta_B$ is the contribution of the BC, that proportional to $\eta_O$ is the contribution of the OMM, and the product is their cross term. The total is the sum over the two sheets at $\Gamma$ and the two copies at $R$, with the numbers $N_a$ and $G_a$ of Table~\ref{tabnodes2}. The RTA does not contain the scattering between the sheets. The overlaps between $\Gamma$ and $R$ do not enter, and the dependence on $\mathcal Y$ is absent. 
For noncollinear fields, the RTA drops the in-scattering term of Eq.~\eqref{eqell} and keeps the LFO. For $\bE\perp\bB$, the sources $h^i_a=D_a^{-1}\,w_{a,i}$ with $i=x,y$ carry no pumping term, since $B_x=B_y=0$. The mean free-path of every sheet is then $\ell_a=\left(q_a/D_a\right)\sin\theta\,\tau/\left(1+i\,\omega_a\,\tau\right)$, and Eq.~\eqref{eqsigperp} gives
\begin{align}
\sigma_\perp-i\,\sigma_H^{\rm B}=\frac{e^2\,\tau}{8\,\pi^2}\sum_a\int_0^\pi d\theta\,\sin^3\theta\,\rho_a\,\frac{\left(q_a/D_a\right)^2}{1+i\,\omega_a\,\tau}\,.
\label{eqrtaperp2}
\end{align}
The sum runs over the two sheets at $\Gamma$ and the two copies at $R$. At $B=0$, each sheet contributes $e^2\,\tau\,k_a^2\,\bar v_a/\left(6\,\pi^2\right)$, which is its value of $\sigma_\parallel$ in Eq.~\eqref{eqrtaquad}. No planar-Hall current flows at $B=0$. For $\eta_B=\eta_O=0$, the integrand is independent of $\theta$ and the result is exact in $B$,
\begin{align}
\sigma_\perp-i\,\sigma_H=\frac{e^2\,\tau}{6\,\pi^2}\sum_a\frac{k_a^2\,\bar v_a}{1+i\,\omega_a\,\tau}\,,\quad
\omega_a=\frac{e\,B\,\bar v_a}{k_a}\,.
\label{eqrtalfo2}
\end{align}
To expand in $B$ with the BC and the OMM, we use the sheet dispersion $\epsilon_a=\mathsf u_a\,k+\mathsf w_a\,k^2$ and the quantities $\bar v_a$, $k_a$, $N_a$, $G_a$, and $v_\nu$ of Eq.~\eqref{eqrtaquad}. Expanding the integrand of Eq.~\eqref{eqrtaperp2} to the third order in $B$ [cf. Appendix~\ref{secapprta3}], we obtain for one sheet
\begin{align}
\sigma^{\rm RTA}_{\perp,a}=\;&\frac{e^2\,\tau\,k_a^2\,\bar v_a}{6\,\pi^2}+\frac{e^4\,\tau\,B^2}{120\,\pi^2\,k_a^2}\Big[\eta_B\,N_a^2\,\bar v_a-\eta_B\,\eta_O\,N_a\,G_a\,v_\nu\,\frac{4\,\bar v_a-\mathsf u_a}{\bar v_a}-\eta_O\,G_a^2\,v_\nu^2\,\frac{\left(\bar v_a-\mathsf u_a\right)^2}{2\,\bar v_a^3}-20\,\tau^2\,k_a^2\,\bar v_a^3\Big]+\mathcal O\!\left(B^4\right),
\label{eqrtaexp2}
\end{align}
and for the Boltzmann part of the Hall conductivity
\begin{align}
\sigma^{\rm B,RTA}_{H,a}=\,&\frac{e^3\,\tau^2\,k_a\,\bar v_a^2}{6\,\pi^2}\,B-\frac{e^5\,\tau^2\,B^3}{240\,\pi^2\,k_a^3\,\bar v_a}\Big[40\,\tau^2\,k_a^2\,\bar v_a^5-6\,\eta_B\,N_a^2\,\bar v_a^3+8\,\eta_B\,\eta_O\,N_a\,G_a\,v_\nu\,\bar v_a\left(3\,\bar v_a-\mathsf u_a\right)-\eta_O\,G_a^2\,v_\nu^2\left(11\,\bar v_a-9\,\mathsf u_a\right)\Big]\nn
&+\mathcal O\!\left(B^5\right).
\label{eqrtaHexp2}
\end{align}
The AH part of Eq.~\eqref{eqsigmw} does not depend on $\tau$, and its expansion is
\begin{align}
\sigma^{\rm AH}_{H,a}=\eta_B\,\eta_O\,N_a\,G_a\,v_\nu\left[\frac{e^3\,B}{24\,\pi^2\,k_a\,\bar v_a}+\frac{e^5\,G_a^2\,v_\nu^2\left(8\,\bar v_a^2-5\,\mathsf u_a\,\bar v_a+\mathsf u_a^2\right)B^3}{320\,\pi^2\,k_a^5\,\bar v_a^5}\right]+\mathcal O\!\left(B^5\right).
\label{eqrtaAH2}
\end{align}
The difference of the transverse and longitudinal conductivities of one sheet, which governs the planar-Hall current of Eq.~\eqref{eqcurr}, follows from Eqs.~\eqref{eqrtaquad} and \eqref{eqrtaexp2} as
\begin{align}
\sigma^{\rm RTA}_{\perp,a}-\sigma^{\rm RTA}_{\parallel,a}=\,\frac{e^4\,\tau\,B^2}{120\,\pi^2\,k_a^2}\Big[-7\,\eta_B\,N_a^2\,\bar v_a-2\,\eta_B\,\eta_O\,N_a\,G_a\,v_\nu\,\frac{\bar v_a-4\,\mathsf u_a}{\bar v_a}-\eta_O\,G_a^2\,v_\nu^2\,\frac{4\,\bar v_a^2+2\,\mathsf u_a\,\bar v_a-\mathsf u_a^2}{\bar v_a^3}-20\,\tau^2\,k_a^2\,\bar v_a^3\Big]+\mathcal O\!\left(B^4\right).
\label{eqrtaph2}
\end{align}
The terms linear in $B$ of $\sigma^{\rm B,RTA}_{H,a}$ do not depend on $\eta_B$ and $\eta_O$, and $\sigma^{\rm RTA}_{\perp,a}$ has no odd orders. The terms with $\tau^2$ inside the brackets are the expansion of Eq.~\eqref{eqrtalfo2}. The totals are the sums over the four sheets with the numbers $N_a$ and $G_a$ of Table~\ref{tabnodes2}, with the copies at $R$ counted twice. For $c_{3/2}=c_R=0$, we have $\bar v_a=\mathsf u_a$, so that the term of Eq.~\eqref{eqrtaexp2} proportional to $\eta_O$ vanishes. With $\mu_\Gamma$ and $\mu_R$ as in Eq.~\eqref{eqrtarsw}, the sheets give
\begin{align}
& \sigma^{\rm RTA}_{\perp,(\Gamma,3/2)}=\frac{e^2\,\tau\,\mu_\Gamma^2}{9\,\pi^2\,v_{3/2}}+\frac{e^4\,\tau\,v_{3/2}^3\,B^2}{\pi^2\,\mu_\Gamma^2}\left(\frac{81\,\eta_B}{320}-\frac{81\,\eta_B\,\eta_O}{320}\right)-\frac{9\,e^4\,\tau^3\,v_{3/2}^3\,B^2}{16\,\pi^2}\,,\nn
& \sigma^{\rm RTA}_{\perp,(\Gamma,1/2)}=\frac{e^2\,\tau\,\mu_\Gamma^2}{3\,\pi^2\,v_{3/2}}+\frac{e^4\,\tau\,v_{3/2}^3\,B^2}{\pi^2\,\mu_\Gamma^2}\left(\frac{\eta_B}{960}-\frac{7\,\eta_B\,\eta_O}{320}\right)-\frac{e^4\,\tau^3\,v_{3/2}^3\,B^2}{48\,\pi^2}\,,\nn
& \sigma^{\rm RTA}_{\perp,(R_i,+)}=\frac{e^2\,\tau\,\mu_R^2}{6\,\pi^2\,v_R}+\frac{e^4\,\tau\,v_R^3\,B^2}{\pi^2\,\mu_R^2}\left(\frac{\eta_B}{30}-\frac{\eta_B\,\eta_O}{20}\right)-\frac{e^4\,\tau^3\,v_R^3\,B^2}{6\,\pi^2}\,,
\label{eqrtaperplin2}
\end{align}
and the differences entering the planar-Hall current are
\begin{align}
& \sigma^{\rm RTA}_{\perp,(\Gamma,3/2)}-\sigma^{\rm RTA}_{\parallel,(\Gamma,3/2)}=\frac{e^4\,\tau\,v_{3/2}^3\,B^2}{\pi^2\,\mu_\Gamma^2}\left(\frac{81\,\eta_B\,\eta_O}{160}-\frac{567\,\eta_B}{320}-\frac{9\,\eta_O}{64}\right)-\frac{9\,e^4\,\tau^3\,v_{3/2}^3\,B^2}{16\,\pi^2}\,,\nn
& \sigma^{\rm RTA}_{\perp,(\Gamma,1/2)}-\sigma^{\rm RTA}_{\parallel,(\Gamma,1/2)}=\frac{e^4\,\tau\,v_{3/2}^3\,B^2}{\pi^2\,\mu_\Gamma^2}\left(\frac{7\,\eta_B\,\eta_O}{160}-\frac{7\,\eta_B}{960}-\frac{49\,\eta_O}{192}\right)-\frac{e^4\,\tau^3\,v_{3/2}^3\,B^2}{48\,\pi^2}\,,\nn
& \sigma^{\rm RTA}_{\perp,(R_i,+)}-\sigma^{\rm RTA}_{\parallel,(R_i,+)}=\frac{e^4\,\tau\,v_R^3\,B^2}{\pi^2\,\mu_R^2}\left(\frac{\eta_B\,\eta_O}{10}-\frac{7\,\eta_B}{30}-\frac{\eta_O}{24}\right)-\frac{e^4\,\tau^3\,v_R^3\,B^2}{6\,\pi^2}\,.
\label{eqrtaphlin2}
\end{align}
The leading terms of the Hall conductivity are $\sigma^{\rm B,RTA}_{H,a}=e^3\,\tau^2\,\mu_a\,\mathsf u_a\,B/\left(6\,\pi^2\right)$ and $\sigma^{\rm AH}_{H,a}=\eta_B\,\eta_O\,e^3\,N_a\,G_a\,v_\nu\,B/\left(24\,\pi^2\,\mu_a\right)$, with $\mu_a=\mu_\Gamma$ for the two sheets at $\Gamma$ and $\mu_a=\mu_R$ for the sheets at $R$. The AH part of the two sheets at $\Gamma$ adds up to $\eta_B\,\eta_O\,e^3\,v_{3/2}\,B/\left(3\,\pi^2\,\mu_\Gamma\right)$, and that of the two copies at $R$ to $\eta_B\,\eta_O\,e^3\,v_R\,B/\left(6\,\pi^2\,\mu_R\right)$. The RTA does not contain the scattering between the sheets, and the overlaps between $\Gamma$ and $R$ do not enter.

The internode scattering between $\Gamma$ and $R$ is included with the relaxation time $\tau_{\rm inter}$ following Ref.~\cite{ips-internode}, in the same way as in Sec.~\ref{secrta}. The RSW node at $\Gamma$ and the double-TSM node at $R$ are the two nodes, and the two copies at $R$ form a single node. The contribution of the node $\nu$ for $\bE\parallel\bB=B\,\boldsymbol{\hat z}$ is given by Eq.~\eqref{eqrtainter6}, and the tensor of the internode current is proportional to $B_i\,B_j$, so that it vanishes for $\bE\perp\bB$. With the numbers of Table~\ref{tabnodes2}, the sums of the sheet vertices of Eq.~\eqref{eqrtab11} over the sheets of each node are
\begin{align}
\Upsilon_\Gamma=-\frac{e\,B}{12\,\pi^2}\,\mathcal E_\Gamma\,,\quad
\Upsilon_R=\frac{e\,B}{12\,\pi^2}\,\mathcal E_R\,,\quad
\mathcal E_\Gamma=12\,\eta_B+\eta_O\left(\frac{3}{2\,S_{3/2}}+\frac{7}{2\,S_{1/2}}\right),\quad
\mathcal E_R=12\,\eta_B+\frac{2\,\eta_O}{S_R}\,,
\label{eqrtavert7}
\end{align}
where $S_{\tilde\psi}=\sqrt{\tilde\psi^2+4\,c_{3/2}\,\mu_\Gamma/v_{3/2}}$ and $S_R=\sqrt{1+4\,c_R\,\mu_R/v_R}$ are the group velocities at the FS in units of $v_{3/2}$ and $v_R$. With $\bar\rho_\Gamma=\rho_{\Gamma,3/2}+\rho_{\Gamma,1/2}$ and $\bar\rho_R=2\,\rho_R$ of Eq.~\eqref{eqfsb02}, the sum over the two nodes is
\begin{align}
\sigma^{\rm inter}_\parallel=\frac{e^4\,B^2}{72\,\pi^2}\sum_{\nu=\Gamma,R}\frac{\mathcal E_\nu^2}{\bar\rho_\nu}\left[\frac{2\,\tau_{\rm inter}\,\bar\rho_{\bar\nu}}{\bar\rho_\Gamma+\bar\rho_R}-\tau\right]+\mathcal O\!\left(B^3\right).
\label{eqrtainter7}
\end{align}
The derivation is given in Appendix~\ref{secrtab8}. The term is of the order $B^2$ and it adds to the intranode expressions of this subsection. It is proportional to $\tau_{\rm inter}-\tau$ for $\bar\rho_\Gamma=\bar\rho_R$ and does not vanish at $\tau_{\rm inter}=\tau$ otherwise, as in Sec.~\ref{secrta}. The OMM part of the vertices is specific to the RTA, and the exact solution contains the OMM through $\rho_a$, $h_a$, and $\omega_a$.

These results are the reference for the solution beyond the RTA of Eq.~\eqref{eqtiltsol}. The derivation of Eqs.~\eqref{eqrtarsw} and \eqref{eqrtaquad} is given in Appendix~\ref{secrtab}. For linear dispersions, the RTA treatment of the RSW node and of the TSM node was carried out in Refs.~\cite{ips-rsw-ph} and \cite{ips-spin1-ph}, respectively, for an arbitrary orientation of $\bB$ relative to $\bE$. Those works give the longitudinal and planar-Hall components up to $\mathcal O\!\left(B^2\right)$, since the odd orders vanish for the isotropic dispersions. They give the intrinsic anomalous-Hall and the Lorentz-force-contributed components up to $\mathcal O\!\left(B^3\right)$. The longitudinal RTA conductivity of the RSW node was also studied in Ref.~\cite{ips-exact-rsw}. Equation~\eqref{eqrtarsw} is therefore the case $\bE\parallel\bB$ of the earlier linear-dispersion results, and the coefficients of $B^2$ of the two sheets at $\Gamma$ agree with the longitudinal conductivity of Ref.~\cite{ips-rsw-ph}. Here, these contributions are collected for the four sheets of the present set-up, with the BC and the OMM tracked separately through $\eta_B$ and $\eta_O$. Equation~\eqref{eqrtaquad} is a general expression for a sheet with the dispersion $\epsilon_a=\mathsf u_a\,k+\mathsf w_a\,k^2$ for $\bE\parallel\bB$. The quadratic expressions of the QTSM node in Ref.~\cite{ips-exact-spin1} follow from it, and it also applies to the sheets of the RSW and TSM nodes with the quadratic corrections $c_{3/2}$ and $c_R$. The term linear in $B$ vanishes and the next correction is of the order $B^4$, so that Eq.~\eqref{eqrtaquad} is complete up to $\mathcal O\!\left(B^3\right)$. For noncollinear fields, Eqs.~\eqref{eqrtaexp2}--\eqref{eqrtaph2} include the quadratic corrections of the sheets, in the same way as Sec.~\ref{secrta3} does for the QTSM nodes. The RTA contribution of the internode scattering between the RSW node at $\Gamma$ and the double-TSM node at $R$ was reported in Ref.~\cite{ips-internode} for linear dispersions, for an arbitrary orientation of $\bB$, and up to $\mathcal O\!\left(B^2\right)$. Eq.~\eqref{eqrtainter7} reproduces it for $c_{3/2}=c_R=0$ and extends it to the quadratic corrections and to the separate tracking of the BC and the OMM. It is not part of Eqs.~\eqref{eqrtarsw} and \eqref{eqrtaquad}, which contain only the intranode scattering.

%%%%%%%%%%%%%%%
\subsection{Results}
\label{secres2}

The analytical results for the RSW node at $\Gamma$ and the two TSM nodes at $R$ are collected here. They concern the limit $B=0$ and the terms that are linear in $B$, for which the sectors $m=0$ and $m=1$ of Eq.~\eqref{eqmomsys2} close in a few unknowns. We abbreviate $k_{\tilde\psi}=k_{\Gamma,\tilde\psi}$ and $\rho_{\tilde\psi}=\rho_{\Gamma,\tilde\psi}$ for $\tilde\psi=3/2,1/2$, while $k_R$ and $\rho_R$ are those of Eq.~\eqref{eqfsb02}. The velocities at the FS for $B=0$ are $w_{\tilde\psi}=v_{3/2}\left(\tilde\psi+2\,c_{3/2}\,k_{\tilde\psi}\right)$ and $w_R=v_R\left(1+2\,c_R\,k_R\right)$, and they satisfy $\rho_a\,w_a=k_a^2$. The scattering amplitudes are $\beta_{\tilde\psi\tilde\psi'}=\beta^{\Gamma}_{\tilde\psi\tilde\psi'}$, $g_{\tilde\psi}=\beta^{\Gamma R_i}_{\tilde\psi}$, and $\beta_R=\beta^R_{\rm intra}$. The amplitude $g_{\tilde\psi}$ is the same for both copies [cf. the discussion below Eq.~\eqref{eqsymP2}]. The two copies are equivalent, and we therefore have $\beta_{R_1R_1}=\beta_{R_2R_2}=\beta_R$. The amplitude between the two copies does not enter, since their overlap vanishes [cf. Eq.~\eqref{eqovtsm}].

\subsubsection{Longitudinal magnetoconductivity for parallel fields}
\label{secresll2}

At $B=0$, the source of the sheet $a$ is $h_a=w_a\cos\theta$ for $\bE\parallel\hat{\boldsymbol z}$, and the weights $\rho_a$ are constants. The out-scattering rates follow from Eq.~\eqref{eqtau}. For two sheets of the same node, only the coefficient $o_0$ of Eq.~\eqref{eqlegendre} contributes, and it gives the term $4\,\beta^2\,o_0\,\rho$ for each pair. For the overlaps between $\Gamma$ and $R$, we integrate over the solid angle of the momentum that is not fixed. The projector $v_+\,v_+^\dagger$ of Eq.~\eqref{eqspintsm} averages to $\left(4\,\pi/3\right)\mathbb I_3$ over the directions at $R$. The projector $\psi_{\Gamma,\tilde\psi}\,\psi_{\Gamma,\tilde\psi}^\dagger$ averages to $\pi\,\mathbb I_4$ over the directions at $\Gamma$, by Schur's lemma. Equation~\eqref{eqyblocks} gives $\mathcal Y_1^\dagger\,\mathcal Y_1+\mathcal Y_2^\dagger\,\mathcal Y_2=\mathbb I_4$ and $\mathcal Y_1\,\mathcal Y_1^\dagger={\rm diag}\left(1,\zeta^2,\eta^2\right)$. The rates of the sheets at $\Gamma$ are therefore independent of $\theta$, whereas the rates of the copies depend on it. They are
\begin{align}
& \tau_{\Gamma,3/2}^{-1}=\beta_{3/2,3/2}^2\,\rho_{3/2}+\beta_{3/2,1/2}^2\,\rho_{1/2}+\frac43\,g_{3/2}^2\,\rho_R\,,\quad
\tau_{\Gamma,1/2}^{-1}=\beta_{3/2,1/2}^2\,\rho_{3/2}+\beta_{1/2,1/2}^2\,\rho_{1/2}+\frac43\,g_{1/2}^2\,\rho_R\,,\nn
& \tau_{R_1}^{-1}(\theta)=\frac43\,\beta_R^2\,\rho_R+\left(g_{3/2}^2\,\rho_{3/2}+g_{1/2}^2\,\rho_{1/2}\right)\left[\frac{\left(1-\cos\theta\right)^2+\eta^2\left(1+\cos\theta\right)^2}{4}+\frac{\zeta^2\,\sin^2\theta}{2}\right],\quad
\tau_{R_2}^{-1}(\theta)=\tau_{R_1}^{-1}(\pi-\theta)\,.
\label{eqtauB02}
\end{align}

At $B=0$, the field does not restrict the proper rotation $g$ in Eq.~\eqref{eqsymg2}. We use $g=C_{2x}$, for which $g_{zz}=-1$. For the component of $\bs\Lambda$ along $\boldsymbol{\hat z}$, this gives $\Lambda_{(\Gamma,\tilde\psi)}(\pi-\theta)=-\Lambda_{(\Gamma,\tilde\psi)}(\theta)$ and $\Lambda_{R_2}(\theta)=-\Lambda_{R_1}(\pi-\theta)$. Let $M_{\tilde\psi,l}=M^{(0)}_{(\Gamma,\tilde\psi),l}$ and $M_{R,l}=M^{(0)}_{(R_1,+),l}$ be the moments of Eq.~\eqref{eqmom2}. The moments of the sheets at $\Gamma$ vanish for even $l$, and the moments of the second copy are $M^{(0)}_{(R_2,+),l}=-\left(-1\right)^l M_{R,l}$. The independent unknowns are thus $M_{\tilde\psi,1}$ and $M_{\tilde\psi,3}$ for $\tilde\psi=3/2,1/2$, together with $M_{R,0}$, $M_{R,1}$, and $M_{R,2}$, which makes seven. The number constraint of Eq.~\eqref{eqnumber} is satisfied identically, since $M_{\tilde\psi,0}=0$ and $M^{(0)}_{(R_1,+),0}+M^{(0)}_{(R_2,+),0}=0$. The solvability condition of Eq.~\eqref{eqsolv} holds because the sources are odd in $\cos\theta$. The system therefore has no null vector at $B=0$.

Let $o^{\tilde\psi}_{ll'}$ be the matrix elements $o^{(0)}_{ll'}$ of Sec.~\ref{secfsker2} for the pair $(\Gamma,\tilde\psi)$ and $R_1$. By Eq.~\eqref{eqsymP1} and $\Theta_{l0}(\pi-\theta)=\left(-1\right)^l\Theta_{l0}(\theta)$, the matrix elements for the pair $(\Gamma,\tilde\psi)$ and $R_2$ are $\left(-1\right)^{l+l'}o^{\tilde\psi}_{ll'}$. For odd $l$, the in-scattering from the two copies into the sheet $(\Gamma,\tilde\psi)$ therefore adds up to $4\,g_{\tilde\psi}^2\sum_{l'}o^{\tilde\psi}_{ll'}\,M_{R,l'}$. The integrals give, for the rows $l=1,3$ and the columns $l'=0,1,2$,
\begin{align}
o^{3/2}_{ll'}=\begin{pmatrix}
\dfrac{\sqrt3\left(1+\zeta^2\right)}{15} & -\dfrac{4-\zeta^2}{20} & \dfrac{\sqrt{15}\left(2-\zeta^2\right)}{300}\\[2ex]
\dfrac{\sqrt7\left(2-3\,\zeta^2\right)}{105} & \dfrac{\sqrt{21}\left(2-3\,\zeta^2\right)}{420} & \dfrac{\sqrt{35}\left(4+3\,\zeta^2\right)}{2100}
\end{pmatrix},\quad
o^{1/2}_{1l'}=\frac13\,o^{3/2}_{1l'}\,,\quad
o^{1/2}_{3l'}=-3\,o^{3/2}_{3l'}\,.
\label{eqovmom2}
\end{align}
For the in-scattering within the node at $\Gamma$, let $o^{(\tilde\psi\tilde\psi')}_l$ be the coefficient $o_l$ of the pair $(\Gamma,\tilde\psi)$ and $(\Gamma,\tilde\psi')$ in Eq.~\eqref{eqlegendre}, and let $o^R_q$ be that of the pair $(R_i,+)$ and $(R_i,+)$. Let $\bar T_p$ and $T_{pq}$ be the integrals
\begin{align}
\bar T_p=\int_0^\pi d\theta\,\sin\theta\,\cos\theta\,\Theta_{p0}(\theta)\,\tau_{R_1}(\theta)\,,\quad
T_{pq}=\int_0^\pi d\theta\,\sin\theta\,\Theta_{p0}(\theta)\,\Theta_{q0}(\theta)\,\tau_{R_1}(\theta)\,.
\label{eqTint2}
\end{align}
Eq.~\eqref{eqsol2} then gives the closed linear system for the seven unknowns. For $l=1,3$ and $\tilde\psi=3/2,1/2$, the equations of the sheets at $\Gamma$ read
\begin{align}
M_{\tilde\psi,l}=\rho_{\tilde\psi}\,\tau_{\Gamma,\tilde\psi}\left[\sqrt{\frac23}\,w_{\tilde\psi}\,\delta_{l1}+\sum_{\tilde\psi'}\frac{4\,\beta_{\tilde\psi\tilde\psi'}^2\,o^{(\tilde\psi\tilde\psi')}_l}{2\,l+1}\,M_{\tilde\psi',l}+4\,g_{\tilde\psi}^2\sum_{l'=0}^{2}o^{\tilde\psi}_{ll'}\,M_{R,l'}\right],
\label{eqsys2G}
\end{align}
and for $p=0,1,2$, the equations of the copy $R_1$ read
\begin{align}
M_{R,p}=\rho_R\left[w_R\,\bar T_p+\sum_{q=0}^{2}T_{pq}\left(\frac{4\,\beta_R^2\,o^R_q}{2\,q+1}\,M_{R,q}+2\sum_{\tilde\psi}g_{\tilde\psi}^2\sum_{l=1,3}o^{\tilde\psi}_{lq}\,M_{\tilde\psi,l}\right)\right].
\label{eqsys2R}
\end{align}
The zero-field conductivity, which is the reference for the magnetoconductivity, follows from Eq.~\eqref{eqsigpar} with $\cos\theta=\sqrt{2/3}\,\Theta_{10}(\theta)$ and $M^{(0)}_{(R_2,+),1}=M_{R,1}$,
\begin{align}
\sigma_\parallel\big\vert_{B=0}=\frac{e^2}{4\,\pi^2}\sqrt{\frac23}\left(w_{3/2}\,M_{3/2,1}+w_{1/2}\,M_{1/2,1}+2\,w_R\,M_{R,1}\right).
\label{eqsigll2}
\end{align}
In contrast to the QTSM and Weyl nodes of Sec.~\ref{secsys}, the mean free-paths are not proportional to $\cos\theta$ for $g_{\tilde\psi}\neq0$. The sheets at $\Gamma$ acquire the component with $l=3$ through the matrix elements $o^{\tilde\psi}_{3l'}$, and the rates of the copies depend on $\theta$ through Eq.~\eqref{eqtauB02}.

\subsubsection{Planar-Hall and Hall response for perpendicular fields and a general angle}
\label{secresph2}

At $B=0$, the sector $m=1$ has the source $h^{(1)}_a=w_a\sin\theta$ and the rates of Eq.~\eqref{eqtauB02}, and it has no frequency $\omega_a$. The rotation $C_{2x}$ has $g_{xx}=+1$. Eq.~\eqref{eqsymg2} therefore gives $\Lambda^{(1)}_{(\Gamma,\tilde\psi)}(\pi-\theta)=\Lambda^{(1)}_{(\Gamma,\tilde\psi)}(\theta)$ and $\Lambda^{(1)}_{R_2}(\theta)=\Lambda^{(1)}_{R_1}(\pi-\theta)$. We fix the sign of the functions $\Theta_{l1}$ by $\Theta_{l1}(\theta)=\sqrt{(2\,l+1)\left(l-1\right)!/\left[2\left(l+1\right)!\right]}\,\sin\theta\,P_l'(\cos\theta)$, which obeys $\Theta_{l1}(\pi-\theta)=\left(-1\right)^{l+1}\Theta_{l1}(\theta)$. Let $\tilde M_{\tilde\psi,l}=M^{(1)}_{(\Gamma,\tilde\psi),l}$ and $\tilde M_{R,l}=M^{(1)}_{(R_1,+),l}$ be the moments of Eq.~\eqref{eqmom2}. The moments of the sheets at $\Gamma$ with even $l$ vanish, and those of the second copy are $M^{(1)}_{(R_2,+),l}=\left(-1\right)^{l+1}\tilde M_{R,l}$. The independent unknowns are $\tilde M_{\tilde\psi,1}$ and $\tilde M_{\tilde\psi,3}$ for $\tilde\psi=3/2,1/2$, together with $\tilde M_{R,1}$ and $\tilde M_{R,2}$. They are six real numbers, in place of the ten complex unknowns at finite $B$.

Let $o^{(1),\tilde\psi}_{ll'}$ be the matrix elements of $O^{(1)}$ between the sheet $(\Gamma,\tilde\psi)$ and $R_1$, defined as the elements $o^{(m)}_{ll'}$ of Sec.~\ref{secfsker2} for $m=1$. By Eq.~\eqref{eqsymP1}, those for the second copy are $\left(-1\right)^{l+l'}o^{(1),\tilde\psi}_{ll'}$. For odd $l$, the in-scattering from the two copies into $(\Gamma,\tilde\psi)$ therefore adds up to $2\,g_{\tilde\psi}^2\sum_{l'}o^{(1),\tilde\psi}_{ll'}\,\tilde M_{R,l'}$. For the rows $l=1,3$ and the columns $l'=1,2$, the integrals give
\begin{align}
o^{(1),3/2}_{ll'}=\begin{pmatrix}
\dfrac{\zeta\left(\sqrt6+2\sqrt2\,\eta\right)}{10} & \dfrac{\zeta\left(2\sqrt{10}\,\eta-\sqrt{30}\right)}{50}\\[2ex]
\dfrac{\zeta\left(\sqrt{21}-3\sqrt7\,\eta\right)}{105} & -\dfrac{\zeta\left(\sqrt{105}+3\sqrt{35}\,\eta\right)}{525}
\end{pmatrix},\quad
o^{(1),1/2}_{1l'}=\frac13\,o^{(1),3/2}_{1l'}\,,\quad
o^{(1),1/2}_{3l'}=-3\,o^{(1),3/2}_{3l'}\,.
\label{eqovmom2perp}
\end{align}
For the in-scattering within a node, the matrices $K^{(1)}$ have the same diagonal elements $4\,\beta^2\,o_l/(2\,l+1)$ as the matrices $K^{(0)}$. Let $\bar T^{(1)}_p$ and $T^{(1)}_{pq}$ be the integrals
\begin{align}
\bar T^{(1)}_p=\int_0^\pi d\theta\,\sin^2\theta\,\Theta_{p1}(\theta)\,\tau_{R_1}(\theta)\,,\quad
T^{(1)}_{pq}=\int_0^\pi d\theta\,\sin\theta\,\Theta_{p1}(\theta)\,\Theta_{q1}(\theta)\,\tau_{R_1}(\theta)\,.
\end{align}
For $l=1,3$ and $\tilde\psi=3/2,1/2$, Eq.~\eqref{eqsol2} gives the equations of the sheets at $\Gamma$ as
\begin{align}
\tilde M_{\tilde\psi,l}=\rho_{\tilde\psi}\,\tau_{\Gamma,\tilde\psi}\left[\frac{2}{\sqrt3}\,w_{\tilde\psi}\,\delta_{l1}+\sum_{\tilde\psi'}\frac{4\,\beta_{\tilde\psi\tilde\psi'}^2\,o^{(\tilde\psi\tilde\psi')}_l}{2\,l+1}\,\tilde M_{\tilde\psi',l}+2\,g_{\tilde\psi}^2\sum_{l'=1}^{2}o^{(1),\tilde\psi}_{ll'}\,\tilde M_{R,l'}\right],
\label{eqsys2Gperp}
\end{align}
and for $p=1,2$, it gives the equations of the copy $R_1$ as
\begin{align}
\tilde M_{R,p}=\rho_R\left[w_R\,\bar T^{(1)}_p+\sum_{q=1}^{2}T^{(1)}_{pq}\left(\frac{4\,\beta_R^2\,o^R_q}{2\,q+1}\,\tilde M_{R,q}+\sum_{\tilde\psi}g_{\tilde\psi}^2\sum_{l=1,3}o^{(1),\tilde\psi}_{lq}\,\tilde M_{\tilde\psi,l}\right)\right].
\label{eqsys2Rperp}
\end{align}
The first integral of Eq.~\eqref{eqsigperp} gives, with $\sin\theta=\left(2/\sqrt3\right)\Theta_{11}(\theta)$ and $M^{(1)}_{(R_2,+),1}=\tilde M_{R,1}$, the zero-field conductivity perpendicular to $\bB$,
\begin{align}
\sigma_\perp\big\vert_{B=0}=\frac{e^2}{4\sqrt3\,\pi^2}\left(w_{3/2}\,\tilde M_{3/2,1}+w_{1/2}\,\tilde M_{1/2,1}+2\,w_R\,\tilde M_{R,1}\right),\quad
\sigma_H\big\vert_{B=0}=0\,.
\label{eqsigperp02}
\end{align}
The Hall conductivity vanishes, because the sector $m=1$ is real at $B=0$. For $g_{3/2}=g_{1/2}=0$, each node is invariant under all rotations. The sector $m=1$ then obeys the same equations as the sector $m=0$, and we have $\sigma_\perp=\sigma_\parallel=\sigma_\Gamma+\sigma_R$ with the expressions of Sec.~\ref{secresbeta2}. For $g_{\tilde\psi}\neq0$, the difference $\sigma_\perp-\sigma_\parallel$ at $B=0$ follows from Eqs.~\eqref{eqsigll2} and \eqref{eqsigperp02}, and Eq.~\eqref{eqcurr} gives the planar-Hall current $j_{\rm PH}=\left(\sigma_\perp-\sigma_\parallel\right)\sin\vartheta\,\cos\vartheta\,E$ at $B=0$ for an oblique $\bE$.

To the leading order in $B$, the Boltzmann part of the Hall conductivity follows from the imaginary part of the sector $m=1$. At this order, the terms that depend on $\eta_B$ and $\eta_O$ do not contribute to it, and only the frequency $\omega_a=e\,B\,w_a/k_a$ acts as a perturbation of the real solution $\ell_a$ at $B=0$. Because the operator of the sector is symmetric with respect to the weight $\rho_a$, we obtain
\begin{align}
\sigma_H^{\rm B}\simeq\frac{e^3\,B}{8\,\pi^2}\sum_a\frac{w_a}{k_a}\int_0^\pi d\theta\,\sin\theta\,\rho_a\,\ell_a^2(\theta)\,,
\label{eqsigHB2}
\end{align}
where $\ell_a$ is the real solution of the sector $m=1$ at $B=0$ and the sum runs over the four sheets. The second copy contributes as much as the first one. The AH part of Eq.~\eqref{eqsigah2} is linear in $B$ as well. The radius of the sheet $a$ changes by $\delta k_a=-\,\eta_O\,\chi_\nu\,e\,v\,G_a\,B\,\cos\theta/\left(2\,k_a\,w_a\right)$ at this order, and only the product of the BC and the OMM contributes,
\begin{align}
\sigma_H^{\rm AH}\simeq\frac{\eta_B\,\eta_O\,e^3\,B}{24\,\pi^2}\left[\frac{9\,v_{3/2}}{2\,k_{3/2}\,w_{3/2}}+\frac{7\,v_{3/2}}{2\,k_{1/2}\,w_{1/2}}+\frac{4\,v_R}{k_R\,w_R}\right].
\label{eqsigHAH2}
\end{align}
It does not depend on the scattering amplitudes. For $c_{3/2}=c_R=0$, the bracket becomes $3/k_{3/2}+7/k_{1/2}+4/k_R$, and the two terms of the sheets at $\Gamma$ add up to $\eta_B\,\eta_O\,e^3\,v_{3/2}\,B/\left(3\,\pi^2\,\mu_\Gamma\right)$ while the two copies give $\eta_B\,\eta_O\,e^3\,v_R\,B/\left(6\,\pi^2\,\mu_R\right)$. These are the values of the RTA [cf. Eq.~\eqref{eqrtaAH2}].

For $g_{3/2}=g_{1/2}=0$, we have $\ell_a=A_a\sin\theta$ with the amplitudes $A_{3/2}$ and $A_{1/2}$ of Sec.~\ref{secresbeta2} for the sheets at $\Gamma$ and $A_R=3\,w_R/\left(2\,\beta_R^2\,\rho_R\right)$ for the copies, and Eq.~\eqref{eqsigHB2} gives
\begin{align}
\sigma_H^{\rm B}\simeq\frac{e^3\,B}{6\,\pi^2}\left(k_{3/2}\,A_{3/2}^2+k_{1/2}\,A_{1/2}^2+2\,k_R\,A_R^2\right).
\label{eqsigHBdec2}
\end{align}
The solution is not a superposition of the results for the components of $\bB$ along and perpendicular to $\bE$, since the quantities of the sheets depend on the full magnitude $B$. The two sectors are therefore solved once at the full $B$, and the angle enters only through the weights in Eq.~\eqref{eqcurr}.

\subsubsection{Expansion to the third order in $B$}
\label{secresB32}

The expansion of Sec.~\ref{secresB33} applies to the sheets of this section. Equations~\eqref{eqexp3}, \eqref{eqkser}, \eqref{eqforms}, \eqref{eqlam1}, \eqref{eqSn}, and \eqref{eqsigcoef} hold with the sums over the four sheets $(\Gamma,3/2)$, $(\Gamma,1/2)$, $R_1$, and $R_2$, and with the kernel $\mathcal K^{(m)}_{aa'}(\theta,\theta')=\sum_{l,l'}\Theta_{lm}(\theta)\,K^{(m)}_{aa',ll'}\,\Theta_{l'm}(\theta')$ of Eq.~\eqref{eqsol2}, which obeys $\mathcal K^{(m)}_{aa'}(\theta,\theta')=\mathcal K^{(m)}_{a'a}(\theta',\theta)$. Let $\tilde\lambda_a=\tilde\psi$ for the sheet $(\Gamma,\tilde\psi)$ and $\tilde\lambda_{R_i}=1$. Let $v_a$ and $c_a$ be $v_{3/2}$ and $c_{3/2}$ for the sheets at $\Gamma$, and $v_R$ and $c_R$ for the sheets at $R$. We define $\tilde\nu_a=\tilde\lambda_a+2\,c_a\,k_a$, such that $w_a=v_a\,\tilde\nu_a$. The numbers $N_a$ and $G_a$ are those of Table~\ref{tabnodes2}. The radius of each sheet is the positive root of the cubic equation below Eq.~\eqref{eqsigah2}, and its expansion is Eq.~\eqref{eqkser} with these parameters. The first-order coefficients of the weights, of the sources, and of the frequency are those of Eq.~\eqref{eqing1}, and the coefficients of the higher orders are polynomials in $\cos\theta$ of the same degrees at most. The sum of Eq.~\eqref{eqsigHAH3} over these four sheets gives the AH part to the third order. Its first term is Eq.~\eqref{eqsigHAH2}. For $c_{3/2}=c_R=0$, the second term becomes $\eta_B\,\eta_O^3\,e^5\,B^3\left(6/k_{3/2}^5+686/k_{1/2}^5+8/k_R^5\right)/\left(160\,\pi^2\right)$. Neither term depends on the scattering amplitudes. The coefficient $\sigma_H^{(1)}$ of Eq.~\eqref{eqsigcoef} reproduces Eq.~\eqref{eqsigHB2}.

The solution $\Lambda_0$ at $B=0$ is that of Secs.~\ref{secresll2} and \ref{secresph2}. At $B=0$, the rotation $C_{2x}$ exchanges the two copies at $R$ and maps each sheet at $\Gamma$ onto itself, and the parities of $\Lambda_1$ follow from Eq.~\eqref{eqsymg2}. For $m=0$, the function $\Lambda_1$ has the opposite parity to $\Lambda_0$, namely $\Lambda_{1,(\Gamma,\tilde\psi)}(\pi-\theta)=\Lambda_{1,(\Gamma,\tilde\psi)}(\theta)$ and $\Lambda_{1,R_2}(\theta)=\Lambda_{1,R_1}(\pi-\theta)$. It contains the zero mode, and the constant drops out of Eq.~\eqref{eqSn} for the reason given in Sec.~\ref{secresB3}. For $m=1$, we write $\Lambda_1=\Lambda_1^{\rm R}+i\,\Lambda_1^{\rm I}$. The imaginary part $\Lambda_1^{\rm I}$ has the symmetry of $\Lambda_0$, and the real part $\Lambda_1^{\rm R}$ has the opposite one, with $\Lambda^{\rm R}_{1,(\Gamma,\tilde\psi)}(\pi-\theta)=-\Lambda^{\rm R}_{1,(\Gamma,\tilde\psi)}(\theta)$ and $\Lambda^{\rm R}_{1,R_2}(\theta)=-\Lambda^{\rm R}_{1,R_1}(\pi-\theta)$.

\subsubsection{Dependence on the scattering amplitudes}
\label{secresbeta2}

For $g_{3/2}=g_{1/2}=0$, the sheets at $\Gamma$ and the copies at $R$ decouple. The kernel of each node then preserves $l$, and the mean free-paths are proportional to $\cos\theta$. For the sheets at $\Gamma$, we write $\Lambda_{(\Gamma,\tilde\psi)}=A_{\tilde\psi}\cos\theta$. The rates of Eq.~\eqref{eqtauB02} lose their terms with $g_{\tilde\psi}$, and the amplitudes obey a $2\times2$ system with the entries
\begin{align}
& N_{3/2,3/2}=\frac25\,\beta_{3/2,3/2}^2\,\rho_{3/2}+\beta_{3/2,1/2}^2\,\rho_{1/2}\,,\quad
N_{1/2,1/2}=\beta_{3/2,1/2}^2\,\rho_{3/2}+\frac{14}{15}\,\beta_{1/2,1/2}^2\,\rho_{1/2}\,,\nn
& N_{3/2,1/2}=-\frac15\,\beta_{3/2,1/2}^2\,\rho_{1/2}\,,\quad
N_{1/2,3/2}=-\frac15\,\beta_{3/2,1/2}^2\,\rho_{3/2}\,.
\end{align}
With $\mathcal N=N_{3/2,3/2}\,N_{1/2,1/2}-N_{3/2,1/2}\,N_{1/2,3/2}$, the amplitudes are
\begin{align}
A_{3/2}=\frac{N_{1/2,1/2}\,w_{3/2}-N_{3/2,1/2}\,w_{1/2}}{\mathcal N}\,,\quad
A_{1/2}=\frac{N_{3/2,3/2}\,w_{1/2}-N_{1/2,3/2}\,w_{3/2}}{\mathcal N}\,,
\end{align}
and the sheets at $\Gamma$ give
\begin{align}
\sigma_\Gamma=\frac{e^2}{6\,\pi^2}\left(k_{3/2}^2\,A_{3/2}+k_{1/2}^2\,A_{1/2}\right).
\end{align}
For each copy at $R$, the rate is $\tau_{R_i}^{-1}=\left(4/3\right)\beta_R^2\,\rho_R$ and the vertex correction is $\left(2/3\right)\beta_R^2\,\rho_R$. The amplitude is therefore $3\,w_R/\left(2\,\beta_R^2\,\rho_R\right)$, and the two copies together give
\begin{align}
\sigma_R=\frac{e^2\,w_R^2}{2\,\pi^2\,\beta_R^2}\,.
\end{align}
For $c_R=0$, we have $w_R=v_R$, and $\sigma_R$ does not depend on $\mu_R$ and on the energy of the nodes at $R$. The zero-field conductivity is continuous at $g_{3/2}=g_{1/2}=0$, and it is $\sigma_\parallel\vert_{B=0}=\sigma_\Gamma+\sigma_R$ in this limit. The limit does not extend to $B\neq0$. The Chern numbers of Table~\ref{tabnodes2} add up to $-4$ for the sheets at $\Gamma$ and to $+4$ for the copies at $R$, and Eq.~\eqref{eqsolv} is therefore violated separately for each set of sheets if $\bE\parallel\bB$.

The amplitudes $\beta_{3/2,3/2}$, $\beta_{3/2,1/2}$, and $\beta_{1/2,1/2}$ enter Eq.~\eqref{eqtauB02} only through the rates of the sheets at $\Gamma$ and through the in-scattering of Eq.~\eqref{eqsys2G}. The amplitude $\beta_R$ enters the rate of the copies and their in-scattering, and the amplitude $\beta^R_{\rm inter}$ does not appear at all. The ratio $\zeta/\eta$ of the block $\mathcal Y$ enters only through the rates of the copies and through Eq.~\eqref{eqovmom2}.

\subsubsection{Numerical results}
\label{secresnum2}

The figures show the exact solution of the sectors $m=0$ and $m=1$ of Eq.~\eqref{eqmomsys2} for $\bB=B\,\boldsymbol{\hat z}$, which is case (a) of Table~\ref{tabgeom2}. The field enters through $\alpha=e\,B/(2\,k_{\Gamma,3/2}^2)$, where $k_{\Gamma,3/2}$ is the radius of the smallest sheet at $B=0$, and both signs of $\alpha$ are shown. The parameters are taken from the fitted tight-binding model of RhSi that is described in Sec.~\ref{secresnum}. To the leading order in the spin-orbit coupling, the velocities are $v_{3/2}=|v_p|/3$ and $v_R=|v_p|/2$ [cf. Sec.~\ref{secfsker2}], and the small spin-orbit terms of the fit are neglected. The second-neighbour hopping gives $c_{3/2}\,v_{3/2}=-v_2/2$ and $c_R\,v_R=v_2/2$. The offset of the nodes is $E_\Gamma-E_R=\Delta=0.41$ eV, and we set $\mu_\Gamma=0.01$ eV and $\mu_R=\mu_\Gamma+\Delta$. The bands with $\tilde\psi<0$ at $\Gamma$ are not cut for $c_{3/2}<0$. For $c_R>0$, the bands with $s\leq0$ at $R$ are cut at a large radius, and these sheets are not included, as stated below Eq.~\eqref{eqeval2}. We take $\mu_\Gamma\,\tau_0=15$ with $\tau_0$ the out-scattering time of the sheet $(\Gamma,3/2)$ at $B=0$, and $\zeta=\eta=1/\sqrt2$ in Eq.~\eqref{eqyblocks}. One amplitude $\beta$ is used for all pairs of sheets within the RSW node and within each copy at $R$, and the amplitude $r_\beta\,\beta$ is used for all pairs of sheets $(\Gamma,\tilde\psi)R_i$. The copies at $R$ are coupled only through the sheets at $\Gamma$. We set $r_\beta=1/2$ unless stated otherwise. All the quantities are normalised to the corresponding values at $B=0$, and the numerical implementation is described in Appendix~\ref{secnum}.

\begin{figure}[t!]
\centering
\subfigure[\label{figrt}]{\includegraphics[width=0.85\textwidth]{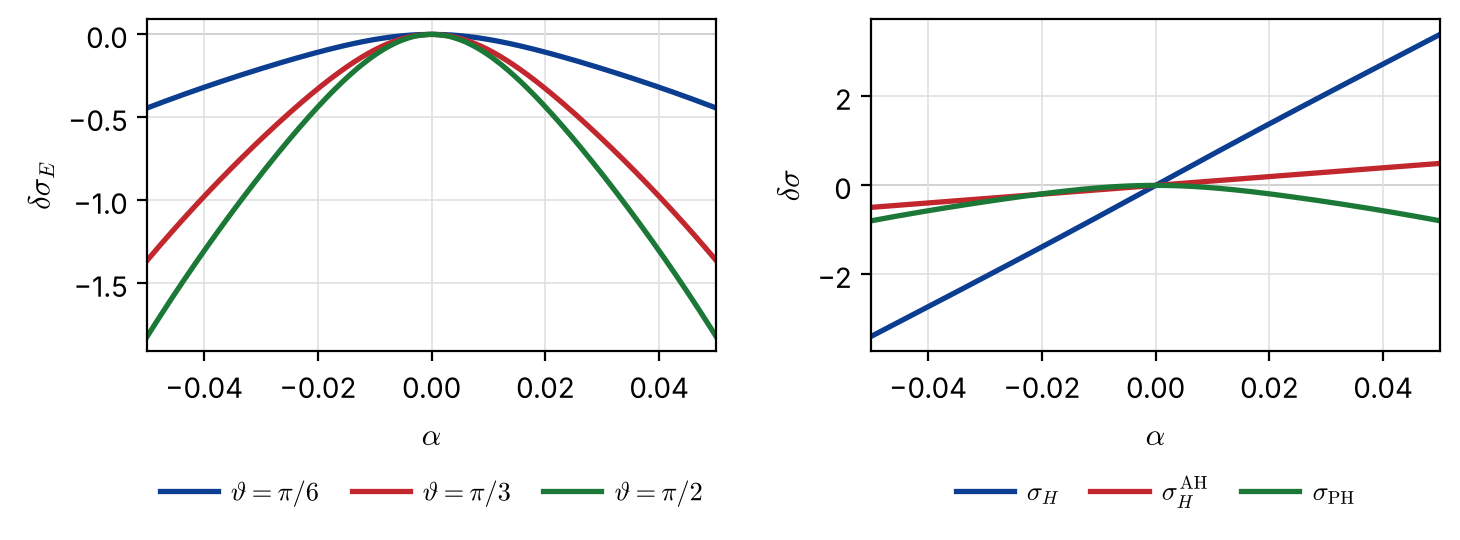}}\\
\subfigure[\label{figrtratio}]{\includegraphics[width=0.85\textwidth]{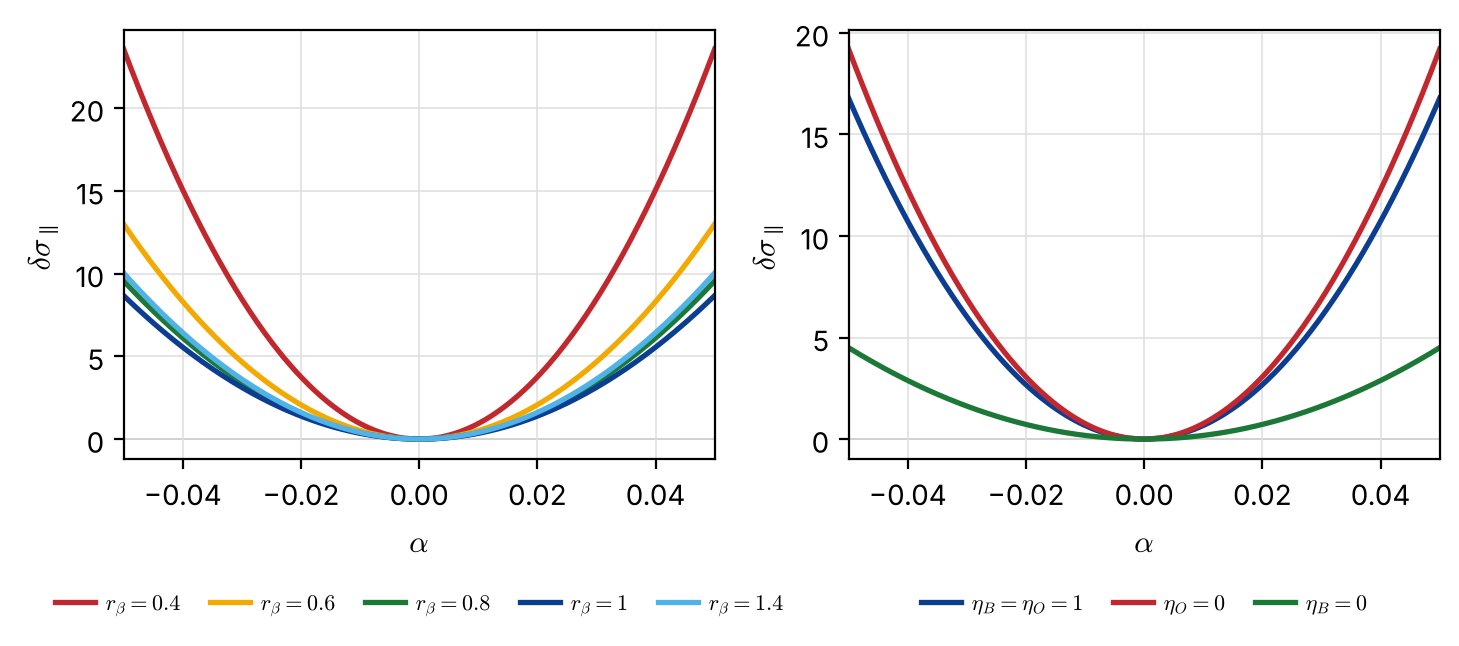}}
\caption{RSW and TSM nodes for $\bB\parallel\hat{\boldsymbol z}$. (a) Left: $\delta\sigma_{E}$ multiplied by $10^3$ for $\bE$ at the angle $\vartheta=\pi/6$ (navy), $\pi/3$ (crimson), and $\pi/2$ (green) to $\bB$. Right: $\sigma_H$ multiplied by $10^2$ (navy), $\sigma_H^{\rm AH}$ multiplied by $10^5$ (crimson), and the planar-Hall change for $\vartheta=\pi/3$ multiplied by $10^3$ (green), in units of $\sigma_\perp$ at $B=0$. (b) For $\bE\parallel\bB$, $\delta\sigma_\parallel$ is multiplied by $10^6$. Left: $r_\beta=0.4$ (crimson), $0.6$ (amber), $0.8$ (green), $1$ (navy), and $1.4$ (sky). Right: $r_\beta=1/2$ with $\eta_B=\eta_O=1$ (navy), $\eta_O=0$ (crimson), and $\eta_B=0$ (green).}
\label{figrtbz}
\end{figure}

By Eq.~\eqref{eqons} and the axisymmetric form of the tensor, $\sigma_\parallel$ and $\sigma_\perp$ are even and $\sigma_H$ is odd in $\alpha$, and the curves of Fig.~\ref{figrt} display this symmetry. The diagonal response of this set-up has therefore no term odd in $B$. For these parameters, $\sigma_\perp$ and $\sigma_\parallel$ differ at $B=0$, since the overlap between $\Gamma$ and $R$ is invariant only under the rotations about the polar axis. A planar-Hall current thus flows at $B=0$ for an oblique $\bE$ [cf. Eq.~\eqref{eqcurr}], and the right panel shows its change relative to the value at $B=0$. The sheets at $R$ carry almost the whole of $\sigma_0$, as in Sec.~\ref{secresnum}, and the changes in units of $\sigma_\perp$ at $B=0$ are therefore small. The change of the current along $\bE$ is negative and grows in magnitude as $\vartheta$ increases. The Hall response is dominated by the Boltzmann part, the AH part is much smaller, and the Hall response increases monotonically with $\alpha$ in the range shown. The change of the planar-Hall current is negative for both signs of $\alpha$.

For $\bE\parallel\bB$, the change of $\sigma_\parallel$ is positive and even in $\alpha$ for all the ratios in Fig.~\ref{figrtratio}. It is not monotonic in $r_\beta$, and the smallest change among the values shown occurs for $r_\beta=1$. No change of sign occurs in this range. The right panel separates the contributions of the BC and the OMM to the longitudinal change. Both are positive, and the combined change is smaller than the BC contribution alone, which shows that the cross term is negative. The OMM contribution alone is the smallest of the three.

\begin{figure}[t!]
\centering
\includegraphics[width=\textwidth]{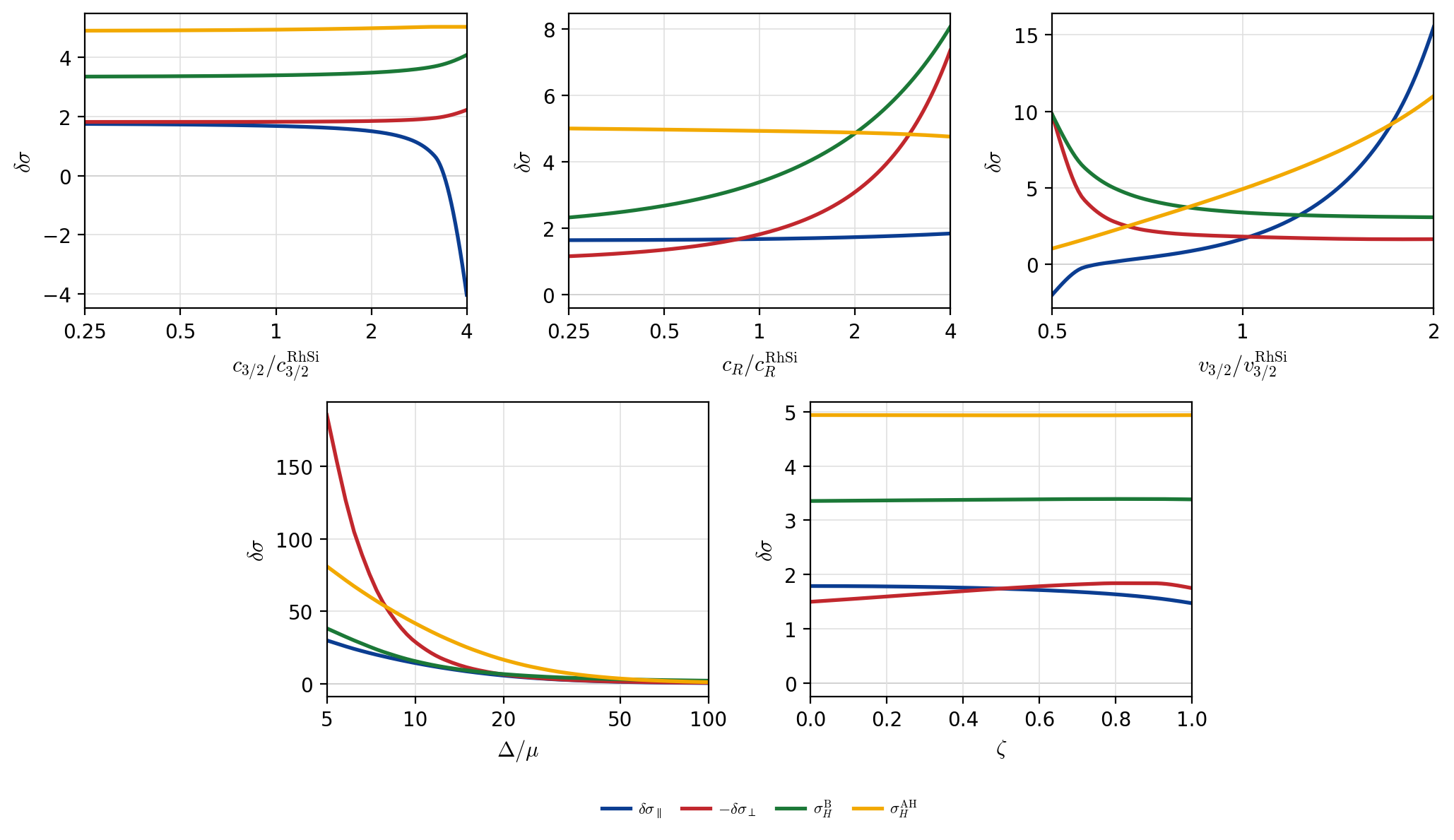}
\caption{RSW and TSM nodes for $\bB\parallel\hat{\boldsymbol z}$ at fixed $\alpha=0.05$. The panels show the dependence on $c_{3/2}$, $c_R$, $v_{3/2}$ (in units of the RhSi values), on $\Delta/\mu$, and on $\zeta$. In every panel, $\delta\sigma_\parallel$ multiplied by $10^5$ (navy), $-\delta\sigma_\perp$ multiplied by $10^3$ (crimson), $\sigma_H^{\rm B}$ multiplied by $10^2$ (green), and $\sigma_H^{\rm AH}$ multiplied by $10^6$ (amber), in units of the values at $B=0$.}
\label{figrtscan}
\end{figure}

Figure~\ref{figrtscan} shows the dependence on the parameters of Sec.~\ref{secresnum2}. In each panel, one of the quadratic coefficients $c_{3/2}$ and $c_R$, the velocity $v_{3/2}$ at fixed quadratic coefficients, the offset $\Delta$, and the ratio $\zeta$ of Eq.~\eqref{eqham2} is varied about the RhSi values, with $\zeta=1/\sqrt2$ as the reference. The other parameters are kept fixed, and the field is fixed at the value $\alpha=0.05$ of the RhSi parameters. The ratio $\zeta$ is not fixed by the symmetry, and it is not obtained from the DFT bandstructure. CoSi is not used in this section, since the RSW node requires strong SOC, whereas the SOC of CoSi is weak \cite{sanchez2019}.

The dependence on $\zeta$ is weak. The change $\delta\sigma_\parallel$ decreases monotonically with $\zeta$ and stays positive, the magnitude of $\delta\sigma_\perp$ has a maximum at an intermediate value of $\zeta$, and $\sigma_H^{\rm B}$ and $\sigma_H^{\rm AH}$ are almost independent of $\zeta$. The assumption on $\zeta$ therefore does not change the conclusions of Sec.~\ref{secresnum2}. The coefficient $c_{3/2}$ has a weak influence, except that $\delta\sigma_\parallel$ decreases rapidly and changes sign at about three times the RhSi value. The coefficient $c_R$ has a weak influence on $\delta\sigma_\parallel$ and a strong one on the magnitude of $\delta\sigma_\perp$ and on $\sigma_H^{\rm B}$, which increase with $c_R$, whereas $\sigma_H^{\rm AH}$ decreases slightly. As $v_{3/2}$ increases, $\delta\sigma_\parallel$ and $\sigma_H^{\rm AH}$ increase, and the magnitude of $\delta\sigma_\perp$ and $\sigma_H^{\rm B}$ decrease. The change $\delta\sigma_\parallel$ is negative below about $0.6$ times the RhSi value of $v_{3/2}$. All four quantities decrease monotonically with $\Delta$.

\label{secresnum2bc}

The cases (b) and (c) of Table~\ref{tabgeom2} are solved with the general solver of Eq.~\eqref{eqmomsys}, which uses the full overlaps of Eq.~\eqref{eqovgr} with the harmonics $0$, $1$, and $2$ and no sector decomposition. The implementation is described in Appendix~\ref{secnum}. The parameters are those of Sec.~\ref{secresnum2}, and we take $\bB=B\left(\sin\theta_B,0,\cos\theta_B\right)$ with $\theta_B=0$ for the case (a), $\theta_B=\pi/2$ for the case (b), and $0<\beta<\pi/2$ for the case (c). The tensor at $B=0$ is uniaxial about the polar axis, since the kernel is invariant under $\mathcal R_z(\delta)$. All the changes are measured from this tensor and are given in units of $\sigma_0=\mathrm{tr}\,\bs\sigma(0)/3$.

For every $\theta_B$, the rotation $C_{2y}\,\mathcal P$ maps $\bB$ onto $-\bB$ and leaves the kernel invariant, and the exchange $\mathcal P$ does not change the total current. It gives $\sigma_{ij}(-\bB)=\eta_i\,\eta_j\,\sigma_{ij}(\bB)$ with $\eta_x=\eta_z=-1$ and $\eta_y=+1$. The diagonal elements and $\sigma_{xz}$ are therefore even in $\alpha$, and $\sigma_{xy}$ and $\sigma_{yz}$ are odd. This is consistent with Eq.~\eqref{eqons}, by which the symmetric part of $\bs\sigma$ is even and the antisymmetric part is odd. The diagonal elements are even in every case. The numerical tensor has these properties and satisfies the relations of Table~\ref{tabgeom2}. In the case (b), the elements $\sigma_{xy}$, $\sigma_{xz}$, $\sigma_{yx}$, and $\sigma_{zx}$ vanish and are not shown.

\begin{figure}[t!]
\centering
\subfigure[\label{figrtb}]{\includegraphics[width=0.85\textwidth]{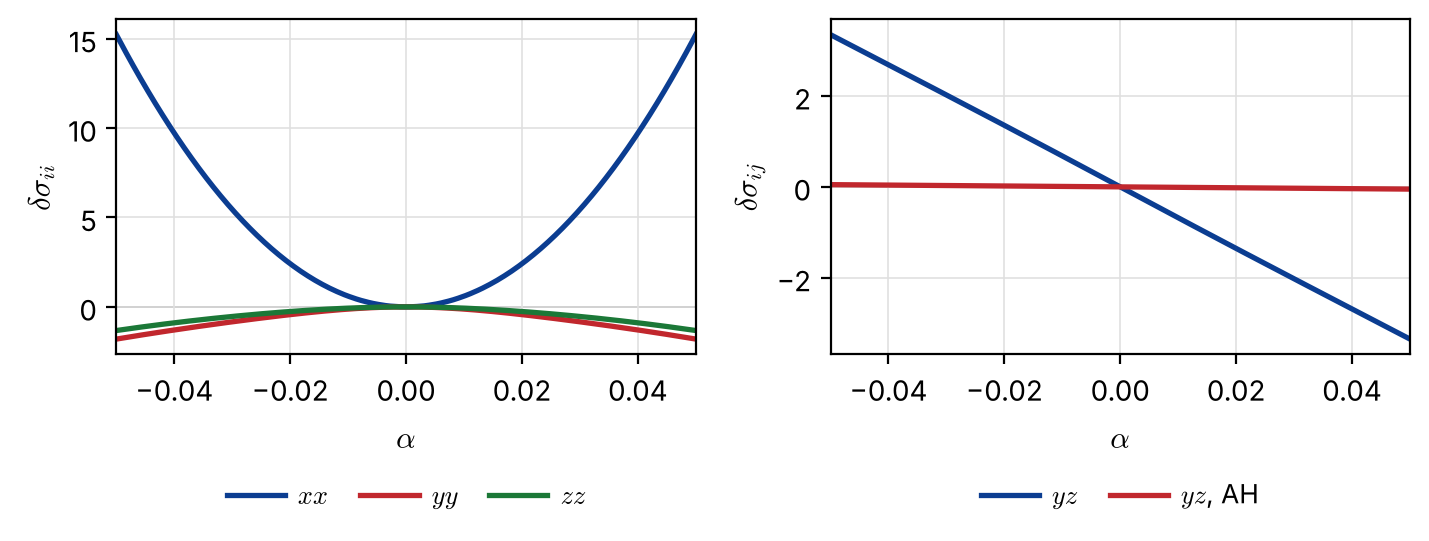}}\\
\subfigure[\label{figrtc}]{\includegraphics[width=\textwidth]{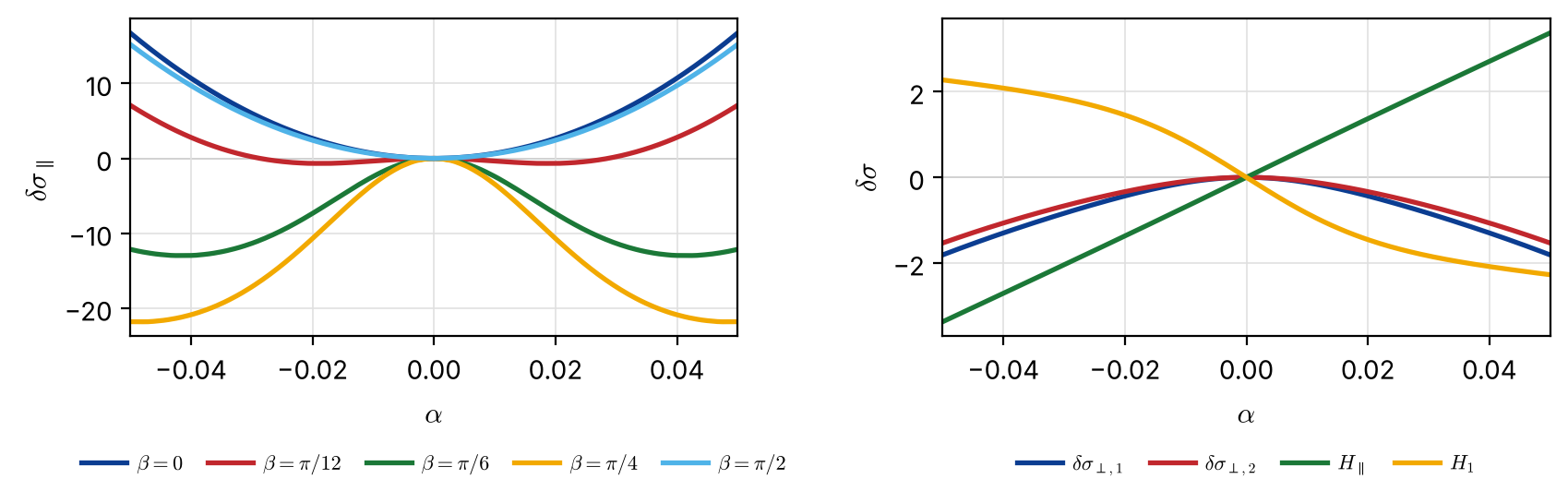}}
\caption{RSW and TSM nodes. (a) Case (b) with $\bB\parallel\hat{\boldsymbol x}$. Left: $\delta\sigma_{xx}$ multiplied by $10^6$ (navy), $\delta\sigma_{yy}$ (crimson), and $\delta\sigma_{zz}$ (green) multiplied by $10^3$. Right: $\delta\sigma_{yz}$ multiplied by $10^2$ (navy) and its AH part multiplied by $10^4$ (crimson), in units of $\sigma_0$. (b) Case (c) with $\bB=B\left(\sin\theta_B,0,\cos\theta_B\right)$. Left: $\delta\sigma_\parallel=\hat{\bB}\cdot\delta\bs\sigma\cdot\hat{\bB}$ multiplied by $10^6$ for $\theta_B=0$ (navy), $\pi/12$ (crimson), $\pi/6$ (green), $\pi/4$ (amber), and $\pi/2$ (sky). Right: for $\theta_B=\pi/4$, the eigenvalues $\delta\sigma_{\perp,1}$ (navy) and $\delta\sigma_{\perp,2}$ (crimson) of the symmetric part of $\delta\bs\sigma$ in the plane perpendicular to $\bB$ multiplied by $10^3$, the component $H_\parallel$ of the Hall vector along $\bB$ (green) multiplied by $10^2$, and its component $H_1$ (amber) along $\left(\cos\theta_B,0,-\sin\theta_B\right)$ multiplied by $10^4$, in units of $\sigma_0$.}
\label{figrtbc}
\end{figure}

In Fig.~\ref{figrtb}, the change $\delta\sigma_{xx}$ for $\bE\parallel\bB$ is positive and grows with $|\alpha|$ without a sign change. The changes $\delta\sigma_{yy}$ and $\delta\sigma_{zz}$ for $\bE\perp\bB$ are negative, and the magnitude of $\delta\sigma_{yy}$ is larger. The Hall element $\sigma_{yz}$ is odd and close to linear in $\alpha$. Its AH part is smaller than the total by almost four orders of magnitude, and the Hall response is therefore of the Boltzmann type.

The sign of $\delta\sigma_\parallel$ in Fig.~\ref{figrtc} depends on the orientation of $\bB$. It is positive for $\theta_B=0$ and $\pi/2$, where the two curves are close, and it is negative for $\theta_B=\pi/6$ and $\pi/4$, with the most negative values at $\theta_B=\pi/4$. For $\theta_B=\pi/12$, it is small, with a shallow negative interval at smaller $|\alpha|$ and a positive value at the largest $|\alpha|$. The change thus depends strongly on the angle between $\bB$ and the polar axis. In the right panel, the two eigenvalues of the symmetric transverse block are negative and close to each other. The Hall vector is dominated by $H_\parallel$, which is close to linear in $\alpha$. Its component $H_1$ perpendicular to $\bB$ is odd and smaller than $H_\parallel$ by more than two orders of magnitude. The component along $\boldsymbol{\hat y}$ vanishes, since $\sigma_{xz}=\sigma_{zx}$ in Table~\ref{tabgeom2}, and it is not shown.

\begin{figure}[t!]
\centering
\includegraphics[width=\textwidth]{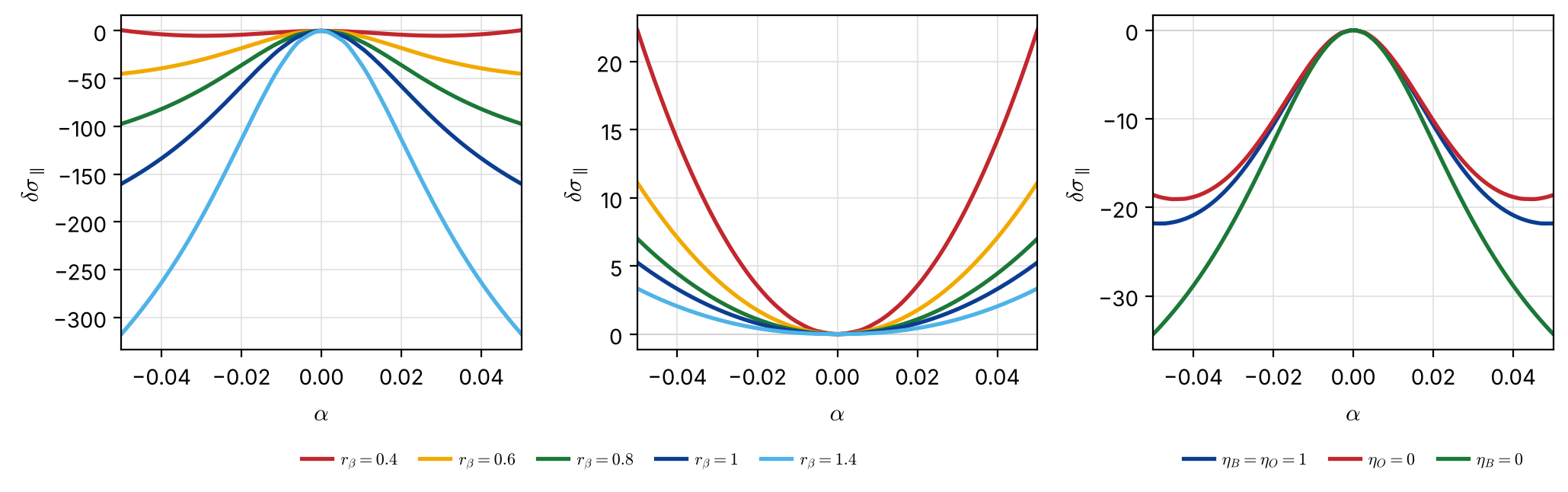}
\caption{RSW and TSM nodes for $\bB=B\left(\sin\theta_B,0,\cos\theta_B\right)$. Left and middle: $\delta\sigma_\parallel$ multiplied by $10^6$ for $\theta_B=\pi/4$ [case (c)] and $\theta_B=\pi/2$ [case (b)], respectively, with $r_\beta=0.4$ (crimson), $0.6$ (amber), $0.8$ (green), $1$ (navy), and $1.4$ (sky). Right: $\theta_B=\pi/4$ and $r_\beta=1/2$ with $\eta_B=\eta_O=1$ (navy), $\eta_O=0$ (crimson), and $\eta_B=0$ (green), multiplied by $10^6$. All changes are in units of $\hat{\bB}\cdot\bs\sigma(0)\cdot\hat{\bB}$.}
\label{figrtcratio}
\end{figure}

In the left panel of Fig.~\ref{figrtcratio}, $\delta\sigma_\parallel$ for $\theta_B=\pi/4$ is negative for $r_\beta\geq0.6$, and its magnitude increases monotonically with $r_\beta$. For $r_\beta=0.4$ it is close to zero, with a shallow negative interval and a small positive value at the largest $|\alpha|$. In the middle panel, $\delta\sigma_\parallel$ for $\theta_B=\pi/2$ is positive for all the ratios, and it decreases monotonically with $r_\beta$. The two orientations therefore have opposite signs and opposite dependences on $r_\beta$ for $r_\beta\geq0.6$. In the right panel, both the BC and the OMM contributions are negative, and the combined change is smaller in magnitude than their sum. The cross term is therefore positive, which is opposite to the sign for the QTSM and Weyl nodes in Fig.~\ref{figqwcratio}.

\label{secresnum2gf}

The orbital overlaps of Eq.~\eqref{eqovgr8c} between $\Gamma$ and $R$ are now used in place of Eq.~\eqref{eqovgr}. The overlap of Eq.~\eqref{eqovgr8c} is summed over the two copies at $R$, and each copy carries one half of it. The intranode overlaps and the vanishing overlap between the copies are unchanged, and the amplitude $r_\beta\,\theta_B$ multiplies the overlaps between $\Gamma$ and $R$ as before. The Chang scenario is the case $\phi=0$ of Eq.~\eqref{eqovgr8ch}, and the Flicker scenario is shown for $\phi=\pi/2$ and $\phi=\pi$, which correspond to $\upsilon=1/8$ and $\upsilon=1/4$, since $\upsilon$ is not specified here. The terms of Eq.~\eqref{eqovgr8c} that are proportional to $\sin\phi$ are odd under the simultaneous inversion of $\hat{\bk}$ and $\hat{\bs q}$. Time reversal exchanges the two copies at $R$ and inverts both momenta, and the overlap that is shared equally between the copies must therefore be even under this inversion. With the odd terms kept, the numerical tensor violates Eq.~\eqref{eqons}, and $\delta\sigma_\parallel$ for $\bB\parallel\hat{\bs c}$ acquires a term linear in $B$. We therefore use the part of Eq.~\eqref{eqovgr8c} that is even in $\phi$, namely $\left[\mathcal O(\phi)+\mathcal O(-\phi)\right]/2$, for which the dependence on $s_p$ drops out. The Chang scenario and the case $\phi=\pi$ are not affected, since $\sin\phi$ vanishes there.

The overlaps of Eq.~\eqref{eqovgr8c} are invariant under the rotations of all the momenta about $\hat{\bs c}=\left(1,1,1\right)/\sqrt3$ for every $\phi$. The set-ups are therefore those of the second case of Table~\ref{tabgeom}, with $\boldsymbol{\hat z}=\hat{\bs c}$ and $\bB=B\left(\sin\theta_B,0,\cos\theta_B\right)$. The tensor at $B=0$ is uniaxial about $\hat{\bs c}$, and its anisotropy is small. For $\bB\perp\hat{\bs c}$, the rotation $\mathcal R_{\hat c}(\pi)$ and Eq.~\eqref{eqons} give the parities of Eq.~\eqref{eqplanar5}. The out-scattering rate of the sheet $(\Gamma,3/2)$ depends on the direction of $\bk$ for these overlaps. The time $\tau_0$ with $\mu_\Gamma\,\tau_0=15$ is therefore defined through the average of this rate over the sheet at $B=0$. All the other parameters are those of Sec.~\ref{secresnum2}, and the solution is obtained with the general solver of Eq.~\eqref{eqmomsys}. The changes are measured from the tensor at $B=0$.

\begin{figure}[t!]
\centering
\subfigure[\label{figrtgf}]{\includegraphics[width=\textwidth]{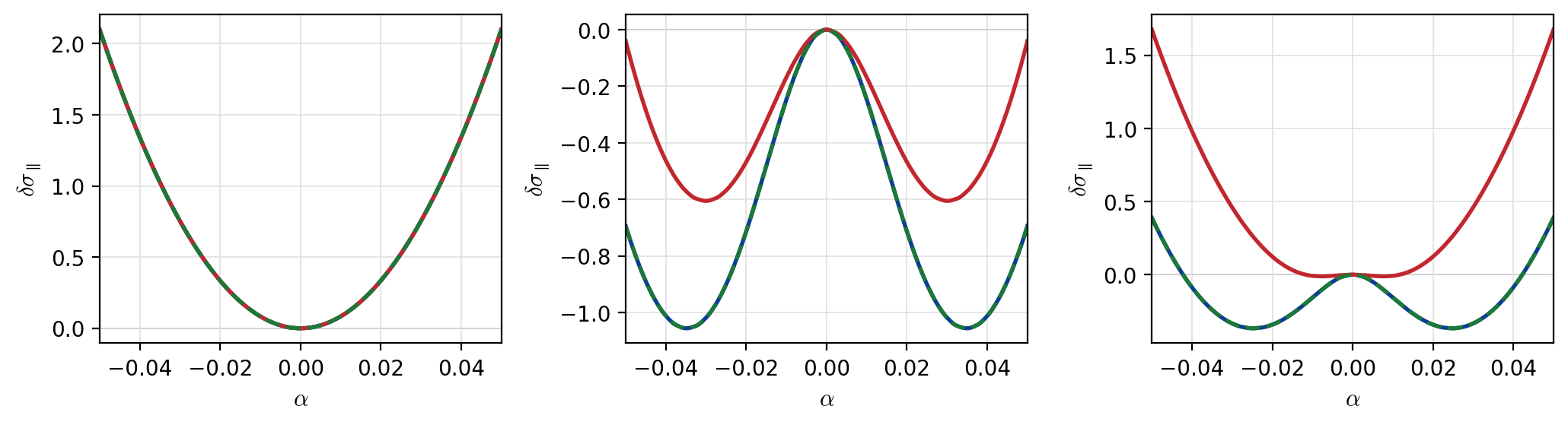}}\\
\subfigure[\label{figrtgf2}]{\includegraphics[width=0.85\textwidth]{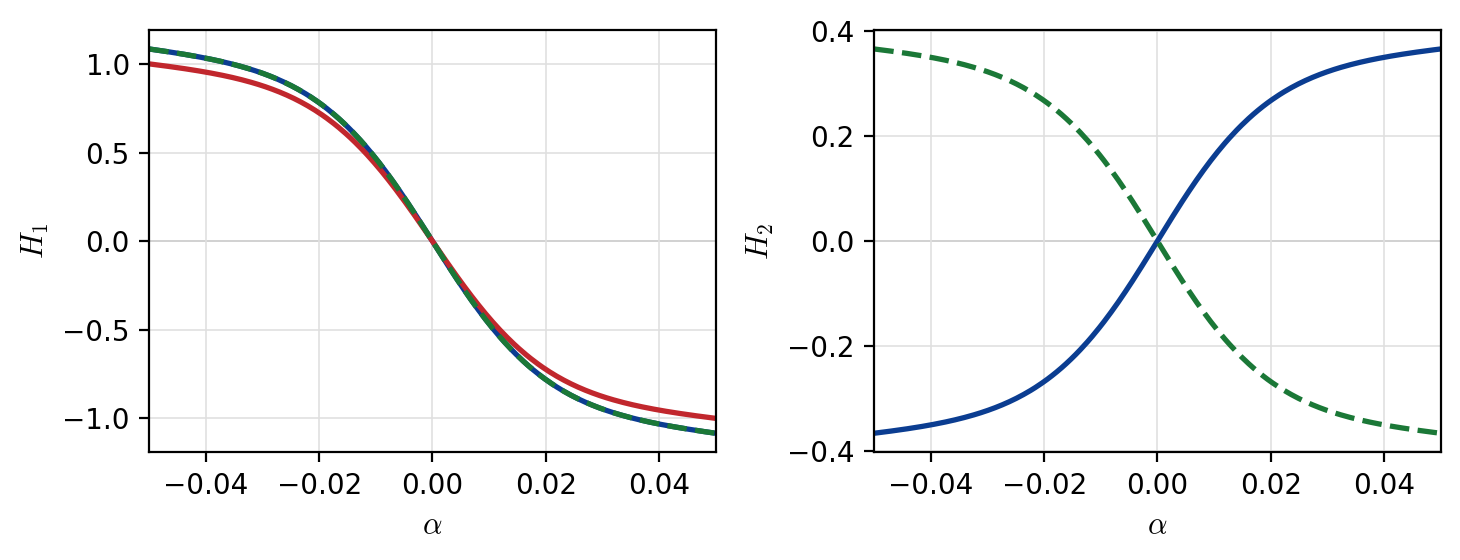}}
\caption{RSW and TSM nodes in the Chang scenario (navy) and in the Flicker scenario with $\phi=\pi/2$ (crimson) and $\phi=\pi$ (green, dashed). (a) Panels from left to right: $\delta\sigma_\parallel=\hat{\bB}\cdot\delta\bs\sigma\cdot\hat{\bB}$ multiplied by $10^5$ for $\theta_B=0$, $\pi/4$, and $\pi/2$, in units of $\hat{\bB}\cdot\bs\sigma(0)\cdot\hat{\bB}$. In the left panel, the three curves coincide. (b) Here $\theta_B=\pi/4$. Left: component $H_1$ of the Hall vector along $\left(\cos\theta_B,0,-\sin\theta_B\right)$. Right: component $H_2$ along $\boldsymbol{\hat y}$. Both are multiplied by $10^4$ and given in units of $\sigma_0$. In the left panel, the curves of the Chang scenario and of $\phi=\pi$ coincide. In the right panel, the component $H_2$ of $\phi=\pi/2$ vanishes and is not drawn.}
\label{figrtscen}
\end{figure}

Fig.~\ref{figrtgf} shows the longitudinal change for three orientations of $\bB$. All the curves are even in $\alpha$, as required by Eq.~\eqref{eqons}. For $\bB\parallel\hat{\bs c}$, the three scenarios give the same positive change. For $\theta_B=\pi/4$, the change is negative in the whole range for all three scenarios. The curves of the Chang scenario and of $\phi=\pi$ coincide, and the change of $\phi=\pi/2$ is smaller in magnitude. For $\theta_B=\pi/2$, the Chang scenario and $\phi=\pi$ again coincide. The change is negative at smaller $|\alpha|$ and turns positive at larger $|\alpha|$, whereas for $\phi=\pi/2$ it remains close to zero at smaller $|\alpha|$ and is positive and larger at larger $|\alpha|$. The sign of $\delta\sigma_\parallel$ thus depends on the orientation of $\bB$ in every scenario. It is positive for $\bB\parallel\hat{\bs c}$ and negative at an oblique angle, in the same way as in Fig.~\ref{figrtc}. The scenarios differ in the magnitude and in the sign at smaller $|\alpha|$ for $\bB\perp\hat{\bs c}$, and this dependence on the position of the orbitals is absent for the overlaps of Eq.~\eqref{eqovgr}. The agreement of $\phi=\pi$ with the Chang scenario follows from the reversal of the cyclic order, which gives the mirror image of the Chang overlaps and leaves the diagonal elements of the tensor unchanged.

Fig.~\ref{figrtgf2} shows the components of the Hall vector $\bs H$ that are perpendicular to $\bB$ for $\theta_B=\pi/4$. The component $H_\parallel$ along $\bB$ dominates in all the scenarios, and it differs only slightly between them. The components $H_1$ and $H_2$ are odd in $\alpha$ and smaller than $H_\parallel$ by more than two orders of magnitude. The component $H_1$ is almost independent of the scenario. The component $H_2$ along $\boldsymbol{\hat y}$ vanishes for the overlaps of Eq.~\eqref{eqovgr}, since $\sigma_{xz}=\sigma_{zx}$ in Table~\ref{tabgeom2}. It is nonzero for the Chang scenario, which has no mirror symmetry, and it changes sign for $\phi=\pi$. For $\phi=\pi/2$, it vanishes, since the part that is even in $\phi$ is invariant under the reversal of the cyclic order. The AH part of the tensor depends on the FS only, and it is the same in all the scenarios.

\section{Two untilted QTSM nodes}
\label{secsys3}

We consider here two conjugate QTSM nodes with the chiralities $\chi=\pm1$ and the same nodal energy~\cite{ips-exact-spin1}, such that
\begin{align}
\mathcal H_\chi(\bk)=v_F\left(k_x\,S_x+k_y\,S_y+\chi\,k_z\,S_z\right)+c\,v_F\,k^2\,\mathbb I_{3\times3}\,.
\label{eqham3}
\end{align}
Time-reversal maps a node at $\mathbf K$ onto a node at $-\mathbf K$ with the same monopole charge, and hence with the same chirality. The two nodes of opposite chiralities in this section are therefore not time-reversal partners. The threefold nodes at the P point and at its time-reversed partner $-$P of the chiral space groups 199 and 214 in the presence of SOC \cite{grushin-multifold} are such partners and have the same chirality. A chiral space group contains no improper operation, and it has no operation that maps one of these nodes onto a node of opposite chirality. An example of an opposite-chirality pair at $\mathbf K=$ P and $\mathbf K'=-$P is provided by the magnetic space group $I\bar{4}'3d'$ (No.~220.91) \cite{cano-multifold-magnetic}. This group does not contain time reversal. It contains the product of time reversal and the mirror $m_{1\bar{1}0}$, which maps P onto $-$P and thus relates the two nodes with opposite chiralities.

The eigenvalues are those of Eq.~\eqref{eqeval} for $\epsilon_{\Gamma,s}$ at both nodes, and we consider $c>0$, $\mu>0$, and $c\,k<1$ as in Eq.~\eqref{eqeval}. The chemical potential cuts the bands $s=1$ and $s=0$ of each node, which gives the four FS sheets $a=(\chi,s)$ with $s\in\{1,0\}$. The FS radii and the FS weights at $B=0$ are those of $k_{\Gamma,1}$, $k_{\Gamma,0}$, $\rho_{\Gamma,1}$, and $\rho_{\Gamma,0}$ in Eq.~\eqref{eqfsb0}, for both nodes. The BC and the OMM are given by Eq.~\eqref{eqbcomm} with $v_\nu=v_F$ and the numbers $\{N_a,G_a\}=\{2,1\}$ for $s=1$ and $\{0,2\}$ for $s=0$. The Chern numbers $\mathcal C_a=-\chi\,N_a$ are $-2$, $0$, $+2$, and $0$ for the sheets $(+,1)$, $(+,0)$, $(-,1)$, and $(-,0)$, and they add up to zero. The quantities $\xi_s$, $D_a$, and $\bw_a$ follow from Eq.~\eqref{eqxiw}.

The scattering amplitudes between the sheets of the same node are denoted by $\beta^{\rm intra}_{ss'}$, and those between the sheets of different nodes by $\beta^{\rm inter}_{ss'}$. We use $\beta^{\rm intra}_{ss'}=\beta^{\rm intra}_{s's}$ and $\beta^{\rm inter}_{ss'}=\beta^{\rm inter}_{s's}$, and the amplitudes do not depend on the chirality, as in Ref.~\cite{ips-exact-spin1}. The squared overlaps are those of Eq.~\eqref{eqoverlap} for the sheets of the QTSM node, with the variable $X=\chi\,\chi'\cos\theta\cos\theta'$ of Appendix~\ref{secappkernel}, which equals $\cos\theta\cos\theta'$ within a node and $-\cos\theta\cos\theta'$ between the nodes. The sheets $(\chi,1)$ have the range of nine functions of Eq.~\eqref{eqfp}, and the sheets $(\chi,0)$ have the six even functions, as described below Eq.~\eqref{eqfp}.

For $\bB=B\,\boldsymbol{\hat z}$ and $\bE\parallel\bB$, each node has three moments on the sheet $(\chi,1)$ and two on the sheet $(\chi,0)$. This gives ten unknowns for the two nodes, and the number constraint of Eq.~\eqref{eqnumber} removes the null vector of the resulting system ~\cite{ips-exact-spin1}. For $\bE\perp\bB$, each node has two complex unknowns on the sheet $(\chi,1)$ and one on the sheet $(\chi,0)$, which gives six unknowns without a null vector.

For $\bB=B\,\boldsymbol{\hat z}$, we write $u=\cos\theta$ and $p=3\,u^2-1$, and the primed symbols refer to the final state. Eq.~\eqref{eqoverlapm} then gives the components of the overlaps with $m=0$ and $m=1$ in the form
\begin{align}
& \mathcal O^{(0)}_{(\chi,1)(\chi',1)}=\frac13+\frac{\chi\,\chi'\,u\,u'}{2}+\frac{p\,p'}{24}\,,\quad
\mathcal O^{(0)}_{(\chi,0)(\chi',0)}=\frac13+\frac{p\,p'}{6}\,,\quad
\mathcal O^{(0)}_{(\chi,1)(\chi',0)}=\frac13-\frac{p\,p'}{12}\,,\nn
& \mathcal O^{(1)}_{(\chi,1)(\chi',1)}=\frac{Y}{2}\left(1+\chi\,\chi'\,u\,u'\right),\quad
\mathcal O^{(1)}_{(\chi,0)(\chi',0)}=2\,\chi\,\chi'\,u\,u'\,Y\,,\quad
\mathcal O^{(1)}_{(\chi,1)(\chi',0)}=-\,\chi\,\chi'\,u\,u'\,Y\,,
\label{eqoverlap3}
\end{align}
with $Y=\sin\theta\,\sin\theta'$. The components with $m=\pm2$ do not enter [cf. Eq.~\eqref{eqsector}], since no source feeds this sector.

%%%%%%%%%%%%%%%
\subsection{Solution for the magnetic field along the polar axis}
\label{secsol3}

Eqs.~\eqref{eqsector}, \eqref{eqlongsol}, and \eqref{eqell} apply with the sheet index $a=(\chi,s)$, and we write $u=\cos\theta$. The two nodes are exchanged by the reflection $k_z\to-k_z$, because $\mathcal H_\chi(k_x,k_y,-k_z)=\mathcal H_{-\chi}(\bk)$, and this reflection leaves $\bB=B\,\boldsymbol{\hat z}$ invariant. Eqs.~\eqref{eqxiw} and \eqref{eqwqd} show that $k_a$, $D_a$, $w_{a,k}$, $q_a$, $\rho_a$, and $\omega_a$ depend on $\chi$ and $u$ only through the product $\chi\,u$. These quantities for the sheet $(-,s)$ at $u$ therefore equal those of the sheet $(+,s)$ at $-u$. The component $w_{a,z}$ and the source $h_a$ change sign instead,
\begin{align}
w_{(-,s),z}(u)=-\,w_{(+,s),z}(-u)\,,\quad
h_{(-,s)}(u)=-\,h_{(+,s)}(-u)\,.
\label{eqmirror}
\end{align}
Eq.~\eqref{eqoverlap3} is invariant under $(\chi,\chi',u,u')\to(-\chi,-\chi',-u,-u')$, and the amplitudes do not depend on the chirality. The equations of the node $-$ are thus the mirror images of those of the node $+$. For $\bE\parallel\bB$, the constant zero mode, whose amplitude is fixed by Eq.~\eqref{eqnumber}, is even under this exchange, whereas the source is odd. The solution is therefore odd, $\Lambda_{(-,s)}(u)=-\Lambda_{(+,s)}(-u)$, and it satisfies the number constraint identically. For $\bE\perp\bB$, the solution is even, $\ell_{(-,s)}(u)=\ell_{(+,s)}(-u)$. In both cases, the two nodes contribute equally to the conductivities, and the quantities of the node $+$ carry the single index $s$ below.

We denote by $\rho^{(n)}_s=\int_{-1}^{1}du\,u^n\,\rho_s$ the moments of the weight and by $M^{(n)}_s=\int_{-1}^{1}du\,u^n\,\rho_s\,\Lambda_s$ those of the solution for $\bE\parallel\bB$, with $n=0,1,2$ for $s=1$ and $n=0,2$ for $s=0$. For moments $X^{(n)}_s$ and amplitudes $\gamma_{ss'}$, the kernel of Eq.~\eqref{eqoverlap3} with $m=0$ generates the polynomial $\sum_{j=0}^{2}\mathfrak K_{s,j}\,u^j$ with the coefficients
\begin{align}
& \mathfrak K_{1,0}(X,\gamma)=\frac{\gamma_{11}}{4}\left(3\,X^{(0)}_1-X^{(2)}_1\right)+\frac{\gamma_{10}}{2}\left(X^{(0)}_0+X^{(2)}_0\right),\quad
\mathfrak K_{1,1}(X,\gamma)=\gamma_{11}\,X^{(1)}_1\,,\nn
& \mathfrak K_{1,2}(X,\gamma)=\frac{\gamma_{11}}{4}\left(3\,X^{(2)}_1-X^{(0)}_1\right)-\frac{\gamma_{10}}{2}\left(3\,X^{(2)}_0-X^{(0)}_0\right),\nn
& \mathfrak K_{0,0}(X,\gamma)=\frac{\gamma_{10}}{2}\left(X^{(0)}_1+X^{(2)}_1\right)+\gamma_{00}\left(X^{(0)}_0-X^{(2)}_0\right),\quad
\mathfrak K_{0,1}(X,\gamma)=0\,,\nn
& \mathfrak K_{0,2}(X,\gamma)=-\frac{\gamma_{10}}{2}\left(3\,X^{(2)}_1-X^{(0)}_1\right)+\gamma_{00}\left(3\,X^{(2)}_0-X^{(0)}_0\right).
\label{eqkappa}
\end{align}
The sum over the final states of both nodes follows from Eq.~\eqref{eqmirror} and from the relations $\rho^{(n)}_{(-,s)}=(-1)^n\,\rho^{(n)}_{(+,s)}$ and $M^{(n)}_{(-,s)}=-(-1)^n\,M^{(n)}_{(+,s)}$. With $\gamma^{\pm}_{ss'}=\left(\beta^{\rm intra}_{ss'}\right)^2\pm\left(\beta^{\rm inter}_{ss'}\right)^2$ and $\gamma^{\pm}_{01}=\gamma^{\pm}_{10}$, the out-scattering rate of Eq.~\eqref{eqtau} and the mean free-path become
\begin{align}
\tau_s^{-1}(u)=\sum_{j=0}^{2}\mathfrak K_{s,j}\!\left(\rho,\gamma^{+}\right)u^j\,,\quad
\Lambda_s(u)=\tau_s(u)\Big[h_s(u)+\sum_{j=0}^{2}\mathfrak K_{s,j}\!\left(M,\gamma^{-}\right)u^j\Big].
\label{eqlampar}
\end{align}
The rate involves the sum of the intranode and internode strengths, whereas the in-scattering term involves their difference, since the solution is odd under the exchange of the nodes. Multiplying Eq.~\eqref{eqlampar} by $u^n\,\rho_s$ and integrating over $u$ closes the problem as five linear equations,
\begin{align}
M^{(n)}_s=\left\langle u^n\,h_s\right\rangle^{\parallel}_s+\sum_{j=0}^{2}\mathfrak K_{s,j}\!\left(M,\gamma^{-}\right)\left\langle u^{n+j}\right\rangle^{\parallel}_s\,,\quad
\left\langle f\right\rangle^{\parallel}_s=\int_{-1}^{1}du\,\rho_s\,\tau_s\,f\,,
\label{eqmomsys3}
\end{align}
for $n=0,1,2$ with $s=1$ and for $n=0,2$ with $s=0$. The null vector of the full problem has the opposite parity, and the matrix of Eq.~\eqref{eqmomsys3} is therefore regular. The conductivity of Eq.~\eqref{eqsigpar}, summed over the two nodes, reads
\begin{align}
\sigma_\parallel=\frac{e^2}{2\,\pi^2}\sum_{s}\Big[\left\langle h_s^2\right\rangle^{\parallel}_s+\sum_{j=0}^{2}\mathfrak K_{s,j}\!\left(M,\gamma^{-}\right)\left\langle u^j\,h_s\right\rangle^{\parallel}_s\Big].
\label{eqsigpar3}
\end{align}

For $B\to0$, the weights and the times are constants, $h_s=w^0_s\,u$ with $w^0_s=\partial_k\epsilon_{\Gamma,s}$ evaluated at $k_{\Gamma,s}$, and the moments $M^{(0)}_s$ and $M^{(2)}_s$ vanish. Eq.~\eqref{eqmomsys3} is then solved in closed form. Let $\rho_s=\rho_{\Gamma,s}$ and $k_s=k_{\Gamma,s}$ be the quantities of Eq.~\eqref{eqfsb0}. The rates are $\tau_1^{-1}=\tfrac{4}{3}\left(\gamma^+_{11}\,\rho_1+\gamma^+_{10}\,\rho_0\right)$ and $\tau_0^{-1}=\tfrac{4}{3}\left(\gamma^+_{10}\,\rho_1+\gamma^+_{00}\,\rho_0\right)$, and the conductivity is
\begin{align}
\sigma_\parallel\big\vert_{B=0}=\frac{e^2}{3\,\pi^2}\left[\frac{k_1^2\,w^0_1\,\tau_1}{1-\tfrac{2}{3}\,\gamma^{-}_{11}\,\rho_1\,\tau_1}+k_0^2\,w^0_0\,\tau_0\right].
\label{eqsigB0}
\end{align}
The denominator is the vertex correction of the sheet $s=1$. The sheet $s=0$ has none, since its overlaps with all the sheets are even in $u$. The same expression holds for $\sigma_\perp$ with $\gamma^{-}_{11}$ replaced by $\gamma^{+}_{11}$. The two results coincide for $\beta^{\rm inter}_{ss'}=0$, as required by isotropy. They differ otherwise, because the factor $\chi\,\chi'$ in Eq.~\eqref{eqoverlap3} makes the internode overlap depend on the orientation of the $z$ axis.

For $\bE\perp\bB$, we write $\mathcal G_s(u)=\left[\tau_s^{-1}(u)+i\,\omega_s(u)\right]^{-1}$ with $\tau_s^{-1}$ of Eq.~\eqref{eqlampar}, and we define the complex moments $\mathcal L^{(n)}_s=\int_{-1}^{1}du\,\sqrt{1-u^2}\,u^n\,\rho_s\,\ell_s$ with $n=0,1$ for $s=1$ and $n=1$ for $s=0$. For the even solution, $\mathcal L^{(n)}_{(-,s)}=(-1)^n\,\mathcal L^{(n)}_{(+,s)}$, and the kernel of Eq.~\eqref{eqoverlap3} with $m=1$ fixes the coefficients $\check\lambda^{(0)}_s$ and $\check\lambda^{(1)}_s$ of $\sin\theta$ and of $\sin\theta\,\cos\theta$ in the in-scattering term of Eq.~\eqref{eqell},
%%%%%%%%%%%%%%%%%%%%%%%
\begin{align}
& \check\lambda^{(0)}_1=\frac{\gamma^+_{11}}{2}\,\mathcal L^{(0)}_1\,,\quad
\check\lambda^{(1)}_1=\frac{\gamma^+_{11}}{2}\,\mathcal L^{(1)}_1-\gamma^+_{10}\,\mathcal L^{(1)}_0\,,\quad
 \check\lambda^{(0)}_0=0\,,\quad
\check\lambda^{(1)}_0=-\,\gamma^+_{10}\,\mathcal L^{(1)}_1+2\,\gamma^+_{00}\,\mathcal L^{(1)}_0\,.
\label{eqlamperp}
\end{align}
Both the rate and the in-scattering term involve $\gamma^+$ in this sector, since the solution is even under the exchange of the nodes. The mean free-path is $\ell_s=\mathcal G_s\,\sin\theta\left[q_s/D_s+\check\lambda^{(0)}_s+\check\lambda^{(1)}_s\,u\right]$. With $\left\langle f\right\rangle^{\perp}_s=\int_{-1}^{1}du\left(1-u^2\right)\rho_s\,\mathcal G_s\,f$, the moments obey three complex linear equations,
%%%%%%%%%%%%
\begin{align}
\begin{pmatrix}
1-\tfrac{1}{2}\,\gamma^+_{11}\left\langle 1\right\rangle^{\perp}_1 & -\tfrac{1}{2}\,\gamma^+_{11}\left\langle u\right\rangle^{\perp}_1 & \gamma^+_{10}\left\langle u\right\rangle^{\perp}_1\\
-\tfrac{1}{2}\,\gamma^+_{11}\left\langle u\right\rangle^{\perp}_1 & 1-\tfrac{1}{2}\,\gamma^+_{11}\left\langle u^2\right\rangle^{\perp}_1 & \gamma^+_{10}\left\langle u^2\right\rangle^{\perp}_1\\
0 & \gamma^+_{10}\left\langle u^2\right\rangle^{\perp}_0 & 1-2\,\gamma^+_{00}\left\langle u^2\right\rangle^{\perp}_0
\end{pmatrix}
\begin{pmatrix}\mathcal L^{(0)}_1\\ \mathcal L^{(1)}_1\\ \mathcal L^{(1)}_0\end{pmatrix}
=
\begin{pmatrix}\left\langle q_1/D_1\right\rangle^{\perp}_1\\ \left\langle u\,q_1/D_1\right\rangle^{\perp}_1\\ \left\langle u\,q_0/D_0\right\rangle^{\perp}_0\end{pmatrix}.
\label{eqmomsysperp}
\end{align}
%%%%%%%%%%%
Eq.~\eqref{eqsigperp}, summed over the two nodes, gives
\begin{align}
\sigma_\perp-i\,\sigma_H^{\rm B}=\frac{e^2}{4\,\pi^2}\sum_{s}\Big[\left\langle \left(q_s/D_s\right)^2\right\rangle^{\perp}_s+\check\lambda^{(0)}_s\left\langle q_s/D_s\right\rangle^{\perp}_s+\check\lambda^{(1)}_s\left\langle u\,q_s/D_s\right\rangle^{\perp}_s\Big].
\label{eqsigperp3}
\end{align}

The Fermi-sea term of Eq.~\eqref{eqsigmw} has only the component along $\bB$, and only the sheets with $s=1$ carry a BC. With Eq.~\eqref{eqmirror}, it becomes
\begin{align}
\sigma_H^{\rm AH}=-\,\frac{\eta_B\,e^2}{2\,\pi^2}\int_{-1}^{1}du\,u\,k_{(+,1)}(u)\,.
\label{eqsigah3}
\end{align}
To the leading order in $B$, the OMM shifts the radius, and $\sigma_H^{\rm AH}\simeq\eta_B\,\eta_O\,e^3\,B/\left[6\,\pi^2\,k_{\Gamma,1}\left(1+2\,c\,k_{\Gamma,1}\right)\right]$. The Hall conductivity is $\sigma_H=\sigma_H^{\rm B}+\sigma_H^{\rm AH}$. The perpendicular and Hall conductivities are new compared with the collinear solution of Ref.~\cite{ips-exact-spin1}.

%%%%%%%%%%%%%%%
\subsection{Symmetries and classification of the set-ups}
\label{secclass3}

For $\bB=B\,\boldsymbol{\hat z}$, the invariance under the rotations about the $z$ axis gives the tensor form below Eq.~\eqref{eqsigperp} for every direction of $\bE$, with the three independent functions $\sigma_\parallel$, $\sigma_\perp$, and $\sigma_H$. The currents follow from Eq.~\eqref{eqcurr}. The reflection of Eq.~\eqref{eqmirror} relates the two nodes and gives equal nodal contributions to each of the three functions. Time reversal yields the Onsager relation of Eq.~\eqref{eqons}, which this tensor form satisfies identically. The internode overlaps of Eq.~\eqref{eqoverlap3} depend on the product $\chi\,\chi'\cos\theta\,\cos\theta'$, and the kernel is therefore invariant only under the rotations about the $z$ axis and not under a general rotation. Eq.~\eqref{eqrot} holds for the intranode scattering alone. For $\beta^{\rm inter}_{ss'}\neq0$, the orientations of $\bB$ other than $\boldsymbol{\hat z}$ do not follow from the reference set-up by a rotation. The set-ups with $\bB$ at a general direction are therefore classified by the symmetries that remain.

The overlaps of Eq.~\eqref{eqoverlap3} depend on $\hat{\bk}\cdot\hat{\bk}'$ within a node and on $\hat{\bk}\cdot\mathsf T\,\hat{\bk}'$ between the nodes, with $\mathsf T=\mathrm{diag}\left(1,1,-1\right)$. A transformation $g$ of the momenta leaves the kernel invariant if $g$ commutes with $\mathsf T$. These transformations are the rotations $\mathcal R_z$ about the $z$ axis, the rotations by $\pi$ about axes in the $xy$ plane, the reflections $\Pi_z$ and $\Pi_{x}$ normal to the $z$ axis and to an axis in the $xy$ plane, the inversion $\mathcal J$, and their products. Proper elements map every node onto itself, and improper elements exchange the two nodes [cf. Eq.~\eqref{eqmirror}]. The field $\bB$ is a pseudovector and transforms as $\det(g)\,g\,\bB$, whereas $\bE$ transforms as a vector. Appendix~\ref{secappsym3} gives the derivation. The inversion fixes $\bB$ for every orientation, and it gives
\begin{align}
\Lambda^j_{(-,s)}(-\hat{\bk})=-\,\Lambda^j_{(+,s)}(\hat{\bk})\,.
\label{eqinv3}
\end{align}
The number constraint of Eq.~\eqref{eqnumber} is then satisfied identically, and only the node $+$ has to be expanded. The rotations $\mathcal R_z$ relate orientations of $\bB$ with the same polar angle $\theta_B$ to the $z$ axis, and we take $\bB=B\left(\sin\theta_B,0,\cos\theta_B\right)$. The classification is collected in Table~\ref{tabgeom3}, and all directions of $\bE$ follow from the reference set-ups.

\begin{table}[t!]
\centering
\small
\renewcommand{\arraystretch}{1.6}
\begin{tabular}{|c|c|c|c|}
\hline
Case & Orientation of $\bB$ & Symmetries & \makecell{Independent elements \\ of $\bs\sigma$} \\
\hline
(a) & $\theta_B=0$ & $\mathcal R_z$, $\Pi_z$, $\mathcal J$ & \makecell{three: \\ $\sigma_\parallel$, $\sigma_\perp$, $\sigma_H$} \\
\hline
(b) & $\theta_B=\pi/2$ & $C_{2x}$, $\Pi_x$, $\mathcal J$ & \makecell{four: $\sigma_{xx}$, $\sigma_{yy}$, $\sigma_{zz}$, \\ $\sigma_{yz}=-\sigma_{zy}$} \\
\hline
(c) & $0<\theta_B<\pi/2$ & $\mathcal J$ & \makecell{six: $\sigma_{xx}$, $\sigma_{yy}$, $\sigma_{zz}$, $\sigma_{xz}=\sigma_{zx}$, \\ $\sigma_{xy}=-\sigma_{yx}$, $\sigma_{yz}=-\sigma_{zy}$} \\
\hline
\end{tabular}
\caption{\label{tabgeom3} Independent geometries of the two conjugate QTSM nodes. The elements of $\bs\sigma$ follow from the symmetries and from the Onsager relation of Eq.~\eqref{eqons}.}
\end{table}

%%%%%%%%%%%%%%%
\subsection{Ansatz for the mean free-path}
\label{secansatz3}

The nodes are untilted, and the FS of every sheet is axisymmetric about $\bB$ for every orientation of $\bB$. The distinct set-ups are the cases (a) to (c) of Table~\ref{tabgeom3}.

For each sheet $a=(\chi,s)$, the exact form of $\Lambda^j_a$ is Eq.~\eqref{eqtiltsol} with the functions $f^a_p$ of Eq.~\eqref{eqfp}. The only unknowns are the moments $\mathcal M^j_{a,p}$. The sheets $(\chi,1)$ carry nine functions and the sheets $(\chi,0)$ carry six, which are listed for the sectors about the $z$ axis in Eq.~\eqref{eqspans}. The out-scattering rate in Eq.~\eqref{eqtiltsol} is of finite rank as well,
\begin{align}
\frac{1}{\tau_a(\hat{\bk})}=\sum_{a'}\sum_{p,p'}f^a_p(\hat{\bk})\,Q^{aa'}_{pp'}\,\mathcal N^{a'}_{p'}\,,\quad
\mathcal N^{a}_{p}=\int d^2\hat{\bk}\,\rho_a\,f^a_p\,,
\label{eqtau3}
\end{align}
where $\mathcal N^{a}_{p}$ are the moments of the weight $\rho_a$ at the given $\bB$.

For case (a), $\bB=B\,\boldsymbol{\hat z}$, the rotations about the $z$ axis make the ansatz closed in the azimuthal sectors for every angle $\vartheta$ between $\bE$ and $\bB$. The components of the node $+$ are
\begin{align}
\Lambda^z_{(+,s)}=\tau_s(\theta)\Big[h_s(\theta)+\sum_{j=0}^{2}\mathfrak K_{s,j}\!\left(M,\gamma^{-}\right)\cos^j\theta\Big]\,,\quad
\Lambda^x_{(+,s)}+i\,\Lambda^y_{(+,s)}=e^{i\,\varphi}\,\sin\theta\,\frac{q_s/D_s+\check\lambda^{(0)}_s+\check\lambda^{(1)}_s\cos\theta}{\tau_s^{-1}(\theta)+i\,\omega_s(\theta)}\,,
\label{eqansatz3a}
\end{align}
where $\mathfrak K_{s,j}$ and $\check\lambda^{(n)}_s$ are given by Eqs.~\eqref{eqkappa} and \eqref{eqlamperp}, and $\tau_s$, $\omega_s$, $q_s$, and $D_s$ by Eqs.~\eqref{eqlampar} and \eqref{eqresalg}. The unknowns are the five real moments $M^{(n)}_s$ of the sector $m=0$ and the three complex moments $\mathcal L^{(n)}_s$ of the sector $m=1$. The mean free-path of the node $-$ follows from the mirror relations of Eq.~\eqref{eqmirror}, which read $\Lambda^z_{(-,s)}(\theta)=-\,\Lambda^z_{(+,s)}(\pi-\theta)$ and $\ell_{(-,s)}(\theta)=\ell_{(+,s)}(\pi-\theta)$.

For the cases (b) and (c), the internode overlaps of Eq.~\eqref{eqoverlap3} depend on the orientation of the $z$ axis, and the rate of Eq.~\eqref{eqtau3} is not axisymmetric about $\bB$. The mirror relations of Eq.~\eqref{eqmirror} are not available. Only the node $+$ is expanded, since inversion fixes $\bB$ for every orientation and gives Eq.~\eqref{eqinv3}. The moments of the node $-$ then follow from Eq.~\eqref{eqmomJ3}, and the ansatz is
\begin{align}
\Lambda^j_{(+,s)}(\hat{\bk})=\hat{\mathcal R}_{(+,s)}\!\left[h^j_{(+,s)}\right]+\sum_{s'}\sum_{p,p'}\Big(Q^{(+,s)(+,s')}_{pp'}-\varepsilon_{p'}\,Q^{(+,s)(-,s')}_{pp'}\Big)\,\mathcal M^j_{(+,s'),p'}\,\hat{\mathcal R}_{(+,s)}\!\left[f^{(+,s)}_p\right],
\label{eqansatz3bc}
\end{align}
with $j=x,y,z$ and the signs $\varepsilon_{p'}$ of Eq.~\eqref{eqmomJ3}. The number constraint of Eq.~\eqref{eqnumber} is satisfied identically, since $\epsilon_0=+1$, and the unknowns are not reduced further by it.

For case (c), $\bB=B\left(\sin\theta_B,0,\cos\theta_B\right)$ with $0<\theta_B<\pi/2$, only inversion fixes $\bB$. All the fifteen moments $\mathcal M^j_{(+,s),p}$ are independent for each $j$, with nine on the sheet $(+,1)$ and six on the sheet $(+,0)$. For case (b), $\bB=B\,\boldsymbol{\hat x}$, the rotation $C_{2x}$ gives $\Lambda^x_{(+,s)}(\pi-\theta,-\varphi)=\Lambda^x_{(+,s)}(\theta,\varphi)$ and $\Lambda^{y,z}_{(+,s)}(\pi-\theta,-\varphi)=-\,\Lambda^{y,z}_{(+,s)}(\theta,\varphi)$. The moments that are odd under this relation vanish [cf. Appendix~\ref{secappsym3}]. This leaves nine unknowns for $j=x$, with five on the sheet $(+,1)$ and four on the sheet $(+,0)$, and six unknowns for $j=y,z$, with four and two, respectively. The limits $\theta_B\to0$ and $\theta_B\to\pi/2$ of case (c) reduce to the cases (a) and (b).

%%%%%%%%%%%%%%%
\subsection{Limit of the relaxation-time approximation}
\label{secrta3}

%% NOTE: the intranode expressions below were rescaled from (2 pi)^3 to (2 pi)^2 in the denominators to match the normalisation of Eq. (eqsigmw).
%% Checked numerically against the RTA evaluation of Eq. (eqsigmw): the Drude and BC parts of s=1 and the Drude and OMM parts of s=0 were exactly 2 pi too small.
%% The s=1 OMM expression matches only if the BC-OMM cross term is included in it.
%% The internode expressions are those of Ref. ips-exact-spin1.

In the RTA, the intranode and internode contributions are computed with the relaxation times $\tau$ and $\tau_{\rm inter}$, and the quantities that depend on $B$ are expanded up to the order $B^2$ \cite{ips-exact-spin1}. The contributions of the Drude, BC, and OMM parts for the intranode scattering of the sheets $s=1$ and $s=0$ are given by
\begin{align}
&  \sigma_{zz}^{\rm dr} \big \vert_{s=1} =
\frac{e^2\, \tau \, \sqrt{ 4 \, c \, \mu  + v_F } 
\left(\sqrt{v_F}-\sqrt{4 \, c \, \mu +v_F}\right)^2} 
{(2\,\pi)^2 \times 6 \, c^2 \, \sqrt v_F}
\,, \quad
%%%%%%%%%%%%%%%%
\sigma_{zz}^{\rm bc} \big \vert_{s=1} =
\frac{64 \,B^2 \,c^2 \,e^4 \,\tau \, v_F^{\frac{3} {2} } 
\;\sqrt{4 \,c \, \mu + v_F}}
{(2\,\pi)^2 \times  15 \left(\sqrt{v_F}-\sqrt{4\, c \,\mu +v_F}\right)^2}\,,
%%%%%%%%%%%%%%%%%%%%%%%%%%%
\nn & \sigma_{zz}^{\rm omm} \big \vert_{s=1} =
\frac{ - \,2 \,B^2 \,c^2 \,e^4 \,\tau  \,v_F^{\frac{3} {2} } 
\left[  12  \,v_F^2
+ 42 \, c  \,\mu  \, v_F + 16  \,c \, \mu  \,
 \sqrt{ v_F \left(4  \,c  \,\mu + v_F\right)}
 + \sqrt{ v_F^3
   \left(4  \, c  \,\mu +v_F\right)} \right ]}
%%%%%%%%%%%%%   
   {(2\,\pi)^2 \times  15 \left(4  \,c  \,\mu +v_F\right)^{\frac{3} {2} } \left(\sqrt{v_F}
  -\sqrt{4  \, c  \, \mu  + v_F} \right )^2} \,,
\end{align}
and
\begin{align}
&  \sigma_{zz}^{\rm dr} \big \vert_{s=0} = 
\frac{4 \,e^2\, \mu ^{\frac{3} {2} } \,\tau }  {(2\,\pi)^2 \times  3 \, \sqrt{c \,v_F} } \,,\quad
\sigma_{zz}^{\rm bc} \big \vert_{s=0} = 0\,,\quad
%%%%%%%%%%%%%
\sigma_{zz}^{\rm omm} \big \vert_{s=0} =
\frac{7\, B^2\, \sqrt{c} \,e^4 \,\tau \, v_F^{5/2}}
{ (2\,\pi)^2 \times  30 \, \mu ^{\frac{3} {2} }}\,.
\end{align}
respectively, with $\sigma^{\rm dr}$, $\sigma^{\rm bc}$, and $\sigma^{\rm omm}$ the parts that are independent of the BC and the OMM, due to the BC, and due to the OMM. The internode contribution of the sheet $s=1$ is
\begin{align}
 \sigma_{zz}^{\rm inter} \big \vert_{s=1} & =
\frac{ B^2 \, e^4 \, \tau  \left(c \,  v_F\right)^{\frac{3} {2} }
 \sqrt{4 \,c \,\mu + v_F}
%%%%%%%%%%
  \left(\sqrt{\frac{v_F} {4 \, c \,\mu +v_F}}+6\right) 
  \left [ \sqrt{c\, \mu \, v_F}
   \left ( \sqrt{\frac{v_F}{4 \, c\, \mu +v_F}}+6\right) + v_F\right ]  
 } 
 {36 \, \pi ^2\,\sqrt \mu 
 \left(  v_F^{\frac{3} {2} } - v_F \, \sqrt{4 \, c \, \mu + v_F}
  + 2 \, c \, \mu\, \sqrt{v_F} + \sqrt{c \,\mu\, v_F} 
  \, \sqrt{4 \, c \, \mu + v_F} \right)}
 \left(\frac{ \tau_{\text{\tiny{inter}}} }{\tau }-1\right)  ,\nn
 %%%%%%%%%%%%%%%%%%% 
 \sigma_{zz}^{\rm inter} \big \vert_{s=0} & =
 \frac{B^2 \, e^4 \, \tau \left(c \, v_F\right)^{\frac{3} {2} } v_F \,
 \sqrt{4 \, c \, \mu + v_F}
  \left(\sqrt{\frac{v_F} {c \, \mu}}+\sqrt{\frac{v_F} {4 \, c \, \mu +v_F}}+6\right)}
 {36 \, \pi ^2 \, \sqrt \mu
  \left( v_F^{\frac{3} {2} } - v_F \, \sqrt{4 \, c \, \mu + v_F}
  + 2 \, c \, \mu \, \sqrt{v_F} + \sqrt{c \,\mu \, v_F}
  \, \sqrt{4 \, c \, \mu + v_F} \right)}
 \left(\frac{ \tau_{\text{\tiny{inter}}} }{\tau }-1\right) .
\end{align}
The intranode expressions for the two sheets are those of Ref.~\cite{ips-exact-spin1}, and they are derived in Appendix~\ref{secrtab} from the general formula of Eq.~\eqref{eqrtaquad} for a sheet with a quadratic dispersion [cf. Appendix~\ref{secrtab5}]. The internode contributions follow from the scattering between the two sheets at the same node [cf. Appendix~\ref{secrtab6}]. The expressions are given for the comparison with the solution beyond the RTA.

The RTA is not expected to reproduce the exact solution, and we do not use it as a quantitative reference for the internode term. In particular, the OMM part of the pumping vertex $\Upsilon_s$ of Eq.~\eqref{eqrtab8} need not appear in the exact solution. The exact solution contains the OMM through the radius $k_a(\theta;B)$, the weights $\rho_a$, the sources $h_a$, and the frequency $\omega_a$. The net pumping of a node is $\sum_s\int_{-1}^{1}du\,\rho_s\,h_s$, and it is the quantity that drives the mode that is uniform over the node. To the first order in $B$, the OMM part of this integral is the integral of a total derivative over the closed sheets, and it vanishes. Only the BC part remains, and it is $-\,\eta_B\,\chi\,e\,B\sum_sN_s$ for the node $\chi$. If the mode that is uniform over the node is relaxed with a single time $\tau_{\rm inter}$ in the exact formalism, the internode term is therefore given by the BC part of $\Upsilon_s$ alone. The OMM is then not lost. It enters the exact solution through $\rho_s$, $h_s$, and $\omega_s$ at the higher orders in $B$, and through the in-scattering with the full kernel, which the RTA does not contain.

For noncollinear fields, the RTA drops the in-scattering term of Eq.~\eqref{eqell} and keeps the LFO. For $\bE\perp\bB$, the sources $h^i_a=D_a^{-1}\,w_{a,i}$ with $i=x,y$ carry no pumping term, since $B_x=B_y=0$. The internode relaxation time $\tau_{\rm inter}$ therefore does not enter the sector $m=1$, and it affects only the longitudinal response. With the constant time $\tau$, the mean free-path is $\ell_a=\left(q_a/D_a\right)\sin\theta\,\tau/\left(1+i\,\omega_a\,\tau\right)$, and Eq.~\eqref{eqsigperp}, summed over the two nodes, gives
\begin{align}
\sigma_\perp-i\,\sigma_H^{\rm B}=\frac{e^2\,\tau}{4\,\pi^2}\sum_s\int_{-1}^{1}du\left(1-u^2\right)\rho_s\,\frac{\left(q_s/D_s\right)^2}{1+i\,\omega_s\,\tau}\,.
\label{eqrtaperp3}
\end{align}
At $B=0$, this reduces to $\sigma_\perp=\sigma_\parallel=\left(e^2\,\tau/3\,\pi^2\right)\sum_s k_s^2\,w^0_s$, which is twice the Drude expressions above for one node. No planar-Hall current flows at $B=0$ in the RTA. For $\eta_B=\eta_O=0$, the integrand is independent of $u$ and the result is exact in $B$,
\begin{align}
\sigma_\perp-i\,\sigma_H=\frac{e^2\,\tau}{3\,\pi^2}\sum_s\frac{k_s^2\,w^0_s}{1+i\,\omega_s\,\tau}\,,\quad
\omega_s=\frac{e\,B\,w^0_s}{k_s}\,.
\label{eqrtalfo3}
\end{align}
To expand in $B$ with the BC and the OMM, we consider a sheet with the dispersion $\epsilon_s=b_a\,k+g_a\,k^2$ measured from its node. Let $\bar v_a=\partial_k\epsilon_s$ at $k_a$, so that $\partial_k^2\epsilon_s=\left(\bar v_a-b_a\right)/k_a$, and let $N_a$, $G_a$, and $v_\nu$ be the numbers and the velocity of Eq.~\eqref{eqbcomm}. Expanding the integrand of Eq.~\eqref{eqrtaperp3} to the third order in $B$ [cf. Appendix~\ref{secapprta3}], we obtain for one node
\begin{align}
\sigma^{\rm RTA}_{\perp,a}=\;&\frac{e^2\,\tau\,k_a^2\,\bar v_a}{6\,\pi^2}+\frac{e^4\,\tau\,B^2}{120\,\pi^2\,k_a^2}\Big[\eta_B\,N_a^2\,\bar v_a-\eta_B\,\eta_O\,N_a\,G_a\,v_\nu\,\frac{4\,\bar v_a-b_a}{\bar v_a} -\,\eta_O\,G_a^2\,v_\nu^2\,\frac{\left(\bar v_a-b_a\right)^2}{2\,\bar v_a^3}-20\,\tau^2\,k_a^2\,\bar v_a^3\Big]+\mathcal O\!\left(B^4\right),
\label{eqrtaexp3}
\end{align}
and for the Boltzmann part of the Hall conductivity
\begin{align}
\sigma^{\rm B,RTA}_{H,a}=\,& \frac{e^3\,\tau^2\,k_a\,\bar v_a^2}{6\,\pi^2}\,B-\frac{e^5\,\tau^2\,B^3}{240\,\pi^2\,k_a^3\,\bar v_a}  \Big[40\,\tau^2\,k_a^2\,\bar v_a^5-6\,\eta_B\,N_a^2\,\bar v_a^3 
+8\,\eta_B\,\eta_O\,N_a\,G_a\,v_\nu\,\bar v_a\left(3\,\bar v_a-b_a\right)-\eta_O\,G_a^2\,v_\nu^2\left(11\,\bar v_a-9\,b_a\right)\Big]
\nn & +\mathcal O\!\left(B^5\right).
\label{eqrtaHexp3}
\end{align}
The terms linear in $B$ do not depend on $\eta_B$ and $\eta_O$, and $\sigma_{\perp,a}^{\rm RTA}$ has no odd orders. The terms with $\tau^2$ inside the brackets are the expansion of Eq.~\eqref{eqrtalfo3}. The AH part of Eq.~\eqref{eqsigah3} does not depend on $\tau$ and has the expansion
\begin{align}
\sigma^{\rm AH}_{H,a}=\eta_B\,\eta_O\,N_a\,G_a\,v_\nu\left[\frac{e^3\,B}{24\,\pi^2\,k_a\,\bar v_a}+\frac{e^5\,G_a^2\,v_\nu^2\left(8\,\bar v_a^2-5\,b_a\,\bar v_a+b_a^2\right)B^3}{320\,\pi^2\,k_a^5\,\bar v_a^5}\right]+\mathcal O\!\left(B^5\right).
\label{eqrtaAH3}
\end{align}
The corresponding expansion of the longitudinal conductivity is Eq.~\eqref{eqrtaquad} with the replacement of $\mathsf u_a$ by $b_a$ in the notation of that equation, and the difference of the transverse and longitudinal conductivities of one node, which governs the planar-Hall current of Eq.~\eqref{eqcurr}, is
\begin{align}
\sigma^{\rm RTA}_{\perp,a}-\sigma^{\rm RTA}_{\parallel,a}= &\,
\frac{e^4\,\tau\,B^2}{120\,\pi^2\,k_a^2}\Big[-7\,\eta_B\,N_a^2\,\bar v_a-2\,\eta_B\,\eta_O\,N_a\,G_a\,v_\nu\,\frac{\bar v_a-4\,b_a}{\bar v_a}
 -\,\eta_O\,G_a^2\,v_\nu^2\,\frac{4\,\bar v_a^2+2\,b_a\,\bar v_a-b_a^2}{\bar v_a^3}-20\,\tau^2\,k_a^2\,\bar v_a^3\Big]+\mathcal O\!\left(B^4\right).
\label{eqrtaph3}
\end{align}
It vanishes at $B=0$. For the two conjugate QTSM nodes, the totals are twice the sums over $s$ of these contributions, with $b_a=v_F$, $\bar v_a=w^0_1$, $N_a=2$, and $G_a=1$ for $s=1$, and with $b_a=0$, $\bar v_a=w^0_0$, $N_a=0$, and $G_a=2$ for $s=0$, and with $v_\nu=v_F$ for both.

%%%%%%%%%%%%%%%
\subsection{Results from beyond-RTA treatement}
\label{secres3}

For $\bB=B\,\boldsymbol{\hat z}$, the angle $\vartheta$ between $\bE$ and $\bB$ enters only through the weights of Eq.~\eqref{eqcurr}, and the problem reduces to the three functions $\sigma_\parallel$, $\sigma_\perp$, and $\sigma_H$. This is case (a) of Table~\ref{tabgeom3}. We abbreviate $k_s=k_{\Gamma,s}$, $\rho_s=\rho_{\Gamma,s}$, and $w^0_s=\partial_k\epsilon_{\Gamma,s}$ at $k_s$, with $w^0_1=v_F\left(1+2\,c\,k_1\right)$ and $w^0_0=2\,c\,v_F\,k_0$. The rates $\tau_s^{-1}$ at $B=0$ are those given below Eq.~\eqref{eqmomsys3}. The vertex corrections of the sheet $s=1$ differ for the two orientations of $\bE$, since the amplitudes enter as $\gamma^-_{11}$ for $\bE\parallel\bB$ and as $\gamma^+_{11}$ for $\bE\perp\bB$. We therefore define
\begin{align}
 \Xi^{\parallel}_1=\tau_1^{-1}-\frac23\,\gamma^{-}_{11}\,\rho_1\,,\quad
\Xi^{\perp}_1=\tau_1^{-1}-\frac23\,\gamma^{+}_{11}\,\rho_1\,,\quad
\Xi^{\parallel}_0=\Xi^{\perp}_0=\tau_0^{-1}\,,\quad
 \mathcal A^{\parallel}_s=\frac{w^0_s}{\Xi^{\parallel}_s}\,,\quad
\mathcal A^{\perp}_s=\frac{w^0_s}{\Xi^{\perp}_s}\,.
\label{eqxiamp3}
\end{align}
The amplitudes $\mathcal A^{\parallel}_s$ and $\mathcal A^{\perp}_s$ are the coefficients of $u$ and of $\sin\theta$ in the mean free-path at $B=0$ for the two orientations of $\bE$, respectively. The intermediate steps are given in Appendix~\ref{secappres3}. 

\subsubsection{Longitudinal magnetoconductivity for parallel fields}
\label{secresll3}

For $\bB\parallel\bE$, the results of Ref.~\cite{ips-exact-spin1} apply. There, the change of the longitudinal magnetoconductivity of the sheets with $s=1$ has contributions from the BC and from the OMM, which have opposite signs, and the OMM can reverse the sign of the total for large values of the ratio $\beta^{\rm inter}_{1,1}/\beta^{\rm intra}_{1,1}$. The sheets with $s=0$ are affected by the OMM alone. The RTA does not show these sign changes. In the notation of Eq.~\eqref{eqxiamp3}, the zero-field limit of Eq.~\eqref{eqsigB0} reads $\sigma_\parallel\vert_{B=0}=\left(e^2/3\,\pi^2\right)\sum_s k_s^2\,\mathcal A^\parallel_s$.

\subsubsection{Planar-Hall and Hall response for perpendicular fields and a general angle}
\label{secresph3}

For $B=0$, the sector $m=1$ obeys the equations of the sector $m=0$ with $\gamma^-_{11}$ replaced by $\gamma^+_{11}$ [cf. Appendix~\ref{secappzero3}]. The conductivities are
\begin{align}
\sigma_\perp\big\vert_{B=0}=\frac{e^2}{3\,\pi^2}\sum_s k_s^2\,\mathcal A^{\perp}_s\,,\quad
\sigma_H\big\vert_{B=0}=0\,,
\label{eqsigperp0}
\end{align}
and the difference of the transverse and longitudinal conductivities is
\begin{align}
\left(\sigma_\perp-\sigma_\parallel\right)\big\vert_{B=0}=\frac{4\,e^2}{9\,\pi^2}\,\frac{\left(\beta^{\rm inter}_{11}\right)^2k_1^4}{\Xi^{\parallel}_1\,\Xi^{\perp}_1}\,.
\label{eqdsig0}
\end{align}
The difference is positive and vanishes for $\beta^{\rm inter}_{11}=0$. By Eq.~\eqref{eqcurr}, a planar-Hall current $j_{\rm PH}=\left(\sigma_\perp-\sigma_\parallel\right)\sin\vartheta\,\cos\vartheta\,E$ flows already at $B=0$ for an oblique $\bE$. It originates from the dependence of the internode overlap on the orientation of the $z$ axis, which was discussed below Eq.~\eqref{eqsigB0}, and not from the magnetic field.

To isolate the effect of the LFO, we first switch off the BC and the OMM, $\eta_B=\eta_O=0$. The FS radii, the weights $\rho_s$, and the rates $\tau_s^{-1}$ are then independent of $u$ and of $B$, and the cyclotron frequency is $\omega_s=e\,B\,w^0_s/k_s$. The sector $m=0$ is annihilated by the LFO, and $\sigma_\parallel$ keeps its zero-field value. The sector $m=1$ is solved exactly by $\ell_s=w^0_s\sin\theta/\left(\Xi^{\perp}_s+i\,\omega_s\right)$ for every $B$ [cf. Appendix~\ref{secapplfo}], which gives
\begin{align}
\sigma_\perp-i\,\sigma_H=\frac{e^2}{3\,\pi^2}\sum_s\frac{k_s^2\,w^0_s}{\Xi^{\perp}_s+i\,\omega_s}\,.
\label{eqlfoexact}
\end{align}
The real and imaginary parts are the Lorentzian and dispersive functions of $\omega_s/\Xi^{\perp}_s$, with the vertex correction of the sheet $s=1$ absorbed in $\Xi^{\perp}_1$. For weak fields, Eq.~\eqref{eqlfoexact} gives
\begin{align}
\sigma_\perp=\sigma_\perp\big\vert_{B=0}-\frac{e^4\,B^2}{3\,\pi^2}\sum_s\left(\mathcal A^{\perp}_s\right)^3+\mathcal O\!\left(B^4\right),\quad
\sigma_H=\frac{e^3\,B}{3\,\pi^2}\sum_s k_s\left(\mathcal A^{\perp}_s\right)^2+\mathcal O\!\left(B^3\right).
\label{eqlfosmall}
\end{align}
For $\omega_s\gg\Xi^{\perp}_s$, the conductivity $\sigma_\perp$ decays as $B^{-2}$, and the Hall conductivity saturates to $\sigma_H\to e\left(k_1^3+k_0^3\right)/\left(3\,\pi^2\,B\right)$, which is independent of the scattering amplitudes. The planar-Hall current then approaches $-\,\sigma_\parallel\sin\vartheta\,\cos\vartheta\,E$, since $\sigma_\parallel$ does not depend on $B$ in this limit. The solution is not a superposition of the responses to the components of $\bB$ along and perpendicular to $\bE$, since the angle enters only through the weights of Eq.~\eqref{eqcurr}.

With the BC and the OMM, the weights, the rates, and the vertex corrections acquire a dependence on $u$ and on $B$. The Hall conductivity to the leading order in $B$ is nevertheless unchanged, since it is determined by the imaginary part of the sector $m=1$ at the first order in $\omega_s$ [cf. Appendix~\ref{secappHall3}]. The Boltzmann part is
\begin{align}
\sigma_H^{\rm B}\simeq\frac{e^3\,B}{3\,\pi^2}\sum_s k_s\left(\mathcal A^{\perp}_s\right)^2\,,
\label{eqsigHB3}
\end{align}
and the AH part of Eq.~\eqref{eqsigah3} adds to it. The total Hall conductivity is
\begin{align}
\sigma_H\simeq\frac{e^3\,B}{3\,\pi^2}\left[\sum_s k_s\left(\mathcal A^{\perp}_s\right)^2+\frac{\eta_B\,\eta_O}{2\,k_1\left(1+2\,c\,k_1\right)}\right].
\label{eqsigH3}
\end{align}
The first term does not depend on $\eta_B$ and $\eta_O$ at this order, and it coincides with the linear term of Eq.~\eqref{eqlfosmall}. The second term requires both the BC and the OMM, and the sheet $(\chi,0)$ does not contribute to it, since its BC vanishes.

The amplitudes enter the results above only through $\Xi^{\parallel}_s$, $\Xi^{\perp}_s$, and $\tau_0^{-1}$. The rates and $\sigma_\perp$ involve the sums $\gamma^+_{ss'}$, and only the vertex correction of $\sigma_\parallel$ involves $\gamma^-_{11}$. The zero-field anisotropy of Eq.~\eqref{eqdsig0} therefore isolates the internode amplitude $\beta^{\rm inter}_{11}$ of the sheets with $s=1$. The zero-field conductivity is isotropic for $\beta^{\rm inter}_{11}=0$, whatever the values of $\beta^{\rm inter}_{10}$ and $\beta^{\rm inter}_{00}$, since these amplitudes enter only through $\gamma^+$. The sheet $s=0$ has no vertex correction in any sector. The Boltzmann Hall term of Eq.~\eqref{eqsigHB3} scales as the inverse square of the scattering strength through $\mathcal A^{\perp}_s$, whereas the AH term does not depend on the amplitudes. The AH term thus dominates for strong scattering, and the Boltzmann term dominates for weak scattering.

\subsubsection{Expansion to the third order in $B$}
\label{secresB33}

The field enters the equations of the sectors only through $\rho_a$, $\tau_a^{-1}$, $\omega_a$, and the sources, and these are analytic in $B$ at $B=0$. By Eq.~\eqref{eqons}, the terms of the Boltzmann conductivities that are allowed by symmetry are
\begin{align}
\sigma_\parallel=\sigma_\parallel^{(0)}+\sigma_\parallel^{(2)}\,B^2+\mathcal O(B^4)\,,\quad
\sigma_\perp=\sigma_\perp^{(0)}+\sigma_\perp^{(2)}\,B^2+\mathcal O(B^4)\,,\quad
\sigma_H^{\rm B}=\sigma_H^{(1)}\,B+\sigma_H^{(3)}\,B^3+\mathcal O(B^5)\,.
\label{eqexp3}
\end{align}
Eq.~\eqref{eqcurr} gives the currents with the same orders. For example, the planar-Hall current is $$ j_{\rm PH}=\left[\sigma_\perp^{(0)}-\sigma_\parallel^{(0)}+\left(\sigma_\perp^{(2)}-\sigma_\parallel^{(2)}\right)B^2\right]\sin\vartheta\,\cos\vartheta\,E+\mathcal O(B^4), $$ where the zero-field difference is given by Eq.~\eqref{eqdsig0}.

Let $\tilde\lambda_a=1$ for the sheets $(\chi,1)$ and $\tilde\lambda_a=0$ for the sheets $(\chi,0)$, and let $v_a=v_F$ and $c_a=c$ for all the sheets. We define $\tilde\nu_a=\tilde\lambda_a+2\,c_a\,k_a$, such that $w_a=v_a\,\tilde\nu_a$ gives $w^0_1$ and $w^0_0$ of Sec.~\ref{secres3}. The numbers $N_a$ and $G_a$ are $2$ and $1$ for $s=1$, and $0$ and $2$ for $s=0$. The symbol $\chi_\nu=\chi$ is the chirality of the node of the sheet $a$. The radius $k_a(\theta;B)$ is the positive root of the cubic equation of Appendix~\ref{secappqtsm} with this chirality, and it is continued from $k_a=k_{\Gamma,s}$ at $B=0$. Its expansion is
%%%%%%%%%%%%%%%
\begin{align}
\label{eqkser}
& k_a(\theta;B)=k_a+\sum_{n=1}^{3}\mathfrak k_a^{(n)}\left(e\,B\right)^n\cos^n\theta+\mathcal O(B^4)\,,\quad
%%%%%%%%%%%%%%
 \mathfrak k_a^{(1)}=-\frac{\eta_O\,\chi_\nu\,G_a}{2\,k_a\,\nu_a}\,,\quad
\mathfrak k_a^{(2)}=-\frac{\eta_O^2\,G_a^2\left(\tilde\lambda_a+3\,c_a\,k_a\right)}{4\,k_a^3\,\nu_a^3}\,,\nn
& \mathfrak k_a^{(3)}=-\frac{\eta_O^3\,\chi_\nu\,G_a^3\left(2\,\tilde\lambda_a^2+11\,c_a\,\tilde\lambda_a\,k_a+16\,c_a^2\,k_a^2\right)}{8\,k_a^5\,\nu_a^5}\,.
\end{align}
The Fermi-sea term of Eq.~\eqref{eqsigmw} is Eq.~\eqref{eqsigah3}, which reads $\sigma_H^{\rm AH}=-\left(\eta_B\,e^2/8\,\pi^2\right)\sum_a\chi_\nu\,N_a\int_0^\pi d\theta\,\sin\theta\,\cos\theta\,k_a(\theta;B)$, where the sum runs over the four sheets. With Eq.~\eqref{eqkser}, it gives the AH part to the third order,
\begin{align}
\sigma_H^{\rm AH}\simeq\frac{\eta_B\,\eta_O\,e^3\,B}{24\,\pi^2}\sum_a\frac{N_a\,G_a}{k_a\,\nu_a}+\frac{\eta_B\,\eta_O^3\,e^5\,B^3}{160\,\pi^2}\sum_a\frac{N_a\,G_a^3\left(2\,\tilde\lambda_a^2+11\,c_a\,\tilde\lambda_a\,k_a+16\,c_a^2\,k_a^2\right)}{k_a^5\,\nu_a^5}\,.
\label{eqsigHAH3}
\end{align}
The sheets with $s=0$ do not contribute, since $N_a=0$, and the two sheets with $s=1$ contribute equally. With $k_1=k_{\Gamma,1}$, the sums are
\begin{align}
\sigma_H^{\rm AH}\simeq\frac{\eta_B\,\eta_O\,e^3\,B}{6\,\pi^2\,k_1\left(1+2\,c\,k_1\right)}+\frac{\eta_B\,\eta_O^3\,e^5\,B^3\left(2+11\,c\,k_1+16\,c^2\,k_1^2\right)}{40\,\pi^2\,k_1^5\left(1+2\,c\,k_1\right)^5}\,.
\end{align}
The first term is the leading term below Eq.~\eqref{eqsigah3}. Neither term depends on the scattering amplitudes.

The Taylor coefficients in $e\,B$ of the weights, of the sources, and of the frequency are obtained by inserting Eq.~\eqref{eqkser} into Eq.~\eqref{eqwqd}. The coefficient of $\left(e\,B\right)^n$ is a polynomial in $\cos\theta$ of the degree $n$ at most for $\rho_a$ and for $q_a/D_a$, of the degree $n+1$ at most for the source $h_a$ of Eq.~\eqref{eqlongsol}, and of the degree $n-1$ at most for $\omega_a$. The terms of the first order are
\begin{align}
& \rho_a=\frac{k_a^2}{w_a}-\chi_\nu\,\frac{\eta_O\,G_a\,\tilde\lambda_a+\eta_B\,N_a\,\nu_a^2}{2\,w_a\,\nu_a^2}\,e\,B\,\cos\theta+\mathcal O(B^2)\,,\quad
\omega_a=\frac{w_a}{k_a}\,e\,B+\mathcal O(B^2)\,,\nn
& \frac{q_a}{D_a}=w_a+\chi_\nu\,\frac{v_a\left[\eta_B\,N_a\,\nu_a^2-2\,\eta_O\,G_a\left(\tilde\lambda_a+3\,c_a\,k_a\right)\right]}{2\,k_a^2\,\nu_a}\,e\,B\,\cos\theta+\mathcal O(B^2)\,,\nn
& h_a=w_a\cos\theta+\chi_\nu\,\frac{v_a\left[\eta_O\,G_a\left(\nu_a-2\left(\tilde\lambda_a+3\,c_a\,k_a\right)\cos^2\theta\right)-\eta_B\,N_a\,\nu_a^2\sin^2\theta\right]}{2\,k_a^2\,\nu_a}\,e\,B+\mathcal O(B^2)\,.
\label{eqing1}
\end{align}
The rate $\tau_a^{-1}$ of Eq.~\eqref{eqtau} follows from the weights, and its coefficients are $\tau_a^{-1,(n)}=\sum_{a'}\int_0^\pi d\theta'\,\sin\theta'\,\rho_{a'}^{(n)}(\theta')\,\mathcal K^{(0)}_{aa'}(\theta,\theta')$ with the kernel of Eq.~\eqref{eqsector}. This kernel obeys $\mathcal K^{(m)}_{aa'}(\theta,\theta')=\mathcal K^{(m)}_{a'a}(\theta',\theta)$.

To obtain the terms of the second and the third order from the solutions at $B=0$, we use a variational form of the sector equation. For two sets of functions $f=\left(f_a(\theta)\right)$ and $g=\left(g_a(\theta)\right)$ on the four sheets, let $\mathcal G^{(m)}_B$ be the bilinear form and $\mathfrak H^{(m)}_B$ the linear form as follows:
\begin{align}
& \mathcal G^{(m)}_B[f,g]=\sum_a\int_0^\pi d\theta\,\sin\theta\,\rho_a\left(\tau_a^{-1}+i\,m\,\omega_a\right)f_a\,g_a
 -\sum_{a,a'}\int_0^\pi d\theta\,\sin\theta\int_0^\pi d\theta'\,\sin\theta'\,\rho_a(\theta)\,\rho_{a'}(\theta')\,\mathcal K^{(m)}_{aa'}(\theta,\theta')\,f_a(\theta)\,g_{a'}(\theta')\,,\nn
& \mathfrak H^{(m)}_B[f]=\sum_a\int_0^\pi d\theta\,\sin\theta\,\rho_a\,h^{(m)}_a\,f_a\,.
\label{eqforms}
\end{align}
%%%%%%%%
$\mathcal G^{(m)}_B$ is symmetric, since it has no complex conjugation. Multiplying Eq.~\eqref{eqsector} by $\sin\theta\,\rho_a\,f_a$ and summing over the sheets gives $\mathcal G^{(m)}_B[f,\Lambda^{(m)}]=\mathfrak H^{(m)}_B[f]$ for every $f$. The functional $\mathcal F[f]=2\,\mathfrak H^{(m)}_B[f]-\mathcal G^{(m)}_B[f,f]$ is therefore stationary at $f=\Lambda^{(m)}$, and its value there is $\mathfrak H^{(m)}_B[\Lambda^{(m)}]$, which gives the conductivities through Eqs.~\eqref{eqsigpar} and \eqref{eqsigperp}. A trial function that deviates from $\Lambda^{(m)}$ at the order $B^2$ changes $\mathcal F$ only at the order $B^4$. Let $\mathcal G_n$ and $\mathfrak H_n$ be the coefficients of $B^n$ in the expansions of the forms, let $\Lambda_0$ be the solution at $B=0$ of Secs.~\ref{secresll3} and \ref{secresph3}, and let $\Lambda_1$ be defined by
\begin{align}
\mathcal G_0[f,\Lambda_1]=\mathfrak H_1[f]-\mathcal G_1[f,\Lambda_0]\quad\text{for every }f\,.
\label{eqlam1}
\end{align}
%%%%%%%%%
This is the first-order term of Eq.~\eqref{eqsector}, which reads
$$ \tau_a^{-1,(0)}\Lambda_{1,a}+\left(\tau_a^{-1,(1)}+i\,m\,\omega_a^{(1)}\right)\Lambda_{0,a}=h_a^{(1)}+\sum_{a'}\int_0^\pi d\theta'\,\sin\theta'\left[\rho_{a'}^{(0)}\,\mathcal K^{(m)}_{aa'}\,\Lambda_{1,a'}+\rho_{a'}^{(1)}\,\mathcal K^{(m)}_{aa'}\,\Lambda_{0,a'}\right] .$$
The trial function $\Lambda_0+B\,\Lambda_1$ gives $\mathfrak H_B[\Lambda^{(m)}]=\sum_{n=0}^{3}S_n\,B^n+\mathcal O(B^4)$ with
\begin{align}
& S_0=\mathfrak H_0[\Lambda_0]\,,\quad
S_1=2\,\mathfrak H_1[\Lambda_0]-\mathcal G_1[\Lambda_0,\Lambda_0]\,,
\quad S_2=2\,\mathfrak H_2[\Lambda_0]+2\,\mathfrak H_1[\Lambda_1]-\mathcal G_2[\Lambda_0,\Lambda_0]-2\,\mathcal G_1[\Lambda_0,\Lambda_1]-\mathcal G_0[\Lambda_1,\Lambda_1]\,,\nn
%%%%%%%%%%%%%%%%
& S_3=2\,\mathfrak H_3[\Lambda_0]+2\,\mathfrak H_2[\Lambda_1]-\mathcal G_3[\Lambda_0,\Lambda_0]-2\,\mathcal G_2[\Lambda_0,\Lambda_1]-\mathcal G_1[\Lambda_1,\Lambda_1]\,.
\label{eqSn}
\end{align}
The coefficients of Eq.~\eqref{eqexp3} are then
%%%%%%%%%%%%%%
\begin{align}
& \sigma_\parallel^{(0)}=\frac{e^2 \,S^{(0)}_0}{4\,\pi^2}\,,\quad
\sigma_\parallel^{(2)}=\frac{e^2\,S^{(0)}_2}{4\,\pi^2}\,,\quad
\sigma_\perp^{(0)}=\frac{e^2 \, \Re S^{(1)}_0}{8\,\pi^2}\,,\quad
\sigma_\perp^{(2)}=\frac{e^2\,\Re S^{(1)}_2}{8\,\pi^2}\,,\quad
\sigma_H^{(1)}=\frac{ -e^2}{8\,\pi^2}\Im S^{(1)}_1\,,
\quad \sigma_H^{(3)}=\frac{ -e^2\, \Im S^{(1)}_3}{8\,\pi^2}\,,
\label{eqsigcoef}
\end{align}
%%%%%%%%%%
where the superscript of $S$ is the value of $m$. By Eq.~\eqref{eqons}, the coefficients $S^{(0)}_1$ and $S^{(0)}_3$ vanish, and so do $\Re S^{(1)}_1$, $\Re S^{(1)}_3$, $\Im S^{(1)}_0$, and $\Im S^{(1)}_2$. At the first order, only the term $i\,\rho_a\,\omega_a^{(1)}\,\Lambda_0^2$ of $\mathcal G_1$ contributes to $\Im S^{(1)}_1$, and the coefficient $\sigma_H^{(1)}$ of Eq.~\eqref{eqsigcoef} reproduces Eq.~\eqref{eqsigHB}.

The two nodes are exchanged by the reflection of Eq.~\eqref{eqmirror}, and this relation holds at every order in $B$. For $m=0$, the solutions obey $\Lambda_{(-,s)}(\theta)=-\Lambda_{(+,s)}(\pi-\theta)$ at every order, and the constant zero mode violates this relation. The function $\Lambda_1$ therefore has no zero mode, and Eq.~\eqref{eqlam1} has a unique solution in this sector. The two nodes contribute equally to the conductivities, and the quantities of the node $+$ suffice. For this node, the function $\Lambda_0=\mathcal A^{\parallel}_s\cos\theta$ is odd and $\Lambda_1$ is even in $\cos\theta$. For $m=1$, the solutions obey $\ell_{(-,s)}(\theta)=\ell_{(+,s)}(\pi-\theta)$, and we write $\Lambda_1=\Lambda_1^{\rm R}+i\,\Lambda_1^{\rm I}$ for the node $+$. The function $\Lambda_0=\mathcal A^{\perp}_s\sin\theta$ and the imaginary part $\Lambda_1^{\rm I}$ are even, and the real part $\Lambda_1^{\rm R}$ is odd in $\cos\theta$.

For $\eta_B=\eta_O=0$, the coefficients of Eq.~\eqref{eqexp3} follow from Eq.~\eqref{eqlfoexact}. The conductivity $\sigma_\parallel$ does not depend on $B$, and $\sigma_\perp-i\,\sigma_H=\left(e^2/3\,\pi^2\right)\sum_s k_s^2\,w^0_s/\left(\Xi^{\perp}_s+i\,\omega_s\right)$ with $\omega_s=e\,B\,w^0_s/k_s$. The expansion in $B$ gives $\sigma_\parallel^{(2)}=0$ and
\begin{align}
\sigma_\perp^{(2)}=-\frac{e^4}{3\,\pi^2}\sum_s\left(\mathcal A^{\perp}_s\right)^3,\quad
\sigma_H^{(1)}=\frac{e^3}{3\,\pi^2}\sum_s k_s\left(\mathcal A^{\perp}_s\right)^2,\quad
\sigma_H^{(3)}=-\frac{e^5}{3\,\pi^2}\sum_s\frac{\left(\mathcal A^{\perp}_s\right)^4}{k_s}\,.
\end{align}
The first two are the coefficients of Eq.~\eqref{eqlfosmall}, and the third is new. Eqs.~\eqref{eqSn} and \eqref{eqsigcoef} reproduce all three. The coefficients $S_2^{(0)}$ and $S_2^{(1)}$ are given in terms of elementary quantities in Appendix~\ref{secappB33}.

\subsubsection{Numerical results}
\label{secresnum3}

The figures show the exact solution of the moment system of Eq.~\eqref{eqmomsys} for $\bB=B\,\boldsymbol{\hat z}$, which is case (a) of Table~\ref{tabgeom3}. The field enters through $\alpha=e\,B/(2\,k_1^2)$, where $k_1$ is the radius of the sheets with $s=1$ at $B=0$. We take $\mu\,\tau_0=15$ with $\tau_0$ the transport time of the sheet $(+,1)$ at $B=0$, $c=0.1\,v_F/\mu$, and one scattering amplitude for all pairs of bands within a node and one for all pairs between the nodes. Their ratio is $r_\beta=\beta^{\rm inter}/\beta^{\rm intra}$, and $r_\beta=1/2$ unless stated otherwise. All the quantities are normalised to the corresponding values at $B=0$, and the numerical implementation is described in Appendix~\ref{secnum}. Both signs of $\alpha$ are shown. By Eq.~\eqref{eqons} and the axisymmetric form of the tensor, $\sigma_\parallel$ and $\sigma_\perp$ are even and $\sigma_H$ is odd in $\alpha$. The curves display this symmetry, and the diagonal response of this set-up has no term odd in $B$.

\begin{figure}[t!]
\centering
\subfigure[\label{figqtsma}]{\includegraphics[width=0.85\textwidth]{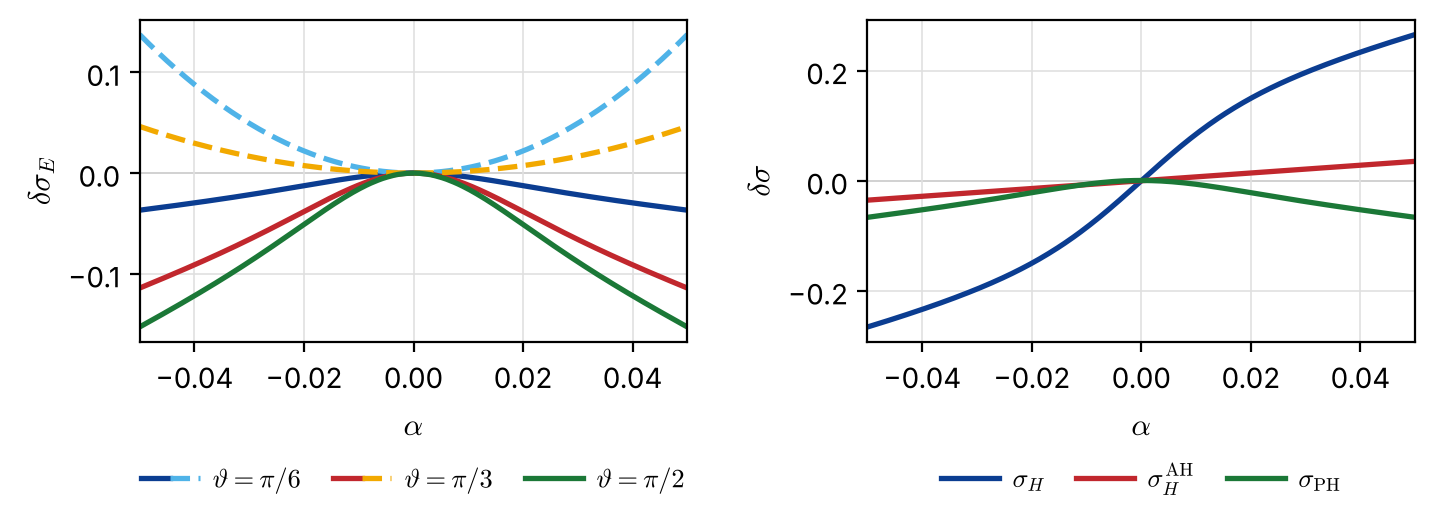}}\\
\subfigure[\label{figqtsmratio}]{\includegraphics[width=0.85\textwidth]{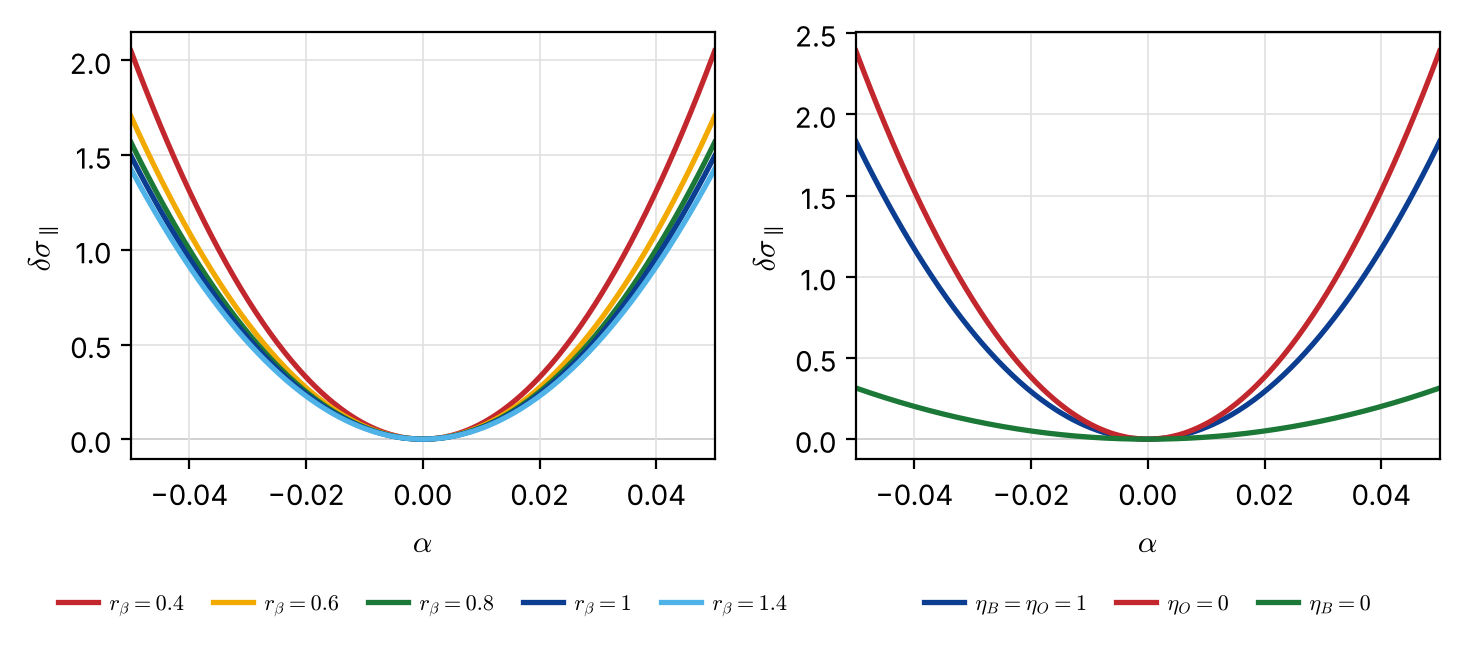}}
\caption{Two conjugate QTSM nodes for $\bB\parallel\hat{\boldsymbol z}$. (a) Left: $\delta\sigma_E$ for $\bE$ at the angle $\vartheta$ to $\bB$. The solid curves are the exact solution for $\vartheta=\pi/6$ (navy), $\pi/3$ (crimson), and $\pi/2$ (green). The dashed curves are the truncated solution multiplied by $10^2$ for $\vartheta=\pi/6$ (sky) and $\pi/3$ (amber). Right: $\sigma_H$ (navy), $\sigma_H^{\rm AH}$ multiplied by $10^2$ (crimson), and the planar-Hall change for $\vartheta=\pi/3$ (green), in units of the value of $\sigma_\perp$ at $B=0$. (b) For $\bE\parallel\bB$, $\delta\sigma_\parallel$ is multiplied by $10^3$. Left: $r_\beta=0.4$ (crimson), $0.6$ (amber), $0.8$ (green), $1$ (navy), and $1.4$ (sky). Right: $r_\beta=1/2$ with $\eta_B=\eta_O=1$ (navy), $\eta_O=0$ (crimson), and $\eta_B=0$ (green).}
\label{figqtsmbz}
\end{figure}

The solution with $\hat{\mathcal R}_a\to\tau_a$, which drops the LFO, is compared with the exact one only in the left panel of Fig.~\ref{figqtsma}. For an oblique $\bE$, the exact change of the current along $\bE$ is negative and grows in magnitude as $\vartheta$ increases. The truncated change has the opposite sign and is smaller by orders of magnitude. It is multiplied by a common factor, and it is not drawn for $\vartheta=\pi/2$, where it is negligibly small. The change is thus the classical LFO contribution, which the truncated solution misses almost entirely. In the right panel, the Hall response is dominated by the Boltzmann part, and the AH part is small and grows linearly with $\alpha$. The planar-Hall change is negative, in accordance with the decrease of $\sigma_\perp$ and the nearly constant $\sigma_\parallel$ in Eq.~\eqref{eqcurr}.

For $\bE\parallel\bB$, the change of $\sigma_\parallel$ is positive and quadratic in $\alpha$ for all the ratios in Fig.~\ref{figqtsmratio}, and it decreases with increasing $r_\beta$. No change of sign occurs in this range, in contrast to the Weyl case. The response for $\bE\perp\bB$ does not depend on $r_\beta$ for either $\sigma_\perp$ or $\sigma_H$, as we find numerically, and it is therefore shown with one value of $r_\beta$ in Fig.~\ref{figqtsma}. The right panel of Fig.~\ref{figqtsmratio} separates the contributions of the BC and the OMM to the longitudinal change. Both are positive, and the combined change is smaller than the BC contribution alone, which shows that the cross term is negative. The OMM contribution alone is the smallest of the three.

\begin{figure}[t!]
\centering
\includegraphics[width=\textwidth]{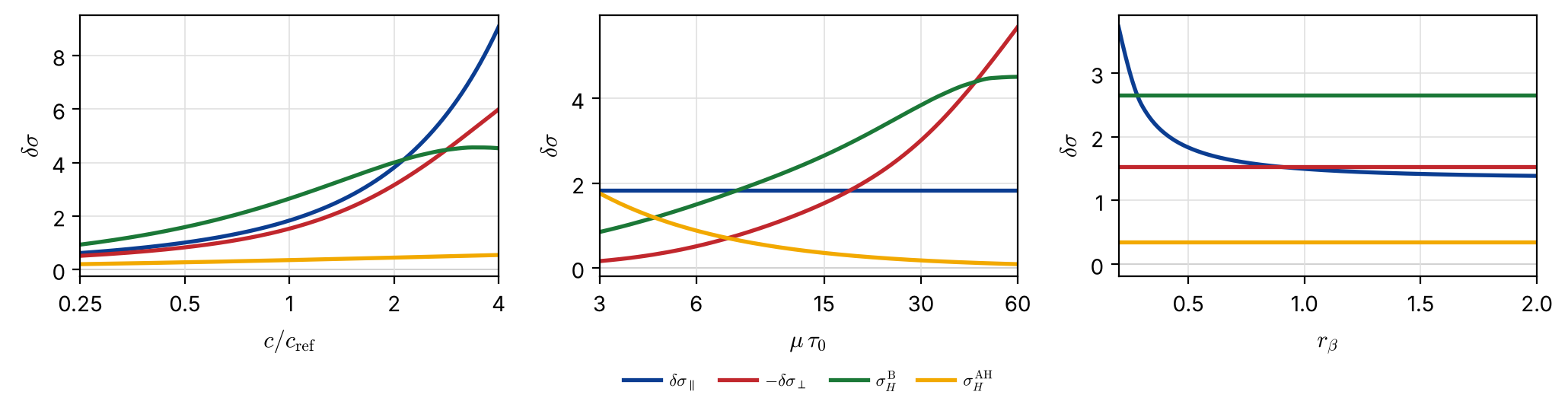}
\caption{Two conjugate QTSM nodes for $\bB\parallel\hat{\boldsymbol z}$ at fixed $\alpha=0.05$. Left, middle, and right: dependence on $c/c_{\rm ref}$ with $c_{\rm ref}=0.1\,v_F/\mu$, on $\mu\,\tau_0$, and on $r_\beta$. In every panel, $\delta\sigma_\parallel$ multiplied by $10^3$ (navy), $-\delta\sigma_\perp$ multiplied by $10$ (crimson), $\sigma_H^{\rm B}$ multiplied by $10$ (green), and $\sigma_H^{\rm AH}$ multiplied by $10^3$ (amber), in units of the values at $B=0$.}
\label{figqtsmscan}
\end{figure}

Figure~\ref{figqtsmscan} shows the dependence on the parameters of this set-up. In each panel, one of $c$, $\mu\,\tau_0$, and $r_\beta$ is varied about the values of Fig.~\ref{figqtsma}, the other two are kept fixed, and the field is fixed at the value $\alpha=0.05$ of $c=0.1\,v_F/\mu$. The changes are in units of the values at $B=0$ of the same parameters.

All four quantities increase with $c$ in the range shown, and $\sigma_H^{\rm B}$ has a shallow maximum near the upper end of the range. The change $\delta\sigma_\parallel$ does not depend on $\mu\,\tau_0$ in the numerical solution. The magnitude of $\delta\sigma_\perp$ and $\sigma_H^{\rm B}$ increase with $\mu\,\tau_0$, where the increase of $\sigma_H^{\rm B}$ saturates at the largest values, and $\sigma_H^{\rm AH}$ decreases monotonically. The dependence on $r_\beta$ is confined to $\delta\sigma_\parallel$, which decreases monotonically with $r_\beta$ as in Fig.~\ref{figqtsmratio}. The quantities $\delta\sigma_\perp$, $\sigma_H^{\rm B}$, and $\sigma_H^{\rm AH}$ do not depend on $r_\beta$ in the numerical solution. No sign change occurs in the ranges shown, and the Boltzmann Hall part remains larger than the AH part throughout.

\label{secresnum3bc}

The cases (b) and (c) of Table~\ref{tabgeom3} are solved with the general solver of Eq.~\eqref{eqmomsys}, which uses the ansatz of Eq.~\eqref{eqansatz3bc} and does not use the azimuthal sectors. The implementation is described in Appendix~\ref{secnum}. The overlaps are the full squared overlaps of Eq.~\eqref{eqoverlap}, with $\hat{\bk}\cdot\hat{\bk}'$ replaced by $\hat{\bk}\cdot\mathsf T\,\hat{\bk}'$ between the nodes, and the other parameters are those of Sec.~\ref{secresnum3}. We take $\bB=B\left(\sin\theta_B,0,\cos\theta_B\right)$, with $\theta_B=0$ for the case (a), $\theta_B=\pi/2$ for the case (b), and $0<\theta_B<\pi/2$ for the case (c). The tensor at $B=0$ is uniaxial about the $z$ axis, and its anisotropy is small. All the changes are measured from this tensor and are given in units of $\sigma_0=\mathrm{tr}\,\bs\sigma(0)/3$.

For a general orientation, Eq.~\eqref{eqons} makes the symmetric part of $\bs\sigma$ even and the antisymmetric part odd in $\alpha$. The diagonal elements are therefore even in every case. The numerical tensor has these properties and satisfies the relations of Table~\ref{tabgeom3}. In the case (b), the elements $\sigma_{xy}$, $\sigma_{xz}$, $\sigma_{yx}$, and $\sigma_{zx}$ vanish and are not shown.

\begin{figure}[t!]
\centering
\subfigure[\label{figqtsmb}]{\includegraphics[width=0.85\textwidth]{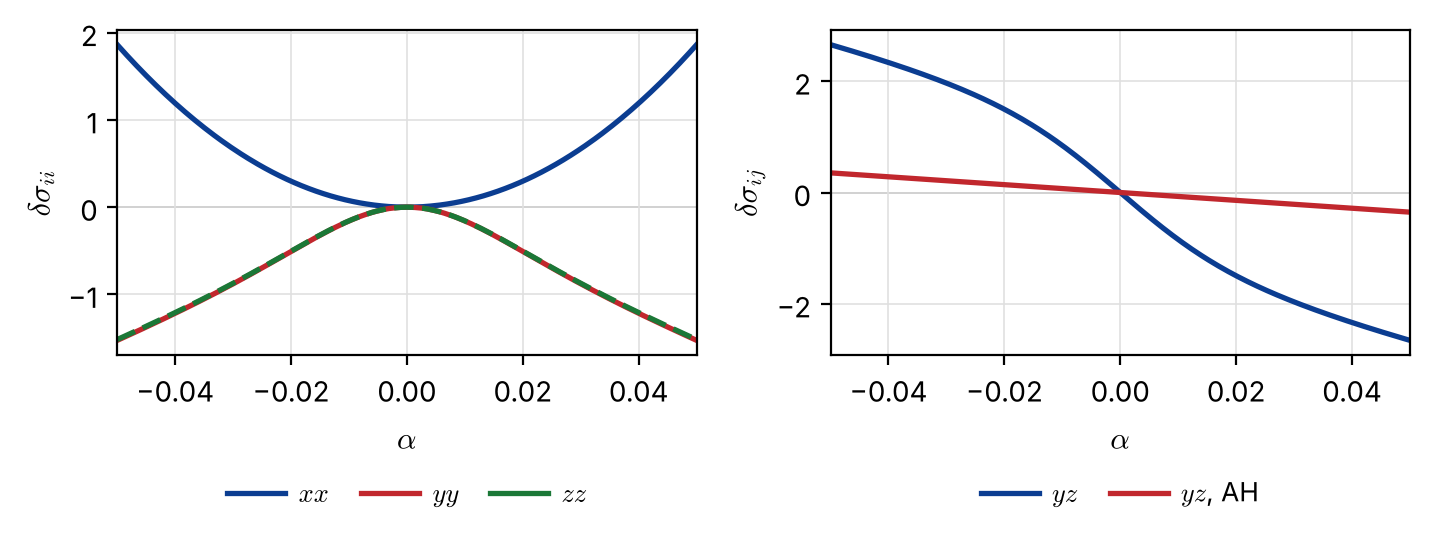}}\\
\subfigure[\label{figqtsmc}]{\includegraphics[width=\textwidth]{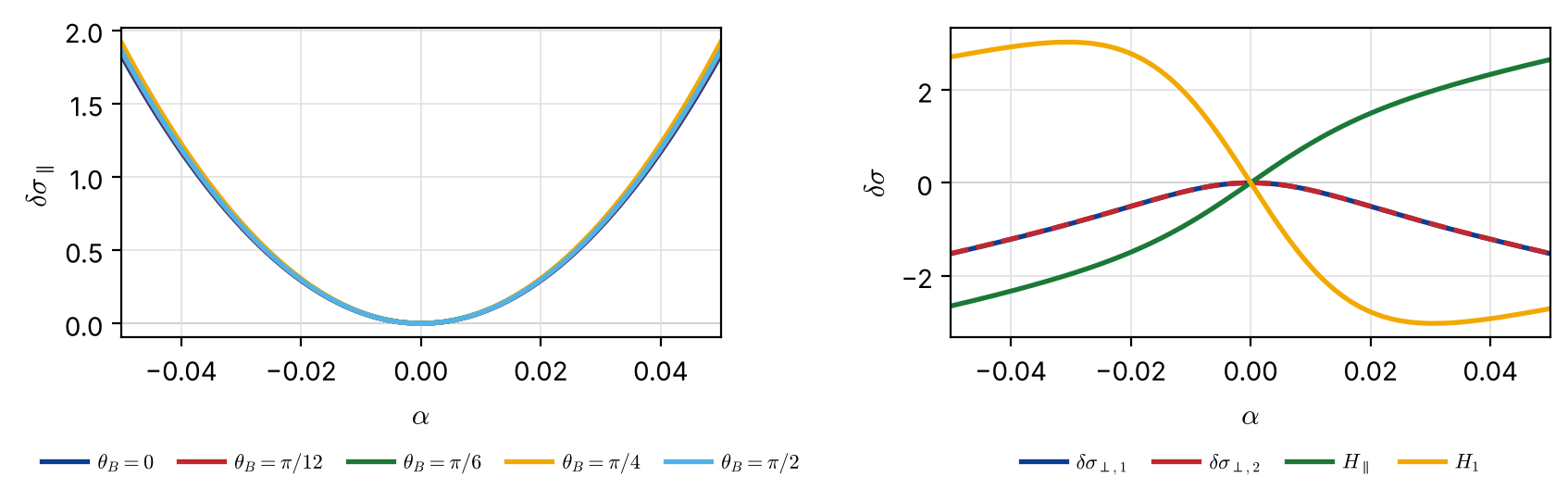}}
\caption{Two conjugate QTSM nodes. (a) Case (b) with $\bB\parallel\hat{\boldsymbol x}$. Left: $\delta\sigma_{xx}$ multiplied by $10^3$ (navy), $\delta\sigma_{yy}$ (crimson), and $\delta\sigma_{zz}$ (green, dashed) multiplied by $10$. Right: $\delta\sigma_{yz}$ multiplied by $10$ (navy) and its AH part multiplied by $10^3$ (crimson), in units of $\sigma_0$. (b) Case (c) with $\bB=B\left(\sin\theta_B,0,\cos\theta_B\right)$. Left: $\delta\sigma_\parallel=\hat{\bB}\cdot\delta\bs\sigma\cdot\hat{\bB}$ multiplied by $10^3$ for $\theta_B=0$ (navy), $\pi/12$ (crimson), $\pi/6$ (green), $\pi/4$ (amber), and $\pi/2$ (sky). Right: for $\theta_B=\pi/4$, the eigenvalues $\delta\sigma_{\perp,1}$ (navy) and $\delta\sigma_{\perp,2}$ (crimson, dashed) of the symmetric part of $\delta\bs\sigma$ in the plane perpendicular to $\bB$ multiplied by $10$, the component $H_\parallel$ of the Hall vector along $\bB$ (green) multiplied by $10$, and its component $H_1$ (amber) along $\left(\cos\theta_B,0,-\sin\theta_B\right)$ multiplied by $10^4$, in units of $\sigma_0$.}
\label{figqtsmbc}
\end{figure}

In Fig.~\ref{figqtsmb}, the change $\delta\sigma_{xx}$ for $\bE\parallel\bB$ is positive and grows with $|\alpha|$ without a sign change. The changes $\delta\sigma_{yy}$ and $\delta\sigma_{zz}$ for $\bE\perp\bB$ are negative and differ only slightly, in contrast to Sec.~\ref{secresnumbc}, where the transverse response depends on the orientation of $\bE$ relative to the crystal axis. The Hall element $\sigma_{yz}$ is odd in $\alpha$, and its magnitude increases sublinearly. Its AH part is smaller than the total by more than two orders of magnitude, and the Hall response is therefore of the Boltzmann type.

The change $\delta\sigma_\parallel$ in Fig.~\ref{figqtsmc} is positive for all the orientations shown, and the curves are close to each other. No sign change occurs, in contrast to Figs.~\ref{figqwc} and \ref{figrtc}. The change increases with $\theta_B$ up to $\pi/4$ and is smaller again at $\pi/2$. In the right panel, the two eigenvalues of the symmetric transverse block are negative and indistinguishable on the scale of the figure. The Hall vector is dominated by $H_\parallel$, which is odd in $\alpha$ and sublinear. The component $H_1$ perpendicular to $\bB$ is odd, smaller than $H_\parallel$ by more than two orders of magnitude, and not monotonic. The component along $\boldsymbol{\hat y}$ vanishes, since $\sigma_{xz}=\sigma_{zx}$ in Table~\ref{tabgeom3}, and it is not shown.

%%%%%%%%%%%%%%%%%
\begin{figure}[t!]
\centering
\includegraphics[width=\textwidth]{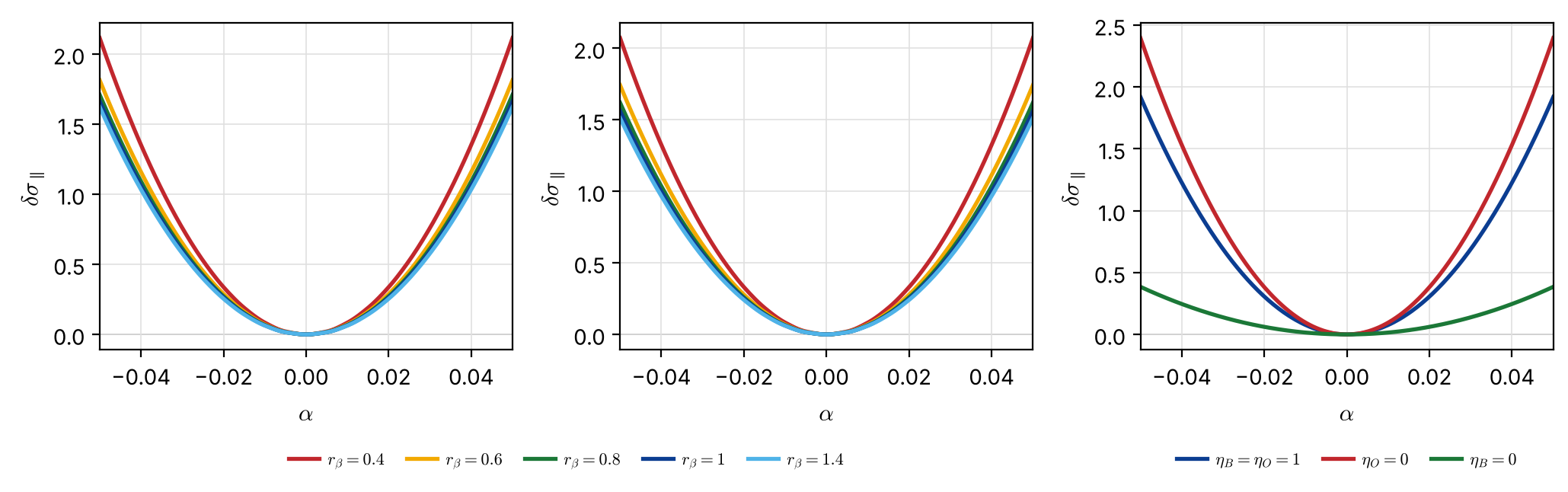}
\caption{Two conjugate QTSM nodes for $\bB=B\left(\sin\theta_B,0,\cos\theta_B\right)$. Left and middle: $\delta\sigma_\parallel$ multiplied by $10^3$ for $\theta_B=\pi/4$ [case (c)] and $\theta_B=\pi/2$ [case (b)], respectively, with $r_\beta=0.4$ (crimson), $0.6$ (amber), $0.8$ (green), $1$ (navy), and $1.4$ (sky). Right: $\theta_B=\pi/4$ and $r_\beta=1/2$ with $\eta_B=\eta_O=1$ (navy), $\eta_O=0$ (crimson), and $\eta_B=0$ (green), multiplied by $10^3$. All changes are in units of $\hat{\bB}\cdot\bs\sigma(0)\cdot\hat{\bB}$.}
\label{figqtsmcratio}
\end{figure}

Figure~\ref{figqtsmcratio} shows the dependence of the longitudinal change on $r_\beta$ and on the BC and the OMM. In the left and middle panels, $\delta\sigma_\parallel$ is positive and decreases monotonically with $r_\beta$ for both orientations, as in Fig.~\ref{figqtsmratio}. The magnitudes for $\theta_B=\pi/2$ are smaller than those for $\theta_B=\pi/4$ at every ratio. No sign change occurs for any ratio, in contrast to Figs.~\ref{figqwcratio} and \ref{figrtcratio}. In the right panel, both the BC and the OMM contributions are positive, and the combined change is smaller than the BC contribution alone. The cross term is therefore negative, and the OMM contribution alone is the smallest of the three, as in the case (a).

%%%%%%%%%%%%%%%
\section{Two tilted QTSM nodes}
\label{secsys4}

A tilt of the two conjugate QTSM nodes of Sec.~\ref{secsys3} is a term $\mathbf t\cdot\bk$ that is proportional to the identity matrix. At a time-reversal-invariant momentum, time reversal maps $\bk$ to $-\bk$ and forbids the tilt term, and the little group of the chiral cubic space groups has no invariant vector, as stated in Sec.~\ref{secintro}. Neither argument applies to a pair of nodes at a momentum $\mathbf K$ and at a momentum $\mathbf K'\neq\mathbf K$ of the BZ. Time reversal then relates the two nodes and does not constrain each node separately. It maps a node with the tilt $\mathbf t$ onto a node of the same chirality with the tilt $-\mathbf t$, since it maps $\bk$ to $-\bk$ and leaves the identity matrix unchanged. The conjugate pair of TSMs of Sec.~\ref{secsys3} \cite{bernevig, ady-spin1} has opposite chiralities, and the two nodes are therefore related by an improper operation, which is the mirror of Eq.~\eqref{eqmirror}, and not by time-reversal alone. The magnetic space group that realises such a pair has the little group of the space group 199 at P, and this little group contains no invariant vector \cite{cano-multifold-magnetic}. A tilt therefore requires a little group with a rotation axis, as discussed below, and we do not specify a space group for the tilted nodes in this work.

A threefold node has no symmetry protection at a generic momentum, as discussed in Sec.~\ref{seclit}. It must therefore sit on a high-symmetry point or line, whose little group contains a rotation about an axis. A rotation by $2\pi/n$ with $n\geq2$ leaves invariant only the vectors along its axis. A tilt that is allowed by this little group is thus directed along the axis, and we call such a tilt axial. We take the axis along $\boldsymbol{\hat z}$, which is the polar axis of Sec.~\ref{secsys3}. Tilts in other directions require a lower symmetry of the little group, for example a mirror plane that contains the tilt, and are not treated here.  

The Hamiltonian of each axially-tilted node is that of Eq.~\eqref{eqham3} with the additional tilt term,
as represented by
\begin{align}
\mathcal H_\chi(\bk)=v_F\left(k_x\,S_x+k_y\,S_y+\chi\,k_z\,S_z\right)+c\,v_F\,k^2\,\mathbb I_{3\times3}+\chi\,t\,v_F\,k_z\,\mathbb I_{3\times3}\,,
\label{eqham4}
\end{align}
with the dimensionless tilt parameter $t$. The sign $\chi$ of the tilt term follows from the mirror relation between the two nodes. The Hamiltonian of Eq.~\eqref{eqham4} obeys $\mathcal H_{-\chi}(k_x,k_y,-k_z)=\mathcal H_\chi(k_x,k_y,k_z)$, and the tilt vectors of the two nodes are $\mathbf t_\chi=\chi\,t\,\boldsymbol{\hat z}$. Since the tilt is proportional to the identity matrix, the eigenstates are those of Sec.~\ref{secsys3}. The BC and the OMM are therefore given by Eq.~\eqref{eqbcomm} with the numbers $\{N_a,G_a\}=\{2,1\}$ for $s=1$ and $\{0,2\}$ for $s=0$, and the Chern numbers are $\mathcal C_a=-\chi\,N_a$. The eigenvalues are
\begin{align}
\epsilon_{\chi,s}(\bk)=v_F\left(s\,k+c\,k^2+\chi\,t\,k\,u\right),\quad s=1,0,-1\,,
\label{eqeval4}
\end{align}
with $u=\cos\theta=k_z/k$. We consider $c>0$, $\mu>0$, and $c\,k<1$ as in Sec.~\ref{secsys3}, and the chemical potential cuts the bands $s=1$ and $s=0$ of each node. This gives the four FS sheets $a=(\chi,s)$. We write $\varpi_{\chi,s}(u)=s+\chi\,t\,u$ and $\bar\mu=c\,\mu/v_F$. For $c>0$ and $\mu>0$, the function $\epsilon_{\chi,s}/v_F=k\left(\varpi_{\chi,s}+c\,k\right)$ is a parabola in $k$ that vanishes at $k=0$. The equation $\epsilon_{\chi,s}=\mu$ therefore has exactly one positive root for every $t$, and every ray from the node crosses the sheet once. At this root, the radial velocity is
\begin{align}
\frac{1}{v_F}\,\partial_k\epsilon_{\chi,s}=\varpi_{\chi,s}+2\,c\,k_{\chi,s}=\sqrt{\varpi_{\chi,s}^2+4\,\bar\mu}>0\,,
\label{eqmono4}
\end{align}
where we have used the radius $k_{\chi,s}$ of Eq.~\eqref{eqfsb04} below. The condition $\partial_k\epsilon_{\chi,s}>0$ on the sheets is thus satisfied for every $t$, and the restriction on $|t|$ comes from the range $c\,k<1$. Multiplying the equation for the radius by $c$ gives $\left(c\,k_{\chi,s}\right)^2+\varpi_{\chi,s}\,c\,k_{\chi,s}=\bar\mu$. Since $k_{\chi,s}$ is the larger root, the condition $c\,k_{\chi,s}<1$ is equivalent to $\bar\mu<1+\varpi_{\chi,s}(u)$. The product $\chi\,u$ takes every value in $\left[-1,1\right]$, and the worst case is $\chi\,t\,u=-|t|$. The sheets $s=1$ and $s=0$ therefore lie inside the range $c\,k<1$ in all directions if
\begin{align}
|t|<2-\frac{c\,\mu}{v_F}\quad (s=1)\,,\qquad |t|<1-\frac{c\,\mu}{v_F}\quad (s=0)\,.
\label{eqtbound4}
\end{align}
At $t=0$, these bounds are the conditions $c\,k_{\Gamma,1}<1$ and $c\,k_{\Gamma,0}<1$. The bound for $s=0$ is the stricter one, and it requires $c\,\mu/v_F<1$. It implies the bound for $s=1$, and it implies $|t|<1$. The two nodes are then of type I, since the slope $1+\chi\,t\,u$ at the node is positive for the band $s=1$ in every direction. We restrict ourselves to this case and do not treat type-II tilts. The band $s=-1$ must not reach $\mu$ inside the range $c\,k<1$. Its radius obeys $\left(c\,k\right)^2+\left(-1+\chi\,t\,u\right)c\,k=\bar\mu$, and the larger root satisfies $c\,k\geq1$ if $\bar\mu\geq\chi\,t\,u$. In all directions, this requires $|t|\leq c\,\mu/v_F$. For $c\,\mu/v_F\leq1/2$, this is stricter than both bounds of Eq.~\eqref{eqtbound4}, and all the tilts used below satisfy $|t|<c\,\mu/v_F$.

At $B=0$, the radius of the sheet $(\chi,s)$ in the direction $u$ is the positive root of $c\,k^2+\left(s+\chi\,t\,u\right)k=\mu/v_F$,
\begin{align}
k_{\chi,s}(u)=\frac{-\left(s+\chi\,t\,u\right)+\sqrt{\left(s+\chi\,t\,u\right)^2+4\,c\,\mu/v_F}}{2\,c}\,.
\label{eqfsb04}
\end{align}
The sheets are no longer spherical, and they depend on $u$ even for $B=0$. The ingredients of the SBE are those of Eq.~\eqref{eqxiw} with $\epsilon_s(k)$ replaced by $\epsilon_{\chi,s}(\bk)$ of Eq.~\eqref{eqeval4}. The tilt adds the constant vector $v_F\,\mathbf t_\chi$ to the velocity,
\begin{align}
\bw_{a}=v_F\left[\left(s+2\,c\,k\right)\hat{\bk}+\mathbf t_\chi\right]+\nabla_{\bk}\left[\eta_O\,\frac{\chi\,e\,v_F\,G_a\,(\bB\cdot\hat{\bk})}{2\,k}\right],
\label{eqw4}
\end{align}
and the density-of-states factor of a sheet is $\rho_a(\hat{\bk})=k_a^2\,D_a/\partial_k\xi_a$ evaluated on the FS. The function $\xi_a$ is invariant under $(\chi,u)\to(-\chi,-u)$, and the mirror relations of Eq.~\eqref{eqmirror} therefore remain valid.

For $\bB=B\,\boldsymbol{\hat z}$, the function
\begin{align}
\xi_a(k,u)=v_F\left(s\,k+c\,k^2+\chi\,t\,k\,u\right)+\eta_O\,\frac{\chi\,e\,v_F\,G_a\,B\,u}{2\,k}
\label{eqxi4}
\end{align}
depends on $\bk$ only through $k$ and $u$, and the vector $\bw_a=\nabla_{\bk}\xi_a$ has no azimuthal component. Its radial and polar components are
\begin{align}
w_{a,k}=\partial_k\xi_a=v_F\left(s+2\,c\,k+\chi\,t\,u\right)-\eta_O\,\frac{\chi\,e\,v_F\,G_a\,B\,u}{2\,k^2}\,,\quad
w_{a,\theta}=\frac{1}{k}\,\partial_\theta\xi_a=-v_F\,\chi\,\sin\theta\left(t+\eta_O\,\frac{e\,G_a\,B}{2\,k^2}\right),
\label{eqwkth4}
\end{align}
where $\theta$ is the polar angle of $\bk$ and $\partial_\theta=-\sin\theta\,\partial_u$. The component along $\boldsymbol{\hat z}$ is $w_{a,z}=u\,w_{a,k}-\sin\theta\,w_{a,\theta}$, which is given in Eq.~\eqref{eqw4z} below. For $B\neq0$, the radius $k_a(\theta)$ is fixed by $\xi_a\!\left(k_a,u\right)=\mu$. After multiplication by $k$, this is the cubic of Eq.~\eqref{eqradius4}, and we take the root that tends to Eq.~\eqref{eqfsb04} for $B\to0$.

The spinors do not depend on the tilt, and the squared overlaps are those of Eq.~\eqref{eqoverlap3} with the variable $X=\chi\,\chi'\cos\theta\cos\theta'$. The amplitudes $\beta^{\rm intra}_{ss'}$ and $\beta^{\rm inter}_{ss'}$ are momentum independent, and the kernel has the finite span of Eq.~\eqref{eqspans}, with nine functions on the sheets $(\chi,1)$ and six on the sheets $(\chi,0)$. The tilt enters only through the weights $\rho_a$ and the FS data, which depend on $u$ at $B=0$ and are no longer constants in the sector with $m=0$. The out-scattering rates, which are constant for the untilted nodes at $B=0$, then depend on the direction of $\bk$. The rates, the weights, and the zero-field solution of the tilted nodes are given in Sec.~\ref{secsol4}.

\subsection{Solution for the magnetic field along the polar axis}
\label{secsol4}

For $\bB=B\,\boldsymbol{\hat z}$, every sheet remains axisymmetric about $\bB$, and the sectors $m=0$ and $m=1$ of Sec.~\ref{secsol3} are retained for the tilted nodes. The radius $k_a(\theta)$ is the positive root of
\begin{align}
c\,v_F\,k^3+v_F\left(s+\chi\,t\,u\right)k^2-\mu\,k+\frac{\eta_O\,\chi\,e\,v_F\,G_a\,B\,u}{2}=0\,,
\label{eqradius4}
\end{align}
which reduces to Eq.~\eqref{eqfsb04} for $B=0$. The phase-space factor, the velocity components, and the function $q_a$ of Eq.~\eqref{eqresalg} are
\begin{align}
& D_a=1-\eta_B\,\frac{\chi\,N_a\,e\,B\,u}{2\,k^2}\,,\quad
w_{a,k}=v_F\left(s+2\,c\,k+\chi\,t\,u\right)-\eta_O\,\frac{\chi\,e\,v_F\,G_a\,B\,u}{2\,k^2}\,,\nn
& w_{a,z}=v_F\left[\left(s+2\,c\,k\right)u+\chi\,t\right]+\eta_O\,\frac{\chi\,e\,v_F\,G_a\,B\left(1-2\,u^2\right)}{2\,k^2}\,,\quad
q_a=v_F\left(s+2\,c\,k\right)-\eta_O\,\frac{\chi\,e\,v_F\,G_a\,B\,u}{k^2}\,,
\label{eqw4z}
\end{align}
with $\rho_a=k^2\,D_a/w_{a,k}$ and $\omega_a=e\,B\,q_a/\left(k\,D_a\right)$. The tilt depends on the radius in $D_a$ and $q_a$ and appears explicitly in $w_{a,k}$ and $w_{a,z}$. The source of the sector $m=0$ is $h_a=\left[w_{a,z}-\eta_B\,\chi\,N_a\,e\,B\,w_{a,k}/(2\,k^2)\right]/D_a$, and that of the sector $m=1$ is $q_a/D_a$. All the quantities depend on $\chi$ and $u$ only through $\chi\,u$, except for $w_{a,z}$ and $h_a$, which change sign under $(\chi,u)\to(-\chi,-u)$. Eq.~\eqref{eqmirror} therefore holds for the tilted nodes, and Eq.~\eqref{eqoverlap3} is invariant under $(\chi,\chi',u,u')\to(-\chi,-\chi',-u,-u')$. The equations of the node $-$ are thus the mirror images of those of the node $+$, as in Sec.~\ref{secsol3}. For $\bE\parallel\bB$, the solution is odd, $\Lambda_{(-,s)}(u)=-\Lambda_{(+,s)}(-u)$, and it satisfies the number constraint identically. For $\bE\perp\bB$, it is even, $\ell_{(-,s)}(u)=\ell_{(+,s)}(-u)$. The two nodes contribute equally to the conductivities, and the quantities of the node $+$ carry the single index $s$ below.

The equations of Sec.~\ref{secsol3} apply with the tilted quantities. The moments $\rho^{(n)}_s$ and $M^{(n)}_s$ are defined with the tilted weights $\rho_s$, and the coefficients $\mathfrak K_{s,j}$ of Eq.~\eqref{eqkappa} are unchanged, since they follow from the kernel alone. The out-scattering rate, the mean free-path, and the conductivity for $\bE\parallel\bB$ are Eqs.~\eqref{eqlampar}, \eqref{eqmomsys3}, and \eqref{eqsigpar3}, with $\rho_s(u)$, $\omega_s(u)$, and $h_s(u)$ built from Eq.~\eqref{eqw4z}. For $\bE\perp\bB$, they are Eqs.~\eqref{eqlamperp}, \eqref{eqmomsysperp}, and \eqref{eqsigperp3}. As in Sec.~\ref{secsys3}, each node has five real moments for $\bE\parallel\bB$ and three complex moments for $\bE\perp\bB$. The null vector of the full problem has the opposite parity to the solution, and the matrix of Eq.~\eqref{eqmomsys3} is regular also for $t\neq0$. The Fermi-sea term is Eq.~\eqref{eqsigah3} with the radius of the tilted sheet $(+,1)$, and the Hall conductivity is $\sigma_H=\sigma_H^{\rm B}+\sigma_H^{\rm AH}$ as for $t=0$.

The explicit forms follow from inserting the coefficients $\mathfrak K_{s,j}$ of Eq.~\eqref{eqkappa} into Eqs.~\eqref{eqmomsys3}, \eqref{eqsigpar3}, and \eqref{eqsigperp3}. The weights $\rho_s$ and the rates $\tau_s$ are not even functions of $u$ for $t\neq0$, and the odd moments do not vanish. For $\bE\parallel\bB$, we define, for the tilted sheets,
\begin{align}
a_r=\left\langle u^r\right\rangle^{\parallel}_1\,,\quad b_r=\left\langle u^r\right\rangle^{\parallel}_0\,,\quad H^{(r)}_s=\left\langle u^r\,h_s\right\rangle^{\parallel}_s\,,\quad G_s=\left\langle h_s^2\right\rangle^{\parallel}_s\,,
\label{eqmomdef4}
\end{align}
with $r=0,\ldots,3$ and $\gamma_{ss'}\equiv\gamma^-_{ss'}$. Then the five moments $M^{(0)}_1$, $M^{(1)}_1$, $M^{(2)}_1$, $M^{(0)}_0$, and $M^{(2)}_0$ obey
\begin{align}
M^{(n)}_1 & =H^{(n)}_1+\gamma_{11}\left[\frac{3\,a_n-a_{n+2}}{4}\,M^{(0)}_1+a_{n+1}\,M^{(1)}_1+\frac{3\,a_{n+2}-a_n}{4}\,M^{(2)}_1\right]
\nn & \quad +\frac{\gamma_{10}}{2}\left[\left(a_n+a_{n+2}\right)M^{(0)}_0+\left(a_n-3\,a_{n+2}\right)M^{(2)}_0\right] \text{ for }  n=0,1,2\,,\nonumber\\
%%%%%%%%%%%%%%%%%%
M^{(n)}_0&=
H^{(n)}_0+\frac{\gamma_{10}}{2}\left[\left(b_n+b_{n+2}\right)M^{(0)}_1
+\left(b_n-3\,b_{n+2}\right)M^{(2)}_1\right]
+\gamma_{00}\left[\left(b_n-b_{n+2}\right)M^{(0)}_0+\left(3\,b_{n+2}-b_n\right)M^{(2)}_0\right] \text{ for } n=0,2\,.
\label{eqmomsys4}
\end{align}
%%%%%%%%%%%%%%
For $t=0$, the odd moments $a_1$, $a_3$, $H^{(0)}_s$ and $H^{(2)}_s$ vanish or decouple by parity, and Eq.~\eqref{eqmomsys4} reduces to Eq.~\eqref{eqmomsys3}. The conductivity is
\begin{align}
\sigma_\parallel&=
\frac{e^2}{2\,\pi^2}\Bigg\{G_1+G_0+\frac{\gamma_{11}}{4}\left[\left(3\,M^{(0)}_1-M^{(2)}_1\right)H^{(0)}_1+4\,M^{(1)}_1\,H^{(1)}_1+\left(3\,M^{(2)}_1-M^{(0)}_1\right)H^{(2)}_1\right]
\nn & \hspace{ 1.75 cm}
 +\frac{\gamma_{10}}{2}\left[\left(M^{(0)}_0+M^{(2)}_0\right)H^{(0)}_1+\left(M^{(0)}_0-3\,M^{(2)}_0\right)H^{(2)}_1
 + \left(M^{(0)}_1+M^{(2)}_1\right)H^{(0)}_0+\left(M^{(0)}_1-3\,M^{(2)}_1\right)H^{(2)}_0\right] 
 %%%%%%%%%
\nn & \hspace{1.75 cm}
+\gamma_{00}\left[\left(M^{(0)}_0-M^{(2)}_0\right)H^{(0)}_0+\left(3\,M^{(2)}_0-M^{(0)}_0\right)H^{(2)}_0\right]\Bigg\}.
\label{eqsigpar4}
\end{align}

For $\bE\perp\bB$, the system is Eq.~\eqref{eqmomsysperp} with $\gamma^+_{ss'}$ and the tilted brackets $\left\langle f\right\rangle^{\perp}_s$. The matrix is that of $t=0$, but the entries with $\left\langle u\right\rangle^{\perp}_s$ no longer vanish. The coefficients of the in-scattering ansatz are
\begin{align}
\check\lambda^{(0)}_1&=\frac{\gamma^+_{11}}{2}\,\mathcal{L}^{(0)}_1\,,\quad
\check\lambda^{(1)}_1=\frac{\gamma^+_{11}}{2}\,\mathcal{L}^{(1)}_1-\gamma^+_{10}\,\mathcal{L}^{(1)}_0\,,
\check\lambda^{(0)}_0 =0\,,\quad
\check\lambda^{(1)}_0=-\gamma^+_{10}\,\mathcal{L}^{(1)}_1+2\,\gamma^+_{00}\,\mathcal{L}^{(1)}_0\,.
\label{eqlam4}
\end{align}
With $Q_s=\left\langle\left(q_s/D_s\right)^2\right\rangle^{\perp}_s$, the transverse and Hall conductivities from the Boltzmann part are
\begin{align}
\sigma_\perp-i\,\sigma_H^{\rm B}&=
\frac{e^2}{4\,\pi^2}\Bigg\{Q_1+Q_0+\frac{\gamma^+_{11}}{2}\,\mathcal{L}^{(0)}_1\left\langle\frac{q_1}{D_1}\right\rangle^{\perp}_1
+\left[\frac{\gamma^+_{11}}{2}\,\mathcal{L}^{(1)}_1-\gamma^+_{10}\,\mathcal{L}^{(1)}_0\right]\left\langle u\,\frac{q_1}{D_1}\right\rangle^{\perp}_1
+\left[-\gamma^+_{10}\,\mathcal{L}^{(1)}_1+2\,\gamma^+_{00}\,\mathcal{L}^{(1)}_0\right]\left\langle u\,\frac{q_0}{D_0}\right\rangle^{\perp}_0\Bigg\}.
\label{eqsigperp4}
\end{align}
The conductivity $\sigma_\perp$ is the real part and $\sigma_H^{\rm B}$ is minus the imaginary part. The anomalous Hall part is
\begin{align}
\sigma_H^{\rm AH}=-\frac{\eta_B\,e^2}{2\,\pi^2}\int_{-1}^{1}du\,u\,k_{(+,1)}(u)\,,
\label{eqsigH4}
\end{align}
where $k_{(+,1)}(u)$ is the radius from Eq.~\eqref{eqradius4}, and $\sigma_H=\sigma_H^{\rm B}+\sigma_H^{\rm AH}$. For $t=0$, the odd brackets vanish and Eqs.~\eqref{eqsigpar3} and \eqref{eqsigperp3} are recovered.

The closed forms for $B\to0$, however, do not carry over. The weights are no longer constants, and the moments $\rho^{(n)}_s$ with odd $n$ enter the rates. At $B=0$, the rate of the sheet $(+,s)$ is given by Eq.~\eqref{eqlampar} with the moments of the tilted weights, and the sheet $(-,s)$ follows from $\tau_{(-,s)}^{-1}(u)=\tau_{(+,s)}^{-1}(-u)$. Here $D_a=1$, and the radial velocity of Eq.~\eqref{eqmono4} gives the weight and its moments,
\begin{align}
\rho_{+,s}(u)=\frac{k_{+,s}^2(u)}{v_F\sqrt{\left(s+t\,u\right)^2+4\,\bar\mu}}\,,\quad
\rho^{(n)}_s=\int_{-1}^{1}du\,u^n\,\rho_{+,s}(u)\,,
\label{eqrho40}
\end{align}
for $n=0,1,2$ with $s=1$ and for $n=0,2$ with $s=0$. Inserting these moments into Eq.~\eqref{eqkappa} with $\gamma^{+}_{ss'}$ gives
\begin{align}
\tau_1^{-1}(u)&=\frac{\gamma^+_{11}}{4}\left(3\,\rho^{(0)}_1-\rho^{(2)}_1\right)+\frac{\gamma^+_{10}}{2}\left(\rho^{(0)}_0+\rho^{(2)}_0\right)+\gamma^+_{11}\,\rho^{(1)}_1\,u
+\left[\frac{\gamma^+_{11}}{4}\left(3\,\rho^{(2)}_1-\rho^{(0)}_1\right)-\frac{\gamma^+_{10}}{2}\left(3\,\rho^{(2)}_0-\rho^{(0)}_0\right)\right]u^2\,,\nn
%%%%%%%%%%%%%%%%%%%%
\tau_0^{-1}(u)&=\frac{\gamma^+_{10}}{2}\left(\rho^{(0)}_1+\rho^{(2)}_1\right)+\gamma^+_{00}\left(\rho^{(0)}_0-\rho^{(2)}_0\right)
+\left[-\frac{\gamma^+_{10}}{2}\left(3\,\rho^{(2)}_1-\rho^{(0)}_1\right)+\gamma^+_{00}\left(3\,\rho^{(2)}_0-\rho^{(0)}_0\right)\right]u^2\,.
\label{eqtau40}
\end{align}
The moment $\rho^{(1)}_1$ is the only odd moment that enters, and it produces the term linear in $u$ of $\tau_1^{-1}$. For $t\to0$, we have $\rho^{(1)}_s=0$ and $\rho^{(0)}_s=3\,\rho^{(2)}_s$, the terms in $u$ and $u^2$ vanish, and the rates reduce to the constants of the untilted nodes.

For $\bE\parallel\hat{\boldsymbol z}$, the zero-field solution keeps the form of Eq.~\eqref{eqlampar},
\begin{align}
\Lambda_s(u)=\tau_s(u)\Big[h_s(u)+\sum_{j=0}^{2}\mathfrak K_{s,j}\!\left(M,\gamma^{-}\right)u^j\Big]\,,\quad
h_s(u)=v_F\left[\left(s+2\,c\,k_{+,s}(u)\right)u+t\right],
\label{eqlam40}
\end{align}
where $h_s=w_{a,z}$ of Eq.~\eqref{eqw4z} at $B=0$ and $\chi=+1$. The five moments $M^{(n)}_s$ obey Eq.~\eqref{eqmomsys3} with $\rho_s$, $\tau_s$, and $h_s$ of Eqs.~\eqref{eqrho40}--\eqref{eqlam40}. For $t=0$, the moments $M^{(0)}_s$ and $M^{(2)}_s$ vanish by parity in $u$. For $t\neq0$, the source $h_s$ has an even part, which contains the constant $v_F\,t$, and the rate $\tau_1^{-1}$ has the odd part $\gamma^+_{11}\,\rho^{(1)}_1\,u$. All five moments are therefore nonzero and coupled. The closed form of Eq.~\eqref{eqsigB0} does not hold, and Eq.~\eqref{eqmomsys3} is solved as a linear system with the tilted weights in the integrals. For $\bE\perp\hat{\boldsymbol z}$, we have $\omega_s=0$ at $B=0$, and the mean free-path of Eq.~\eqref{eqansatz3a} becomes $\ell_s=\tau_s(u)\,\sin\theta\left[v_F\left(s+2\,c\,k_{+,s}(u)\right)+\check\lambda^{(0)}_s+\check\lambda^{(1)}_s\,u\right]$. The source $q_s$ does not contain $t$ explicitly, since the tilt cancels in the combination of $w_{a,k}$ and $w_{a,\theta}$ that defines it. The tilt enters this sector only through $k_{+,s}(u)$ and $\tau_s(u)$.

\subsection{Symmetries and classification of the set-ups}
\label{secclass4}

The dispersion of Eq.~\eqref{eqeval4} and the kernel are invariant under the rotations $\mathcal R_z$ about the $z$ axis, the reflection $\Pi_z$ normal to the $z$ axis, and the inversion $\mathcal J$, and under their products. The inversion maps the node $\chi$ onto the node $-\chi$, and it leaves the tilt term invariant, since $\chi\,k_z$ is invariant under $(\chi,k_z)\to(-\chi,-k_z)$. The same holds for $\Pi_z$. A rotation by $\pi$ about an axis in the $xy$ plane maps each node onto itself and reverses $k_z$. It reverses the sign of the tilt term, and it is therefore not a symmetry for $t\neq0$. Since the reflection $\Pi_x$ normal to an in-plane axis is the product of this rotation and $\mathcal J$, it is not a symmetry either. The set-ups differ from those of Table~\ref{tabgeom3} for this reason. The field $\bB$ transforms as $\det(g)\,g\,\bB$, and we take $\bB=B\left(\sin\theta_B,0,\cos\theta_B\right)$ as in Sec.~\ref{secclass3}. The classification is collected in Table~\ref{tabgeom4}.

Time reversal maps the node $\chi$ onto a node of the same chirality at $-\bk$ with the reversed tilt, and it therefore maps the pair with the tilt $t$ onto the pair with the tilt $-t$. The quantities $\xi_a$, $D_a$, and $\rho_a$ are unchanged under $\bk\to-\bk$, $\bB\to-\bB$, and $t\to-t$, whereas $\bw_a$ and $\Lhat$ change sign, and the in-scattering kernel is symmetric. The proof of Eq.~\eqref{eqons} in Appendix~\ref{secappsym3} then gives the Onsager relation of the pair in the form
\begin{align}
\sigma_{ij}\!\left(\bB;t\right)=\sigma_{ji}\!\left(-\bB;-t\right).
\label{eqons4}
\end{align}
Eq.~\eqref{eqons} at a fixed tilt does not hold for $t\neq0$. In a crystal with time-reversal symmetry, the pair is accompanied by its time-reversed partner, which has the tilt $-t$. By Eq.~\eqref{eqons4}, the conductivity of the two pairs is $\sigma^{\rm tot}_{ij}(\bB)=\sigma_{ij}(\bB;t)+\sigma_{ji}(-\bB;t)$, and it obeys Eq.~\eqref{eqons}. The terms of a single pair that violate Eq.~\eqref{eqons} cancel in this sum. We present the response of a single pair, which contains the information on both.

A rotation $C_{2\hat{\bs n}}$ by $\pi$ about an axis $\hat{\bs n}$ in the $xy$ plane maps every node onto itself and leaves the kernel invariant, and it reverses $k_z$ and with it the tilt. It is a proper rotation, and it gives $\bs\sigma\!\left(C_{2\hat{\bs n}}\bB;-t\right)=C_{2\hat{\bs n}}\,\bs\sigma\!\left(\bB;t\right)C_{2\hat{\bs n}}^{\rm T}$. For $\bB\perp\hat{\bs n}$, we have $C_{2\hat{\bs n}}\bB=-\bB$, and Eq.~\eqref{eqons4} eliminates the tilt,
\begin{align}
\sigma_{ij}=\eta_i\,\eta_j\,\sigma_{ji}\,,
\label{eqplanar4}
\end{align}
where $i$ and $j$ label the axes of a frame with one axis along $\hat{\bs n}$, and $\eta_i=1$ for that axis and $\eta_i=-1$ for the other two. This is the invariance under time reversal combined with $C_{2\hat{\bs n}}$, which maps the pair onto itself and fixes $\bB$. Every orientation of $\bB$ has such an axis. For $\bB\nparallel\hat{\boldsymbol z}$ it is $\hat{\bs n}\parallel\hat{\boldsymbol z}\times\bB$, and for $\bB\parallel\hat{\boldsymbol z}$ every axis in the $xy$ plane serves. For $\bB=B\left(\sin\theta_B,0,\cos\theta_B\right)$, it is the $y$ axis, and Eq.~\eqref{eqplanar4} gives $\sigma_{xy}=-\sigma_{yx}$, $\sigma_{yz}=-\sigma_{zy}$, and $\sigma_{xz}=\sigma_{zx}$. These are the relations of the case (c) of Table~\ref{tabgeom3}. They no longer follow from a unitary symmetry, since $C_{2y}$ reverses the tilt, and they follow from its combination with time reversal, as for the tilted WSMs of Ref.~\cite{ips-exact-noncollinear}.

\begin{table}[t!]
\centering
\small
\renewcommand{\arraystretch}{1.6}
\begin{tabular}{|c|c|c|c|}
\hline
Case & Orientation of $\bB$ & Symmetries & \makecell{Independent elements \\ of $\bs\sigma$} \\
\hline
(a) & $\theta_B=0$ & $\mathcal R_z$, $\Pi_z$, $\mathcal J$ & \makecell{three: \\ $\sigma_\parallel$, $\sigma_\perp$, $\sigma_H$} \\
\hline
(b) & $\theta_B=\pi/2$ & \makecell{$\mathcal J$, $\mathcal R_z(\pi)$, \\ $\mathcal T\,C_{2y}$} & \makecell{six: $\sigma_{xx}$, $\sigma_{yy}$, $\sigma_{zz}$, $\sigma_{xy}=-\sigma_{yx}$, \\ $\sigma_{xz}=\sigma_{zx}$, $\sigma_{yz}=-\sigma_{zy}$} \\
\hline
(c) & $0<\theta_B<\pi/2$ & $\mathcal J$, $\mathcal T\,C_{2y}$ & \makecell{six, as in (b), \\ without parity rule} \\
\hline
\end{tabular}
\caption{\label{tabgeom4} Independent geometries of two axially-tilted conjugate QTSM nodes, where $\mathcal T$ denotes time reversal. The elements of $\bs\sigma$ follow from the symmetries and from the Onsager relation of Eq.~\eqref{eqons4}.}
\end{table}

For the case (a), the field is fixed by $\mathcal R_z$, $\Pi_z$, and $\mathcal J$, and the tensor has the form below Eq.~\eqref{eqsigperp}. No unitary element maps $B\,\boldsymbol{\hat z}$ onto $-B\,\boldsymbol{\hat z}$ at a fixed tilt, since $C_{2x}$ and $\Pi_x$ reverse the tilt. The relation $\bs\sigma\!\left(-\bB;-t\right)=C_{2x}\,\bs\sigma\!\left(\bB;t\right)C_{2x}^{\rm T}$ then shows that $\sigma_\parallel$ and $\sigma_\perp$ contain terms that are odd in $B$ and odd in $t$ \cite{amit_magneto}, and that $\sigma_H$ contains a term that is even in $B$ and odd in $t$. The latter is the Fermi-sea AH term of Eq.~\eqref{eqsigmw}, which does not vanish at $B=0$ because the radii $k_a$ are not even in $u$.

For the case (b), the rotation $\mathcal R_z(\pi)$ maps $\bB$ onto $-\bB$ and leaves the kernel, the dispersion, and the tilt invariant. It gives $\sigma_{ij}(-\bB)=\left[\mathcal R_z(\pi)\right]_{ii}\left[\mathcal R_z(\pi)\right]_{jj}\sigma_{ij}(\bB)$. The diagonal elements and $\sigma_{xy}$ are therefore even in $B$, and $\sigma_{xz}$ and $\sigma_{yz}$ are odd. In contrast to Table~\ref{tabgeom3}, the elements $\sigma_{xy}$ and $\sigma_{xz}$ do not vanish for the case (b), since $C_{2x}$ is broken by the tilt. For the case (c), $\mathcal R_z(\pi)$ maps $\bB$ onto $\left(-\sin\theta_B,0,\cos\theta_B\right)$ and not onto $-\bB$. Only $\mathcal J$ and $\mathcal T\,C_{2y}$ fix $\bB$, no relation connects $\bB$ and $-\bB$, and the diagonal elements contain terms odd in $B$ as in the case (a).

For every case, $\sigma_\parallel=\hat{\bB}\cdot\bs\sigma\,\hat{\bB}$ is the response to $\bE\parallel\bB$. For the cases (b) and (c), the symmetric part $\bs\sigma^{\rm S}=\left(\bs\sigma+\bs\sigma^{\rm T}\right)/2$ has no $xy$ and $yz$ elements, and $\boldsymbol{\hat y}$ is a principal axis. We denote by $\sigma_{\perp,1}$ and $\sigma_{\perp,2}$ the eigenvalues of $\bs\sigma^{\rm S}$ in the plane perpendicular to $\bB$. The antisymmetric part is $\sigma^{\rm A}_{ij}=-\epsilon_{ijl}\,H_l$ with the Hall vector $\bs H$, which has $H_y=0$ and lies in the plane spanned by $\bB$ and $\boldsymbol{\hat z}$. Its components are $H_\parallel=\hat{\bB}\cdot\bs H$ and $H_1=\hat{\bs e}_1\cdot\bs H$ with $\hat{\bs e}_1=\left(\cos\theta_B,0,-\sin\theta_B\right)$. At $B=0$, the vector $\bs H$ points along $\boldsymbol{\hat z}$ and is the AH term. In the sum of the pair and its partner, the symmetric part is twice the part of $\bs\sigma^{\rm S}$ that is even in $B$, and the Hall vector is twice the part of $\bs H$ that is odd in $B$. The terms odd in $B$ of $\bs\sigma^{\rm S}$ and the terms even in $B$ of $\bs H$, which include the AH term at $B=0$, belong to a single pair and cancel in $\bs\sigma^{\rm tot}$.

\subsection{Ansatz and linear system}
\label{secansatz4}

The exact form of $\Lambda^j_a$ is Eq.~\eqref{eqtiltsol} with the functions $f^a_p$ of Eq.~\eqref{eqfp}, and the unknowns are the moments $\mathcal M^j_{a,p}$ of Eq.~\eqref{eqmomsys}. The sheets $(\chi,1)$ carry nine functions and the sheets $(\chi,0)$ carry six, as in Eq.~\eqref{eqspans}, since neither the spinors nor the overlaps depend on the tilt. The tilt enters through $k_a(\hat{\bk})$, $\bw_a$, and $\rho_a$, and hence through the rates $\tau_a^{-1}$ and the resolvent $\hat{\mathcal R}_a$. At $B=0$, the weight $\rho_a=k_a^2/\partial_k\xi_a$ is not even in $\hat{\bk}$ for $t\neq0$, and the moments of the odd functions $f^a_p$ do not vanish. The rates of Eq.~\eqref{eqtau3} therefore depend on the direction of $\hat{\bk}$ already at $B=0$. The inversion $\mathcal J$ maps the node $\chi$ onto $-\chi$, fixes $\bB$ for every orientation, and leaves the tilt term invariant. Eq.~\eqref{eqinv3} then holds for the tilted nodes, and the moments of the node $-$ follow from those of the node $+$ by Eq.~\eqref{eqmomJ3}. The number constraint holds identically. Only the sheets of the node $+$ are expanded, which leaves fifteen moments for each $j$.

The case $\bB\parallel\hat{\boldsymbol z}$ is solved in Sec.~\ref{secsol4}.

For the other orientations, the FS is not axisymmetric about $\bB$. The component $\hat{\bB}\cdot\bk$ is conserved along the cyclotron motion of Eq.~\eqref{eqeom}, because $\hat{\bB}\cdot\left(\bw_a\times\bB\right)=0$, and the orbits are the intersections of a sheet with the planes $\hat{\bB}\cdot\bk=\mathfrak o$. Without $\bE$, we have $\dot\bk=-e\,D_a^{-1}\,\bw_a\times\bB$, and the operator $-e\,D_a^{-1}\Lhat$ is the derivative $\partial_{\mathfrak s}$ with respect to the time $\mathfrak s$ along the orbit. The period and the cyclotron frequency of the orbit are
\begin{align}
T_a(\mathfrak o)=\oint\frac{D_a\,dl}{e\,B\,|\bw_{a,\perp\bB}|}\,,\quad
\bar\omega_a(\mathfrak o)=\frac{2\,\pi}{T_a(\mathfrak o)}\,,
\label{eqorbit4}
\end{align}
where $dl$ is the line element of the orbit in momentum space and $\bw_{a,\perp\bB}$ is the component of $\bw_a$ perpendicular to $\bB$. With the orbit angle $\psi=\bar\omega_a(\mathfrak o)\,\mathfrak s$, Eq.~\eqref{eqboltzop} becomes $\left[\tau_a^{-1}(\mathfrak o,\psi)+\bar\omega_a(\mathfrak o)\,\partial_\psi\right]\Lambda=u$ on each orbit, and the resolvent is
\begin{align}
\hat{\mathcal R}_a[u](\mathfrak o,\psi)=\frac{1}{\bar\omega_a(\mathfrak o)}\int_0^\infty d\phi\;u(\mathfrak o,\psi-\phi)\,\exp\!\left[-\frac{1}{\bar\omega_a(\mathfrak o)}\int_{\psi-\phi}^{\psi}\tau_a^{-1}(\mathfrak o,\psi')\,d\psi'\right].
\label{eqresorb4}
\end{align}
It converges since $\tau_a^{-1}>0$, and it reduces to $\tau_a$ for $\bar\omega_a\to0$. For $\bB\parallel\hat{\boldsymbol z}$, the orbit is the ring of constant $\theta$, we have $\bar\omega_a=\omega_a(\theta)$, and the rate is constant along the ring. The integral then gives the algebraic resolvent of Eq.~\eqref{eqresalg} for each $m$. For $t=0$, the orbits are the rings about $\bB$, but the rates depend on the angle around them, and Eq.~\eqref{eqresorb4} is not algebraic. For $t\neq0$ and $\bB\nparallel\hat{\boldsymbol z}$, the orbits are not circles, and $\bar\omega_a$ and $\tau_a^{-1}$ both vary along them.

We evaluate the resolvent by the Galerkin expansion of Eq.~\eqref{eqansatzc} in the spherical harmonics $Y_{lm}$ with $l\leq l_{\max}$, with the polar axis along $\boldsymbol{\hat z}$. The LFO acts on the harmonics through Eq.~\eqref{eqlfomw}, and the matrix elements are
\begin{align}
\left(\mathsf F_a\right)_{lm,\,l'm'}=\int d^2\hat{\bk}\,\rho_a\,Y^*_{lm}\left[\tau_a^{-1}\,Y_{l'm'}-D_a^{-1}\left(V_\theta\,\partial_\theta+\frac{V_\varphi}{\sin\theta}\,\partial_\varphi\right)Y_{l'm'}\right],
\label{eqgalerkin4}
\end{align}
with $e=1$ and with the weight $\rho_a\,d^2\hat{\bk}$ also in $\mathsf U_{lm}$. The weight is the measure for which $D_a^{-1}\Lhat$ is antisymmetric, and the symmetric part of $\mathsf F_a$ is then positive definite. The matrix is therefore invertible for every field, and the truncation is stable. For $\bB\parallel\hat{\boldsymbol z}$, $\mathsf F_a$ is diagonal in $m$, and the expansion reproduces Eq.~\eqref{eqresalg}. For $t=0$, it reproduces the solution of Sec.~\ref{secsys3} for every orientation of $\bB$.

\subsection{Limit of the relaxation-time approximation}
\label{secrta4}

In the RTA, the in-scattering term of Eq.~\eqref{eqell} is dropped, and the out-scattering rate is replaced by a constant $1/\tau$. This is the intranode RTA of Sec.~\ref{secrta3} for $\tau_{\rm inter}=\tau$, for which the internode correction proportional to $\tau_{\rm inter}/\tau-1$ vanishes. For $\bB=B\,\boldsymbol{\hat z}$, Eqs.~\eqref{eqsigpar3} and \eqref{eqsigperp3} hold with $M^{(n)}_s=0$, $\check\lambda^{(n)}_s=0$, and $\tau_s=\tau$, and they give
\begin{align}
\sigma_\parallel=\frac{e^2\,\tau}{2\,\pi^2}\sum_s\int_{-1}^{1}du\,\rho_s\,h_s^2\,,\quad
\sigma_\perp-i\,\sigma_H^{\rm B}=\frac{e^2\,\tau}{4\,\pi^2}\sum_s\int_{-1}^{1}du\left(1-u^2\right)\rho_s\,\frac{\left(q_s/D_s\right)^2}{1+i\,\omega_s\,\tau}\,,
\label{eqrta4}
\end{align}
with $\rho_s$, $D_s$, $q_s$, $\omega_s$, and $h_s$ of Eq.~\eqref{eqw4z} for the node $+$. For $t=0$, the second expression is Eq.~\eqref{eqrtaperp3}, and the first one reproduces the intranode terms of Sec.~\ref{secrta3} to the order $B^2$.

At $B=0$, we have $D_s=1$. We define $\Delta_s(u)=\left(s+t\,u\right)^2+4\,\bar\mu$, so that Eq.~\eqref{eqfsb04} gives $2\,c\,k_{+,s}=-\left(s+t\,u\right)+\sqrt{\Delta_s}$ and $s+2\,c\,k_{+,s}=\sqrt{\Delta_s}-t\,u$. Eqs.~\eqref{eqmono4} and \eqref{eqw4z} then give
\begin{align}
\rho_s=\frac{k_{+,s}^2}{v_F\sqrt{\Delta_s}}\,,\quad
h_s=v_F\left[u\sqrt{\Delta_s}+t\left(1-u^2\right)\right],\quad
q_s=v_F\left(\sqrt{\Delta_s}-t\,u\right),
\label{eqrtab04}
\end{align}
and the conductivities of Eq.~\eqref{eqrta4} at $B=0$ are
\begin{align}
\sigma_\parallel\big\vert_{B=0}=\frac{e^2\,\tau\,v_F}{2\,\pi^2}\sum_s\int_{-1}^{1}du\,\frac{k_{+,s}^2\left[u\sqrt{\Delta_s}+t\left(1-u^2\right)\right]^2}{\sqrt{\Delta_s}}\,,\quad
\sigma_\perp\big\vert_{B=0}=\frac{e^2\,\tau\,v_F}{4\,\pi^2}\sum_s\int_{-1}^{1}du\,\frac{\left(1-u^2\right)k_{+,s}^2\left(\sqrt{\Delta_s}-t\,u\right)^2}{\sqrt{\Delta_s}}\,.
\label{eqrtadr4}
\end{align}
Both are even in $t$. We write $\tilde\nu_s=\sqrt{s^2+4\,\bar\mu}=w^0_s/v_F$, with $k_s=k_{\Gamma,s}$ and $w^0_s$ of Sec.~\ref{secsol3}. Expanding to the order $t^2$, we obtain
\begin{align}
& \sigma_\parallel\big\vert_{B=0}=\frac{e^2\,\tau}{3\,\pi^2}\sum_s k_s^2\,w^0_s\left(1+a_s\,t^2\right),\quad
\sigma_\perp\big\vert_{B=0}=\frac{e^2\,\tau}{3\,\pi^2}\sum_s k_s^2\,w^0_s\left(1+b_s\,t^2\right),\nn
& a_1=\frac{3\left(5\,\tilde\nu_1^2-2\,\tilde\nu_1-1\right)}{10\,\tilde\nu_1^4}\,,\quad
b_1=\frac{\left(3\,\tilde\nu_1-1\right)\left(5\,\tilde\nu_1+1\right)}{10\,\tilde\nu_1^4}\,,\quad
a_0=b_0=\frac{3}{2\,\tilde\nu_0^2}\,.
\label{eqrtat24}
\end{align}
The tensor at $B=0$ is uniaxial. Its anisotropy comes from the sheet $s=1$ only, and it is
\begin{align}
\sigma_\perp\big\vert_{B=0}-\sigma_\parallel\big\vert_{B=0}=\frac{e^2\,\tau}{3\,\pi^2}\,k_1^2\,w^0_1\,\frac{2\,\tilde\nu_1+1}{5\,\tilde\nu_1^4}\,t^2+\mathcal O\!\left(t^4\right).
\label{eqrtaan4}
\end{align}

To the first order in $B$ and in $t$, the tilt generates terms that vanish for $t=0$. The coefficients below are obtained by expanding the integrals of Eq.~\eqref{eqrta4} with the radius of Eq.~\eqref{eqradius4}, where $N_s=2$ and $G_s=1$ for $s=1$, and $N_s=0$ and $G_s=2$ for $s=0$. The parts of $\sigma_\parallel$ and $\sigma_\perp$ that are odd in $B$ are
\begin{align}
\sigma_\parallel^{\rm odd}&=\frac{e^3\,\tau\,v_F\,B\,t}{2\,\pi^2}\left[-\frac{\eta_B\left(32\,\tilde\nu_1^3+14\,\tilde\nu_1^2\right)+\eta_O\left(3-11\,\tilde\nu_1^2\right)}{15\,\tilde\nu_1^3}+\frac{22\,\eta_O}{15\,\tilde\nu_0}\right],\nn
\sigma_\perp^{\rm odd}&=\frac{e^3\,\tau\,v_F\,B\,t}{4\,\pi^2}\left[-\frac{2\left[\eta_B\left(4\,\tilde\nu_1^3-2\,\tilde\nu_1^2\right)+\eta_O\left(1-7\,\tilde\nu_1^2\right)\right]}{15\,\tilde\nu_1^3}+\frac{28\,\eta_O}{15\,\tilde\nu_0}\right],
\label{eqrtaodd4}
\end{align}
and the part of the Boltzmann Hall conductivity that is even in $B$, which vanishes at $B=0$ in the RTA, is
\begin{align}
\sigma_H^{\rm B,\,even}=\frac{e^4\,\tau^2\,v_F^3\,B^2\,t}{4\,\pi^2\,\mu}\left[-\frac{\left(\tilde\nu_1^2-1\right)\left(8\,\eta_B\,\tilde\nu_1-9\,\eta_O\right)}{15\,\tilde\nu_1}+\frac{6\,\eta_O\,\tilde\nu_0}{5}\right].
\label{eqrtaHev4}
\end{align}
These terms are linear in $\eta_B$ and $\eta_O$ with no cross term at this order. They are odd in $t$, as required by the parity rules of Sec.~\ref{secclass4}. The AH conductivity does not depend on $\tau$, and its terms at $B=0$ and to the first order in $B$ are those of Sec.~\ref{secres4}.

For $\tau_{\rm inter}\neq\tau$, the longitudinal response has the internode term of Sec.~\ref{secrta3}. We generalise it to the tilted sheets by keeping the form of Eq.~\eqref{eqrtab7}, in which the velocity $\bs v_s$ and the energy $\epsilon_s=v_F\left(s\,k+c\,k^2+t\,k\,u\right)$ of the node $+$ are those of the tilted sheets. The BC and the OMM do not depend on $t$, because the tilt is proportional to the unit matrix. The BC term of $\Upsilon_s$ does not depend on $t$ either, since $\int dk\,\partial_k\epsilon_s\,\delta(\epsilon_s-\mu)=1$ in every direction. The OMM term is evaluated like that of Appendix~\ref{secrtab6}, with $m^s_z\,v_{s,z}=-\,e\,v_F\,G_s\,u\,v_{s,z}/(2\,k)$ and the tilted velocity $v_{s,z}=v_F\left[\left(s+2\,c\,k\right)u+t\right]$. For the node $\chi=+$, we obtain $\Upsilon_s=-\,e\,B\,a_s(t)/\left(12\,\pi^2\right)$ with
\begin{align}
a_1(t)=6+\frac{1}{\tilde\nu_1}+\frac{\tilde\nu_1^2-3}{10\,\tilde\nu_1^5}\,t^2\,,\quad
a_0(t)=\frac{2}{\tilde\nu_0}+\frac{t^2}{5\,\tilde\nu_0^3}\,,\quad
\rho_0(t)=\frac{1}{2\,\pi^2}\sum_s\frac{k_s^2}{w^0_s}\left(1+d_s\,t^2\right),
\label{eqrtaa4}
\end{align}
where $d_1=\left(\tilde\nu_1+1\right)^2/\left(2\,\tilde\nu_1^4\right)$ and $d_0=1/\left(2\,\tilde\nu_0^2\right)$, with $\tilde\nu_s$ of Eq.~\eqref{eqrtat24} and the terms of the order $t^2$ taken from the expansion of the integrals over $u$. The terms linear in $t$ vanish. For the node $-$, we have $\Upsilon_s\to-\Upsilon_s$. The internode term of the two nodes is then
\begin{align}
\sigma_\parallel^{\rm inter}=\frac{e^4\,B^2\left(\tau_{\rm inter}-\tau\right)}{36\,\pi^2}\,\frac{\left[a_1(t)+a_0(t)\right]^2}{\sum_s\left(k_s^2/w^0_s\right)\left(1+d_s\,t^2\right)}+\mathcal O\!\left(B^3\right).
\label{eqrtainter4}
\end{align}
For $t=0$, this is the sum of the internode expressions of Sec.~\ref{secrta3} over $s$ and over the two nodes. To the order $t^2$, we have $\sigma_\parallel^{\rm inter}=\sigma_\parallel^{\rm inter}\vert_{t=0}\left(1+\mathcal C_{\rm inter}\,t^2\right)$, where
\begin{align}
\mathcal C_{\rm inter}=\frac{2\sum_s a_s^{(2)}}{\sum_s a_s^{(0)}}-\frac{\sum_s\left(k_s^2/w^0_s\right)d_s}{\sum_s k_s^2/w^0_s}\,,
\label{eqrtacint4}
\end{align}
and $a_s^{(0)}$ and $a_s^{(2)}$ are the coefficients of $t^0$ and $t^2$ in Eq.~\eqref{eqrtaa4}. Since $\Upsilon_s$ is proportional to $B$, the internode term is of the order $B^2$, and it is even in $t$ at this order. Its part odd in $B$ starts at the order $B^3$. The linear magnetochiral term of Eq.~\eqref{eqrtaodd4} is hence independent of $\tau_{\rm inter}$. As in Sec.~\ref{secrta3}, the OMM part of $\Upsilon_s$ is specific to the RTA, and the exact solution contains the OMM through $\rho_s$, $h_s$, and $\omega_s$.

\subsection{Results}
\label{secres4}

The tilt enters the solution of Sec.~\ref{secsol3} only through the FS data. For $\bB\parallel\hat{\boldsymbol z}$, by Eq.~\eqref{eqons4} and the rotation $C_{2x}$, the tensor at $B=0$ obeys $\sigma_{ij}(0;t)=\sigma_{ji}(0;-t)$ and $\bs\sigma(0;-t)=C_{2x}\,\bs\sigma(0;t)\,C_{2x}^{\rm T}$. The elements $\sigma_\parallel(0)$ and $\sigma_\perp(0)$ are therefore even in $t$, and $\sigma_H(0)$ is odd. The tensor at $B=0$ is uniaxial about $\boldsymbol{\hat z}$ for every $t$. The element $\sigma_H(0)$ is the Fermi-sea AH term, which is given by Eq.~\eqref{eqsigah3} with the radius of the tilted sheet $(+,1)$. To leading order in $t$, this radius is $k_{(+,1)}(u)\simeq k_{\Gamma,1}\left[1-t\,u/\left(1+2\,c\,k_{\Gamma,1}\right)\right]$, and
\begin{align}
\sigma_H^{\rm AH}\big\vert_{B=0}\simeq\frac{\eta_B\,e^2\,t\,k_{\Gamma,1}}{3\,\pi^2\left(1+2\,c\,k_{\Gamma,1}\right)}\,.
\label{eqsigah4}
\end{align}
The OMM term of Sec.~\ref{secsol3} is $\eta_B\,\eta_O\,e^3\,B/\left[6\,\pi^2\,k_{\Gamma,1}\left(1+2\,c\,k_{\Gamma,1}\right)\right]$. The ratio of the tilt term to the OMM term is $t/\left(\eta_O\,\alpha\right)$ with $\alpha=e\,B/(2\,k_{\Gamma,1}^2)$, and the tilt term dominates the AH conductivity for $\alpha<t$. Neither term depends on the scattering amplitudes.

For $t\to0$, the solution reduces to that of Sec.~\ref{secsol3}, and the RTA limit is given in Sec.~\ref{secrta4}.

\subsubsection{Longitudinal magnetoconductivity for parallel fields}
\label{secresll4}

For $\bB\parallel\bE\parallel\hat{\boldsymbol z}$, the tilt couples all five moments of Eq.~\eqref{eqmomsys4}, and the conductivity is given by Eq.~\eqref{eqsigpar4}. By Eq.~\eqref{eqons4} and the rotation $C_{2x}$, we have $\sigma_\parallel(B;t)=\sigma_\parallel(-B;-t)$. Hence $\sigma_\parallel$ is even in $t$ at $B=0$, and its part odd in $B$ is odd in $t$. In the RTA, this part is the linear magnetochiral term of Eq.~\eqref{eqrtaodd4}. Beyond the RTA, it arises from the odd moments $a_1$, $a_3$, $H^{(0)}_s$, and $H^{(2)}_s$ of Eq.~\eqref{eqmomdef4}, which vanish for $t=0$. It has no closed form, because Eq.~\eqref{eqmomsys4} must be solved with the weights and rates at nonzero $B$.

At $B=0$, an analytical treatment is possible. The weights and sources depend on $t$ and $u$ only through the product $t\,u$, and we set $v_F=1$ in the remainder of this subsubsection. With $\tilde\nu_s$ of Eq.~\eqref{eqrtat24} and $\rho_s=k_s^2/\tilde\nu_s$ at $t=0$, the expansions to the order $t^2$ read
\begin{align}
\rho_s(u)&=\rho_s\left[1+\mathfrak p_s\,t\,u+\mathfrak q_s\,t^2\,u^2\right],\quad \mathfrak p_s=-\frac{2\,\tilde\nu_s+s}{\tilde\nu_s^2}\,,\quad \mathfrak q_s=\frac{3\left(\tilde\nu_s+s\right)^2}{2\,\tilde\nu_s^4}\,,\nonumber\\
h_s(u)&=\tilde\nu_s\,u+t\left[1-\left(1-\frac{s}{\tilde\nu_s}\right)u^2\right]+t^2\,\frac{\tilde\nu_s^2-s^2}{2\,\tilde\nu_s^3}\,u^3\,.
\label{eqexp4}
\end{align}
For $t=0$, the only nonzero moment is $M^{(1)}_1=\tfrac{2}{3}\,\rho_1\,\mathcal A^\parallel_1$, and $\sigma_\parallel$ reduces to the zero-field limit of Eq.~\eqref{eqsigB0}. The term of the order $t$ in $h_s$ is even in $u$, and the odd part of the weights is $\mathfrak p_s$. Together they drive the even moments $M^{(0)}_s$ and $M^{(2)}_s$ at the order $t$ through a linear system of four equations. Its matrix is the even block of Eq.~\eqref{eqmomsys3} with the amplitudes $\gamma^-_{ss'}$, and its source terms are fixed by $\mathfrak p_s$ and $1-s/\tilde\nu_s$ of Eq.~\eqref{eqexp4}. These moments feed back into $M^{(1)}_1$ and into $\sigma_\parallel$ at the order $t^2$, and the term linear in $t$ vanishes. We solved this perturbation series and compared it with the numerical solution of Eq.~\eqref{eqmomsys4} at small $t$, and the coefficient of $t^2$ agrees with the exact solution. The first-order moment system and the explicit coefficient of $t^2$ are given in Appendix~\ref{secappll4}.

The result has the form $\sigma_\parallel\vert_{B=0}=\sigma_\parallel^{(0)}\left(1+\mathcal C_\parallel\,t^2\right)$, with a coefficient $\mathcal C_\parallel$ that is a rational function of $\tilde\nu_1$, $\tilde\nu_0$, and the amplitudes $\gamma^\pm_{ss'}$, which is given in Appendix~\ref{secappll4}. It differs from the RTA coefficients $a_s$ of Eq.~\eqref{eqrtat24} in two respects. The vertex correction through $\gamma^-_{ss'}$ dresses the amplitudes $\mathcal A^\parallel_s$, and the rates $\tau_s(u)$ of Eq.~\eqref{eqtau40} depend on $u$ through the odd moment $\rho^{(1)}_1$, whereas the RTA uses a constant $\tau$. The dependence on the scattering amplitudes is analysed in Sec.~\ref{secresbeta4}, and that on $B$ is shown numerically in Sec.~\ref{secresnum4}.

\subsubsection{Planar-Hall and Hall response for perpendicular fields and a general angle}
\label{secresph4}

For $\bE\perp\bB$ with $\bB\parallel\hat{\boldsymbol z}$, the sector $m=1$ is governed by Eq.~\eqref{eqmomsysperp} with the tilted weights, the complex rates $\tau_s^{-1}(u)+i\,\omega_s(u)$, and the in-scattering coefficients of Eq.~\eqref{eqlam4}. The transverse and Hall conductivities are given by Eq.~\eqref{eqsigperp4}. Eq.~\eqref{eqons4} and the rotation $C_{2x}$ imply that the Boltzmann Hall conductivity is odd under the simultaneous reversal of $B$ and $t$. Its part odd in $B$ is therefore even in $t$, and its part even in $B$ is odd in $t$, in agreement with Eq.~\eqref{eqrtaHev4}. At $B=0$, we have $\omega_s=0$, and the Boltzmann part $\sigma_H^{\rm B}$ vanishes. The Hall conductivity is then the AH term of Eq.~\eqref{eqsigah4}, which is odd in $t$ and does not depend on the scattering amplitudes.

At $B=0$, the transverse conductivity can be treated in the same way as the longitudinal one. We keep $v_F=1$ and use $\tilde\nu_s$, $\rho_s(u)$, and $h_s(u)$ of Eq.~\eqref{eqexp4}. The source of the sector $m=1$ is $q_s(u)=\sqrt{\left(s+t\,u\right)^2+4\,\bar\mu}-t\,u$, which expands as
\begin{align}
q_s(u)=\tilde\nu_s-\left(1-\frac{s}{\tilde\nu_s}\right)t\,u+\frac{\tilde\nu_s^2-s^2}{2\,\tilde\nu_s^3}\,t^2\,u^2+\mathcal O\!\left(t^3\right).
\label{eqexpq4}
\end{align}
For $t=0$, the only nonzero moment is $\mathcal L^{(0)}_1$, and $\sigma_\perp$ reduces to Eq.~\eqref{eqsigperp0}. The parity of this sector is opposite to that of the sector $m=0$. The moments $\mathcal L^{(1)}_1$ and $\mathcal L^{(1)}_0$ are driven at the order $t$ by the odd parts of $q_s(u)$ and of the weights, and they feed back into $\mathcal L^{(0)}_1$ and into $\sigma_\perp$ at the order $t^2$. The term linear in $t$ vanishes. The coefficient of $t^2$ was obtained by solving the perturbation series with $\gamma^+_{ss'}$ in both the rates and the in-scattering terms. Hence $\sigma_\perp\vert_{B=0}=\sigma_\perp^{(0)}\left(1+\mathcal C_\perp\,t^2\right)$, where $\mathcal C_\perp$ is a rational function of $\tilde\nu_1$, $\tilde\nu_0$, and $\gamma^+_{ss'}$ [cf. Appendix~\ref{secappph4}].

The zero-field anisotropy follows from $\mathcal C_\perp$ and $\mathcal C_\parallel$. It is the sum of two contributions. The first is the internode term of Eq.~\eqref{eqdsig0}, which requires $\beta^{\rm inter}_{11}\neq0$. The second is generated by the tilt at the order $t^2$. It remains for $\beta^{\rm inter}_{ss'}=0$, where $\gamma^+_{ss'}=\gamma^-_{ss'}$, as in the RTA of Eq.~\eqref{eqrtat24}, since the tilt makes the zero-field tensor uniaxial. The planar-Hall current $j_{\rm PH}=\left(\sigma_\perp-\sigma_\parallel\right)\sin\vartheta\,\cos\vartheta\,E$ at $B=0$ is therefore no longer a probe of the internode amplitude alone.

For nonzero $B$ and $\eta_B=\eta_O=0$, the weights and rates are independent of $B$, but they depend on $u$ through $t$, and $\omega_s(u)=e\,B\,q_s(u)/k_{+,s}(u)$. The closed form of Eq.~\eqref{eqlfoexact} then no longer holds, since $\tau_s^{-1}(u)+i\,\omega_s(u)$ is not constant. The Hall conductivity to the first order in $B$ reduces to Eq.~\eqref{eqsigHB3} for $t\to0$. Its dependence on $t$ is even, and the first correction is of the order $t^2$. The part even in $B$ is absent for $\eta_B=\eta_O=0$ and requires the BC or the OMM, in agreement with the terms linear in $\eta_B$ and $\eta_O$ in Eq.~\eqref{eqrtaHev4}. The total Hall conductivity at the leading orders is the sum of the Boltzmann part and of the AH part of Eq.~\eqref{eqsigah4}. Its tilt contribution at $B=0$ dominates over the OMM term of Eq.~\eqref{eqsigH3} for $\alpha<t$, as stated below Eq.~\eqref{eqsigah4}.

\subsubsection{Expansion to the third order in $B$}
\label{secresB34}

For $t\neq0$, the weights, the rates, the frequencies, and the sources remain analytic in $B$ at $B=0$, but the parity rules of Sec.~\ref{secclass4} replace those of Eq.~\eqref{eqexp3}. By Eq.~\eqref{eqons4} and the rotation $C_{2x}$, we have $\sigma_\parallel(B;t)=\sigma_\parallel(-B;-t)$, $\sigma_\perp(B;t)=\sigma_\perp(-B;-t)$, and $\sigma_H(B;t)=-\,\sigma_H(-B;-t)$. Writing $\sigma_\parallel=\sum_n\sigma_\parallel^{(n)}(t)\,B^n$ and likewise for $\sigma_\perp$ and $\sigma_H^{\rm B}$, the coefficients obey
\begin{align}
\sigma_{\parallel,\perp}^{(n)}(-t)=\left(-1\right)^n\sigma_{\parallel,\perp}^{(n)}(t)\,,\quad
\sigma_H^{(n)}(-t)=-\left(-1\right)^n\sigma_H^{(n)}(t)\,.
\label{eqparity4}
\end{align}
The odd powers of $B$ therefore appear in $\sigma_\parallel$ and $\sigma_\perp$ with odd powers of $t$, and the even powers of $B$ appear in $\sigma_H$ with odd powers of $t$. The leading terms are the linear magnetochiral conductivity and the Hall conductivity of the order $B^2\,t$ of Eq.~\eqref{eqrtaHev4}. For the planar-Hall current, the same rules give terms odd in $B$ that are absent for the untilted nodes.

The variational construction of Eqs.~\eqref{eqforms} to \eqref{eqsigcoef} carries over, with the tilted weights, rates, and sources in the forms $\mathcal G^{(m)}_B$ and $\mathfrak H^{(m)}_B$. The coefficients $S^{(m)}_n$ of Eq.~\eqref{eqSn} now exist for every $n\leq3$, and the coefficients $S^{(0)}_1$ and $S^{(0)}_3$ of the sector $m=0$ no longer vanish. They are odd in $t$ by Eq.~\eqref{eqparity4}. The leading terms of the order $t\,B$ in $\sigma_\parallel$ and $\sigma_\perp$ have closed forms, which are given in Appendix~\ref{secappmix4}. A closed form of the coefficient $S^{(0)}_3$ and of the terms of higher order in $t$ is too lengthy to be informative, and we do not carry it out. These coefficients follow from the linear systems of Eqs.~\eqref{eqmomsys4} and \eqref{eqmomsysperp} with the tilted moments and can be obtained numerically whenever needed, as in Sec.~\ref{secresnum4}. The Hall conductivity of the Boltzmann part at the orders $t^2B$ and $t\,B^2$ is given in Appendix~\ref{secappmixH4}.

The AH part is the exception, because it involves only the radius of the sheet $(+,1)$. For the tilted node, the cubic equation of Appendix~\ref{secappqtsm} has the same form with $\tilde\lambda_a\to s+t\,u$, and the expansion of Eq.~\eqref{eqkser} holds at every $u$ with coefficients that depend on $u$ through $k=k_{+,1}(u)$ and $A=1+t\,u$. With $\tilde\nu=A+2\,c\,k$ and $b=e\,B$, the radius is $k_{+,1}(u;B)=k+\sum_{n=1}^{3}\eta_O^n\,\mathfrak k_n\left(b\,u\right)^n+\mathcal O(B^4)$, where
\begin{align}
\mathfrak k_1=-\frac{1}{2\,k\,\tilde\nu}\,,\quad
\mathfrak k_2=-\frac{A+3\,c\,k}{4\,k^3\,\tilde\nu^3}\,,\quad
\mathfrak k_3=-\frac{2\,A^2+11\,c\,A\,k+16\,c^2\,k^2}{8\,k^5\,\tilde\nu^5}\,.
\label{eqkhat4}
\end{align}
Inserting this into Eq.~\eqref{eqsigH4} and expanding in $t$ gives, with $k_1=k_{\Gamma,1}$ and $\tilde\nu_1=1+2\,c\,k_1$,
\begin{align}
\sigma_H^{\rm AH}&\simeq\frac{\eta_B\,e^2\,t\,k_1}{3\,\pi^2\,\tilde\nu_1}\left[1+\frac{3\left(\tilde\nu_1+1\right)}{10\,\tilde\nu_1^4}\,t^2\right]+\frac{\eta_B\,\eta_O\,e^3\,B}{6\,\pi^2\,k_1\,\tilde\nu_1}\left[1-\frac{9\left(\tilde\nu_1-1\right)}{10\,\tilde\nu_1^4}\,t^2\right]\nonumber\\
&\quad+\frac{\eta_B\,\eta_O^2\,e^4\,B^2\,t}{40\,\pi^2\,k_1^3\,\tilde\nu_1^5}\left[8\,\tilde\nu_1^2-9\,\tilde\nu_1+3-\frac{25\left(\tilde\nu_1-1\right)^3}{2\,\tilde\nu_1^4}\,t^2\right]\nonumber\\
&\quad+\frac{\eta_B\,\eta_O^3\,e^5\,B^3}{80\,\pi^2\,k_1^5\,\tilde\nu_1^5}\left[8\,\tilde\nu_1^2-5\,\tilde\nu_1+1+\frac{5\left(128\,\tilde\nu_1^4-325\,\tilde\nu_1^3+345\,\tilde\nu_1^2-175\,\tilde\nu_1+35\right)}{14\,\tilde\nu_1^4}\,t^2\right]+\mathcal O\!\left(t^4,\,B^4\right).
\label{eqahser4}
\end{align}
The terms of the orders $B^0$, $B^1$, and $B^3$ at $t=0$ are those of Eqs.~\eqref{eqsigah4} and \eqref{eqsigHAH3}, and the order of the powers of $t$ agrees with Eq.~\eqref{eqparity4}. The new term is the one of the order $B^2\,t$, which requires the BC and the squared OMM, and it is the AH counterpart of the even-in-$B$ Boltzmann Hall term of Eq.~\eqref{eqrtaHev4}. None of the terms depends on the scattering amplitudes, and the sheets with $s=0$ do not contribute. The derivation of Eq.~\eqref{eqahser4} is given in Appendix~\ref{secappah4}.

\subsubsection{Dependence on the scattering amplitudes}
\label{secresbeta4}

The tilt does not change which amplitudes enter each sector. For $\bE\parallel\bB$, the rates $\tau_s(u)$ of Eq.~\eqref{eqtau40} and the weights $\left\langle f\right\rangle^{\parallel}_s$ involve the sums $\gamma^+_{ss'}$, whereas the in-scattering terms of Eqs.~\eqref{eqmomsys4} and \eqref{eqsigpar4} involve the differences $\gamma^-_{ss'}$. For $\bE\perp\bB$, Eqs.~\eqref{eqlam4} and \eqref{eqsigperp4} contain $\gamma^+_{ss'}$ only. For $t\neq0$, all three amplitudes $\gamma^-_{11}$, $\gamma^-_{10}$, and $\gamma^-_{00}$ enter $\sigma_\parallel$ already at $B=0$, through the even moments that the tilt drives at the order $t$ [cf. Eq.~\eqref{eqappm4}]. For $t=0$, only $\gamma^-_{11}$ enters at $B=0$, as in Eq.~\eqref{eqsigB0}. The AH conductivity of Eqs.~\eqref{eqsigah4} and \eqref{eqahser4} does not depend on the amplitudes.

The dependence on the overall strength follows from a scaling. Let $\gamma^\pm_{ss'}\to\mathfrak f\,\gamma^\pm_{ss'}$ for all the amplitudes. Then $\tau_s(u)\to\tau_s(u)/\mathfrak f$ and $\left\langle f\right\rangle^{\parallel}_s\to\left\langle f\right\rangle^{\parallel}_s/\mathfrak f$ for every $f$, so that $H^{(r)}_s$, $G_s$, and $M^{(n)}_s$ of Eqs.~\eqref{eqmomdef4} and \eqref{eqmomsys4} scale as $1/\mathfrak f$, and the products $\gamma_{ss'}\,a_n$ and $\gamma_{ss'}\,b_n$ are invariant. Eq.~\eqref{eqsigpar4} then gives $\sigma_\parallel\to\sigma_\parallel/\mathfrak f$ for every $B$ and $t$. The relative change $\delta\sigma_\parallel$ is therefore independent of the scattering strength, and in particular of $\mu\,\tau_0$. In the sector $m=1$, the frequency $\omega_s$ does not scale, and $\delta\sigma_\perp$ and $\sigma_H^{\rm B}$ depend on the strength through $\omega_s\,\tau_s$.

For the ratio $r_\beta$ of the figures, we have $\beta^{\rm inter}_{ss'}=r_\beta\,\beta^{\rm intra}_{ss'}$ for all the pairs of bands, so that $\gamma^\pm_{ss'}=\left(1\pm r_\beta^2\right)\left(\beta^{\rm intra}_{ss'}\right)^2$. The time $\tau_0$ is the rate of the sheet $(+,1)$ at $B=0$ and $t=0$, and it depends on $\gamma^+_{ss'}$ only. Fixing $\tau_0$ therefore fixes $\gamma^+_{ss'}$ and leaves the amplitudes in the in-scattering of the sector $m=0$ as
\begin{align}
\gamma^-_{ss'}=\tilde\delta\,\gamma^+_{ss'}\,,\quad
\tilde\delta=\frac{1-r_\beta^2}{1+r_\beta^2}\,,
\label{eqkappa4}
\end{align}
with $\tilde\delta\in\left(-1,1\right]$. Two consequences follow. The sector $m=1$, and hence $\sigma_\perp$, $\sigma_H^{\rm B}$, and $\sigma_H^{\rm AH}$ for $\bB\parallel\hat{\boldsymbol z}$, do not depend on $r_\beta$ at fixed $\tau_0$, for every $B$ and $t$. The longitudinal conductivity depends on $r_\beta$ only through $\tilde\delta$, and the scan over $r_\beta$ in Fig.~\ref{figtiltratio} is a scan over the vertex correction of the sector $m=0$ alone.

For $r_\beta=1$, we have $\tilde\delta=0$, and the in-scattering of the sector $m=0$ vanishes. Eq.~\eqref{eqmomsys4} gives $M^{(n)}_s=H^{(n)}_s$, and Eq.~\eqref{eqsigpar4} reduces to
\begin{align}
\sigma_\parallel\Big\vert_{r_\beta=1}=\frac{e^2}{2\,\pi^2}\sum_s\left\langle h_s^2\right\rangle^{\parallel}_s\,.
\label{eqsigpar4r1}
\end{align}
This is the RTA with the rate $\tau_s(u)$ of Eq.~\eqref{eqtau40} in place of the constant $\tau$ of Sec.~\ref{secrta4}. For $r_\beta<1$, the factor $\tilde\delta$ is positive, and for $r_\beta>1$ it is negative, which changes the sign of the vertex correction.

At $B=0$ and $t=0$, Eq.~\eqref{eqsigB0} and its counterpart for $\sigma_\perp$ give
\begin{align}
\left(\sigma_\perp-\sigma_\parallel\right)\Big\vert_{B=0}=\frac{e^2}{3\,\pi^2}\,k_1^2\,w^0_1\,\tau_1\,\frac{\left(1-\tilde\delta\right)\tilde\varsigma_1}{\left(1-\tilde\varsigma_1\right)\left(1-\tilde\delta\,\tilde\varsigma_1\right)}\,,\quad
\tilde\varsigma_1=\frac23\,\gamma^+_{11}\,\rho_1\,\tau_1\,,
\label{eqdsig04}
\end{align}
where $0<\tilde\varsigma_1\leq1/2$ by the expression of $\tau_1^{-1}$ below Eq.~\eqref{eqmomsys3}. The difference is proportional to $1-\tilde\delta=2\,r_\beta^2/\left(1+r_\beta^2\right)$ at fixed $\gamma^+_{ss'}$, and it vanishes for $r_\beta=0$, in agreement with Eq.~\eqref{eqdsig0}. For $t\neq0$, a second contribution remains at $r_\beta=0$, as discussed in Sec.~\ref{secresph4}.

\subsubsection{Numerical results}
\label{secresnum4}

The figures show the exact solution for the system of Table~\ref{tabgeom4}. The parameters are those of Sec.~\ref{secresnum3}, with $\mu\,\tau_0=15$, $c=0.1\,v_F/\mu$, and $r_\beta=1/2$ unless stated otherwise. The scattering amplitudes are fixed by $\tau_0$ of the sheet $(+,1)$ at $B=0$ and $t=0$, and they do not change with $t$. The field enters through $\alpha=e\,B/(2\,k_1^2)$, where $k_1$ is the radius of the sheets with $s=1$ at $B=0$ and $t=0$, and $t$ is in units of $v_F$. All tilts obey $|t|<c\,\mu/v_F$. The change $\delta\sigma$ is relative to $B=0$ and in units of the value of the same element at $B=0$ for the case (a), and in units of $\sigma_0=\mathrm{tr}\,\bs\sigma(0)/3$ for the cases (b) and (c). Both signs of $\alpha$ are shown. For $\bB\parallel\hat{\boldsymbol z}$, the solution coincides with that of the sectors of Sec.~\ref{secsol3}, and for $t=0$ with that of Sec.~\ref{secsys3}.

\begin{figure}[t!]
\centering
\subfigure[\label{figtilta}]{\includegraphics[width=\textwidth]{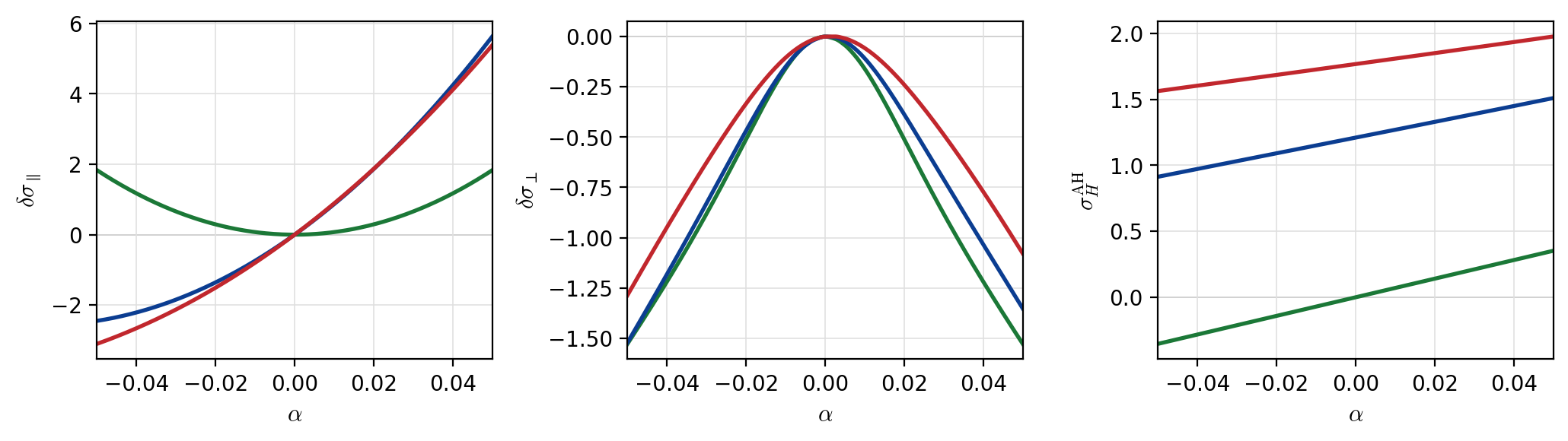}}\\
\subfigure[\label{figtiltratio}]{\includegraphics[width=0.85\textwidth]{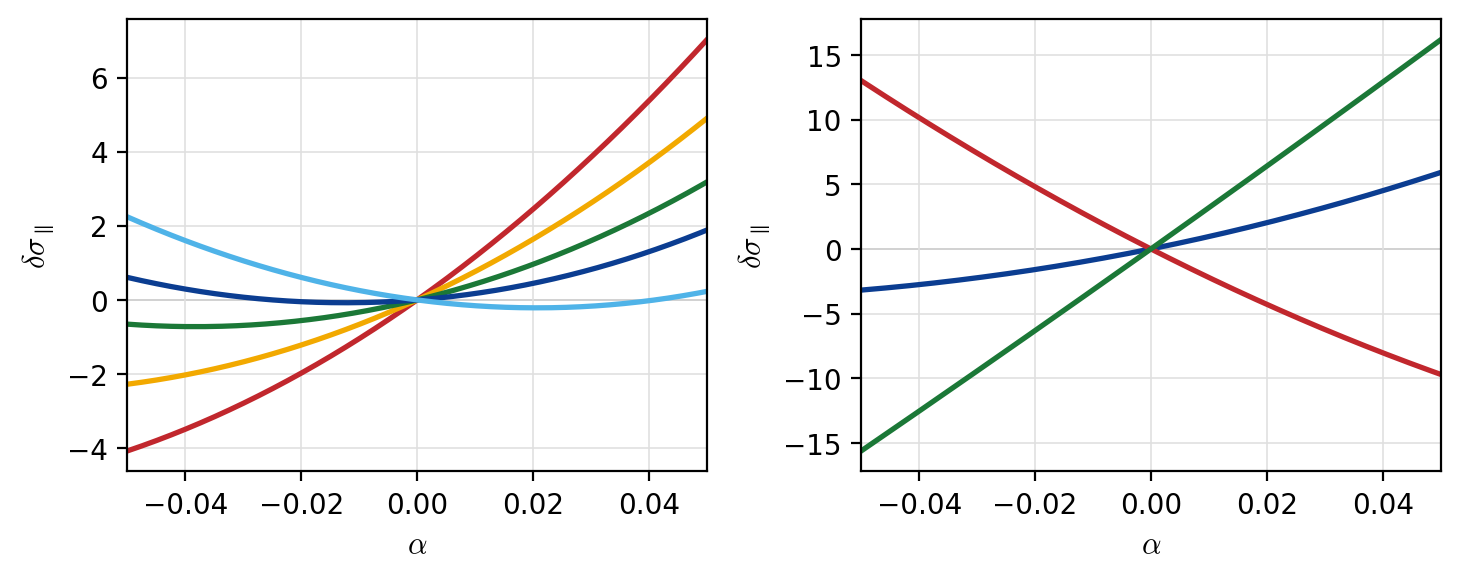}}
\caption{Two axially-tilted conjugate QTSM nodes for $\bB\parallel\hat{\boldsymbol z}$ [case (a)]. (a) The curves are for $t=0$ (green), $0.04$ (navy), and $0.08$ (crimson). Left: $\delta\sigma_\parallel$ multiplied by $10^3$. Middle: $\delta\sigma_\perp$ multiplied by $10$. Right: $\sigma_H^{\rm AH}$ multiplied by $10^3$, in units of $\sigma_\perp$ at $B=0$. (b) For $\bE\parallel\bB$ and $t=0.06$, $\delta\sigma_\parallel$ is multiplied by $10^3$. Left: $r_\beta=0.4$ (crimson), $0.6$ (amber), $0.8$ (green), $1$ (navy), and $1.4$ (sky). Right: $r_\beta=1/2$ with $\eta_B=\eta_O=1$ (navy), $\eta_O=0$ (crimson), and $\eta_B=0$ (green).}
\label{figtiltbz}
\end{figure}

Fig.~\ref{figtilta} shows the three functions of the case (a). For $t=0$, the curves are even in $\alpha$ and reproduce Sec.~\ref{secresnum3}. For $t\neq0$, $\delta\sigma_\parallel$ acquires a term odd in $\alpha$, which grows with $t$ and is the linear magnetochiral term of Sec.~\ref{secclass4}. It increases $\delta\sigma_\parallel$ for $\alpha>0$ and decreases it for $\alpha<0$. For the larger tilt, $\delta\sigma_\parallel$ is negative in a range of $\alpha<0$. The asymmetry of $\delta\sigma_\perp$ is weaker. Its magnitude is larger for $\alpha<0$, and the tilt increases it for $\alpha<0$ and reduces it for $\alpha>0$. In the right panel, the AH conductivity is linear in $\alpha$ with a nonzero intercept, which is the tilt term of Eq.~\eqref{eqsigah4}. The intercept grows with $t$, and the slope, which is the OMM term, decreases slightly. The Hall response is dominated by the Boltzmann part as for $t=0$, and the planar-Hall response follows from the first two panels through Eq.~\eqref{eqcurr}. It is dominated by the classical difference $\sigma_\perp-\sigma_\parallel$, as in Sec.~\ref{secresnum3}, and it is not shown. In the sum of the pair and its time-reversed partner, only the parts of $\delta\sigma_\parallel$ and $\delta\sigma_\perp$ that are even in $\alpha$ remain. The odd parts are properties of a single pair.

Fig.~\ref{figtiltratio} shows the dependence of $\delta\sigma_\parallel$ on $r_\beta$ and on the BC and the OMM. At fixed $\tau_0$, only $\sigma_\parallel$ depends on $r_\beta$, and it does so through $\tilde\delta$ of Eq.~\eqref{eqkappa4}. In the left panel, the even part of the change is positive and depends only weakly on $r_\beta$. The odd part decreases monotonically with $r_\beta$ and changes its sign at a ratio of order unity. The sign of the term linear in $B$ is therefore controlled by the ratio, as the sign of the even term is for $t=0$ in Fig.~\ref{figqtsmratio}. In the right panel, the odd part has opposite signs for the BC and for the OMM, and the OMM contribution is larger. The combined odd part is smaller than the OMM contribution alone, and the cross term is small. The even part is positive for both, and it is dominated by the BC.

\begin{figure}[t!]
\centering
\includegraphics[width=\textwidth]{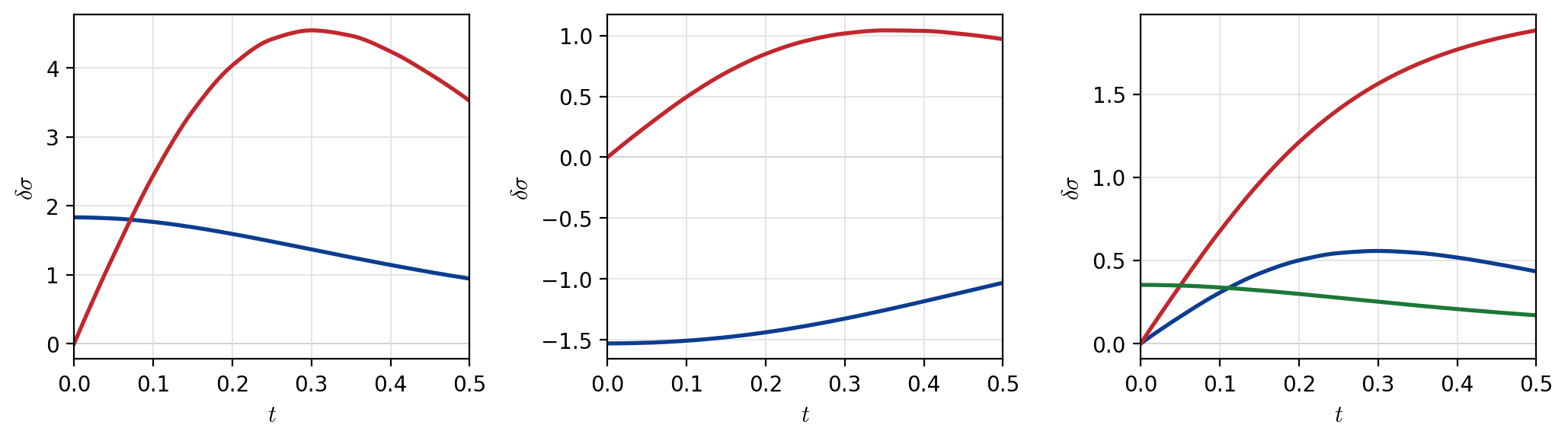}
\caption{Two axially-tilted conjugate QTSM nodes for $\bB\parallel\hat{\boldsymbol z}$ at $\alpha=0.05$ as functions of $t$ for $0\leq t\leq0.09$. Left: even (navy) and odd (crimson) parts of $\delta\sigma_\parallel$ multiplied by $10^3$. Middle: even part of $\delta\sigma_\perp$ multiplied by $10$ (navy) and odd part multiplied by $10^2$ (crimson). Right: even part of $\sigma_H^{\rm B}$ multiplied by $10^2$ (navy), $\sigma_H^{\rm AH}$ at $B=0$ multiplied by $10^3$ (crimson), and odd part of $\sigma_H^{\rm AH}$ multiplied by $10^3$ (green), in units of $\sigma_\perp$ at $B=0$.}
\label{figtiltscan}
\end{figure}

Fig.~\ref{figtiltscan} shows the dependence on $t$ at fixed $\alpha$. The even part of $\delta\sigma_\parallel$ is positive and decreases monotonically with $t$. The odd part vanishes at $t=0$ and grows almost linearly with $t$. The same holds for the odd part of $\delta\sigma_\perp$, which is positive, whereas the even part of $\delta\sigma_\perp$ is negative and decreases in magnitude with $t$. In the right panel, the Boltzmann Hall conductivity acquires a part even in $\alpha$, which vanishes at $t=0$ and increases with $t$. It is the counterpart of the AH term at $B=0$, which also increases with $t$, whereas the odd part of the AH conductivity decreases slightly. The terms odd in $\alpha$ for the diagonal elements and even in $\alpha$ for the Hall element are all odd in $t$, as required by Sec.~\ref{secclass4}.

\begin{figure}[t!]
\centering
\subfigure[\label{figtiltb}]{\includegraphics[width=0.85\textwidth]{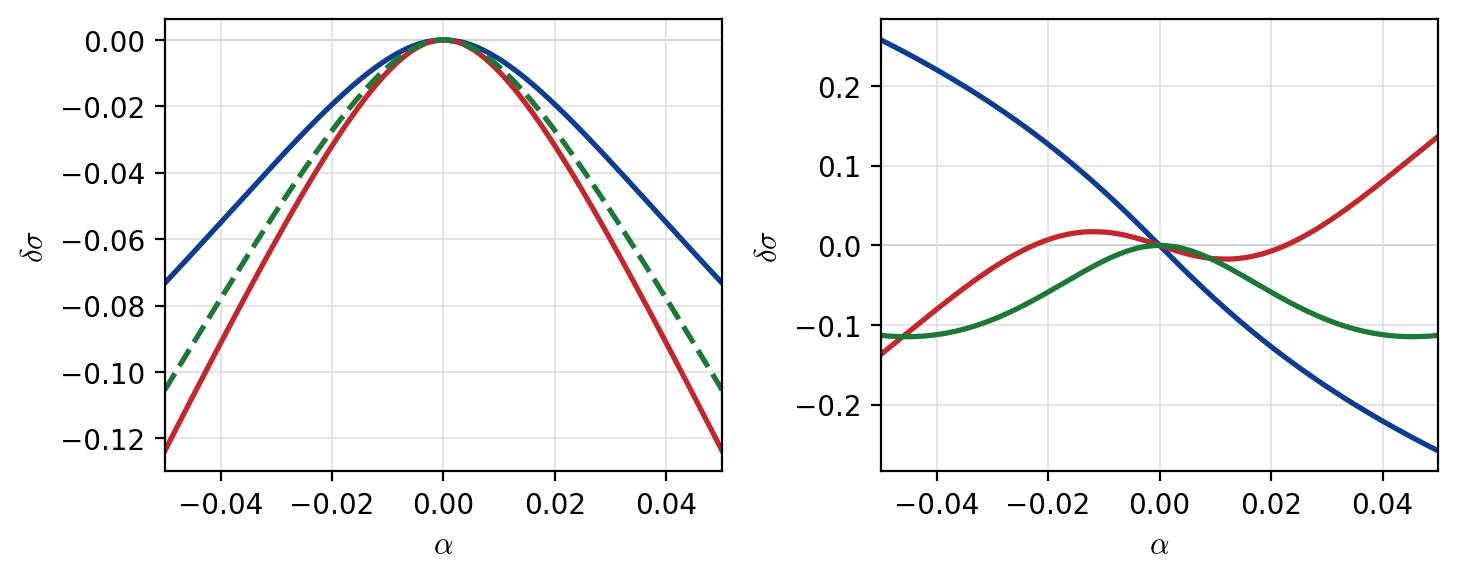}}\\
\subfigure[\label{figtiltc}]{\includegraphics[width=0.85\textwidth]{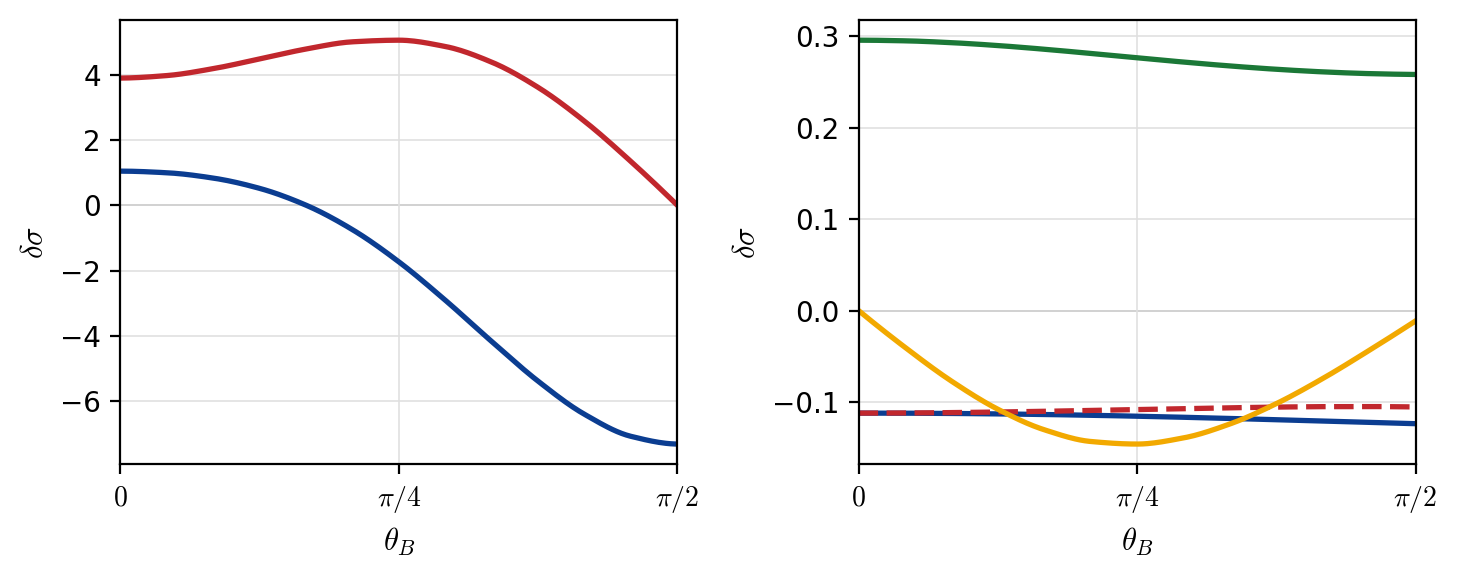}}
\caption{Two axially-tilted conjugate QTSM nodes with $t=0.08$. (a) Case (b) with $\bB\parallel\hat{\boldsymbol x}$. Left: $\delta\sigma_{xx}$ multiplied by $10^3$ (navy), $\delta\sigma_{yy}$ multiplied by $10$ (crimson), and $\delta\sigma_{zz}$ multiplied by $10$ (green, dashed). Right: $\delta\sigma_{yz}$ multiplied by $10$ (navy), $\delta\sigma_{xz}$ multiplied by $10^3$ (crimson), and $\delta\sigma_{xy}$ multiplied by $10^3$ (green), in units of $\sigma_0$. (b) Case (c) with $\bB=B\left(\sin\theta_B,0,\cos\theta_B\right)$ and $\alpha=0.05$. Left: even (navy) and odd (crimson) parts of $\delta\sigma_\parallel=\hat{\bB}\cdot\delta\bs\sigma\cdot\hat{\bB}$ multiplied by $10^3$. Right: eigenvalues $\delta\sigma_{\perp,1}$ (navy) and $\delta\sigma_{\perp,2}$ (crimson, dashed) of the symmetric part of $\delta\bs\sigma$ in the plane perpendicular to $\bB$ multiplied by $10$, the component $H_\parallel$ of the Hall vector along $\bB$ multiplied by $10$ (green), and its component $H_1$ along $\left(\cos\theta_B,0,-\sin\theta_B\right)$ multiplied by $10^3$ (amber), in units of $\sigma_0$.}
\label{figtiltbc}
\end{figure}

For the case (b), we take $\bB=B\,\boldsymbol{\hat x}$ and $t=0.08$, and the tensor is obtained from the general solver. In Fig.~\ref{figtiltb}, the change $\delta\sigma_{xx}$ for $\bE\parallel\bB$ is positive and even in $\alpha$, as the parity rule of Sec.~\ref{secclass4} requires. It has the sign of the untilted nodes in Sec.~\ref{secresnum3}, and its magnitude is smaller than that of the transverse elements by more than two orders of magnitude. The tilt perpendicular to $\bB$ removes the axisymmetry about $\bB$. The changes $\delta\sigma_{yy}$ and $\delta\sigma_{zz}$ are negative and even in $\alpha$, and they differ only slightly. In the right panel, the Hall element $\sigma_{yz}$ is odd in $\alpha$ and large. The elements $\sigma_{xz}$ and $\sigma_{xy}$ are smaller by more than two orders of magnitude. They vanish for $t=0$ and are due to the tilt, since $C_{2x}$ is broken. The element $\sigma_{xz}$ is odd in $\alpha$ and is not monotonic, and $\sigma_{xy}$ is even in $\alpha$.

Fig.~\ref{figtiltc} shows the case (c) as a function of $\theta_B$. The even part of $\delta\sigma_\parallel$ is positive at all angles. It is the longitudinal change of the case (a) at $\theta_B=0$, it changes weakly at small $\theta_B$, and it decreases to its smallest value at $\theta_B=\pi/2$, which is the case (b). This is the sign of the untilted nodes (Sec.~\ref{secresnum3}). The odd part is positive, has a maximum at an intermediate angle, and vanishes at $\theta_B=\pi/2$, where the parity rule of the case (b) makes $\delta\sigma_\parallel$ even in $B$. In the right panel, the two eigenvalues of the symmetric transverse block are negative. They coincide at $\theta_B=0$, where the axisymmetry is restored, and they split with increasing $\theta_B$. The Hall vector is dominated by $H_\parallel$, which is positive and decreases with $\theta_B$. The component $H_1$ is negative and smaller than $H_\parallel$ by more than two orders of magnitude. It vanishes at $\theta_B=0$ and is not monotonic. The component along $\boldsymbol{\hat y}$ vanishes by Eq.~\eqref{eqplanar4} and is not shown.

%%%%%%%%%%%%%%%
\section{Comparison with earlier work}
\label{seclit}

Here we compare the exact solution with the RTA, including the phenomenological modelling used to interpret the measurements on CoSi \cite{claudia-multifold}. We then discuss the inadequacies of the earlier works that our treatment fills in. Ref.~\cite{ahmad-sharma-2025} treats conjugate nodes of pseudospin 1 and 3/2 beyond the RTA for collinear fields, and Ref.~\cite{ahmad-phc-2026} extends the treatment of pseudospin-1 nodes to noncollinear fields and tilted nodes.

\subsection{Relaxation-time approximation and the exact solution}

\begin{figure}[t!]
\centering
\includegraphics[width=\textwidth]{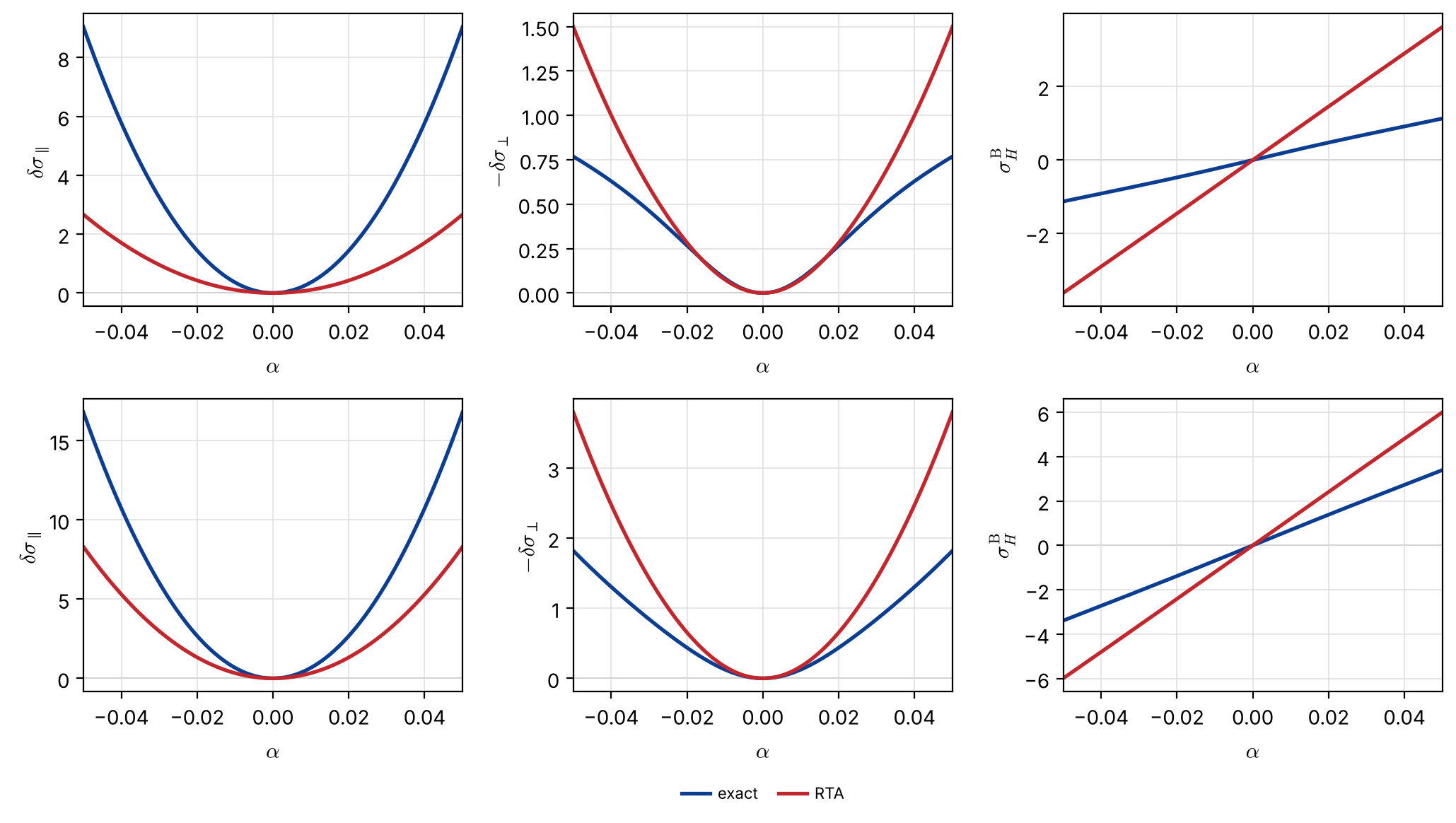}
\caption{Exact solution (navy) and RTA (crimson) for $\bB\parallel\hat{\boldsymbol z}$ and $r_\beta=1/2$. Top: QTSM and Weyl nodes. Bottom: RSW and TSM nodes. Left: $\delta\sigma_\parallel$ multiplied by $10^6$. Middle: $-\delta\sigma_\perp$ multiplied by $10^3$. Right: $\sigma_H^{\rm B}$ multiplied by $10^2$. The two curves of a panel have the same scale factor.}
\label{figcmp}
\end{figure}

The RTA of Sec.~\ref{secrta} drops the in-scattering term and relaxes all the sheets with the time $\tau=\tau_0$ of the reference sheet, which fixes $\beta_{\rm intra}$ in the exact solution. Fig.~\ref{figcmp} compares the two for the systems of Secs.~\ref{secsys} and \ref{secsys2}, with the parameters of Figs.~\ref{figqw} and \ref{figrt} and $\bB\parallel\hat{\boldsymbol z}$. Every curve is in units of the value of its own solution at $B=0$, and the Hall response is in units of $\sigma_\perp$ at $B=0$. Table~\ref{tabrta} gives the ratios of the RTA to the exact solution at the largest field.

\begin{table}[t!]
\centering
\small
\renewcommand{\arraystretch}{1.5}
\begin{tabular}{|c|c|c|c|c|}
\hline
System & $\sigma_\parallel(0)$ & $\delta\sigma_\parallel$ & $\delta\sigma_\perp$ & $\sigma_H^{\rm B}$ \\
\hline
QTSM and Weyl nodes & $3.33$ & $0.30$ & $1.95$ & $3.21$ \\
\hline
RSW and TSM nodes & $1.81$ & $0.49$ & $2.09$ & $1.77$ \\
\hline
\end{tabular}
\caption{\label{tabrta} Ratios of the RTA to the exact solution for $\bB\parallel\hat{\boldsymbol z}$, $r_\beta=1/2$, and $\alpha=0.05$. The first column is the ratio of the conductivities at $B=0$. The other columns are the ratios of the changes, each measured in units of the value of its own solution at $B=0$.}
\end{table}

The RTA overestimates the conductivity at $B=0$ in both systems, since the sheets at $R$, which carry almost the whole conductivity, have shorter out-scattering times than the reference sheet. It does not depend on $r_\beta$, because the internode amplitude enters only through the in-scattering term and through the fixed time $\tau_0$. The longitudinal change has the same sign and the same quadratic dependence on $\alpha$ in the two solutions, but the RTA underestimates it, by a larger factor for the QTSM and Weyl nodes. The magnitudes of the transverse change and of the Hall response are overestimated. The ratios of Table~\ref{tabrta} differ between the two systems and between the quantities. For $\bE\parallel\bB$, the RTA solution is $\Lambda=\tau\,h$, and its relative longitudinal change is therefore independent of $\tau$. No rescaling of $\tau$ removes the difference in the column of $\delta\sigma_\parallel$. The AH conductivity depends only on the FS [cf. Eq.~\eqref{eqsigah2}] and is the same in the two solutions in absolute units, and it differs in Fig.~\ref{figcmp} only through the normalisation. The in-scattering term thus changes the longitudinal, the transverse, and the Hall responses quantitatively at the field strengths of the figure.

\begin{figure}[t!]
\centering
\includegraphics[width=0.5\textwidth]{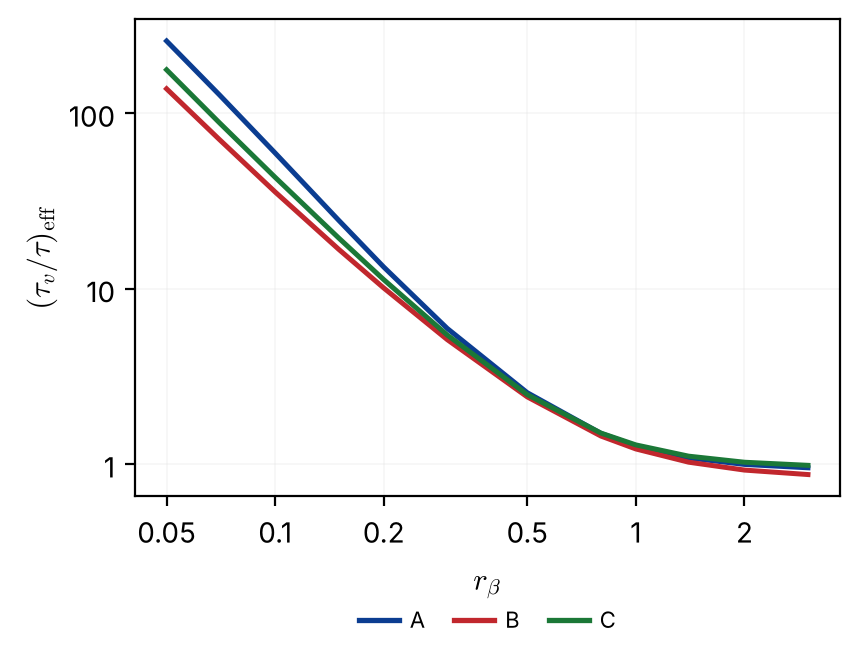}
\caption{QTSM and Weyl nodes: effective ratio $\left(\tau_v/\tau\right)_{\rm eff}$ versus $r_\beta$. A (navy): linear dispersions, $k_F^R=5\,k_F^\Gamma$, and $v_R=v_\Gamma$. B (crimson): the same with $v_R=v_\Gamma/\sqrt3$. C (green): the RhSi parameters of Sec.~\ref{secresnum}.}
\label{figxeff}
\end{figure}

The interpretation of the longitudinal magnetoconductance of CoSi in Ref.~\cite{claudia-multifold} goes beyond the RTA with a single time. It uses the intranode time $\tau$, the internode time $\tau_v$, and a shift $\delta\mu\propto\bE\cdot\bB$ of the chemical potential, and it keeps the BC and the OMM of the linear band of the threefold node at $\Gamma$ and of the two Weyl nodes at $R$. For $\bE$ at the angle $\theta$ to $\bB$, the change of the conductivity is
\begin{align}
\frac{\sigma-\sigma_0}{\sigma_0}=\frac{\left(\check a+\check b\,\tau_v/\tau\right)\cos^2\theta-\check c}{20\left(1+2\,r_2\right)}\,\frac{e^2\,B^2}{\hbar^2\left(k_F^\Gamma\right)^4}\,,
\label{eqlitbald}
\end{align}
with $\check a=-9+2\,r_1$, $\check b=10\left(3+r_1\right)$, $\check c=2+4\,r_1$, $r_1=\left(k_F^\Gamma/k_F^R\right)^2 v_R/v_\Gamma$, and $r_2=\left(k_F^R/k_F^\Gamma\right)^2 v_R/v_\Gamma$. For $\tau_v=\tau$ and $\theta=0$, the contribution of the Weyl nodes reduces to the RTA expression of Eq.~\eqref{eqrtaweyl}, up to the factor $2$ of the spin degeneracy that is kept in Ref.~\cite{claudia-multifold}. The ratio $\tau_v/\tau$ is a fitting parameter there, whereas in the exact solution the internode scattering enters through the amplitude $r_\beta\,\beta$ and the overlaps. We therefore define an effective ratio $\left(\tau_v/\tau\right)_{\rm eff}$ as the value for which Eq.~\eqref{eqlitbald} at $\theta=0$ reproduces the coefficient of $B^2$ of the exact $\sigma_\parallel$, which is obtained from two small values of $\alpha$ by extrapolation. Fig.~\ref{figxeff} shows it as a function of $r_\beta$ for the system of Sec.~\ref{secsys}.

For the three parameter sets, the effective ratio decreases monotonically with $r_\beta$. For small $r_\beta$, the product $\left(\tau_v/\tau\right)_{\rm eff}\,r_\beta^2$ is approximately constant, and the effective ratio is therefore proportional to $r_\beta^{-2}$. This is the dependence expected of an internode time that is set by the internode rate, which is proportional to $r_\beta^2$. For $r_\beta\gtrsim1$, the effective ratio is of order unity, which corresponds to the RTA with equal times. The sets A and B have the ratio $k_F^R/k_F^\Gamma=5$ of that reference, and the set C has the larger ratio of Sec.~\ref{secresnum}. Eq.~\eqref{eqlitbald} is not an exact solution, and this mapping relates only its $B^2$ coefficient at $\theta=0$ to the exact one. It shows that the large ratio $\tau_v/\tau\approx10^2$ fitted in Ref.~\cite{claudia-multifold} corresponds, within the exact treatment, to weak internode amplitudes $r_\beta\ll1$ for all three sets.

Four consequences for the interpretation of the measurements follow from the comparison of Eq.~\eqref{eqlitbald} with the exact solution. First, the dependence on $\theta$ in Eq.~\eqref{eqlitbald} follows from the axisymmetry about $\bB$. For isotropic nodes, the symmetric part of the tensor has the form of Eq.~\eqref{eqrot}, and the change of the current along $\bE$ is $\delta\sigma_\perp\sin^2\theta+\delta\sigma_\parallel\cos^2\theta$ for every strength of the scattering. The $\cos^2\theta$ dependence therefore does not identify the chiral anomaly by itself, which enters through the coefficient. For the second-case overlaps, the dependence on the orientation of $\bB$ relative to $\hat{\bs c}$ in Fig.~\ref{figqwbc} is a correction to this form. Second, the expansion of the distribution function in Ref.~\cite{claudia-multifold} is truncated at the second order in $\tau$. The term of the third order, which gives the classical $B^2$ dependence of the response for $\bE\perp\bB$, is therefore absent from Eq.~\eqref{eqlitbald}. By Eq.~\eqref{eqrtaweylperp}, the ratio of the classical term of a Weyl node to its BC term is $20\left(\mu_R\,\tau\right)^2$ in the units $\hbar=1$, which is large for $\mu_R\,\tau\gg1$. Third, the finding of Ref.~\cite{claudia-multifold} that the OMM reduces but does not overcome the BC contribution to the longitudinal magnetoconductance agrees with Figs.~\ref{figqwratio} and \ref{figrtratio}. In both, the OMM contribution alone is smaller than the BC contribution alone, and the combined change is smaller than the BC contribution alone.

Fourth, the cubic-in-$B$ term of the Hall resistivity, which Ref.~\cite{claudia-multifold} attributes to the BC and the OMM of the linear band at $\Gamma$, is dominated by the Boltzmann response of the sheets in the exact solution. For $\bB$ perpendicular to the plane of the current, the resistivity of the in-plane block is $\tilde\rho_{xy}=\sigma_H/\left(\sigma_\perp^2+\sigma_H^2\right)$. Its terms beyond the linear one in $B$ arise already from the dependences of $\sigma_\perp$ and $\sigma_H^{\rm B}$ on $B$ for the Boltzmann response of several sheets with different radii and times. We extract the coefficients of $\alpha$ and $\alpha^3$ of $\tilde\rho_{xy}$ from a fit for $\alpha\leq0.02$, for the parameter set A of Fig.~\ref{figxeff} with $r_\beta=0.07$ and $0.5$ and for the RhSi parameters with $r_\beta=1/2$. The AH conductivity contributes a negligible part of the cubic coefficient in all three cases, and the cubic coefficient with and without the AH part is practically the same. The RTA underestimates the ratio of the cubic to the linear coefficient considerably. In this model, the cubic term of $\tilde\rho_{xy}$ is therefore not a signature of the BC and the OMM. It follows from the classical response of the sheets at $\Gamma$ and $R$, and the in-scattering term changes its size appreciably. 

%%%%%%%%%%%%%%%%%%
\subsection{Inconsistent orbital magnetic moment}

Refs.~\cite{ahmad-sharma-2025} and \cite{ahmad-phc-2026} use an OMM of the pseudospin-1 node that is twice the value given by the starting expression of Ref.~\cite{ips-exact-spin1}. With $\boldsymbol m^s(\bk)=-\left(i\,e/2\right)\left[\nabla_{\bk}\psi_s\right]^\dagger\times\left[\left(\mathcal H-\varepsilon^s\right)\nabla_{\bk}\psi_s\right]$, the upper band of the pseudospin-1 node has the magnitude $e\,v_F/(2\,k)$, which is Eq.~\eqref{eqbcomm} with $G_a=1$, whereas the two works quote $e\,v_F/k$. For the pseudospin-3/2 node, Ref.~\cite{ahmad-sharma-2025} quotes the weights $7/4$ and $3/4$ of Ref.~\cite{ips-exact-rsw} and Table~\ref{tabnodes2}. The OMM weights of its pseudospin-1 and pseudospin-3/2 nodes are therefore inconsistent with each other. The BC is not affected, but the balance of the OMM and BC contributions of the pseudospin-1 node is wrong by a factor of two. This balance fixes the sign reversals of the longitudinal magnetoconductivity and of the planar Hall conductivity, the critical amplitudes $\tilde\alpha_c$ including their angular dependences in Ref.~\cite{ahmad-phc-2026}, and the comparison with the pseudospin-1/2 node. The quantitative pseudospin-1 results of both works are therefore likely to change when recomputed with the correct OMM.

To estimate the size of this error, we use the model of these works, which has two conjugate pseudospin-1 nodes with a linear dispersion and a single FS sheet per node. This is the system of Sec.~\ref{secsys3} without the sheets $(\chi,0)$ and without the quadratic term. The other parameters are those of Sec.~\ref{secresnum3}, with $\mu\,\tau_0=15$, $\alpha=0.05$, and $r_\beta=1/2$ unless stated otherwise. The ratio of the internode to the intranode strength of Refs.~\cite{ahmad-sharma-2025, ahmad-phc-2026} enters the rate linearly [cf. Eq.~\eqref{eqlittr}] and therefore corresponds to $r_\beta^2$. We solve the model for $\bE\parallel\bB\parallel\hat{\boldsymbol z}$ with the OMM of Eq.~\eqref{eqbcomm}, with the doubled OMM, and with the BC alone. Ref.~\cite{ahmad-sharma-2025} does not specify the magnetic field of its figures in terms of the model parameters, since neither the Fermi energy, nor the Fermi velocity, nor the impurity strength is given a value, and the absolute scale of its curves cannot be reproduced. The calculation therefore isolates the effect of the doubled OMM within the present solver.

\begin{figure}[t!]
\centering
\subfigure[\label{figlitomm}]{\includegraphics[width=0.85\textwidth]{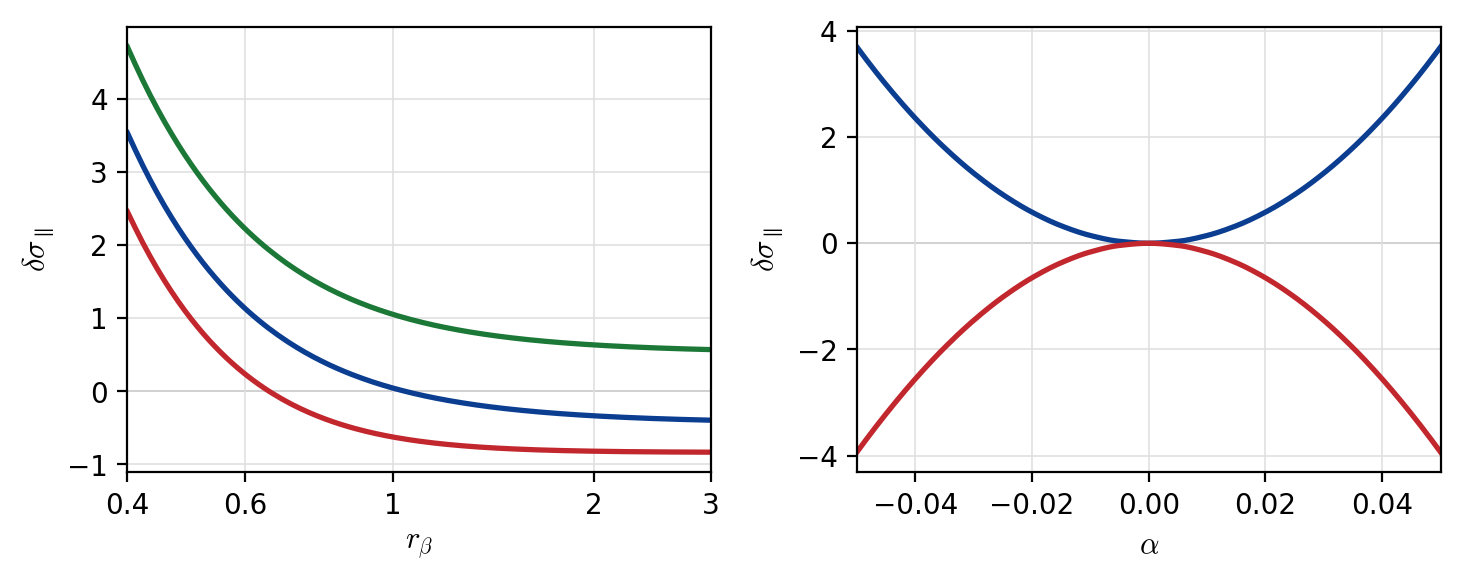}}\\
\subfigure[\label{figlitlfo}]{\includegraphics[width=\textwidth]{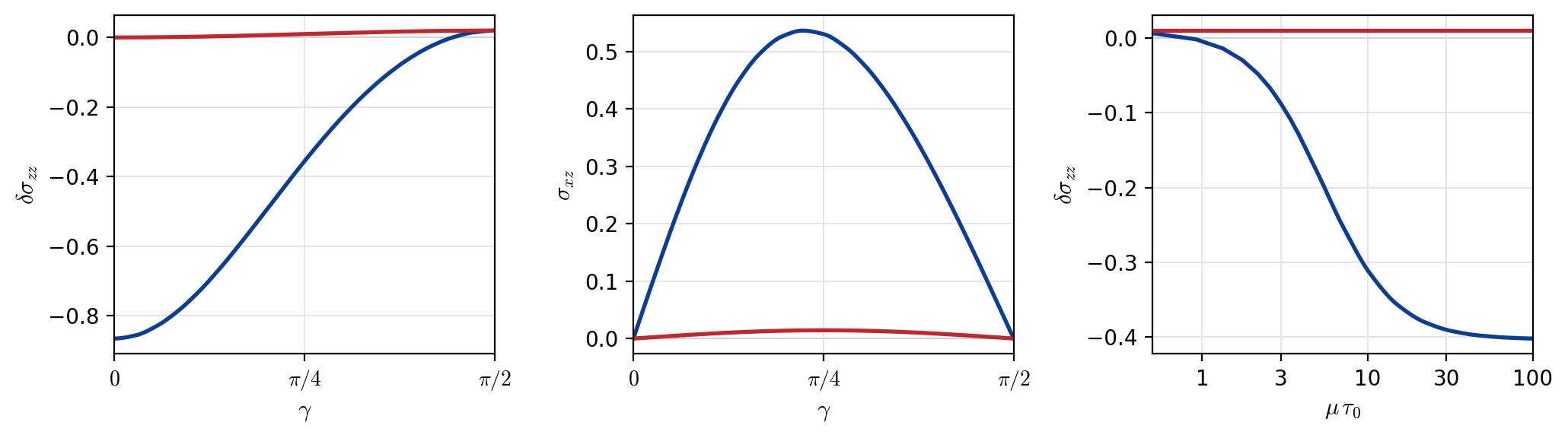}}
\caption{Two conjugate pseudospin-1 nodes with one FS sheet per node. (a) Here $\bE\parallel\bB\parallel\hat{\boldsymbol z}$. Left: $\delta\sigma_\parallel$ multiplied by $10^2$ versus $r_\beta$ for $\alpha=0.05$, with the correct OMM (navy), the doubled OMM (crimson), and the BC alone (green). Right: $\delta\sigma_\parallel$ multiplied by $10^3$ versus $\alpha$ for $r_\beta=0.8$, with the correct OMM (navy) and the doubled OMM (crimson). (b) Here $\bE\parallel\hat{\boldsymbol z}$ and $\bB=B\left(\cos\tilde\vartheta,0,\sin\tilde\vartheta\right)$. The exact solution is shown in navy, and the truncated solution with $\hat{\mathcal R}_a\to\tau_a$ in crimson. Left: $\delta\sigma_{zz}$ versus $\tilde\vartheta$. Middle: $\sigma_{xz}$ versus $\tilde\vartheta$. Right: $\delta\sigma_{zz}$ versus $\mu\,\tau_0$ for $\tilde\vartheta=\pi/4$.}
\label{figlit}
\end{figure}

Fig.~\ref{figlitomm} shows the longitudinal change in units of the value of the same solution at $B=0$. With the BC alone, the change is positive and decreases with increasing $r_\beta$. The OMM reduces it for every $r_\beta$, and the doubled OMM reduces it further. With the OMM, the change reverses its sign at a critical internode amplitude, which is considerably smaller for the doubled OMM than for the correct one (Table~\ref{tablit}) and does not depend on $\mu\,\tau_0$. For the amplitudes between the two critical values, the change is positive and quadratic in $\alpha$ with the correct OMM, and negative and quadratic with the doubled OMM. The doubled OMM therefore gives the wrong sign of the longitudinal magnetoconductivity in this window, and it changes the magnitude by a factor of order unity outside it. It also doubles the AH conductivity, which requires both the BC and the OMM. The transverse and the Boltzmann Hall responses are practically unchanged, since the classical LFO contribution controls them (Table~\ref{tablit}). In the two-sheet system of Sec.~\ref{secsys3}, the longitudinal change stays positive for all the internode amplitudes that we have examined, because the sheets with $s=0$ remove the reversal of the single-sheet model. The sign reversal is therefore a property of the model with an inert flat band, and its location is fixed by the balance of the BC and OMM terms.

%%%%%%%%%%%%%%%
\subsection{Shortcomings of earlier beyond-RTA treatments}

Ref.~\cite{ahmad-sharma-2025} considers two conjugate pseudospin-1 nodes for $\bE\parallel\bB\parallel\hat{\boldsymbol z}$, with momentum-independent amplitudes and the ratio $\tilde\alpha$ of the internode to the intranode amplitude as the control parameter. The flat band of each node is inert, and every node carries a single FS sheet. This is the system of two conjugate QTSM nodes with the sheet $(\chi,0)$ omitted, because the quadratic correction to the dispersion is not taken into account. Their overlap in Eq.~(14) is the azimuthal average of $\left[(1\pm \hat{\bk}\cdot\hat{\bk}')/2\right]^2$, where the upper sign refers to intranode and the lower sign to internode scattering. The in-scattering term therefore lies in the span $\{1,\cos\theta,\cos^2\theta\}$ for each node, which is the span of Eq.~\eqref{eqspans} for $m=0$. Their ansatz in Eq.~(20) uses the four functions $1$, $\cos^4 \!\left( \theta/2 \right)$, $\sin^4 \!\left( \theta/2 \right)$, and $\sin^2\theta$ in the factor multiplying the out-scattering time. These functions are linearly dependent, since
\begin{align}
\cos^4 \!\left( \frac{\theta}{2} \right)+\sin^4 \!\left( \frac{\theta}{2} \right)+\frac{1}{2}\,\sin^2\theta=1\,,
\label{eqlitdep}
\end{align}
and they span the same space as $\{1,\cos\theta,\cos^2\theta\}$. The ansatz thus has the structure of the first line of Eq.~\eqref{eqansatz3a} for a single sheet per node, which is appropriate for this geometry because the LFO vanishes on $\Lambda^z$. However, it carries four coefficients per node for a space of dimension three. A shift of the coefficients of the four functions by $\{\delta,-\delta,-\delta,-\delta/2\}$ leaves $\Lambda_\chi$ unchanged, and the eight coefficients therefore determine only six independent numbers, which are the moments $M^{(n)}_\chi$ of Eq.~\eqref{eqlongsol} for the two nodes. This redundancy has no physical meaning and is unrelated to the zero mode of Sec.~\ref{seczero}. The moment system has dimension six, and its only null vector is the zero mode, which makes the rank five. The number constraint of Eq.~\eqref{eqnumber} closes the system, and the solvability condition of Eq.~\eqref{eqsolv}, which is the continuity equation, holds because the pumping rates of the two conjugate nodes cancel. Equating the coefficients of four linearly dependent functions is sufficient but not necessary, since only three conditions per node are independent. The paper states eight equations for eight unknowns and imposes the constraint of its Eq.~(21) in addition, which on its face value makes the system overdetermined. It does not explain that one combination of the equations is redundant by the continuity equation, which makes the system consistent and the constraint an independent equation. The solution quoted in Appendix~B of Ref.~\cite{ahmad-sharma-2025} fixes the redundancy by $\lambda_\pm=c_\pm=0$, and its Appendix~A works with the six remaining unknowns. The redundancy therefore does not affect the results for $\bE\parallel\bB$. The relation $b_\pm=a_\pm$ printed in that appendix should read $b_\pm=-a_\pm$. At $B=0$, the source is odd under $\theta\to\pi-\theta$, and the solution is odd as well, since the even sector is homogeneous and contains only the zero mode removed by the number constraint. For $\lambda_\pm=c_\pm=0$, this requires $a_\pm+b_\pm=0$. The solution with $a_\pm=v_F\left(\tilde\alpha-1\right)/\left(3\,\tilde\alpha+1\right)$ then reproduces its Eq.~(B6). The collinear problem was solved properly in Ref.~\cite{ips-exact-spin1} for the QTSM nodes, including the second sheet and the interband scattering within a node, with the independent moments $M^{(n)}_a$ of Eq.~\eqref{eqlongsol}.

Their zero-field conductivities in Eq.~(25) follow from the same kernel. For $B=0$, the transport rate of a single sheet is proportional to
\begin{align}
\tau_{\rm tr}^{-1}\propto\int\frac{d^2\hat{\bk}'}{4\pi}\left(1-C\right)\left[\mathcal O_{\rm intra}(C)+\tilde\alpha\,\mathcal O_{\rm inter}(C)\right],
\label{eqlittr}
\end{align}
since the upper bands of the two conjugate nodes have the same velocity direction, and the in-scattering term enters with the same sign for both nodes. For pseudospin 1, with $\mathcal O_{\rm intra}=(1+C)^2/4$ and $\mathcal O_{\rm inter}=(1-C)^2/4$, Eq.~\eqref{eqlittr} gives $(1+3\,\tilde\alpha)/6$. For pseudospin 1/2, with $\mathcal O_{\rm intra}=(1+C)/2$ and $\mathcal O_{\rm inter}=(1-C)/2$, it gives $(1+2\,\tilde\alpha)/3$. These are the dependences on $\tilde\alpha$ of their zero-field conductivities. The BC that they quote for the four bands of the pseudospin-3/2 node agrees with that of the sheets $(\Gamma,\tilde\psi)$ in Table~\ref{tabnodes2}. The sign reversal of the longitudinal magnetoconductivity that they find only with the OMM is the mechanism of Ref.~\cite{timm} for Weyl nodes.

The ansatz of Ref.~\cite{ahmad-sharma-2025} has the form of the out-scattering time multiplied by a polynomial in $\cos\theta$, for the pseudospin-1 nodes in Eq.~(20) and for the bands of the pseudospin-3/2 nodes in Appendix~D. This form is exact only when the LFO acts trivially, which holds for $\bE\parallel\bB$ in the geometry that they consider. It corresponds to the replacement $\hat{\mathcal R}_a\to\tau_a$ in Eq.~\eqref{eqtiltsol}, which is the truncation of Refs.~\cite{PhysRevB.103.115146, sharma-nandy-tewari-2020}. For a field with a component perpendicular to $\bE$, the LFO couples the components of the mean free-path, and the function multiplying the in-scattering moments is no longer the out-scattering time. The Hall and planar-Hall responses are then missed. They make no claim for these responses, and their ansatz cannot be carried over to them. The overlaps are also written as functions of the two polar angles after the azimuthal terms are dropped. This is exact for $\bE\parallel\bB$ because $\Lambda^z$ has no azimuthal dependence. For a general direction of $\bE$ or $\bB$, the terms with $\varphi-\varphi'$ generate the sectors $m\neq0$, and the harmonics of Eq.~\eqref{eqoverlapm} are required. Finally, the channels are studied one at a time, with all the amplitudes except the intranode-intraband one set to zero, and with the amplitudes of the two conjugate nodes varied together. The sign of the response that follows is a property of the isolated channel, and it does not hold for a crystal in which all the channels are present with ratios fixed by the microscopic model.

Ref.~\cite{ahmad-phc-2026} considers two conjugate pseudospin-1 nodes with $\bE\parallel\hat{\boldsymbol z}$ and $\bB=B\left(\cos\tilde\vartheta,0,\sin\tilde\vartheta\right)$, with and without tilts along $\boldsymbol{\hat x}$ or $\boldsymbol{\hat z}$ that act on the dispersive bands only. The overlap is the same as in Ref.~\cite{ahmad-sharma-2025}, and the in-scattering term therefore has the span of the harmonics with $l\leq2$ also for the tilted nodes. The LFO term of the kinetic equation is dropped at the outset, on the ground that it is of order $\omega_c\,\tau$ relative to the anomaly term. The ansatz below its Eq.~(21) is then the out-scattering time multiplied by a combination of ten angular functions, with the unknowns $\tilde\lambda$, $\tilde a_{1,2}$, $\tilde b_{1,2}$, $\tilde c_{1,2}$, and $\tilde d_{1,2,3}$ for each node. This is again the replacement $\hat{\mathcal R}_a\to\tau_a$. The justification is not sufficient, because the LFO contribution of Eq.~\eqref{eqrtalfo3} exceeds the BC and OMM corrections by a factor set by $\left(\mu\,\tau\right)^2$, and not by $\omega_c\,\tau$ alone. The classical LFO contribution therefore dominates for noncollinear fields in the semiclassical regime, and it is absent from the longitudinal, the planar-Hall, and the ordinary Hall conductivities of Ref.~\cite{ahmad-phc-2026}. The magnitudes, the signs, and the critical amplitudes $\tilde\alpha_c(\tilde\vartheta)$ obtained there do not reflect the actual scenario.

To estimate the effect, we use the model in question for $\bE\parallel\hat{\boldsymbol z}$ and $\bB=B\left(\cos\tilde\vartheta,0,\sin\tilde\vartheta\right)$, without the tilts. The exact solution is the Galerkin solution of Eq.~\eqref{eqansatz3bc}, and the truncated solution replaces $\hat{\mathcal R}_a$ by $\tau_a$. The parameters are $r_\beta=1/2$, $\alpha=0.05$, and $\mu\,\tau_0=15$ unless stated otherwise. Both $\delta\sigma_{zz}$ and $\sigma_{xz}$ are in units of $\sigma_{zz}$ at $B=0$, and $\sigma_{xz}$ vanishes at $B=0$ for this orientation. Fig.~\ref{figlitlfo} compares the two solutions. For noncollinear fields, the exact change of $\sigma_{zz}$ is negative and large over most of the range of angles, and its magnitude decreases as $\bB$ turns towards $\bE$. This is the classical LFO contribution. The exact change reverses its sign close to $\tilde\vartheta=\pi/2$ and merges with the truncated one for $\bB\parallel\bE$, where the LFO has no effect. The truncated change is small and positive, and it vanishes for $\bB\perp\bE$. It therefore has the opposite sign to the exact change except in a narrow range of angles near $\bB\parallel\bE$. The exact planar-Hall conductivity has the same sign as the truncated one, and it is much larger. Both vanish for $\bB\parallel\bE$ and for $\bB\perp\bE$ (see Table~\ref{tablit}). In the right panel, the truncated change does not depend on $\mu\,\tau_0$, whereas the magnitude of the exact change grows with $\mu\,\tau_0$ and saturates for large $\mu\,\tau_0$. The two solutions are comparable only for $\mu\,\tau_0$ of order unity, and the error of the truncation increases with the purity of the sample, in agreement with the dependence on $\left(\mu\,\tau\right)^2$ stated above.

The ten functions of the ansatz are not independent either, since
\begin{align}
\cos^2\theta+\sin^2\theta\,\cos^2\varphi+\sin^2\theta\,\sin^2\varphi=1\,,
\label{eqlitdep2}
\end{align}
and they span a nine-dimensional space, which is the space of the harmonics with $l\leq2$ generated by the kernel. The ansatz thus carries one redundant coefficient per node, which leaves the mean free-path unchanged. The count of $20$ equations for $20$ unknowns that is stated for the two nodes corresponds to equating the coefficients of the ten functions, which is sufficient but not necessary, since only nine conditions per node are independent. A shift along the redundant direction changes the left-hand side of the matched equations but not their right-hand side, because the in-scattering term depends on the coefficients only through $\Lambda_\chi$. The redundant directions are therefore not null vectors of the matched system, and its only null vector is the zero mode, which the number constraint of its Eq.~(23) removes. The paper states $20$ equations for $20$ unknowns and imposes the constraint in addition, which on its face value makes the system overdetermined. It does not explain that one combination of the $20$ equations is redundant by the continuity equation, which makes the system consistent and the constraint an independent equation. The kernel of its Eq.~(16) also lacks the term $\tfrac12\,\sin^2\theta\,\sin^2\theta'\,\sin\varphi\,\cos\varphi\,\sin\varphi'\,\cos\varphi'$ of the expansion of $\left[\left(1\pm C\right)/2\right]^2$, which arises from $\cos^2(\varphi-\varphi')$ and generates the coefficient $\tilde d_3$ of the ansatz. Whether this is a misprint or whether it enters the numerical solution cannot be decided from the text. We remove the term from the kernel of the model, for the exact and for the truncated solution, for $r_\beta=1/2$, $1$, and $2$, and for all the angles $\tilde\vartheta$. The relative changes of $\delta\sigma_{zz}$ and $\sigma_{xz}$ are negligible (cf. Table~\ref{tablit}), and the term is therefore immaterial for the untilted model. The BC is quoted as $-\chi\,\bk/k^3$ in the text, which corresponds to the monopole charge $2$, and as $-\chi\,\bk/(2\,k^3)$ below the equations of motion, which is the value for a Weyl node. The effect of the two quoted values on the numerical results cannot be assessed from the text. The moments of Eq.~\eqref{eqlongsol} and the Galerkin form of Eq.~\eqref{eqansatzc} contain only independent unknowns and give the exact solution for the same kernel.

\begin{table}[t!]
\centering
\small
\renewcommand{\arraystretch}{1.6}
\begin{tabular}{|c|c|c|c|c|}
\hline
Limitation & Set-up & Quantity & Exact & Limited \\
\hline
Doubled OMM & $\bE\parallel\bB$, $r_\beta=1/2$ & $10^{2}\,\delta\sigma_\parallel$ & 1.96 & 0.97 \\
\hline
Doubled OMM & $\bE\parallel\bB$, $r_\beta=0.8$ & $10^{3}\,\delta\sigma_\parallel$ & 3.70 & $-$3.93 \\
\hline
Doubled OMM & $\bE\parallel\bB$ & critical $r_\beta$ & 1.04 & 0.65 \\
\hline
Doubled OMM & $\bE\perp\bB$, $r_\beta=1/2$ & $10^{3}\,\sigma_H^{\rm AH}/\sigma_\perp(0)$ & 1.67 & 3.38 \\
\hline
Doubled OMM & $\bE\perp\bB$, $r_\beta=1/2$ & $10^{2}\,\sigma_H^{\rm B}/\sigma_\perp(0)$ & 29.96 & 29.90 \\
\hline
Dropped LFO & $\bE\parallel\hat{\boldsymbol z}$, $\tilde\vartheta=\pi/12$ & $10^{2}\,\delta\sigma_{zz}$ & $-$78.66 & 0.12 \\
\hline
Dropped LFO & $\bE\parallel\hat{\boldsymbol z}$, $\tilde\vartheta=\pi/4$ & $10^{2}\,\delta\sigma_{zz}$ & $-$35.56 & 0.97 \\
\hline
Dropped LFO & $\bE\parallel\hat{\boldsymbol z}$, $\gamma=5\,\pi/12$ & $10^{2}\,\delta\sigma_{zz}$ & $-$2.49 & 1.83 \\
\hline
Dropped LFO & $\bE\parallel\hat{\boldsymbol z}$, $\gamma=\pi/4$ & $10^{2}\,\sigma_{xz}/\sigma_{zz}(0)$ & 53.01 & 1.44 \\
\hline
Missing kernel term & $\bE\parallel\hat{\boldsymbol z}$, $0<\gamma<\pi/2$ & \makecell{largest relative change \\ of $\delta\sigma_{zz}$ and $\sigma_{xz}$} & $0$ & $3\times10^{-4}$ \\
\hline
\end{tabular}
\caption{\label{tablit} Effect of the limitations of Refs.~\cite{ahmad-sharma-2025, ahmad-phc-2026} in the model of two conjugate pseudospin-1 nodes with one FS sheet per node, for $\alpha=0.05$ and $\mu\,\tau_0=15$. The column Exact is the solution of this work. The column Limited is the same solver with the limitation imposed, which is the doubled OMM, the dropped LFO, or the kernel without the term $\tfrac12\,\sin^2\theta\,\sin^2\theta'\,\sin\varphi\,\cos\varphi\,\sin\varphi'\,\cos\varphi'$.}
\end{table}

Ref.~\cite{ahmad-phc-2026} places the two nodes at the P point of the space groups 199, 214, and 220 and assigns tilts to them. A symmetry analysis does not support this assignment for the pair of opposite chiralities of that work. In none of these space groups does the little group at P admit a tilt, since it contains the twelve rotations of the space group 199, which leave no vector invariant. In the space groups 199 and 214, the nodes at P and $-$P have the same chirality, since time reversal and the proper rotations preserve it. In the space group 220, the improper operations of the little group force the flux of the BC of each band through a sphere around P to vanish, and this space group therefore hosts no chiral node at P. The unitary little group at P of the magnetic group 220.91 is that of the space group 199, and it excludes a tilt as well. A pair of nodes with opposite chiralities and non-vanishing tilts of the kind assigned in Ref.~\cite{ahmad-phc-2026} is thus not obtained in any of these cases. The derivation, with the counts of the operations, is given in Appendix~\ref{secappP}.

The two conjugate RSW nodes of Ref.~\cite{ahmad-sharma-2025} cannot occur in a chiral crystal. The fourfold degeneracy of the RSW node is enforced by the time-reversal-symmetric corepresentation of the little group of a high-symmetry momentum, which is the chiral tetrahedral group in the space group 198 and the octahedral group in the space group 212 \cite{prl119206401, grushin-multifold}. The RSW node therefore sits at $\Gamma$, whereas $R$ hosts the sixfold node and $M$ the double spin-1/2 node \cite{grushin-multifold, gao2022}. At a generic momentum, the little group is trivial and two bands cross only at a twofold Weyl point. Two RSW nodes at conjugate generic momenta would therefore have no symmetry protection and would require a fine-tuned Hamiltonian, and the BZ contains a single RSW node at $\Gamma$. The crystal thus contains a single RSW node at $\Gamma$, compensated by the TSM nodes at $R$, which is the system of Sec.~\ref{secsys2}. It differs from the conjugate-pair description in four respects. First, the two ends of the internode channel carry different pseudospins, and the kernel has the degree three on the sheets at $\Gamma$ and the degree two on those at $R$ [cf. Eq.~\eqref{eqspans2}]. Second, the sheets are not related by an exchange of equivalent nodes, since there are two nested sheets at $\Gamma$ and one sheet on each of the two copies at $R$. Third, the overlaps between $\Gamma$ and $R$ are fixed by the unitary block of Eq.~\eqref{eqyblocks}, whose ratio $\zeta/\eta$ is a parameter of the microscopic model, and they differ for the two copies [cf. Eq.~\eqref{eqovgr2}]. Fourth, the two copies at $R$ are coupled only through the sheets at $\Gamma$, since their mutual overlap vanishes. Appendix~D of Ref.~\cite{ahmad-sharma-2025} remarks that unequal numbers of bands in the two valleys are allowed, but no such system is solved there. The quadratic correction, which turns the flat band of each TSM node into a dispersive band, and the interband scattering within a node are not included for the pseudospin-1 nodes either. Since the tilt is forbidden at $\Gamma$ and $R$ (see the Introduction), Ref.~\cite{ahmad-phc-2026} describes a different physical situation from the chiral crystals, in addition to the incomplete ansatz.

The exact solution for general orientations of $\bE$ and $\bB$ is given neither in Refs.~\cite{timm, ips-exact-spin1, ahmad-sharma-2025}, which are confined to collinear fields, nor in Ref.~\cite{ahmad-phc-2026}, whose noncollinear treatment has the shortcomings above. The solution of this work covers the longitudinal, the transverse, the Hall, and the planar-Hall responses with the LFO kept.

%%%%%%%%%%%%%%%
\section{Summary and outlook}
\label{secsum}

We have solved the linearised SBE beyond the RTA for four set-ups, which are two conjugate QTSM nodes without and with a type-I axial tilt, a QTSM node with two Weyl nodes, and an RSW node with two TSM nodes. The nodes of the last two set-ups are untilted, as required by the symmetry of the crystal. The orientations of $\bE$ and $\bB$ are arbitrary, and the BC, the OMM, and the LFO are all kept. The momentum-independent amplitudes make the scattering kernel of finite rank, which reduces the problem to a linear system for a few moments of the mean free-path vector. The number constraint removes the zero mode and closes the system.

The FS of every sheet of the untilted nodes is axisymmetric about $\bB$, and the LFO is therefore diagonal in the azimuthal harmonic. For the tilted nodes, this holds only for $\bB$ along the polar axis. In both cases, the resolvent is algebraic for $\bB$ along the polar axis, and it is evaluated by a Galerkin expansion otherwise. The symmetries of each set-up fix the independent geometries, the independent elements of the conductivity tensor, and the form of the ansatz [cf. Tables~\ref{tabgeom3}, \ref{tabgeom4}, \ref{tabgeom}, and \ref{tabgeom2}]. For the tilted pair, the Onsager relation takes the form of Eq.~\eqref{eqons4}. The terms of the diagonal elements that are odd in $B$ and the term of the Hall element that is even in $B$ are then odd in $t$. They belong to a single pair and cancel in the sum with its time-reversed partner. The relations between $\sigma_{ij}$ and $\sigma_{ji}$ follow from time reversal combined with a rotation by $\pi$ about an axis perpendicular to the polar axis [cf. Eq.~\eqref{eqplanar4}]. For $\bB$ perpendicular to the polar axis, the tilt makes the elements $\sigma_{xy}$ and $\sigma_{xz}$ nonzero.

The results are available in closed form at zero field and to the leading orders in $B$, and numerically for the full field dependence. The amplitudes enter the two sectors differently. The sum of the intranode and internode strengths governs the rates and the sector of $\bE\perp\bB$, whereas their difference governs the in-scattering for $\bE\parallel\bB$. For the tilted pair with $\bB$ along the polar axis, this implies that the dependence on the ratio of the amplitudes is confined to the longitudinal conductivity at fixed scattering time, and that the relative longitudinal change does not depend on the overall scattering strength. The AH part depends only on the radius of the sheet with $s=1$, and it does not depend on the amplitudes. For the untilted pair, it requires both the BC and the OMM. For the tilted pair, the tilt generates a term that requires only the BC and does not vanish at zero field. For $\bE\parallel\bB$ without a tilt, the longitudinal change is positive and quadratic in $B$ for the ratios of the amplitudes that we have scanned, and it does not change sign, in contrast to the Weyl case. The BC and OMM contributions have the same sign, and their cross term is negative. For the tilted pair, the tilt adds a term that is odd in $B$ for $\bE\parallel\bB$, and its sign is controlled by the ratio of the amplitudes. For a noncollinear $\bE$, the change of the current along $\bE$ is dominated by the classical LFO contribution, and the Hall response is dominated by the Boltzmann part.

A planar-Hall current flows already at zero field for an oblique $\bE$ in the systems whose internode overlap depends on the orientation of the polar axis. For two conjugate QTSM nodes without a tilt, it vanishes when the internode amplitude between the sheets of the upper band vanishes. For the tilted pair, the tilt makes the zero-field tensor uniaxial at the order $t^2$ even without the internode scattering. The planar-Hall current at zero field is therefore no longer a probe of the internode amplitude alone. For the QTSM and Weyl nodes and for the RSW and TSM nodes, the sheets at $R$ carry almost the whole zero-field conductivity for the parameters of RhSi, and the changes relative to it are therefore small. For the first of these two systems, the dependence on the quadratic coefficient is weak. When the overlap of the Chang scenario replaces its rotational average, the response depends on the orientation of $\bB$ and $\bE$ relative to the crystal axes, and the sign of the longitudinal change depends on the ratio of the amplitudes. For the RSW and TSM nodes, the longitudinal change is positive and even in $B$ for the ratios that we have scanned, and the cross term of the BC and the OMM is again negative.

The RTA contains the internode scattering through the pumping of the nodes by the BC and the OMM. We have extended its expression for conjugate nodes to the tilted pair, and to nodes of different type and DOS at $\Gamma$ and at $R$. This term is of the order $B^2$, and it acts only for $\bE\parallel\bB$. For the tilted pair, it is even in $t$ at this order, and the linear magnetochiral term of the RTA does not depend on the internode relaxation time. The OMM part of the pumping vertex is specific to the RTA, since the exact solution contains the OMM through the weights, the sources, and the frequencies.

The comparison with the earlier works shows that their beyond-RTA treatments are confined to conjugate pairs and to collinear fields, and that for noncollinear fields the truncated ansatz removes the classical LFO contribution. Their pseudospin-1 results should be recomputed with the correct OMM and with the LFO kept, which is a direct application of the present framework. The pair of conjugate RSW nodes of Ref.~\cite{ahmad-sharma-2025} does not occur in chiral crystals.

The framework also offers a route to the thermal and thermoelectric responses beyond the RTA. In the RTA, the magnetothermoelectric and magnetothermal transport of WSMs has been studied extensively \cite{lundgren14_thermoelectric, ips-kush, das-agarwal_omm, ips-rsw-ph, ips-kush-review, ips-ruiz}. It includes the Nernst and thermal Hall effects and the contributions of the thermal and gravitational chiral anomalies \cite{das20_thermal, spivak16_chiral}, which have been probed experimentally through the thermopower and the thermal conductivity \cite{hirschberger16_chiral, stockert17_thermopower, gca-expt}. In these treatments, the chiral-anomaly-induced thermoelectric conductivity obeys the Mott relation, whereas the longitudinal thermal conductivity violates the Wiedemann-Franz law \cite{sharma16_Nernst, das20_thermal, sekine19_quantum}. Both are also broken in the hydrodynamic regime, where inelastic electron-electron collisions separate the charge and heat channels \cite{gca_andy, messica22_heat}. At finite temperature, elastic scattering conserves the electron number on every energy shell, and the zero mode and the number constraint of Eq.~\eqref{eqnumber} acquire an energy label. The solution with the energy-resolved moments entering the Sommerfeld expansion would show whether the Mott relation and the Wiedemann-Franz law survive when the in-scattering term and the full LFO are kept for multifold nodes. It would also supply the elastic-scattering reference for the violations in the hydrodynamic regime. Further directions are the coherences between the two degenerate copies at $R$, which are kept in the matrix treatment of Appendix~\ref{secappdeg}, the tilts that are not directed along a rotation axis, and the type-II tilts.

%%%%%%%%%%%%%%%%%%%%%%%

\bibliography{ref_ex}

\newpage

\appendix

\setcounter{page}{1}

\section*{Supplemental Material for ``Beyond-RTA magnetoconductivity of multifold semimetals under generic orientations of electromagnetic fields''}

In this supplemental material, we provide the details of our derivations and arguments presented in the main text.  

\section{Spinors, overlaps, FS ingredients, and degenerate copies}
\label{secappingr}

\subsection{Spinors and FS ingredients of the QTSM node}
\label{secappqtsm}

The spin-1 matrices are
\begin{align}
S_x=\frac{1}{\sqrt{2}}
\begin{pmatrix}
0&1&0\\1&0&1\\0&1&0
\end{pmatrix},\quad
S_y=\frac{1}{\sqrt{2}}
\begin{pmatrix}
0&-i&0\\i&0&-i\\0&i&0
\end{pmatrix},\quad
S_z=
\begin{pmatrix}
1&0&0\\0&0&0\\0&0&-1
\end{pmatrix}.
\end{align}
With $\bk=k\left(\sin\theta\,\cos\varphi,\,\sin\theta\,\sin\varphi,\,\cos\theta\right)$, the normalised eigenvectors of the positive-energy bands for $\chi_\Gamma=+1$ are
\begin{align}
\psi_{s=1}=
\begin{pmatrix}
e^{-2\,i\,\varphi}\cos^2 \!\left(\frac{\theta}{2}\right) \\
\frac{e^{-i\,\varphi}\sin\theta}{\sqrt2} \\
\sin^2 \!\left(\frac{\theta}{2}\right)
\end{pmatrix},\quad
\psi_{s=0}=
\begin{pmatrix}
-\frac{e^{-2\,i\,\varphi}\sin\theta}{\sqrt2} \\
e^{-i\,\varphi}\cos\theta \\
\frac{\sin\theta}{\sqrt2}
\end{pmatrix}.
\end{align}
Since the quadratic correction multiplies the identity matrix, these spinors are those of the TSM, and the BC and OMM of Eq.~\eqref{eqbcomm} are unchanged by it. They are the eigenstates of the spin-1 projection along $\hat{\bs n}$ with the eigenvalues $s=1$ and $s=0$, and their squared overlaps depend only on the angle between $\hat{\bs n}$ and $\hat{\bs n}'$ [cf. Eq.~\eqref{eqoverlap}].

For $\bB=B\,\boldsymbol{\hat z}$, the FS radius is the positive root of $\xi_s(k\,\hat{\bk})=\mu$ that is continuously connected to the radius of Eq.~\eqref{eqfsb0}. The equation for the root reads
\begin{align}
& c\,v_F\,k^3+v_F\,k^2-\mu\,k+\frac{\eta_O\,\chi_\Gamma\,e\,v_F\,B\cos\theta}{2}=0\quad\text{for }(\Gamma,1)\,,\nn
& c\,v_F\,k^3-\mu\,k+\eta_O\,\chi_\Gamma\,e\,v_F\,B\cos\theta=0\quad\text{for }(\Gamma,0)\,.
\end{align}
The phase-space factor and the group velocity have the components
\begin{align}
& D_a=1-\eta_B\,\frac{\chi_\nu\,N_a\,e\,B\cos\theta}{2\,k^2}\,,\quad
w_{a,k}=\partial_k\epsilon_s-\eta_O\,\frac{\chi_\nu\,e\,v_\nu\,G_a\,B\cos\theta}{2\,k^2}\,,\nn
& \sin\theta\;w_{a,\theta}=-\eta_O\,\frac{\chi_\nu\,e\,v_\nu\,G_a\,B\sin^2\theta}{2\,k^2}\,,\quad
q_a\equiv\frac{w_{a,\perp}}{\sin\theta}=\partial_k\epsilon_s-\eta_O\,\frac{\chi_\nu\,e\,v_\nu\,G_a\,B\cos\theta}{k^2}\,,
\label{eqwqd}
\end{align}
with $w_{a,z}=\cos\theta\,w_{a,k}-\sin\theta\,w_{a,\theta}$. All quantities are evaluated at $k=k_a(\theta)$. The derivative $\partial_k\epsilon_s$ equals $v_F\left(1+2\,c\,k\right)$ for $(\Gamma,1)$, $2\,c\,v_F\,k$ for $(\Gamma,0)$, and $\tilde v_F$ for the sheets at $R$. The FS weight is $\rho_a=k^2\,D_a/w_{a,k}$.

\subsection{Spinors and FS ingredients of the Weyl nodes}
\label{secappweyl}

The Hamiltonian of a Weyl node with the chirality $\chi_R$ is $\tilde v_F\,k\,\hat{\bs n}\cdot\bs\sigma-\Delta$ with $\hat{\bs n}=\left(\sin\theta\,\cos\varphi,\sin\theta\,\sin\varphi,\chi_R\cos\theta\right)$. The upper-band spinor is the spin-1/2 coherent state along $\hat{\bs n}$, and the squared overlap is $\mathcal O_{R_iR_i}$ of Eq.~\eqref{eqoverlap}. For $\bB=B\,\boldsymbol{\hat z}$, the FS radius is
\begin{align}
k_R(\theta)=\frac{\mu+\Delta+\sqrt{\left(\mu+\Delta\right)^2-2\,\eta_O\,\chi_R\,e\,\tilde v_F^2\,B\cos\theta}}{2\,\tilde v_F}\,,
\end{align}
and the quantities of Eq.~\eqref{eqwqd} apply with $N_a=G_a=1$ and $\partial_k\epsilon_s=\tilde v_F$.

\subsection{Spinors and FS ingredients of the RSW node}
\label{secapprsw}

With $k_\pm=k_x\pm i\,k_y$, the spin-3/2 matrices enter the RSW Hamiltonian of Eq.~\eqref{eqham2} through
\begin{align}
\bk\cdot\bs J=\begin{pmatrix}
\frac{3}{2}\,k_z & \frac{\sqrt3}{2}\,k_- & 0 & 0 \\
\frac{\sqrt3}{2}\,k_+ & \frac{1}{2}\,k_z & k_- & 0 \\
0 & k_+ & -\frac{1}{2}\,k_z & \frac{\sqrt3}{2}\,k_- \\
0 & 0 & \frac{\sqrt3}{2}\,k_+ & -\frac{3}{2}\,k_z
\end{pmatrix}.
\end{align}
The eigenvalue of $\hat{\bk}\cdot\bs J$ is the helicity $\tilde\psi$. For $\bB=B\,\boldsymbol{\hat z}$, the quantities of Eq.~\eqref{eqwqd} apply to the sheets of Eq.~\eqref{eqham2} with the numbers $N_a$ and $G_a$ of Table~\ref{tabnodes2}. The derivative $\partial_k\epsilon_s$ equals $\tilde\psi\,v_{3/2}$ for the sheets at $\Gamma$ and $v_R$ for the sheets at $R$, and $v_\nu$ is replaced by $v_{3/2}$ and $v_R$, respectively.

\subsection{Spinors and overlaps of the RSW and TSM nodes}
\label{secappsp2}

The spin-1 matrices at $R$ are the left-handed set $\bs\Lambda=\left(S_x,S_y,-S_z\right)$ with the matrices $S_a$ of Appendix~\ref{secappqtsm}. The eigenvectors of $\bk\cdot\bs\Lambda$ follow from those of the QTSM node by $\theta\to\pi-\theta$, and the eigenvector of the band $s=1$ is
\begin{align}
v_+(\theta,\varphi)=\Big(e^{-2\,i\,\varphi}\sin^2\!\Big(\frac{\theta}{2}\Big),\ \frac{e^{-i\,\varphi}\sin\theta}{\sqrt2},\ \cos^2\!\Big(\frac{\theta}{2}\Big)\Big)^{\rm T}.
\label{eqspintsm}
\end{align}
Its components are labelled by the eigenvalues $j_R=+1,0,-1$ of $S_z$, and the component $j_R$ carries the phase $e^{-i\,(1+j_R)\,\varphi}$.

The spinors of the upper bands at $\Gamma$ are four-component vectors, whose components are labelled by the eigenvalues $j_\Gamma=3/2,1/2,-1/2,-3/2$ of $J_z$. For $\mu>E_\Gamma$, the relevant bands are $\tilde\psi=3/2$ and $\tilde\psi=1/2$,
\begin{align}
& \psi_{\Gamma,3/2}=\Big(e^{-3\,i\,\varphi}\cos^3\!\Big(\frac{\theta}{2}\Big),\ \sqrt3\,e^{-2\,i\,\varphi}\cos^2\!\Big(\frac{\theta}{2}\Big)\sin\Big(\frac{\theta}{2}\Big),\ \sqrt3\,e^{-i\,\varphi}\cos\Big(\frac{\theta}{2}\Big)\sin^2\!\Big(\frac{\theta}{2}\Big),\ \sin^3\!\Big(\frac{\theta}{2}\Big)\Big)^{\rm T},\nn
& \psi_{\Gamma,1/2}=\Big(-\sqrt3\,e^{-3\,i\,\varphi}\cos^2\!\Big(\frac{\theta}{2}\Big)\sin\Big(\frac{\theta}{2}\Big),\ e^{-2\,i\,\varphi}\cos\Big(\frac{\theta}{2}\Big)\Big[3\cos^2\!\Big(\frac{\theta}{2}\Big)-2\Big],\nn
& \qquad\qquad -\,e^{-i\,\varphi}\sin\Big(\frac{\theta}{2}\Big)\Big[3\sin^2\!\Big(\frac{\theta}{2}\Big)-2\Big],\ \sqrt3\cos\Big(\frac{\theta}{2}\Big)\sin^2\!\Big(\frac{\theta}{2}\Big)\Big)^{\rm T}.
\label{eqspinrsw}
\end{align}
The component $j_\Gamma$ carries the phase $e^{-i\,(3/2+j_\Gamma)\,\varphi}$, and these spinors agree with the eigenvectors of Ref.~\cite{ips-exact-rsw} up to the convention for the components. Let $C=\hat{\bk}\cdot\hat{\bk}'$ be the cosine of the angle between the momenta. The squared overlaps between the sheets at $\Gamma$ are
\begin{align}
\mathcal O_{(\Gamma,3/2)(\Gamma,3/2)}=\frac{\left(1+C\right)^3}{8}\,,\quad
\mathcal O_{(\Gamma,1/2)(\Gamma,1/2)}=\frac{\left(1+C\right)\left(3\,C-1\right)^2}{8}\,,\quad
\mathcal O_{(\Gamma,3/2)(\Gamma,1/2)}=\frac{3\left(1+C\right)^2\left(1-C\right)}{8}\,,
\label{eqovrsw}
\end{align}
and those within a copy at $R$ and between the two copies are
\begin{align}
\mathcal O_{(R_i,+)(R_i,+)}=\frac{\left(1+C\right)^2}{4}\,,\quad
\mathcal O_{(R_1,+)(R_2,+)}=0\,.
\label{eqovtsm}
\end{align}

The spinors of $R_1$ and $R_2$ are $\left(v_+,0,0,0\right)^{\rm T}$ and $\left(0,0,0,v_+\right)^{\rm T}$ in the six-dimensional space. The amplitude from $(\Gamma,\tilde\psi)$ into $R_i$ is $A_i=v_+^\dagger(\theta',\varphi')\,\mathcal Y_i\,\psi_{\Gamma,\tilde\psi}(\theta,\varphi)$ with $\mathcal Y_i$ of Eq.~\eqref{eqyblocks}, and the squared overlap is $\mathcal O_{(\Gamma,\tilde\psi)(R_i,+)}=|A_i|^2$. Let $r_p$, with $p=1,\dots,4$, be the real functions of $\theta$ that multiply the phases in the components of $\psi_{\Gamma,\tilde\psi}$ [cf. Eq.~\eqref{eqspinrsw}]. The phases combine into the factors $e^{i\,(\varphi'-2\,\varphi)}\,e^{i\,j_R\,(\varphi'-\varphi)}$ for $R_1$ and $e^{i\,(\varphi'-\varphi)}\,e^{i\,j_R\,(\varphi'-\varphi)}$ for $R_2$. The global factors drop out of the modulus, and with the real amplitudes
\begin{align}
& R_1:\ t_{+1}=\sin^2\!\Big(\frac{\theta'}{2}\Big)\,r_1\,,\quad t_{0}=\zeta\,\frac{\sin\theta'}{\sqrt2}\,r_2\,,\quad t_{-1}=\eta\,\cos^2\!\Big(\frac{\theta'}{2}\Big)\,r_3\,,\nn
& R_2:\ t_{+1}=\eta\,\sin^2\!\Big(\frac{\theta'}{2}\Big)\,r_2\,,\quad t_{0}=\zeta\,\frac{\sin\theta'}{\sqrt2}\,r_3\,,\quad t_{-1}=\cos^2\!\Big(\frac{\theta'}{2}\Big)\,r_4\,,
\label{eqovgr}
\end{align}
the overlap is $\mathcal O_{(\Gamma,\tilde\psi)(R_i,+)}=\big|\sum_{j=-1}^{1}t_j\,e^{i\,j\,(\varphi'-\varphi)}\big|^2$. It depends only on $\varphi-\varphi'$ and has the decomposition of Eq.~\eqref{eqoverlapm} with
\begin{align}
\mathcal O^{(0)}=t_{+1}^2+t_0^2+t_{-1}^2\,,\quad
\mathcal O^{(1)}=2\left(t_{+1}\,t_0+t_0\,t_{-1}\right),\quad
\mathcal O^{(2)}=2\,t_{+1}\,t_{-1}\,.
\end{align}
The components $r_p$ are polynomials of degree $3$ in $\cos(\theta/2)$ and $\sin(\theta/2)$, and the amplitudes $t_j$ are polynomials of degree $2$ in the half angles at $R$. The harmonic $m=2$ is
\begin{align}
& \mathcal O^{(2)}_{(\Gamma,3/2)(R_1,+)}=\frac{\sqrt3\,\eta}{16}\,\sin^2\theta\,\sin^2\theta'\left(1+\cos\theta\right),\quad
\mathcal O^{(2)}_{(\Gamma,3/2)(R_2,+)}=\frac{\sqrt3\,\eta}{16}\,\sin^2\theta\,\sin^2\theta'\left(1-\cos\theta\right),\nn
& \mathcal O^{(2)}_{(\Gamma,1/2)(R_1,+)}=-\frac{\sqrt3\,\eta}{16}\,\sin^2\theta\,\sin^2\theta'\left(3\cos\theta+1\right),\quad
\mathcal O^{(2)}_{(\Gamma,1/2)(R_2,+)}=\frac{\sqrt3\,\eta}{16}\,\sin^2\theta\,\sin^2\theta'\left(3\cos\theta-1\right).
\label{eqovgr2}
\end{align}
These differ for the two copies, and the exchange of $R_1$ and $R_2$ is therefore not a symmetry of the kernel. Only the four orbitals of the RSW node enter, and the $6\times2$ block of $\mathcal U$ that belongs to the Weyl node at $\Gamma$ gives no contribution.

\subsection{Fermi surfaces and scattering kernel}
\label{secappkernel}

For two sheets of the same pseudospin, $\hat{\bs n}\cdot\hat{\bs n}'=X+Y\cos(\varphi-\varphi')$ with $X=\chi_\nu\,\chi_{\nu'}\cos\theta\cos\theta'$ and $Y=\sin\theta\sin\theta'$. The squared overlaps of Eq.~\eqref{eqoverlap} then have the azimuthal decomposition $\mathcal O_{aa'}=\mathcal O^{(0)}_{aa'}+\mathcal O^{(1)}_{aa'}\cos(\varphi-\varphi')+\mathcal O^{(2)}_{aa'}\cos\!\left[2\,(\varphi-\varphi')\right]$, with
\begin{align}
& \mathcal O^{(0)}_{(\Gamma,1)(\Gamma,1)}=\frac14\left(1+2\,X+X^2+\frac{Y^2}{2}\right),\quad
\mathcal O^{(1)}_{(\Gamma,1)(\Gamma,1)}=\frac12\,Y\left(1+X\right),\quad
\mathcal O^{(2)}_{(\Gamma,1)(\Gamma,1)}=\frac{Y^2}{8}\,,\nn
& \mathcal O^{(0)}_{(\Gamma,0)(\Gamma,0)}=X^2+\frac{Y^2}{2}\,,\quad
\mathcal O^{(1)}_{(\Gamma,0)(\Gamma,0)}=2\,X\,Y\,,\quad
\mathcal O^{(2)}_{(\Gamma,0)(\Gamma,0)}=\frac{Y^2}{2}\,,\nn
& \mathcal O^{(0)}_{(\Gamma,1)(\Gamma,0)}=\frac12\left(1-X^2-\frac{Y^2}{2}\right),\quad
\mathcal O^{(1)}_{(\Gamma,1)(\Gamma,0)}=-\,X\,Y\,,\quad
\mathcal O^{(2)}_{(\Gamma,1)(\Gamma,0)}=-\frac{Y^2}{4}\,,\nn
& \mathcal O^{(0)}_{R_iR_i}=\frac12\left(1+X\right),\quad
\mathcal O^{(1)}_{R_iR_i}=\frac12\,Y\,,\quad
\mathcal O^{(2)}_{R_iR_i}=0\,.
\label{eqoverlapm}
\end{align}
The azimuthal integrals needed are
\begin{align}
& \int_0^{2\pi}d\varphi'\,e^{i\,m\,\varphi'}=2\,\pi\,\delta_{m,0}\,,\quad
\int_0^{2\pi}d\varphi'\,\cos \!\left(\varphi-\varphi'\right) e^{i\,m\,\varphi'}=\pi\,e^{i\,m\,\varphi}\left(\delta_{m,1}+\delta_{m,-1}\right),\nn
& \int_0^{2\pi}d\varphi'\,\cos\!\left[2\left(\varphi-\varphi'\right)\right] e^{i\,m\,\varphi'}=\pi\,e^{i\,m\,\varphi}\left(\delta_{m,2}+\delta_{m,-2}\right).
\label{eqazint}
\end{align}
With $\mathcal W_{aa'}=\left(\beta_{aa'}^2/\pi\right)\mathcal O_{aa'}$, the kernel acting on $e^{i\,m\,\varphi'}$, with the factor $e^{i\,m\,\varphi}$ stripped off, becomes
\begin{align}
m=0:\ \ 2\,\beta_{aa'}^2\,\mathcal O^{(0)}_{aa'}\,,\quad
m=\pm1:\ \ \beta_{aa'}^2\,\mathcal O^{(1)}_{aa'}\,,\quad
m=\pm2:\ \ \beta_{aa'}^2\,\mathcal O^{(2)}_{aa'}\,.
\label{eqkernel}
\end{align}
It vanishes for $|m|\geq3$. The out-scattering rate follows from contracting the $m=0$ kernel with $\Lambda=1$ in Eq.~\eqref{eqcoll},
\begin{align}
\frac{1}{\tau_a(\theta)}=\sum_{a'}2\,\beta_{aa'}^2\int_0^\pi d\theta'\,\sin\theta'\,\rho_{a'}(\theta')\,\mathcal O^{(0)}_{aa'}(\theta,\theta')\,.
\label{eqtau}
\end{align}
For two Weyl nodes, Eqs.~\eqref{eqkernel} and \eqref{eqtau} reduce to the kernel and the rate of the companion work \cite{ips-exact-noncollinear}.

The overlaps of Eq.~\eqref{eqovgr4} between the sheets at $\Gamma$ and at $R$ are summed over the two copies at $R$. In the lattice axes, they depend on the azimuths of $\hat{\bk}$ and $\hat{\bs q}$ separately. In a frame with the polar axis along $\hat{\bs c}=(1,1,1)/\sqrt3$, they depend only on the difference of the azimuths. We choose $\hat{\bs e}_1=\left(2,-1,-1\right)/\sqrt6$ and $\hat{\bs e}_2=\left(0,1,-1\right)/\sqrt2$ with $\hat{\bs e}_1\times\hat{\bs e}_2=\hat{\bs c}$, and let $\left(\theta,\varphi\right)$ and $\left(\theta',\varphi'\right)$ be the polar and azimuthal angles of $\hat{\bk}$ and $\hat{\bs q}$ in the frame $\left(\hat{\bs e}_1,\hat{\bs e}_2,\hat{\bs c}\right)$. Another choice of $\hat{\bs e}_1$ shifts $\varphi$ and $\varphi'$ by the same angle. Let $\mathcal R_{\hat{\bs c}}$ be the permutation matrix with $\mathcal R_{\hat{\bs c}}\,\boldsymbol{\hat x}=\boldsymbol{\hat y}$, $\mathcal R_{\hat{\bs c}}\,\boldsymbol{\hat y}=\boldsymbol{\hat z}$, and $\mathcal R_{\hat{\bs c}}\,\boldsymbol{\hat z}=\boldsymbol{\hat x}$, which is the rotation by $2\,\pi/3$ about $\hat{\bs c}$. The sum in Eq.~\eqref{eqovgr4} is $\hat q_x\,\hat k_y+\hat q_y\,\hat k_z+\hat q_z\,\hat k_x=\hat{\bk}\cdot\mathcal R_{\hat{\bs c}}\,\hat{\bs q}$, and the Rodrigues formula for this rotation gives
\begin{align}
\hat{\bk}\cdot\mathcal R_{\hat{\bs c}}\,\hat{\bs q}&=-\frac12\,\hat{\bk}\cdot\hat{\bs q}+\frac32\left(\hat{\bs c}\cdot\hat{\bk}\right)\left(\hat{\bs c}\cdot\hat{\bs q}\right)+\frac{\sqrt3}{2}\,\hat{\bs c}\cdot\left(\hat{\bs q}\times\hat{\bk}\right)\nn
&=\cos\theta\,\cos\theta'-\frac12\,\sin\theta\,\sin\theta'\,\cos\!\left(\varphi-\varphi'\right)+\frac{\sqrt3}{2}\,\sin\theta\,\sin\theta'\,\sin\!\left(\varphi-\varphi'\right).
\label{eqrodrigues}
\end{align}
The overlaps of Eq.~\eqref{eqovgr4} then become
\begin{align}
\mathcal O_{(\Gamma,1)(R,+)}=\frac12\left(1-\cos\theta\,\cos\theta'\right)+\frac12\,\sin\theta\,\sin\theta'\,\cos\!\left(\varphi-\varphi'+\frac{\pi}{3}\right),\quad
\mathcal O_{(\Gamma,0)(R,+)}=\frac12\,,
\label{eqovgr4ph}
\end{align}
where the phase $\pi/3$ follows from the identity $\cos\!\left(\varphi-\varphi'\right)-\sqrt3\,\sin\!\left(\varphi-\varphi'\right)=2\,\cos\!\left(\varphi-\varphi'+\pi/3\right)$ applied to Eq.~\eqref{eqrodrigues}. The overlap of the sheet $(\Gamma,1)$ contains the harmonics $m=0$ and $|m|=1$ only. It is not even under $\varphi-\varphi'\to-\left(\varphi-\varphi'\right)$, which reflects the absence of improper symmetries.

Let $\mathcal K^{(m)}_{aa'}(\theta,\theta')$ be the kernel of the sector $m$ for the scattering from the sheet $a'=(R,+)$ into the sheet $a=(\Gamma,s)$, defined as in Eq.~\eqref{eqkernel}, and let $\beta$ be the amplitude of this pair. The integral $\int_0^{2\pi}d\varphi'\,\cos\!\left(\varphi-\varphi'+\pi/3\right)e^{i\,m\,\varphi'}=\pi\,e^{i\,m\,\varphi}\,e^{i\,m\,\pi/3}\left(\delta_{m,1}+\delta_{m,-1}\right)$ replaces the third integral of Eq.~\eqref{eqazint}, and we obtain
\begin{align}
&\mathcal K^{(0)}_{(\Gamma,1)(R,+)}=\beta^2\left(1-\cos\theta\,\cos\theta'\right),\quad
\mathcal K^{(\pm1)}_{(\Gamma,1)(R,+)}=\frac12\,\beta^2\,\sin\theta\,\sin\theta'\,e^{\pm i\,\pi/3},\quad
\mathcal K^{(0)}_{(\Gamma,0)(R,+)}=\beta^2\,,
\label{eqkerngr}
\end{align}
while all the other sectors vanish. For the opposite direction, the phase of the sectors $m=\pm1$ is complex-conjugated, since $\mathcal W_{aa'}(\hat{\bk},\hat{\bk}')=\mathcal W_{a'a}(\hat{\bk}',\hat{\bk})$. The sectors with different $m$ remain decoupled in this frame. If the exchange of the two copies at $R$ is a symmetry, each copy carries half of the overlaps of Eq.~\eqref{eqovgr4ph}, and the kernels of a single copy are half of those of Eq.~\eqref{eqkerngr}.

The matrices $Q^{aa'}$ of Eq.~\eqref{eqtilteq} follow from the expansion of $\mathcal W_{aa'}$ in the functions of Eq.~\eqref{eqfp}. Let $f_p$ and $f'_{p'}$ be these functions for the unit vectors $\hat{\bs n}$ and $\hat{\bs n}'$ of the sheets $a$ and $a'$. The symbol $\sum_{ij}f_{ij}\,f'_{ij}$ denotes the contraction of two symmetric traceless tensors, and the corresponding block of $Q^{aa'}$ is the identity on the space of these tensors. With $\hat{\bs n}\cdot\hat{\bs n}'=\sum_i f_i\,f'_i$ and $\left(\hat{\bs n}\cdot\hat{\bs n}'\right)^2=\sum_{ij}f_{ij}\,f'_{ij}+1/3$, the kernels between the sheets of the same node are
\begin{align}
&\mathcal W_{(\Gamma,1)(\Gamma,1)}=\frac{\beta_{11}^2}{\pi}\left[\frac13+\frac12\sum_i f_i\,f'_i+\frac14\sum_{ij}f_{ij}\,f'_{ij}\right],\quad
\mathcal W_{(\Gamma,0)(\Gamma,0)}=\frac{\beta_{00}^2}{\pi}\left[\frac13+\sum_{ij}f_{ij}\,f'_{ij}\right],\nn
&\mathcal W_{(\Gamma,1)(\Gamma,0)}=\frac{\beta_{10}^2}{\pi}\left[\frac13-\frac12\sum_{ij}f_{ij}\,f'_{ij}\right],\quad
\mathcal W_{R_iR_i}=\frac{\beta_{R_iR_i}^2}{\pi}\left[\frac12+\frac12\sum_i f_i\,f'_i\right].
\label{eqwintra}
\end{align}
These matrices are built from $\hat{\bs n}\cdot\hat{\bs n}'$ and do not depend on the orientation of the frame, whereas the kernels between $\Gamma$ and $R$ do. Let $\beta_s=\beta^{\Gamma R}_s$, let $\chi_R$ be the chirality of the nodes at $R$, and let $\chi_\Gamma=+1$ for the QTSM node. Then $\hat k_i=f_i$ and $\hat q_z=\chi_R\,f'_z$ for the polar axis $\boldsymbol{\hat z}$ of the frame in use. In the lattice axes, Eq.~\eqref{eqovgr4} gives
\begin{align}
\mathcal W_{(\Gamma,1)(R,+)}=\frac{\beta_1^2}{\pi}\left[\frac12\,f_0\,f'_0-\frac12\left(f_y\,f'_x+f_z\,f'_y+\chi_R\,f_x\,f'_z\right)\right],\quad
\mathcal W_{(\Gamma,0)(R,+)}=\frac{\beta_0^2}{\pi}\,\frac12\,f_0\,f'_0\,,
\label{eqwgrlat}
\end{align}
and in the frame with the polar axis along $\hat{\bs c}$, Eq.~\eqref{eqovgr4ph} and the identities $\sin\theta\,\sin\theta'\cos\!\left(\varphi-\varphi'\right)=f_x\,f'_x+f_y\,f'_y$ and $\sin\theta\,\sin\theta'\sin\!\left(\varphi-\varphi'\right)=f_y\,f'_x-f_x\,f'_y$ give
\begin{align}
\mathcal W_{(\Gamma,1)(R,+)}=\frac{\beta_1^2}{\pi}\left[\frac12\,f_0\,f'_0-\frac{\chi_R}{2}\,f_z\,f'_z+\frac14\left(f_x\,f'_x+f_y\,f'_y\right)-\frac{\sqrt3}{4}\left(f_y\,f'_x-f_x\,f'_y\right)\right],\quad
\mathcal W_{(\Gamma,0)(R,+)}=\frac{\beta_0^2}{\pi}\,\frac12\,f_0\,f'_0\,.
\label{eqwgrc}
\end{align}
The entries of $Q^{aa'}$ are the coefficients of the products $f_p\,f'_{p'}$ in Eqs.~\eqref{eqwintra}, \eqref{eqwgrlat}, and \eqref{eqwgrc}. All entries with a function $f_{ij}$ vanish for the pairs between $\Gamma$ and $R$, and the matrices of the opposite pairs are the transposes, $Q^{a'a}_{p'p}=Q^{aa'}_{pp'}$. The sheets at $R$ have no functions $f_{ij}$, and the sheet $(\Gamma,0)$ has no functions $f_i$ in its range.

The decompositions of the overlaps of Eqs.~\eqref{eqovgr4} and \eqref{eqovgrx} are both obtained in the frame with the polar axis along $\hat{\bs c}$. Only the block of the projection between the three triplet orbitals of the QTSM node and the four orbitals at $R$ enters, because the singlet provides no FS sheet. The cyclic sums in Eq.~\eqref{eqovgrx} are
\begin{align}
& \sum_a\hat q_a\,\hat k_{a+1}=\cos\theta\,\cos\theta'-\frac12\,\sin\theta\,\sin\theta'\cos\!\left(\varphi-\varphi'\right)+\frac{\sqrt3}{2}\,\sin\theta\,\sin\theta'\sin\!\left(\varphi-\varphi'\right),\quad
\sum_a\hat q_a=\sqrt3\,\cos\theta'\,,\nn
& \sum_a\hat q_a\left(\hat k_{a+1}-\hat k_{a+2}\right)=\sqrt3\,\sin\theta\,\sin\theta'\sin\!\left(\varphi-\varphi'\right),\nn
& \sum_a\hat q_a\left(\hat k_a^2-\hat k_{a+1}\,\hat k_{a+2}\right)=\frac{\sqrt3}{2}\,\sin^2\theta\,\cos\theta'+\sqrt3\,\sin\theta\,\cos\theta\,\sin\theta'\cos\!\left(\varphi-\varphi'\right).
\label{eqcycsums}
\end{align}
The first sum is Eq.~\eqref{eqrodrigues}. The second follows from $\sum_a\hat q_a\,\hat k_{a+2}=\hat{\bk}\cdot\mathcal R_{\hat{\bs c}}^{\rm T}\hat{\bs q}$ and the antisymmetric part of the rotation, $\left(\mathcal R_{\hat{\bs c}}-\mathcal R_{\hat{\bs c}}^{\rm T}\right)\hat{\bs q}=\sqrt3\,\hat{\bs c}\times\hat{\bs q}$, together with $\hat{\bs c}\cdot\left(\hat{\bs q}\times\hat{\bk}\right)=\sin\theta\,\sin\theta'\sin\!\left(\varphi-\varphi'\right)$. The third follows by inserting the components of $\hat{\bk}$ and $\hat{\bs q}$ in the frame. Eq.~\eqref{eqovgrx} then has the decomposition $\mathcal O_{(\Gamma,s)(R,+)}=\mathcal O^{(0)}_{(\Gamma,s)(R,+)}+2\,\Re\!\left[g_{(\Gamma,s)(R,+)}\,e^{i\,(\varphi-\varphi')}\right]$ with
\begin{align}
& \mathcal O^{(0)}_{(\Gamma,1)(R,+)}=\frac12\left(1-\cos\theta\,\cos\theta'\right)-\frac{\sqrt3}{8}\,s_p\sin\phi\,\sin^2\theta\,\cos\theta'\,,\nn
& g_{(\Gamma,1)(R,+)}=\frac18\,\sin\theta\,\sin\theta'\left(1-\sqrt3\,s_p\sin\phi\,\cos\theta+i\,\sqrt3\cos\phi\right),\nn
& \mathcal O^{(0)}_{(\Gamma,0)(R,+)}=\frac12-\frac{\sqrt3}{4}\,s_p\sin\phi\,\cos^2\theta\,\cos\theta'\,,\nn
& g_{(\Gamma,0)(R,+)}=\frac{\sqrt3}{4}\,s_p\sin\phi\,\cos\theta\,\sin\theta\,\sin\theta'\,.
\label{eqovgrxph}
\end{align}
The overlaps of both scenarios therefore contain the harmonics $m=0$ and $|m|=1$ only. The Chang scenario is the case $\phi=0$, for which $g_{(\Gamma,1)(R,+)}=\frac14\sin\theta\,\sin\theta'\,e^{i\,\pi/3}$ reproduces Eq.~\eqref{eqovgr4ph}. For $\phi=\pi$, the phase of $g_{(\Gamma,1)(R,+)}$ is $e^{-i\,\pi/3}$, and the overlap of the sheet $(\Gamma,0)$ is $1/2$, which is the reversed cyclic order stated below Eq.~\eqref{eqovgrx}. The kernels of the sectors follow as in Eq.~\eqref{eqkerngr},
\begin{align}
\mathcal K^{(0)}_{(\Gamma,s)(R,+)}=2\,\beta_s^2\,\mathcal O^{(0)}_{(\Gamma,s)(R,+)}\,,\quad
\mathcal K^{(+1)}_{(\Gamma,s)(R,+)}=2\,\beta_s^2\,g_{(\Gamma,s)(R,+)}\,,\quad
\mathcal K^{(-1)}_{(\Gamma,s)(R,+)}=2\,\beta_s^2\,g^*_{(\Gamma,s)(R,+)}\,,
\label{eqkerngrx}
\end{align}
with $\beta_s=\beta^{\Gamma R}_s$, and all the other sectors vanish. As for Eq.~\eqref{eqkerngr}, the sectors $m=\pm1$ are complex-conjugated for the opposite direction, and the sectors with different $m$ remain decoupled in the Flicker scenario as well. The terms proportional to $\sin\phi$ are quadratic in $\hat{\bk}$ and linear in $\hat{\bs q}$. They give entries of $Q^{aa'}$ between the functions $f_{ij}$ at $\Gamma$ and $f'_i$ at $R$ for both sheets at $\Gamma$. The statement above that all entries with a function $f_{ij}$ vanish for the pairs between $\Gamma$ and $R$ therefore holds for the Chang scenario only.

The kernels of the RSW and TSM nodes are treated analogously. The overlaps of Eqs.~\eqref{eqovrsw} and \eqref{eqovtsm} are polynomials in $C$, and their expansions in the Legendre polynomials, $\mathcal O=\sum_{l}o_l\,P_l(C)$, have the coefficients
\begin{align}
& \left\{o_0,o_1,o_2,o_3\right\}_{(\Gamma,3/2)(\Gamma,3/2)}=\left\{\tfrac14,\tfrac{9}{20},\tfrac14,\tfrac{1}{20}\right\},\quad
\left\{o_0,o_1,o_2,o_3\right\}_{(\Gamma,1/2)(\Gamma,1/2)}=\left\{\tfrac14,\tfrac{1}{20},\tfrac14,\tfrac{9}{20}\right\},\nn
& \left\{o_0,o_1,o_2,o_3\right\}_{(\Gamma,3/2)(\Gamma,1/2)}=\left\{\tfrac14,\tfrac{3}{20},-\tfrac14,-\tfrac{3}{20}\right\},\quad
\left\{o_0,o_1,o_2\right\}_{(R_i,+)(R_i,+)}=\left\{\tfrac13,\tfrac12,\tfrac16\right\}.
\label{eqlegendre}
\end{align}
By the addition theorem, $P_l(C)=\left(2/(2\,l+1)\right)\sum_{m=-l}^{l}\Theta_{lm}(\theta)\,\Theta_{lm}(\theta')\,e^{i\,m\,(\varphi-\varphi')}$ with the functions $\Theta_{lm}$ used for the moments of Eq.~\eqref{eqmomsys}. The azimuthal harmonics of the overlap between two sheets of the same node are therefore $\mathcal O^{(0)}=\sum_l\left(2\,o_l/(2\,l+1)\right)\Theta_{l0}(\theta)\,\Theta_{l0}(\theta')$ and $\mathcal O^{(m)}=\sum_l\left(4\,o_l/(2\,l+1)\right)\Theta_{lm}(\theta)\,\Theta_{lm}(\theta')$ for $m\geq1$. The matrices $K^{(m)}_{aa'}$ of Eq.~\eqref{eqsol2} are then diagonal in $l$ with the entries $4\,\beta_{aa'}^2\,o_l/(2\,l+1)$ for every $m$, and the kernel between the sheets of the same node has the range $l\leq3$ for the RSW node and $l\leq2$ for the TSM node.

For the overlaps between $\Gamma$ and $R$, the harmonics $\mathcal O^{(m)}_{(\Gamma,\tilde\psi)(R_i,+)}$ with $m=0,1,2$ follow from the amplitudes of Eq.~\eqref{eqovgr}. Let
\begin{align}
o^{(m)}_{ll'}=\int_0^\pi d\theta\,\sin\theta\int_0^\pi d\theta'\,\sin\theta'\;\Theta_{lm}(\theta)\,\mathcal O^{(m)}_{(\Gamma,\tilde\psi)(R_i,+)}(\theta,\theta')\,\Theta_{l'm}(\theta')\,.
\end{align}
The matrix of Eq.~\eqref{eqsol2} between the sheet $(\Gamma,\tilde\psi)$ and the copy $R_i$ is $K^{(m)}_{ll'}=c_m\left(\beta^{\Gamma R_i}_{\tilde\psi}\right)^2o^{(m)}_{ll'}$ with $c_0=2$ and $c_{1}=c_{2}=1$, and the matrix between $R_i$ and $(\Gamma,\tilde\psi)$ is its transpose. The range of the kernel follows from a multipole argument. As a function of the direction at $\Gamma$, the overlap is $\mathcal O_{(\Gamma,\tilde\psi)(R_i,+)}=\psi_{\Gamma,\tilde\psi}^\dagger\left(\mathcal Y_i^\dagger\,v_+v_+^\dagger\,\mathcal Y_i\right)\psi_{\Gamma,\tilde\psi}$, where the matrix in brackets is Hermitian and $4\times4$. It is a sum of the multipole operators of the spin 3/2 with the rank $l\leq3$. By the Wigner--Eckart theorem, the expectation value of such an operator in the eigenstate $\psi_{\Gamma,\tilde\psi}$ of $\hat{\bk}\cdot\bs J$ is proportional to $Y_{lm}(\hat{\bk})$ with the same $l$. As a function of the direction at $R$, the overlap is $\mathcal O_{(\Gamma,\tilde\psi)(R_i,+)}=v_+^\dagger\left(\mathcal Y_i\,\psi_{\Gamma,\tilde\psi}\psi_{\Gamma,\tilde\psi}^\dagger\,\mathcal Y_i^\dagger\right)v_+$, where the matrix in brackets is Hermitian and $3\times3$ and contains the multipoles of the spin 1 with the rank $l\leq2$. The expectation value in the eigenstate $v_+$ of $\hat{\bk}'\cdot\bs\Lambda$ is then a combination of $Y_{lm}(\hat{\bk}')$ with $l\leq2$. The ranges are therefore $l\leq3$ at $\Gamma$ and $l\leq2$ at $R$.

The out-scattering rate of the sheet $a$ is given by Eq.~\eqref{eqtau}, since the overlaps depend only on $\varphi-\varphi'$. The contribution of the copies at $R$ to the rate of the sheet $(\Gamma,\tilde\psi)$ is
\begin{align}
\frac{1}{\tau_{(\Gamma,\tilde\psi)}(\theta)}\ni\sum_i2\left(\beta^{\Gamma R_i}_{\tilde\psi}\right)^2\int_0^\pi d\theta'\,\sin\theta'\,\rho_{R_i}(\theta')\,\mathcal O^{(0)}_{(\Gamma,\tilde\psi)(R_i,+)}(\theta,\theta')\,,
\end{align}
and the expression for the sheet $(R_i,+)$ follows by exchanging the roles of $\theta$ and $\theta'$ and using the weights $\rho_{\Gamma,\tilde\psi}$.

%%%%%%%%%%%%%%%
\subsection{Degenerate copies at the \texorpdfstring{$R$}{R} point}
\label{secappdeg}

The copies $R_1$ and $R_2$ have the same dispersion, and every band $s$ of the node at $R$ is therefore twofold degenerate. Its Berry connection is the matrix $\hat{\bs A}^{mn}=i\,\langle u_m|\nabla_{\bk}u_n\rangle$ in the copy space, and the curvature is $\hat{\bs\Omega}=\nabla_{\bk}\times\hat{\bs A}-i\,\hat{\bs A}\times\hat{\bs A}$. The blocks of the Hamiltonian at $R$ [cf. Eqs.~\eqref{eqham} and \eqref{eqham2}] are independent of $\bk$ in the copy basis. The connection is therefore diagonal, $\hat{\bs A}=\mathrm{diag}\!\left(\bs A^{R_1},\bs A^{R_2}\right)$, the commutator term vanishes, and the curvature reduces to $\hat{\bs\Omega}=\bOm^a\,\mathbb{I}$ with the abelian curvature of Eq.~\eqref{eqbcomm}. The OMM, the phase-space factor, and the velocity are likewise proportional to $\mathbb{I}$. The field terms of the SBE are thus the scalar ones for each copy, and all differences from the scalar treatment arise from the collision integral.

We distinguish two cases, according to whether the overlap between $R_1$ and $R_2$ vanishes. If it vanishes, as for the block-diagonal Hamiltonians of Eqs.~\eqref{eqham} and \eqref{eqham2}, the direct scattering between the copies is absent and $\beta^{R}_{\rm inter}=0$. The copies remain coupled through the sheets at $\Gamma$, whose spinors overlap with both copies. The scattering from a sheet $a'$ into the coherence of the copies carries the factor $\langle u_{R_1}(\bk)|\hat V|u_{a'}(\bk')\rangle\langle u_{a'}(\bk')|\hat V^\dagger|u_{R_2}(\bk)\rangle$, which is nonzero for the block $\mathcal Y$ of Eq.~\eqref{eqyblocks}, and the coherences feed back into the populations at $\Gamma$. With the amplitudes $A_1$ and $A_2$ of Appendix~\ref{secappsp2}, the in-scattering from the band at $R$ into the sheet $(\Gamma,\tilde\psi)$ involves $\left(A_1^*,A_2^*\right)\hat\Lambda_R\left(A_1,A_2\right)^{\rm T}=|A_1|^2\hat\Lambda_{11}+|A_2|^2\hat\Lambda_{22}+2\,\Re\!\left(A_1^*\,A_2\,\hat\Lambda_{12}\right)$, and the last term is the contribution of the coherence. It does not vanish for the overlaps of Eq.~\eqref{eqovgr}, since $A_1^*\,A_2\neq0$ in general. The scalar equations of the main text keep only the populations of $R_1$ and $R_2$ and therefore drop this term, whereas the matrix treatment below includes it.

If the overlap between $R_1$ and $R_2$ is nonzero, the same formalism applies. The matrix treatment, which keeps the coherences, covers both cases. The degenerate band $a$ at $R$ carries the distribution matrix
\begin{align}
\hat\Lambda_a=\sum_{\mu=0}^{3}\Lambda^{\mu}_a\,\tilde\tau_\mu\,,\quad
\tilde\tau_0=\mathbb{I}\,,
\label{eqlammat}
\end{align}
with the Pauli matrices $\tilde\tau_{1,2,3}$ in the copy space. The scalar sheets $R_1$ and $R_2$ correspond to the populations $\Lambda^0_a\pm\Lambda^3_a$, and $\Lambda^1_a$ and $\Lambda^2_a$ are the coherences. Let $\mathcal V^{mm'}_{\bk\bk'}=\langle u_m(\bk)|\hat V|u_{m'}(\bk')\rangle$ be the matrix element of the impurity potential between the spinors of the bands $a$ and $a'$, where $m$ and $m'$ are copy indices and a sheet at $\Gamma$ has a single index. The gain term of the collision integral maps a matrix $X$ of the band $a'$ onto $\mathcal W_{aa'}(X)_{mn}=\sum_{m'n'}\overline{\mathcal V^{mm'}_{\bk\bk'}\left(\mathcal V^{nn'}_{\bk\bk'}\right)^{*}}\,X_{m'n'}$, which is normalised as the scalar kernel of Eq.~\eqref{eqkernel}. The bar denotes the disorder average. The loss term is the anticommutator with the matrix $\hat\Gamma_a$, which follows from the same matrix elements by summing over the final states at the Fermi energy. In the basis of the $\tilde\tau_\mu$, the two superoperators are the $4\times4$ matrices
\begin{align}
\mathcal W^{\mu\nu}_{aa'}=\frac12\,\mathrm{Tr}\!\left[\tilde\tau_\mu\,\mathcal W_{aa'}\!\left(\tilde\tau_\nu\right)\right],\quad
\Gamma^{\mu\nu}_a=\frac14\,\mathrm{Tr}\!\left[\tilde\tau_\mu\left(\hat\Gamma_a\,\tilde\tau_\nu+\tilde\tau_\nu\,\hat\Gamma_a\right)\right].
\label{eqkmat}
\end{align}
For a nondegenerate sheet, only the entry $\mu=\nu=0$ exists, and it is the scalar overlap of Eq.~\eqref{eqoverlap}. The field operator is proportional to $\mathbb{I}$ in the copy space, and it acts on the azimuthal dependence of each component. Eq.~\eqref{eqsector} therefore holds for the four-component column $\Lambda^{(m)}_a$ of the sector $m$ in the form
\begin{align}
\left[\Gamma_a(\theta)+i\,\omega_a(\theta)\,\hat M^{(m)}\right]\Lambda^{(m)}_a(\theta)=h_a\,\boldsymbol e_0+\sum_{a'}\int_0^\pi d\theta'\,\sin\theta'\,\rho_{a'}(\theta')\,\mathcal W^{(m)}_{aa'}(\theta,\theta')\,\Lambda^{(m)}_{a'}(\theta')\,,
\label{eqdegsys}
\end{align}
with $\boldsymbol e_0=(1,0,0,0)^{\rm T}$, $\mathcal W^{(m)}_{aa'}$ the azimuthal harmonic of $\mathcal W^{\mu\nu}_{aa'}$ as in Eq.~\eqref{eqkernel}, and $\hat M^{(m)}=\mathrm{diag}\left(m,m,m-\varsigma,m+\varsigma\right)$ in the basis $\left(\Lambda^0_a,\Lambda^3_a,\Lambda_{12,a},\Lambda_{21,a}\right)$ with $\Lambda_{12}=\Lambda^1-i\,\Lambda^2$ and $\Lambda_{21}=\Lambda^1+i\,\Lambda^2$. The integer $\varsigma$ is the angular momentum of the coherence between the copies, and it is determined at the end of this subsection. The source acts only on $\Lambda^0_a$, and the components $\Lambda^{1,2,3}_a$ are generated by the off-diagonal entries of the kernel. The azimuthal decomposition and the finite-rank closure of the main text apply to each entry. The velocity is proportional to $\mathbb{I}$, and the current involves only the trace $\mathrm{Tr}\,\hat\Lambda_a=2\,\Lambda^0_a$. Eq.~\eqref{eqsigmw} therefore holds with $\Lambda_a$ replaced by $\mathrm{Tr}\,\hat\Lambda_a$ for the degenerate bands.

If the exchange of $R_1$ and $R_2$ is a symmetry of the kernel, which does not hold for the overlaps of Eq.~\eqref{eqovgr2}, it maps $\tilde\tau_1\to\tilde\tau_1$ and $\tilde\tau_{2,3}\to-\tilde\tau_{2,3}$. The source is even, the components $\Lambda^2_a$ and $\Lambda^3_a$ decouple, and only $\Lambda^0_a$ and $\Lambda^1_a$ remain. The number of moments of each degenerate band then doubles with respect to the exchange-symmetric scalar sheet.

The matrix elements of Eq.~\eqref{eqkmat} follow from the amplitudes of the two models. The impurity potential is a scalar in the orbital space, and its disorder average gives $\overline{\mathcal V^{mm'}\left(\mathcal V^{nn'}\right)^*}=\left(\beta^2/\pi\right)\langle u_m|u_{m'}\rangle\langle u_{n'}|u_n\rangle$ in the normalisation of Eq.~\eqref{eqkernel}. The copy basis does not depend on the direction of $\hat{\bs q}$, and the overlap between two directions at $R$ is therefore diagonal in the copy index in both models, $\langle u_m(\hat{\bs q})|u_n(\hat{\bs q}')\rangle=\delta_{mn}\,\langle u(\hat{\bs q})|u(\hat{\bs q}')\rangle$. Let $a_m(\hat{\bk},\hat{\bs q})=\langle u_m(\hat{\bs q})|\psi_{\Gamma,\tilde\psi}(\hat{\bk})\rangle$ be the amplitude from the sheet $(\Gamma,\tilde\psi)$ into the copy $m$, which is $A_m$ of Appendix~\ref{secappsp2} for the six-dimensional model. We define the real four-vector
\begin{align}
n_0=|a_1|^2+|a_2|^2=\mathcal O_{(\Gamma,\tilde\psi)(R,+)}\,,\quad
n_1=2\,\Re\!\left(a_1^*\,a_2\right),\quad
n_2=2\,\Im\!\left(a_1^*\,a_2\right),\quad
n_3=|a_1|^2-|a_2|^2\,,
\label{eqnvec}
\end{align}
which is $n_\mu=a^\dagger\tilde\tau_\mu\,a$ and satisfies $n_1^2+n_2^2+n_3^2=n_0^2$, since $a$ is a single spinor in the copy space. The scattering from $(\Gamma,\tilde\psi)$ into $R$ has the matrix $\mathcal W_{R(\Gamma,\tilde\psi)}(\mathscr X)_{mn}=\left(\beta^2/\pi\right)a_m\,a_n^*\,\mathscr X$ for a scalar $\mathscr X$, and the scattering from $R$ into $(\Gamma,\tilde\psi)$ has $\mathcal W_{(\Gamma,\tilde\psi)R}(X)=\left(\beta^2/\pi\right)a^\dagger X\,a$. With Eq.~\eqref{eqkmat}, the entries are
\begin{align}
\mathcal W^{0\nu}_{(\Gamma,\tilde\psi)R}=\frac{\beta^2}{\pi}\,n_\nu\,,\quad
\mathcal W^{\mu0}_{R(\Gamma,\tilde\psi)}=\frac{\beta^2}{2\,\pi}\,n_\mu\,,\quad
\mathcal W^{\mu\nu}_{RR}=\frac{\beta_{RR}^2}{\pi}\,\mathcal O_{RR}\,\delta^{\mu\nu}\,,
\label{eqwmat}
\end{align}
and all the other entries of these blocks vanish. The factor $1/2$ in the second entry is the normalisation of the trace in Eq.~\eqref{eqkmat}, and the first two entries satisfy $\mathcal W^{0\mu}_{(\Gamma,\tilde\psi)R}=2\,\mathcal W^{\mu0}_{R(\Gamma,\tilde\psi)}$, which is compensated by $\mathrm{Tr}\,\hat\Lambda_R=2\,\Lambda^0_R$. The scalar overlap $\mathcal O_{RR}$ of the two directions at $R$ is that of Eq.~\eqref{eqoverlapm} for the four-orbital model and that of Eq.~\eqref{eqovtsm} for the six-dimensional model. The loss matrix of the degenerate band follows from the same amplitudes,
\begin{align}
\hat\Gamma_R(\hat{\bs q})=\sum_{\tilde\psi}\frac{\beta_{\tilde\psi}^2}{2\,\pi}\int d^2\hat{\bk}\,\rho_{\Gamma,\tilde\psi}\sum_{\mu=0}^3n_\mu(\hat{\bk},\hat{\bs q})\,\tilde\tau_\mu+\frac{\beta_{RR}^2}{\pi}\int d^2\hat{\bs q}'\,\rho_R\,\mathcal O_{RR}\,\mathbb{I}\,,
\label{eqgamhat}
\end{align}
since $a\,a^\dagger=\frac12\sum_\mu n_\mu\tilde\tau_\mu$. For $\hat\Gamma_R=\Gamma^0\,\mathbb{I}+\sum_j\Gamma^j\tilde\tau_j$, Eq.~\eqref{eqkmat} gives $\Gamma^{00}=\Gamma^0$, $\Gamma^{0j}=\Gamma^{j0}=\Gamma^j$, and $\Gamma^{jk}=\Gamma^0\,\delta^{jk}$. The scalar rate of the sheets $(\Gamma,\tilde\psi)$ is not modified, since its loss into $R$ involves only $n_0=\mathcal O_{(\Gamma,\tilde\psi)(R,+)}$ summed over the copies.

For the six-dimensional model, the amplitudes of Appendix~\ref{secappsp2} are $A_1=e^{i\left(\varphi'-2\,\varphi\right)}T_1$ and $A_2=e^{i\left(\varphi'-\varphi\right)}T_2$ with $T_i=\sum_jt^{(i)}_j\,e^{-i\,j\,(\varphi-\varphi')}$, where $\mathfrak t^{(i)}_j$ are the real amplitudes of Eq.~\eqref{eqovgr} for the copy $R_i$. The populations are $n_0=|T_1|^2+|T_2|^2$ and $n_3=|T_1|^2-|T_2|^2$, and each has the harmonics $\mathcal O^{(m)}_{1}\pm\mathcal O^{(m)}_{2}$ of Eq.~\eqref{eqovgr} for $m=0,1,2$. The coherence is $n_1+i\,n_2=2\,A_1^*A_2$ with
\begin{align}
& A_1^*A_2=e^{i\,\varphi'}\sum_{p=-2}^{2}\Psi_p\,e^{i\,(p+1)\,(\varphi-\varphi')}\,,\quad
 \Psi_{-2}=\mathfrak t^{(1)}_{-1}\,\mathfrak t^{(2)}_{+1}\,,\quad
\Psi_{-1}=\mathfrak t^{(1)}_{-1}\,\mathfrak t^{(2)}_{0}+\mathfrak t^{(1)}_{0}\,\mathfrak t^{(2)}_{+1}\,,\quad
\Psi_{0}=\sum_jt^{(1)}_j\,\mathfrak t^{(2)}_j\,,\nn
& \Psi_{1}=\mathfrak t^{(1)}_{+1}\,\mathfrak t^{(2)}_{0}+\mathfrak t^{(1)}_{0}\,\mathfrak t^{(2)}_{-1}\,,\quad
\Psi_{2}=\mathfrak t^{(1)}_{+1}\,\mathfrak t^{(2)}_{-1}\,.
\label{eqzp}
\end{align}
The coefficients $\Psi_p$ are real. The contribution of the coherence to the gain term of the sheet $(\Gamma,\tilde\psi)$ is $\left(\beta^2/\pi\right)\left(A_1^*A_2\,\Lambda_{12}+{\rm c.c.}\right)$. If $\Lambda^0$ and $\Lambda^3$ of the band at $R$ are proportional to $e^{i\,m\,\varphi'}$, the integral over $\varphi'$ shows that the sector $m$ of the sheet $(\Gamma,\tilde\psi)$ receives the coherence $\Lambda_{12}\propto e^{i\,(m-1)\,\varphi'}$ with the kernel $2\,\beta^2\,\Psi_{m-1}$, which is nonzero for $-1\leq m\leq3$, and the coherence $\Lambda_{21}\propto e^{i\,(m+1)\,\varphi'}$ with the kernel $2\,\beta^2\,\Psi_{-m-1}$, which is nonzero for $-3\leq m\leq1$. The kernels for the populations are $c_m\,\beta^2\left(\mathcal O^{(m)}_1\pm\mathcal O^{(m)}_2\right)$ with $c_0=2$ and $c_1=c_2=1$ as in Eq.~\eqref{eqkernel}, and the upper sign belongs to $\Lambda^0$. The azimuthal average of Eq.~\eqref{eqgamhat} contains the populations $\mathcal O^{(0)}_1\pm\mathcal O^{(0)}_2$ and the coherence $\bar n_1+i\,\bar n_2=2\,e^{i\,\varphi'}\,\Psi_{-1}$.

For the four-orbital model, the copy space is the space of the operators that commute with the matrices $\Sigma^R_a$ of Eq.~\eqref{eqsigalg}, since the upper pair of bands at $R$ is the product of the spin-1/2 doublet of $\Sigma^R_a$ with this space. The matrices $\check\tau_1=\tilde\tau_x\,\tilde\mu_y$, $\check\tau_2=\tilde\tau_y$, and $\check\tau_3=\tilde\tau_z\,\tilde\mu_y$, written with the Pauli matrices of Eq.~\eqref{eqtb198}, commute with the matrices $\Sigma^R_a$ and $\Sigma^\Gamma_a$, and satisfy $\check\tau_1\,\check\tau_2=i\,\check\tau_3$ together with its cyclic permutations. They act as $\tilde\tau_\mu$ in the copy space and do not depend on $\hat{\bs q}$. The connection is therefore diagonal, as in the first paragraph of this subsection. With the projector $P_R(\hat{\bs q})=\left(1-s_p\,\hat{\bs q}\cdot\bs\Sigma^R\right)/2$ on the upper pair and the phase matrix $U=\mathrm{diag}\left(e^{i\phi},1,1,1\right)$ of the Flicker scenario, $n_\mu=\psi_{\Gamma,\tilde\psi}^\dagger\,U^\dagger P_R(\hat{\bs q})\,\check\tau_\mu\,U\,\psi_{\Gamma,\tilde\psi}$ with $\check\tau_0=\mathbb{I}$, and $U=\mathbb{I}$ for the Chang scenario. We use the components $n_\parallel=\left(n_1+n_2+n_3\right)/\sqrt3$ and $n_\pm=\left(2\,n_1-n_2-n_3\right)/\sqrt6\pm i\left(n_2-n_3\right)/\sqrt2$, which are those along the axis $\left(1,1,1\right)/\sqrt3$ and in the plane perpendicular to it in the copy space. In the frame of Eq.~\eqref{eqcycsums}, they have the decomposition
\begin{align}
n_0=\mathcal O^{(0)}+2\,\Re\!\left(g\,e^{i\,(\varphi-\varphi')}\right),\quad
n_\parallel=\mathcal O^{(0)}_\parallel+2\,\Re\!\left(g_\parallel\,e^{i\,(\varphi-\varphi')}\right),\quad
n_+=e^{-i\,\varphi'}\left(\varpi_0+\varpi_1\,e^{-i\,(\varphi-\varphi')}+\varpi_2\,e^{-2\,i\,(\varphi-\varphi')}\right),\quad
n_-=n_+^*\,,
\label{eqnharm}
\end{align}
where $\mathcal O^{(0)}$ and $g$ are those of Eq.~\eqref{eqovgrxph}. For the sheet $(\Gamma,1)$, the remaining coefficients are
\begin{align}
& \mathcal O^{(0)}_\parallel=-\frac{s_p}{2}\cos\theta\left(1-\cos\theta\,\cos\theta'\right)+\frac{\sqrt3}{8}\sin\phi\,\sin^2\theta\,,\nn
& g_\parallel=\frac18\,\sin\theta\,\sin\theta'\left[\sqrt3\sin\phi-s_p\cos\theta\left(1+i\,\sqrt3\cos\phi\right)\right],\nn
& \varpi_0=\frac{s_p}{16}\,\sin^2\theta\,\sin\theta'\left[1-3\cos\phi-i\,\sqrt3\left(1+\cos\phi\right)\right],\quad
\varpi_2=\frac{s_p}{4}\,\sin^2\theta\,\sin\theta'\,e^{2\,\pi\,i/3}\,,\nn
& \varpi_1=-\frac{s_p}{8}\,\sin\theta\left(1-\cos\theta\,\cos\theta'\right)\left[1+3\cos\phi+i\,\sqrt3\left(\cos\phi-1\right)\right]-\frac{\sqrt3}{8}\,\sin\theta\,\sin\phi\left(1-i\,\sqrt3\right)\left(\cos\theta-\cos\theta'\right),
\label{eqnh1}
\end{align}
and for the sheet $(\Gamma,0)$ they are
\begin{align}
& \mathcal O^{(0)}_\parallel=\frac{\sqrt3}{4}\sin\phi\,\cos^2\theta-\frac{s_p}{2}\cos\theta'\left(2\cos^2\theta-1\right),\quad
g_\parallel=\frac{s_p}{4}\,\cos\theta\,\sin\theta\,\sin\theta'\left(1+i\,\sqrt3\cos\phi\right),\nn
& \varpi_0=\frac{s_p}{8}\,\cos^2\theta\,\sin\theta'\left[1-3\cos\phi-i\,\sqrt3\left(1+\cos\phi\right)\right],\quad
\varpi_2=-\frac{s_p}{2}\,\sin^2\theta\,\sin\theta'\,e^{2\,\pi\,i/3}\,,\nn
& \varpi_1=-\frac{s_p}{4}\,\cos\theta\,\cos\theta'\,\sin\theta\left[1+3\cos\phi+i\,\sqrt3\left(\cos\phi-1\right)\right]+\frac{\sqrt3}{4}\,\cos\theta\,\sin\theta\,\sin\phi\left(1-i\,\sqrt3\right).
\label{eqnh0}
\end{align}
For the Chang scenario, $\phi=0$, the vector $\left(n_\parallel,n_+,n_-\right)$ of the sheet $(\Gamma,1)$ takes the form $-s_p\,\mathcal O_{(\Gamma,1)(R,+)}\left(\cos\theta,\,\sin\theta\,e^{-i\,\varphi},\,\sin\theta\,e^{i\,\varphi}\right)$, which depends on the direction $\hat{\bk}$ at $\Gamma$ only through a unit vector in the copy space. The component $n_+$ is proportional to $e^{-i\,\varphi'}$ in Eq.~\eqref{eqnharm}, and the coherence of the copies at $R$ therefore carries an angular momentum with respect to the axis $\hat{\bs c}$. With $\Lambda_\parallel$ and $\Lambda_\mp=\Lambda'_1\mp i\,\Lambda'_2$ the components of $\hat\Lambda_R$ along the axes of the copy space defined above, the gain term of the sheet $(\Gamma,\tilde\psi)$ is $\left(\beta^2/\pi\right)\left[n_0\,\Lambda^0+n_\parallel\Lambda_\parallel+\frac12\left(n_+\Lambda_-+n_-\Lambda_+\right)\right]$. If $\Lambda^0$ and $\Lambda_\parallel$ are proportional to $e^{i\,m\,\varphi'}$, the sector $m$ of the sheet $(\Gamma,\tilde\psi)$ receives these with the kernels $2\,\beta^2$ times $\mathcal O^{(0)}$ and $\mathcal O^{(0)}_\parallel$ for $m=0$, and $2\,\beta^2$ times $g$ and $g_\parallel$, or their complex conjugates, for $m=\pm1$, as in Eq.~\eqref{eqkerngrx}. The sector receives $\Lambda_-\propto e^{i\,(m+1)\,\varphi'}$ with the kernel $\beta^2\,\varpi_{|m|}$ for $m=0,-1,-2$, and $\Lambda_+\propto e^{i\,(m-1)\,\varphi'}$ with the kernel $\beta^2\,\varpi_{|m|}^*$ for $m=0,1,2$. The kernels of the opposite direction follow from Eq.~\eqref{eqwmat} with the complex-conjugated phases.

In both models, the sector equation of the degenerate band involves four components whose azimuthal numbers are $\left(m,m,m-\varsigma,m+\varsigma\right)$ for $\left(\Lambda^0,\Lambda^3,\Lambda_{12},\Lambda_{21}\right)$, with $\varsigma=+1$ for the six-dimensional model, and $\left(m,m,m+1,m-1\right)$ for $\left(\Lambda^0,\Lambda_\parallel,\Lambda_-,\Lambda_+\right)$, which is $\varsigma=-1$, for the four-orbital model in the frame with the polar axis along $\hat{\bs c}$. The field term of Eq.~\eqref{eqdegsys} acts with the azimuthal number of each component.

%%%%%%%%%%%%%%%
\section{Symmetries, classification, and ansatz for the mean free-path}
\label{secappsymans}

\subsection{Symmetries and classification of the set-ups}
\label{secappsym}

Rotations about $\hat{\bB}$ map every FS sheet onto itself, since the sheets are axisymmetric about $\hat{\bB}$ [cf. Eq.~\eqref{eqwqd}]. The LFO of Eq.~\eqref{eqlfoaxi} is proportional to $\partial_\varphi$, commutes with these rotations, and is therefore diagonal in the azimuthal harmonic $m$ [cf. Eq.~\eqref{eqresalg}]. The collision operator commutes with the rotations if the overlaps depend only on $\varphi-\varphi'$. This holds for the overlaps between the sheets of the same node and for the overlaps between $\Gamma$ and $R$ of Eq.~\eqref{eqovgr}. It fails for the overlaps with terms $m\neq m'$ such as Eq.~\eqref{eqovgr4}, for which the sectors are coupled.

Let $\mathcal P$ be the exchange of the sheets $R_1$ and $R_2$, that is, of the two nodes at $R$. If the amplitudes and the overlaps of the two nodes are equal, the collision operator and the operator $D^{-1}\Lhat$ commute with $\mathcal P$, and the source, the null vector with the components $\rho_a$, and the number constraint are $\mathcal P$-invariant. The solution is unique once the constraint is imposed and is therefore even, $\Lambda^j_{R_1}=\Lambda^j_{R_2}$. This symmetry is used after Eq.~\eqref{eqansatzc}. It is absent for the RSW and TSM nodes, since the overlaps with $\Gamma$ differ for the two copies [cf. Eq.~\eqref{eqovgr2}].

%%%%%%%%%%%%%%%%%
\subsubsection{RSW and TSM nodes}
\label{secappsym2}

The overlaps between two sheets of the same node depend only on $C$, and the overlap between $R_1$ and $R_2$ vanishes. We write $\mathcal P$ for the exchange of $R_1$ and $R_2$, with $\mathcal P(\Gamma,\tilde\psi)=(\Gamma,\tilde\psi)$. Let $\mathcal U_\Gamma(g)$ and $\mathcal U_R(g)$ be the unitary representations of a proper rotation $g$ at $\Gamma$ and at $R$, and let $\mathcal T_\Gamma$ and $\mathcal T_R$ be the antiunitary operators of time reversal. The overlap $\mathcal O_{(\Gamma,\tilde\psi)(R_i,+)}(\hat{\bk},\hat{\bk}')$ is the squared modulus of the matrix element of the block $\mathcal Y_i$ of Eq.~\eqref{eqyblocks} between the spinors of the two sheets. The overlap of Eq.~\eqref{eqovgr} is invariant under $\mathcal R_z(\delta)$ for every $\delta$, and it satisfies, for $g=C_{2x}$ and $g=C_{2y}$,
\begin{align}
\mathcal O_{(\Gamma,\tilde\psi)(R_i,+)}(g\,\hat{\bk},g\,\hat{\bk}')=\mathcal O_{(\Gamma,\tilde\psi)(R_{\bar\imath},+)}(\hat{\bk},\hat{\bk}')\,,\quad \bar\imath=3-i\,,
\label{eqsymP1}
\end{align}
and for the inversion of both momenta,
\begin{align}
\mathcal O_{(\Gamma,\tilde\psi)(R_i,+)}(-\hat{\bk},-\hat{\bk}')=\mathcal O_{(\Gamma,\tilde\psi)(R_{\bar\imath},+)}(\hat{\bk},\hat{\bk}')\,.
\label{eqsymP2}
\end{align}
The helicity projectors transform covariantly, $P(g\,\hat{\bk})=\mathcal U(g)\,P(\hat{\bk})\,\mathcal U^\dagger(g)$, and Eq.~\eqref{eqsymP1} follows from $\mathcal U_R^\dagger(g)\,\mathcal Y_1\,\mathcal U_\Gamma(g)=i\,\mathcal Y_2$ up to a phase, which is verified for the block of Eq.~\eqref{eqyblocks} with $\zeta^2+\eta^2=1$. For Eq.~\eqref{eqsymP2}, time reversal maps the helicity $\tilde\psi$ at $\hat{\bk}$ onto the helicity $\tilde\psi$ at $-\hat{\bk}$, since $\bk$ and $\bs J$ both change sign, and the blocks satisfy $\mathcal Y_2=\mathcal T_R\,\mathcal Y_1\,\mathcal T_\Gamma^{-1}$. An antiunitary operator preserves the modulus of the matrix element. The overlap with $R_1$ at $(-\hat{\bk},-\hat{\bk}')$ therefore equals that with $R_2$ at $(\hat{\bk},\hat{\bk}')$, and the same argument gives $\beta^{\Gamma R_1}_{\tilde\psi}=\beta^{\Gamma R_2}_{\tilde\psi}$ for non-magnetic scatterers. The elements $\mathcal R_z(\delta)$, $\mathcal P\,C_{2x}$, $\mathcal P\,C_{2y}$, and $\mathcal P\,\mathcal J_k$, where $\mathcal J_k$ is the inversion of the momenta, thus leave the kernel invariant. The exchange $\mathcal P$ alone and the rotations about axes in the $xy$ plane by generic angles are not symmetries.

The BC, the OMM, and the dispersion are identical for $R_1$ and $R_2$, and $\mathcal P$ acts trivially on the single-particle part of the SBE. For a proper $g$ with $g\,\bB=\bB$, the arguments below Eq.~\eqref{eqsymg3} in Appendix~\ref{secappsym3} apply, and the kernel is invariant under $(g,\mathcal P)$. The solution is unique once the number constraint is imposed, and it satisfies
\begin{align}
\Lambda^i_{\mathcal P a}(g\,\hat{\bk})=\sum_jg_{ij}\,\Lambda^j_a(\hat{\bk})\,,\quad\bs\sigma=g\,\bs\sigma\,g^{\rm T}\,.
\label{eqsymg2}
\end{align}
Let $\varepsilon_p$ be defined by $f_p(g\,\hat{\bk})=\varepsilon_p\,f_p(\hat{\bk})$ for the real harmonics $f_p$. Changing the integration variable in the moments, we find for diagonal $g$
\begin{align}
M^j_{\mathcal P a,p}=g_{jj}\,\varepsilon_p\,M^j_{a,p}\,.
\label{eqmomrel2}
\end{align}
For the sheets at $\Gamma$, this is a condition on a single sheet, $M^j_{(\Gamma,\tilde\psi),p}=g_{jj}\,\varepsilon_p\,M^j_{(\Gamma,\tilde\psi),p}$, and the moments with $g_{jj}\,\varepsilon_p=-1$ vanish. For the copies, $M^j_{R_2,p}=g_{jj}\,\varepsilon_p\,M^j_{R_1,p}$. For $g=C_{2x}=\mathrm{diag}(1,-1,-1)$ and the real harmonics $f_{lm}\propto P_l^m(\cos\theta)\,\{\cos m\varphi,\sin m\varphi\}$, we have $\varepsilon_p=(-1)^{l+m}$ for the cosine-type and $-(-1)^{l+m}$ for the sine-type functions. For $l\leq3$, there are eight functions with $\varepsilon_p=+1$ and eight with $\varepsilon_p=-1$. For $l\leq2$, there are five with $\varepsilon_p=+1$ and four with $\varepsilon_p=-1$.

The same relations with $g=C_{2y}=\mathrm{diag}(-1,1,-1)$ map $\bB=B\,(\sin\beta,0,\cos\beta)$ onto $-\bB$ and, combined with Eq.~\eqref{eqsymP2}, give the Onsager relation of Eq.~\eqref{eqons} for all the orientations of $\bB$. In the proof of Appendix~\ref{secappsym3}, the third property holds for the RSW and TSM nodes with the exchange of the copies, because of Eq.~\eqref{eqsymP2}. The identity $\sigma_{ij}(\bB)=\sigma_{ji}(-\bB)$ and the transformation of Eq.~\eqref{eqsymg2} with $g=C_{2y}$ give $\sigma_{xy}=-\sigma_{yx}$, $\sigma_{yz}=-\sigma_{zy}$, and $\sigma_{xz}=\sigma_{zx}$ for $\bB$ in the $xz$ plane. For $\bB=B\,\boldsymbol{\hat x}$, the relation with $g=C_{2x}$ removes $\sigma_{xy}$, $\sigma_{xz}$, $\sigma_{yx}$, and $\sigma_{zx}$.

\subsubsection{Two conjugate QTSM nodes}
\label{secappsym3}

Let $\hat{\bs n}^a_{\hat{\bk}}=\mathsf T_\chi\,\hat{\bk}$ be the unit vector of the sheet $a=(\chi,s)$ with $\mathsf T_+=\mathbb{I}$ and $\mathsf T_-=\mathsf T$. The overlaps depend only on $\hat{\bs n}^a_{\hat{\bk}}\cdot\hat{\bs n}^{a'}_{\hat{\bk}'}=\hat{\bk}\cdot\mathsf T_\chi\,\mathsf T_{\chi'}\,\hat{\bk}'$. Let $(g,\pi)$ be a transformation $g$ of the momenta that maps the node $\chi$ onto the node $\pi(\chi)$. The kernel is invariant if $g^{\rm T}\,\mathsf T_{\pi(\chi)}\,\mathsf T_{\pi(\chi')}\,g=\mathsf T_\chi\,\mathsf T_{\chi'}$ for all $\chi$ and $\chi'$. Since $\mathsf T_{-\chi}\,\mathsf T_{-\chi'}=\mathsf T_\chi\,\mathsf T_{\chi'}$ and $\mathsf T^2=\mathbb{I}$, this condition reduces to $g^{\rm T}\,\mathsf T\,g=\mathsf T$ for both choices of $\pi$. It means that $g$ commutes with $\mathsf T$. The Hamiltonian must also be mapped onto itself by a rotation of the pseudospin, which requires that $\mathsf T_{\pi(\chi)}\,g\,\mathsf T_\chi$ be a proper rotation. For $g$ commuting with $\mathsf T$, its determinant is $\det(g)$ for $\pi=\mathrm{id}$ and $-\det(g)$ for the exchange of the nodes. Proper elements therefore keep every node, and improper elements exchange the nodes. A rotation about an axis in the $xy$ plane by a generic angle is not a symmetry.

The BC and the OMM of Eq.~\eqref{eqbcomm} are proportional to $\chi\,\hat{\bk}$ divided by powers of $k$. For a proper $g$ that keeps the node, $\bOm^a(g\,\bk)=g\,\bOm^a(\bk)$. For an improper $g$ that exchanges the nodes, $\bOm^{-\chi}(g\,\bk)=-\,g\,\bOm^{\chi}(\bk)$, and $\boldsymbol m^a$ obeys the same relations. With $\bB\to\det(g)\,g\,\bB$, the quantities $\xi_s$ and $D_a$ are therefore mapped onto those of the image sheet, and $\bw_a$ transforms as a vector. Eq.~\eqref{eqboltzop} is thus invariant if $\bE\to g\,\bE$, and for the transformations that fix $\bB$ we obtain
\begin{align}
\Lambda^j_{\pi(a)}(g\,\hat{\bk})=\sum_{j'}g_{jj'}\,\Lambda^{j'}_a(\hat{\bk})\,,\quad
\bs\sigma=g\,\bs\sigma\,g^{\rm T}\,.
\label{eqsymg3}
\end{align}
The inversion has $g=-\mathbb{I}$ and $\det(g)=-1$, which leaves $\bB$ unchanged, and it gives Eq.~\eqref{eqinv3}.

The elements that fix $\bB$ follow from $\det(g)\,g\,\bB=\bB$. For $\bB\parallel\hat{\boldsymbol z}$, they are all the rotations $\mathcal R_z$, the reflection $\Pi_z$, and $\mathcal J$, and Eq.~\eqref{eqsymg3} gives the azimuthal sectors and Eq.~\eqref{eqmirror}. For $\bB\parallel\hat{\boldsymbol x}$, they are $C_{2x}$, the reflection $\Pi_x$ normal to $\bB$, and $\mathcal J$, and the relation $\bs\sigma=g\,\bs\sigma\,g^{\rm T}$ with $g=C_{2x}$ removes $\sigma_{xy}$, $\sigma_{xz}$, $\sigma_{yx}$, and $\sigma_{zx}$. The rotation $\mathcal R_z(\pi)$ maps $\bB$ onto $-\bB$ and leaves the kernel invariant. Together with the Onsager relation of Eq.~\eqref{eqons}, it gives $\sigma_{ij}(\bB)=\left[\mathcal R_z(\pi)\right]_{ii}\left[\mathcal R_z(\pi)\right]_{jj}\,\sigma_{ji}(\bB)$ and hence $\sigma_{yz}=-\sigma_{zy}$. For $\bB$ in the $xz$ plane at a generic angle, only $\mathcal J$ fixes $\bB$. The rotation $C_{2y}=\mathrm{diag}\left(-1,1,-1\right)$ maps $\bB$ onto $-\bB$, and the same argument gives $\sigma_{xy}=-\sigma_{yx}$, $\sigma_{yz}=-\sigma_{zy}$, and $\sigma_{xz}=\sigma_{zx}$. The three diagonal elements remain independent, as listed in Table~\ref{tabgeom3}.

The moments are $\mathcal M^j_{a,p}=\int d^2\hat{\bk}\,\rho_a\,f^a_p\,\Lambda^j_a$ with $f^a_p=f_p\!\left(\hat{\bs n}^a_{\hat{\bk}}\right)$ of Eq.~\eqref{eqfp}. Under $\mathcal J$, we have $\hat{\bs n}^{-}_{-\hat{\bk}}=\mathcal R_z(\pi)\,\hat{\bk}$ and $\rho_-(-\hat{\bk})=\rho_+(\hat{\bk})$, and Eq.~\eqref{eqinv3} gives
\begin{align}
\mathcal M^j_{(-,s),p}=-\,\varepsilon_p\,\mathcal M^j_{(+,s),p}\,,
\label{eqmomJ3}
\end{align}
where $\varepsilon_p=+1$ for the functions $1$, $\hat n_z$, $\hat n_x\,\hat n_y$, $\hat n_x^2-\hat n_y^2$, and $3\,\hat n_z^2-1$, and $\varepsilon_p=-1$ for $\hat n_x$, $\hat n_y$, $\hat n_x\,\hat n_z$, and $\hat n_y\,\hat n_z$. The number constraint $\sum_a\mathcal M^j_{a,0}=0$ holds identically, since $\epsilon_0=+1$. For case (b), Eq.~\eqref{eqsymg3} with $g=C_{2x}$ gives $\mathcal M^j_{(+,s),p}=s_j\,\varepsilon^{x}_p\,\mathcal M^j_{(+,s),p}$ with $s_x=+1$ and $s_y=s_z=-1$, where $\varepsilon^{x}_p=+1$ for $1$, $\hat n_x$, $\hat n_y\,\hat n_z$, $\hat n_x^2-\hat n_y^2$, and $3\,\hat n_z^2-1$, and $\varepsilon^{x}_p=-1$ for $\hat n_y$, $\hat n_z$, $\hat n_x\,\hat n_y$, and $\hat n_x\,\hat n_z$. The sheet $(+,1)$ keeps five moments for $j=x$ and four for $j=y,z$, and the sheet $(+,0)$ keeps four and two, respectively.

The relation of Eq.~\eqref{eqons} follows from time reversal, and the proof uses three properties. First, $\xi_s(\bk;\bB)=\xi_s(-\bk;-\bB)$ and $D_a(\bk;\bB)=D_a(-\bk;-\bB)$, which follow from Eq.~\eqref{eqxiw} and the oddness of $\bOm^a$ in $\bk$. The FS of the reversed field is thus the inversion of the original FS, and $\bw_a$ changes sign. Second, the operator $D_a^{-1}\Lhat$ is antisymmetric with respect to the measure $\rho_a\,d^2\hat{\bk}$, so that its transpose equals its value at the reversed field. Third, the collision operator is symmetric and invariant under the inversion of all the momenta combined with the exchange of the sheets that are mapped onto each other by time reversal. The last property holds for the overlaps of the sheets of the same node, which depend only on $C$. It is not assumed for the overlaps between $\Gamma$ and $R$. For the reference set-up, the relation follows from the axisymmetric form of the tensor, which satisfies Eq.~\eqref{eqons} identically.

\subsection{Ansatz for the mean free-path}
\label{secappcases}

The unknowns of the solver for the reference set-up of Table~\ref{tabgeom}, $\bB=B\,\boldsymbol{\hat z}$, follow from the spans of Eq.~\eqref{eqspans} in the sectors that carry a source. The sector $m=0$ carries the source of $\bE\parallel\bB$, and the sectors $m=\pm1$ carry that of $\bE\perp\bB$. The sector $m=\pm2$ has no source and does not contribute. The two sheets at $R$ have the same mean free-path by the exchange symmetry of Appendix~\ref{secappsym} and are counted once. The numbers of the unknown moments are collected in Table~\ref{tabunknowns}. Only the sector $m=0$ has a null vector, which is removed by the number constraint of Eq.~\eqref{eqnumber}.

\begin{table}[t!]
\centering
\small
\renewcommand{\arraystretch}{1.6}
\begin{tabular}{|c|c|c|c|}
\hline
Sheet & $m=0$ (real) & $m=1$ (complex) & General $\bB$ (real) \\
\hline
$(\Gamma,1)$ & $3$ & $2$ & $9$ \\
\hline
$(\Gamma,0)$ & $2$ & $1$ & $6$ \\
\hline
$(R,+)$ & $2$ & $1$ & $4$ \\
\hline
Total & $7$ & $4$ & $19$ \\
\hline
\end{tabular}
\caption{\label{tabunknowns} The numbers of the unknown moments for the QTSM and Weyl nodes, with the exchange of the two nodes at $R$ used.}
\end{table}

For the RSW and TSM nodes, the exchange symmetry is absent, and the two copies at $R$ are counted separately. For $\bB=B\,\boldsymbol{\hat z}$, the spans of Eq.~\eqref{eqspans2} in the sectors $m=0$ and $m=1$ give the numbers of case (a) in Table~\ref{tabunknowns2b}, and only the sector $m=0$ has a null vector, which is removed by the number constraint.

For the other orientations of $\bB$ (Table~\ref{tabgeom2}), the exchange of the copies is combined with a rotation, and the symmetry relations of Appendix~\ref{secappsym2} reduce the unknowns. The harmonics of the sheets at $\Gamma$ are those with $l\leq3$ (sixteen real functions), and those of the copies at $R$ are those with $l\leq2$ (nine real functions). The numbers of Table~\ref{tabunknowns2b} follow from the dimension of the eigenspace of the symmetry operator. For case (b), the $\Gamma$ sheets keep the eight functions with $g_{jj}\,\varepsilon_p=+1$, and the copy $R_2$ follows from $R_1$, which keeps all nine functions. The null vector of the number constraint has the weights $\rho_a$ and is invariant under every symmetry of the kernel. It therefore lies in the symmetric sector, that is, in the sector $j=x$ for case (b) and in the sector of every $j$ for case (c). For $j=y,z$ in case (b), the constraint holds identically, and there is no null vector.

\begin{table}[t!]
\centering
\small
\renewcommand{\arraystretch}{1.6}
\begin{tabular}{|c|c|c|c|c|c|}
\hline
Sheet & \makecell{(a), $m=0$ \\ (real)} & \makecell{(a), $m=1$ \\ (complex)} & \makecell{(b), $j=x$ \\ (real)} & \makecell{(b), $j=y,z$ \\ (real)} & \makecell{(c), each $j$ \\ (real)} \\
\hline
$(\Gamma,3/2)$ & $4$ & $3$ & $8$ & $8$ & $16$ \\
\hline
$(\Gamma,1/2)$ & $4$ & $3$ & $8$ & $8$ & $16$ \\
\hline
$(R_1,+)$ & $3$ & $2$ & $9$ & $9$ & $9$ \\
\hline
$(R_2,+)$ & $3$ & $2$ & $0$ & $0$ & $9$ \\
\hline
Total & $14$ & $10$ & $25$ & $25$ & $50$ \\
\hline
\end{tabular}
\caption{\label{tabunknowns2b} The numbers of the unknown moments for the RSW and TSM nodes for the cases of Table~\ref{tabgeom2}, where case (a) is $\bB=B\,\boldsymbol{\hat z}$. For case (b), the moments of $R_2$ follow from those of $R_1$ by Eq.~\eqref{eqmomrel2}. The null vector is removed by the number constraint for (a) with $m=0$, (b) with $j=x$, and (c).}
\end{table}

For the two conjugate QTSM nodes, the spans of Eq.~\eqref{eqspans} give the numbers of Table~\ref{tabunknowns3}. For $\bB=B\,\boldsymbol{\hat z}$, the mirror reflection of Eq.~\eqref{eqmirror} relates the node $-$ to the node $+$, and the number constraint of Eq.~\eqref{eqnumber} holds identically in the sector $m=0$. The sector $m=1$ has no null vector. For cases (b) and (c), the two nodes are related by Eq.~\eqref{eqinv3}, and the constraint again holds identically.

\begin{table}[t!]
\centering
\small
\renewcommand{\arraystretch}{1.6}
\begin{tabular}{|c|c|c|c|c|c|}
\hline
Sheet & \makecell{(a), $m=0$ \\ (real)} & \makecell{(a), $m=1$ \\ (complex)} & \makecell{(b), $j=x$ \\ (real)} & \makecell{(b), $j=y,z$ \\ (real)} & \makecell{(c) \\ (real)} \\
\hline
$(+,1)$ & $3$ & $2$ & $5$ & $4$ & $9$ \\
\hline
$(+,0)$ & $2$ & $1$ & $4$ & $2$ & $6$ \\
\hline
Total & $5$ & $3$ & $9$ & $6$ & $15$ \\
\hline
\end{tabular}
\caption{\label{tabunknowns3} The numbers of the unknown moments for two conjugate QTSM nodes, with the moments of the node $-$ following from those of the node $+$. The cases are those of Table~\ref{tabgeom3}.}
\end{table}

\section{Derivation of the expressions in the RTA limit}
\label{secrtab}

The expressions for $\bE\parallel\bB$ follow from one formula for a sheet with the dispersion $\epsilon_a=\mathsf u_a\,k+\mathsf w_a\,k^2$ measured from the node. The linear coefficient $\mathsf u_a$ contains the band index, for example $\mathsf u_a=s\,v_F$ for the QTSM sheets $s=1,0$. The linear, Weyl, RSW, TSM, and QTSM expressions are special cases of this formula.

In the RTA, the collision operator of Eq.~\eqref{eqboltzop} is replaced by the constant rate $\tau^{-1}$. For $\bE\parallel\bB=B\,\boldsymbol{\hat z}$, the LFO annihilates the sector $m=0$, and hence $\Lambda_a=\tau\,h_a$ with $h_a$ of Eq.~\eqref{eqlongsol}. Equation~\eqref{eqsigpar} gives the contribution of a sheet as
\begin{align}
\sigma_{\parallel,a}^{\rm RTA}=\frac{e^2\,\tau}{4\,\pi^2}\int_0^\pi d\theta\,\sin\theta\;\rho_a\,h_a^2\,,\quad
\rho_a=\frac{k^2\,D_a}{w_{a,k}}\,,\quad
h_a=\frac{1}{D_a}\left[w_{a,z}-\eta_B\,\chi_\nu\,N_a\,e\,B\,\frac{w_{a,k}}{2\,k^2}\right].
\label{eqrtab1}
\end{align}
The second term in $h_a$ is $\eta_B\,e\,B\left(\bOm^a\cdot\bw_a\right)$ of Eq.~\eqref{eqkinlin}, since $\bOm^a\cdot\bw_a=-\chi_\nu\,N_a\,w_{a,k}/(2\,k^2)$ by Eq.~\eqref{eqbcomm}. From Eqs.~\eqref{eqxiw} and \eqref{eqwqd}, the ingredients are
\begin{align}
& D_a=1-\eta_B\,\frac{\chi_\nu\,N_a\,e\,B\cos\theta}{2\,k^2}\,,\quad
w_{a,k}=\partial_k \epsilon_a-\eta_O\,\frac{\chi_\nu\,e\,v_\nu\,G_a\,B\cos\theta}{2\,k^2}\,,\nn
& w_{a,z}=\cos\theta\,w_{a,k}+\eta_O\,\frac{\chi_\nu\,e\,v_\nu\,G_a\,B\sin^2\theta}{2\,k^2}\,,
\label{eqrtab2}
\end{align}
where the expression for $w_{a,z}$ follows from $w_{a,z}=\cos\theta\,w_{a,k}-\sin\theta\,w_{a,\theta}$. All quantities are evaluated at the radius $k=k_a(\theta)$ of the FS.

Let $\bar v_a=\mathsf u_a+2\,\mathsf w_a\,k_a$ be the group velocity at the FS for $B=0$, with $\epsilon_a(k_a)=\mu_a$. With $\tilde\zeta=\eta_O\,\chi_\nu\,e\,v_\nu\,G_a\cos\theta/2$, the condition $\xi_s=\mu_a$ of Eq.~\eqref{eqxiw} reads $\epsilon_a(k)+\tilde\zeta\,B/k=\mu_a$. We insert $k=k_a+k^{(1)}\,B+k^{(2)}\,B^2$ and use $\epsilon_a(k_a+\delta)=\mu_a+\bar v_a\,\delta+\mathsf w_a\,\delta^2$. The terms of the first and second order in $B$ give
\begin{align}
\bar v_a\,k^{(1)}+\frac{\tilde\zeta}{k_a}=0\,,\quad
\bar v_a\,k^{(2)}+\mathsf w_a\left(k^{(1)}\right)^2-\frac{\tilde\zeta\,k^{(1)}}{k_a^2}=0\,,
\end{align}
and the solutions are
\begin{align}
k^{(1)}=-\frac{\tilde\zeta}{k_a\,\bar v_a}\,,\quad
k^{(2)}=\frac{1}{\bar v_a}\left[\frac{\tilde\zeta\,k^{(1)}}{k_a^2}-\mathsf w_a\left(k^{(1)}\right)^2\right].
\label{eqrtab3}
\end{align}
The first of them reproduces the radius that is quoted in Sec.~\ref{secrta} for $\mathsf w_a=0$.

\subsection{Expansion of the integrand and integration for parallel fields}
\label{secrtab3}

We insert $k=k_a+k^{(1)}\,B+k^{(2)}\,B^2$ of Eq.~\eqref{eqrtab3} in Eq.~\eqref{eqrtab2}, where $\partial_k\epsilon_a=\mathsf u_a+2\,\mathsf w_a\,k$, and we expand $\rho_a\,h_a^2$ of Eq.~\eqref{eqrtab1} to the second order in $B$. The relations $\chi_\nu^2=1$, $\eta_B^2=\eta_B$, and $\eta_O^2=\eta_O$ simplify the products. The result is a polynomial in $\cos\theta$,
\begin{align}
\rho_a\,h_a^2=k_a^2\,\bar v_a\cos^2\theta+\chi_\nu\,e\,B\left(\check\alpha_1\cos\theta+\check\alpha_3\cos^3\theta\right)+\frac{e^2\,B^2}{k_a^2\,\bar v_a}\left(\check\beta_0+\check\beta_2\cos^2\theta+\check\beta_4\cos^4\theta\right)+\mathcal O\!\left(B^3\right),
\label{eqrtab4}
\end{align}
with the coefficients
\begin{align}
& \check\alpha_1=\eta_O\,G_a\,v_\nu-\eta_B\,N_a\,\bar v_a\,,\quad
\check\alpha_3=\frac{\eta_O\,G_a\,v_\nu\left(\mathsf u_a-6\,\bar v_a\right)+\eta_B\,N_a\,\bar v_a^2}{2\,\bar v_a}\,,\nn
& \check\beta_0=\frac14\left(\eta_O\,G_a^2\,v_\nu^2-2\,\eta_B\,\eta_O\,N_a\,G_a\,v_\nu\,\bar v_a+\eta_B\,N_a^2\,\bar v_a^2\right),\nn
& \check\beta_2=-\frac12\left[\eta_O\,G_a^2\,v_\nu^2+\eta_B\,\eta_O\,N_a\,G_a\,v_\nu\left(\mathsf u_a-4\,\bar v_a\right)+\eta_B\,N_a^2\,\bar v_a^2\right],\nn
& \check\beta_4=-\frac{1}{8\,\bar v_a^2}\left[\eta_O\,G_a^2\,v_\nu^2\left(\mathsf u_a-\bar v_a\right)^2-2\,\eta_B\,\eta_O\,N_a\,G_a\,v_\nu\,\bar v_a^2\left(\mathsf u_a-4\,\bar v_a\right)-2\,\eta_B\,N_a^2\,\bar v_a^4\right].
\label{eqrtab5}
\end{align}
The terms of the first order in $B$ are odd in $\cos\theta$ and vanish in the integration. With $\int_0^\pi d\theta\,\sin\theta=2$, $\int_0^\pi d\theta\,\sin\theta\cos^2\theta=2/3$, and $\int_0^\pi d\theta\,\sin\theta\cos^4\theta=2/5$, the combination of the second order is
\begin{align}
2\,\check\beta_0+\frac23\,\check\beta_2+\frac25\,\check\beta_4=\frac{\bar v_a}{30}\left[8\,\eta_B\,N_a^2\,\bar v_a-\eta_B\,\eta_O\,N_a\,G_a\,v_\nu\,\frac{7\,\mathsf u_a+2\,\bar v_a}{\bar v_a}+\eta_O\,G_a^2\,v_\nu^2\,\frac{7\,\bar v_a^2+6\,\mathsf u_a\,\bar v_a-3\,\mathsf u_a^2}{2\,\bar v_a^3}\right].
\end{align}
Multiplying with $e^2\,\tau/(4\,\pi^2)$ and with the prefactor $e^2\,B^2/(k_a^2\,\bar v_a)$ of Eq.~\eqref{eqrtab4} gives the second term of Eq.~\eqref{eqrtaquad}, and the first term of Eq.~\eqref{eqrtab4} gives the Drude term $e^2\,\tau\,k_a^2\,\bar v_a/(6\,\pi^2)$.

For $\mathsf w_a\to0$, we have $\bar v_a=\mathsf u_a$ and $k_a=\mu_a/\mathsf u_a$. Eq.~\eqref{eqrtaquad} then gives
\begin{align}
\sigma_{\parallel,a}^{\rm RTA}=\frac{e^2\,\tau\,\mu_a^2}{6\,\pi^2\,\mathsf u_a}+\frac{e^4\,\tau\,v_\nu^3\,B^2}{\pi^2\,\mu_a^2}\left(\frac{N_a^2\,r_a^3}{15}\,\eta_B-\frac{3\,N_a\,G_a\,r_a^2}{40}\,\eta_B\,\eta_O+\frac{G_a^2\,r_a}{24}\,\eta_O\right),\quad
r_a=\frac{\mathsf u_a}{v_\nu}\,,
\label{eqrtab6}
\end{align}
which is Eq.~\eqref{eqrtagen}. For a Weyl node, $r_a=1$ and $N_a=G_a=1$ give the coefficients $1/15$, $-3/40$, and $1/24$ of Eq.~\eqref{eqrtaweyl}. For the sheets of the RSW and TSM nodes, the values of Table~\ref{tabnodes2} and the ratios $r_a=3/2$, $1/2$, and $1$ for $(\Gamma,3/2)$, $(\Gamma,1/2)$, and $(R_i,+)$ give the coefficients of Eq.~\eqref{eqrtarsw}. The Drude terms follow from $e^2\,\tau\,\mu_a^2/(6\,\pi^2\,\mathsf u_a)$ with $\mathsf u_a=3\,v_{3/2}/2$, $v_{3/2}/2$, and $v_R$.

\subsection{Expansion of the RTA integrals for the transverse and Hall conductivities}
\label{secapprta3}

The derivation is written for a single sheet with generic $b_a$, $\bar v_a$, $N_a$, $G_a$, and $v_\nu$. It therefore applies also to the sheets of Secs.~\ref{secsys} and \ref{secsys2}, with $b_a=\mathsf u_a$ in the notation of Eq.~\eqref{eqrtaquad} for the latter. For one node, the RTA gives $\sigma_{\perp,a}-i\,\sigma^{\rm B}_{H,a}=\left(e^2\,\tau/8\,\pi^2\right)\int_{-1}^{1}du\left(1-u^2\right)F_a$ with $F_a=\rho_a\left(q_a/D_a\right)^2/\left(1+i\,\omega_a\,\tau\right)$. We write $\epsilon_s(k)=b_a\,k+\left(\bar v_a-b_a\right)k^2/(2\,k_a)$ and $\mu_a=k_a\left(\bar v_a+b_a\right)/2$, and we use $\eta_B^2=\eta_B$ and $\eta_O^2=\eta_O$. The radius solves $\epsilon_s(k)+\eta_O\,e\,v_\nu\,G_a\,B\,u/(2\,k)=\mu_a$, and its expansion $k=k_a+\sum_{n=1}^{3}k^{(n)}B^n$ follows order by order,
\begin{align}
k^{(1)}=-\,\frac{\eta_O\,e\,v_\nu\,G_a\,u}{2\,k_a\,\bar v_a}\,,\quad
k^{(2)}=-\,\frac{\eta_O\,e^2\,v_\nu^2\,G_a^2\,u^2\left(3\,\bar v_a-b_a\right)}{8\,k_a^3\,\bar v_a^3}\,,
\quad k^{(3)}=-\,\frac{\eta_O\,e^3\,v_\nu^3\,G_a^3\,u^3\left(8\,\bar v_a^2-5\,b_a\,\bar v_a+b_a^2\right)}{16\,k_a^5\,\bar v_a^5}\,.
\end{align}
With $\partial_k\epsilon_s=\bar v_a+\left(\bar v_a-b_a\right)\left(k-k_a\right)/k_a$ inserted into Eq.~\eqref{eqwqd}, the quantities $D_a$, $w_{a,k}$, and $q_a$ are, to the second order,
\begin{align}
& D_a=1-\frac{\eta_B\,N_a\,e\,u\,B}{2\,k_a^2}-\frac{\eta_B\,\eta_O\,N_a\,G_a\,e^2\,v_\nu\,u^2\,B^2}{2\,k_a^4\,\bar v_a}\,,\nn
& w_{a,k}=\bar v_a-\frac{\eta_O\,G_a\,e\,v_\nu\,u\left(2\,\bar v_a-b_a\right)B}{2\,k_a^2\,\bar v_a}-\frac{\eta_O\,G_a^2\,e^2\,v_\nu^2\,u^2\left(7\,\bar v_a^2-4\,b_a\,\bar v_a+b_a^2\right)B^2}{8\,k_a^4\,\bar v_a^3}\,,\nn
& q_a=\bar v_a-\frac{\eta_O\,G_a\,e\,v_\nu\,u\left(3\,\bar v_a-b_a\right)B}{2\,k_a^2\,\bar v_a}-\frac{\eta_O\,G_a^2\,e^2\,v_\nu^2\,u^2\left(11\,\bar v_a^2-4\,b_a\,\bar v_a+b_a^2\right)B^2}{8\,k_a^4\,\bar v_a^3}\,.
\end{align}
The weight $\rho_a=k^2\,D_a/w_{a,k}$ and the frequency $\omega_a=e\,B\,q_a/\left(k\,D_a\right)$ then give the series $\rho_a\left(q_a/D_a\right)^2=\sum_n f_n\,B^n$ and $\omega_a=\sum_{n\geq1}\omega_n\,B^n$ with
\begin{align}
& f_0=k_a^2\,\bar v_a\,,\quad
f_1=\frac{e\,u}{2\,\bar v_a}\left[\eta_B\,N_a\,\bar v_a^2-\eta_O\,G_a\,v_\nu\left(6\,\bar v_a-b_a\right)\right],\nn
& f_2=-\frac{e^2\,u^2}{8\,k_a^2\,\bar v_a^3}\left[\eta_O\,G_a^2\,v_\nu^2\left(\bar v_a-b_a\right)^2+2\,\eta_B\,\eta_O\,N_a\,G_a\,v_\nu\,\bar v_a^2\left(4\,\bar v_a-b_a\right)-2\,\eta_B\,N_a^2\,\bar v_a^4\right],\nn
& \omega_1=\frac{e\,\bar v_a}{k_a}\,,\quad
\omega_2=\frac{e^2\,u}{2\,k_a^3\,\bar v_a}\left[\eta_B\,N_a\,\bar v_a^2-\eta_O\,G_a\,v_\nu\left(2\,\bar v_a-b_a\right)\right],\nn
& \omega_3=-\frac{e^3\,u^2}{8\,k_a^5\,\bar v_a^3}\left[\eta_O\,G_a^2\,v_\nu^2\left(12\,\bar v_a^2-5\,b_a\,\bar v_a+b_a^2\right)-2\,\eta_B\,\eta_O\,N_a\,G_a\,v_\nu\,b_a\,\bar v_a^2-2\,\eta_B\,N_a^2\,\bar v_a^4\right].
\end{align}
The expansion $\left(1+i\,\omega_a\,\tau\right)^{-1}=1-i\,\omega_a\,\tau-\omega_a^2\,\tau^2+i\,\omega_a^3\,\tau^3+\dots$ gives, for the orders that survive the integral over $u$ (the terms odd in $u$ vanish),
\begin{align}
& \sigma_{\perp,a}^{(2)}=\frac{e^2\,\tau}{8\,\pi^2}\int_{-1}^{1}du\left(1-u^2\right)\left[f_2-\tau^2\,\omega_1^2\,f_0\right],\quad
 \sigma_{H,a}^{{\rm B},(1)}=\frac{e^2\,\tau^2}{8\,\pi^2}\,\omega_1\,
 f_0\int_{-1}^{1}du\left(1-u^2\right),\nn
& \sigma_{H,a}^{{\rm B},(3)}=\frac{e^2\,\tau}{8\,\pi^2}\int_{-1}^{1}du\left(1-u^2\right)\left[\tau\left(\omega_3\,f_0+\omega_2\,f_1+\omega_1\,f_2\right)-\tau^3\,\omega_1^3\,f_0\right].
\end{align}
With $\int_{-1}^{1}du\left(1-u^2\right)=4/3$ and $\int_{-1}^{1}du\left(1-u^2\right)u^2=4/15$, these are the terms of Eqs.~\eqref{eqrtaexp3} and \eqref{eqrtaHexp3}. The zeroth order is $\left(e^2\,\tau/8\,\pi^2\right)\left(4/3\right)f_0$. For the AH part, Eq.~\eqref{eqsigah3} gives $\sigma^{\rm AH}_{H,a}=-\left(\eta_B\,e^2/8\,\pi^2\right)N_a\int_{-1}^{1}du\,u\,k(u)$, and the terms $k^{(1)}$ and $k^{(3)}$ with $\int_{-1}^{1}du\,u^2=2/3$ and $\int_{-1}^{1}du\,u^4=2/5$ give Eq.~\eqref{eqrtaAH3}. The difference of Eq.~\eqref{eqrtaph3} follows by subtracting Eq.~\eqref{eqrtaquad}.

\subsection{Intranode expressions of the QTSM node}
\label{secrtab5}

The sheet $s=1$ has $\mathsf u_a=v_F$, $\mathsf w_a=c\,v_F$, $v_\nu=v_F$, $\mu_a=\mu$, and $\left(N_a,G_a\right)=(2,1)$. With $S=\sqrt{4\,c\,\mu+v_F}$, we have
\begin{align}
\bar v_a=\sqrt{v_F}\,S\,,\quad
k_a=\frac{S-\sqrt{v_F}}{2\,c\,\sqrt{v_F}}\,,\quad
\frac{\bar v_a}{k_a^2}=\frac{4\,c^2\,v_F^{3/2}\,S}{\left(S-\sqrt{v_F}\right)^2}\,.
\end{align}
The Drude term $e^2\,\tau\,k_a^2\,\bar v_a/(6\,\pi^2)$ and the BC term $e^4\,\tau\,B^2\,\eta_B\,8\,N_a^2\,\bar v_a/(120\,\pi^2\,k_a^2)$ then give $\sigma^{\rm dr}_{zz}\big\vert_{s=1}$ and $\sigma^{\rm bc}_{zz}\big\vert_{s=1}$ of Sec.~\ref{secrta3}. The expression $\sigma^{\rm omm}_{zz}\big\vert_{s=1}$ is the sum of the two terms of Eq.~\eqref{eqrtaquad} that contain $\eta_O$, namely
\begin{align}
\frac{e^4\,\tau\,B^2}{120\,\pi^2\,k_a^2}\left[-\,\frac{2\,v_F\left(7\,v_F+2\,\bar v_a\right)}{\bar v_a}+\frac{v_F^2\left(7\,\bar v_a^2+6\,v_F\,\bar v_a-3\,v_F^2\right)}{2\,\bar v_a^3}\right],
\end{align}
and it therefore contains the cross term with the BC.

The sheet $s=0$ has $\mathsf u_a=0$, $\mathsf w_a=c\,v_F$, $v_\nu=v_F$, and $\left(N_a,G_a\right)=(0,2)$. Hence $k_a=\sqrt{\mu/(c\,v_F)}$ and $\bar v_a=2\,c\,v_F\,k_a$, which give $k_a^2\,\bar v_a=2\,\mu^{3/2}/\sqrt{c\,v_F}$. The BC term vanishes, and the Drude and the OMM terms are
\begin{align}
\sigma^{\rm dr}_{zz}\big\vert_{s=0}=\frac{e^2\,\tau\,\mu^{3/2}}{3\,\pi^2\,\sqrt{c\,v_F}}\,,\quad
\sigma^{\rm omm}_{zz}\big\vert_{s=0}=\frac{e^4\,\tau\,B^2\,G_a^2\,v_F^2\,7}{240\,\pi^2\,k_a^2\,\bar v_a}=\frac{7\,e^4\,\tau\,B^2\,v_F^{2}\,\sqrt{c\,v_F}}{120\,\pi^2\,\mu^{3/2}}\,.
\end{align}
These are the expressions of Sec.~\ref{secrta3}, and all of them refer to one node.

\subsection{Internode contribution of the QTSM node}
\label{secrtab6}

The internode scattering of the pair of conjugate QTSM nodes is included with the relaxation time $\tau_{\rm inter}$ following Ref.~\cite{ips-internode}. For two nodes with the same dispersions and without an energy offset, the contribution of the sheet $s$ of one node to the conductivity along $\bB$ is
\begin{align}
\sigma^{\rm inter}_{zz}\big\vert_s=\frac{e^2\left(\tau_{\rm inter}-\tau\right)}{\rho_0}\,\Upsilon_s\sum_{s'}\Upsilon_{s'}\,,\quad
\rho_0=\sum_s\int\frac{d^3k}{(2\,\pi)^3}\,\delta\!\left(\epsilon_s-\mu\right)=\sum_s\frac{k_s^2}{2\,\pi^2\,w^0_s}\,,
\label{eqrtab7}
\end{align}
where $w^0_s=\partial_k\epsilon_s(k_s)$ and
\begin{align}
\Upsilon_s=B\int\frac{d^3k}{(2\,\pi)^3}\left[e\left(\bOm^s\cdot\bs v_s\right)\delta\!\left(\epsilon_s-\mu\right)+m^s_z\,v_{s,z}\,n_F''\!\left(\epsilon_s\right)\right],\quad
\bs v_s=\nabla_{\bk}\epsilon_s\,.
\label{eqrtab8}
\end{align}
The first term of $\Upsilon_s$ is the BC and gives, with $\bOm^s\cdot\bs v_s=-\chi\,N_s\,w^0_s/(2\,k^2)$ and $\int dk\,\delta(\epsilon_s-\mu)\,w^0=1$, the value $-\,\chi\,N_s\,e\,B/(4\,\pi^2)$. In the second term, $m^s_z\,v_{s,z}=-\chi\,e\,v_F\,G_s\cos^2\theta\,w^0/(2\,k)$, and $\int_0^\pi d\theta\,\sin\theta\cos^2\theta=2/3$. Changing the variable from $k$ to $\epsilon_s$ and integrating by parts, we use $\int d\epsilon\,k(\epsilon)\,n_F''(\epsilon)=1/w^0_s$ at $T=0$, which gives $-\,\chi\,e\,B\,v_F\,G_s/(12\,\pi^2\,w^0_s)$. The sum is
\begin{align}
\Upsilon_s=-\,\chi\,e\,B
\left ( \frac{N_s}{4\,\pi^2}+\frac{v_F\,G_s}{12\,\pi^2\,w^0_s}\right ).
\end{align}
For the sheets $s=1$ and $s=0$ of the QTSM node, $\Upsilon_s=-\chi\,e\,B\,a_s/(12\,\pi^2)$ with
\begin{align}
a_1=6+\frac{v_F}{w^0_1}=6+\sqrt{\frac{v_F}{4\,c\,\mu+v_F}}\,,\quad
a_0=\frac{2\,v_F}{w^0_0}=\sqrt{\frac{v_F}{c\,\mu}}\,.
\end{align}
With $\rho_0=\left(2\,\pi^2\right)^{-1}\left(k_1^2/w^0_1+k_0^2/w^0_0\right)$ and the radii and velocities of Appendix~\ref{secrtab5}, one finds $k_1^2/w^0_1+k_0^2/w^0_0=\mathcal D/\left(2\,c^2\,v_F^2\,S\right)$, where $\mathcal D$ is the common denominator of the expressions in Sec.~\ref{secrta3},
\begin{align}
\mathcal D=v_F^{3/2}-v_F\,\sqrt{4\,c\,\mu+v_F} +2\,c\,\mu\,\sqrt{v_F}
+\sqrt{c\,\mu\,v_F}\, \sqrt{4\,c\,\mu+v_F}\,.
\end{align}
Eq.~\eqref{eqrtab7} then reads
\begin{align}
\sigma^{\rm inter}_{zz}\big\vert_s=\frac{e^4\,B^2\,c^2\,v_F^2\, \sqrt{4\,c\,\mu+v_F}}
{36\,\pi^2\,\mathcal D}\,a_s\left(a_1+a_0\right)\left(\tau_{\rm inter}-\tau\right).
\label{eqrtab9}
\end{align}
These lead to the expressions of Sec.~\ref{secrta3}.

\subsection{Internode contribution of the QTSM and Weyl nodes}
\label{secrtab7}

The general internode formula of Ref.~\cite{ips-internode} applies to two nodes with different DOS. Let $\varrho_\nu$ be the DOS of the node $\nu$ at $B=0$, and let $\varrho_G=\left(\varrho_\Gamma+\varrho_R\right)/2$. The intranode collision term of each node relaxes its distribution with the time $\tau$ towards the local chemical potential of the node. The internode term relaxes it with the time $\tau_{\rm inter}$ towards the chemical potential $\mu_G$, which is the mean of the two local chemical potentials. To the order $B^2$ and for $\bE\parallel\bB=B\,\boldsymbol{\hat z}$, the contribution of the node $\nu$ is
\begin{align}
\sigma^{\rm inter}_{\parallel,\nu}=\frac{e^2\left[\tau_{\rm inter}\,\varrho_{\bar\nu}-\tau\,\varrho_G\right]}{\varrho_G\,\varrho_\nu}\,\Upsilon_\nu^2\,,\quad
\Upsilon_\nu=\sum_{a\in\nu}\Upsilon_a\,,
\label{eqrtab10}
\end{align}
where $\Upsilon_a$ is the vertex of Eq.~\eqref{eqrtab8} of the sheet $a$. For $\varrho_\Gamma=\varrho_R$, it reduces to Eq.~\eqref{eqrtab7} summed over the sheets, and the two nodes then contribute equally. The two copies at $R$ have the same dispersion and the same chirality. They are one node of the DOS $\varrho_R=2\,\varrho_{R_i}$ and of the vertex $\Upsilon_R=2\,\Upsilon_{R_i}$.

The vertex of a sheet follows as in Appendix~\ref{secrtab6}, with the BC and the OMM of Eq.~\eqref{eqbcomm} and $m^a_z\,v_{a,z}=-\chi_\nu\,e\,v_\nu\,G_a\cos^2\theta\,\bar v_a/\left(2\,k\right)$. The BC part is $-\chi_\nu\,\eta_B\,N_a\,e\,B/\left(4\,\pi^2\right)$, and the OMM part is $-\chi_\nu\,\eta_O\,e\,B\,v_\nu\,G_a/\left(12\,\pi^2\,\bar v_a\right)$. Hence
\begin{align}
\Upsilon_a=-\chi_\nu\,e\,B\left[\frac{\eta_B\,N_a}{4\,\pi^2}+\frac{\eta_O\,v_\nu\,G_a}{12\,\pi^2\,\bar v_a}\right].
\label{eqrtab11}
\end{align}
The sheets $(\Gamma,1)$ and $(\Gamma,0)$ have $\chi_\Gamma=1$, $\left(N_a,G_a\right)=(2,1)$ and $(0,2)$, $v_\nu=v_F$, and $\bar v_a=w_1$ and $w_0$. The sheets at $R$ have $\chi_{R_i}=-1$, $N_a=G_a=1$, and $v_\nu=\bar v_a=\tilde v_F$, and each of them gives $\Upsilon_{R_i}=e\,B\left(3\,\eta_B+\eta_O\right)/\left(12\,\pi^2\right)$. The sums are Eq.~\eqref{eqrtavert6}. At $B=0$, the weight of a sheet is $\rho_a=k_a^2/\bar v_a$, so that $\varrho_\nu=\bar\rho_\nu/\left(2\,\pi^2\right)$ with $\bar\rho_\Gamma=\rho_1+\rho_0$ and $\bar\rho_R=2\,\rho_R$. Inserting these into Eq.~\eqref{eqrtab10} and writing $\left[\tau_{\rm inter}\,\varrho_{\bar\nu}-\tau\,\varrho_G\right]/\varrho_G=2\,\tau_{\rm inter}\,\varrho_{\bar\nu}/\left(\varrho_\Gamma+\varrho_R\right)-\tau$ gives Eqs.~\eqref{eqrtainter6} and \eqref{eqrtainter6b}. For $c=0$ and $\eta_B=\eta_O=1$, only the sheet $(\Gamma,1)$ is present, and $\mathcal E_\Gamma=7$ and $\mathcal E_R=8$. These are the values of the pseudospin-1 node at $\Gamma$ and the two Weyl nodes at $R$ in Ref.~\cite{ips-internode}.

\subsection{Internode contribution of the RSW and TSM nodes}
\label{secrtab8}

The contribution of each node is Eq.~\eqref{eqrtab10}, with the node $\Gamma$ of the RSW sheets $(\Gamma,3/2)$ and $(\Gamma,1/2)$ and the node $R$ of the two copies $(R_i,+)$, which have the same dispersion and the same chirality. The vertex of a sheet is Eq.~\eqref{eqrtab11} with $\chi_\Gamma=1$, $\chi_{R_i}=-1$, and $v_\nu=v_{3/2}$ or $v_R$. For the RSW sheets, $\bar v_a=v_{3/2}\,S_{\tilde\psi}$ with $\left(N_a,G_a\right)=(3,3/2)$ for $\tilde\psi=3/2$ and $(1,7/2)$ for $\tilde\psi=1/2$. Each copy at $R$ has $\bar v_a=v_R\,S_R$ and $\left(N_a,G_a\right)=(2,1)$. Hence
\begin{align}
\Upsilon_\Gamma=-\frac{e\,B}{12\,\pi^2}\left[3\,\eta_B\left(3+1\right)+\eta_O\left(\frac{3}{2\,S_{3/2}}+\frac{7}{2\,S_{1/2}}\right)\right],\quad
\Upsilon_R=\frac{2\,e\,B}{12\,\pi^2}\left[6\,\eta_B+\frac{\eta_O}{S_R}\right],
\end{align}
which is Eq.~\eqref{eqrtavert7}. The weights at $B=0$ are $\rho_a=k_a^2/\bar v_a$ of Eq.~\eqref{eqfsb02}, and the DOS are $\varrho_\Gamma=\bar\rho_\Gamma/\left(2\,\pi^2\right)$ and $\varrho_R=\bar\rho_R/\left(2\,\pi^2\right)$ with $\bar\rho_\Gamma=\rho_{\Gamma,3/2}+\rho_{\Gamma,1/2}$ and $\bar\rho_R=2\,\rho_R$. The steps from Eq.~\eqref{eqrtab10} to Eq.~\eqref{eqrtainter7} are those of Appendix~\ref{secrtab7}. For $c_{3/2}=c_R=0$ and $\eta_B=\eta_O=1$, we have $S_{3/2}=3/2$, $S_{1/2}=1/2$, and $S_R=1$, which give $\mathcal E_\Gamma=20$ and $\mathcal E_R=14$. The two sheets at $\Gamma$ then have the same vertex $\Upsilon_a=-10\,e\,B/\left(12\,\pi^2\right)$, in agreement with the equal contributions of the bands $s=1/2$ and $s=3/2$ in Ref.~\cite{ips-internode}, and Eq.~\eqref{eqrtainter7} reproduces the expressions of that reference for the two nodes.

\section{Derivation of the analytical results}
\label{secappresall}

\subsection{Ingredients of the second-order coefficients}
\label{secappingB}
\label{secappB}

The expansions are taken in $e\,B$. We denote by $\tilde S_2^{(m)}$ the coefficient of $\left(e\,B\right)^2$, which gives $S_2^{(m)}=e^2\,\tilde S_2^{(m)}$, and Eq.~\eqref{eqsigcoef} becomes $\sigma_\parallel^{(2)}=e^4\,\tilde S_2^{(0)}/\left(4\,\pi^2\right)$ and $\sigma_\perp^{(2)}=e^4\,\Re\tilde S_2^{(1)}/\left(8\,\pi^2\right)$.

Let $k_a$, $w_a$, $\tilde\nu_a$, $\tilde\lambda_a$, $v_a$, $c_a$, $N_a$, $G_a$, and $\chi_\nu$ be those of Sec.~\ref{secresB3} for the sheets of the QTSM and Weyl nodes and of Sec.~\ref{secresB33} for the node $+$ of the two conjugate QTSM nodes, and let $\rho_a=k_a^2/w_a$. The coefficients $\varrho^{(n)}_a$ of the weights, $q^{(n)}_a$ of the source of the sector $m=1$, and $h^{(10)}_a$ and $h^{(21)}_a$ of the source of the sector $m=0$ are defined by
\begin{align}
& \rho_a(\theta;B)=\rho_a+\varrho^{(1)}_a\,e\,B\cos\theta+\varrho^{(2)}_a\left(e\,B\right)^2\cos^2\theta+\mathcal O(B^3)\,,\nn
& \frac{q_a}{D_a}=w_a+q^{(1)}_a\,e\,B\cos\theta+q^{(2)}_a\left(e\,B\right)^2\cos^2\theta+\mathcal O(B^3)\,,\nn
& h_a=w_a\cos\theta+\left(h^{(10)}_a+q^{(1)}_a\cos^2\theta\right)e\,B+\left(h^{(21)}_a\cos\theta+q^{(2)}_a\cos^3\theta\right)\left(e\,B\right)^2+\mathcal O(B^3)\,.
\label{eqingB3}
\end{align}
They follow from inserting Eq.~\eqref{eqkser} into Eq.~\eqref{eqwqd}. The coefficients of $\cos^2\theta$ in the first order and of $\cos^3\theta$ in the second order of the source $h_a$ are the coefficients $q^{(1)}_a$ and $q^{(2)}_a$ of $q_a/D_a$. With $\chi_\nu^2=1$, the coefficients are
\begin{align}
& \varrho^{(1)}_a=-\chi_\nu\,\frac{\eta_O\,G_a\,\tilde\lambda_a+\eta_B\,N_a\,\nu_a^2}{2\,w_a\,\nu_a^2}\,,\quad
\varrho^{(2)}_a=\frac{\eta_O\,G_a\left[3\,\eta_O\,G_a\left(\nu_a-\tilde\lambda_a\right)^2-2\,\eta_B\,N_a\,\nu_a^2\left(2\,\nu_a-\tilde\lambda_a\right)\right]}{8\,v_a\,k_a^2\,\nu_a^5}\,,\nn
& q^{(1)}_a=\chi_\nu\,\frac{v_a\left[\eta_B\,N_a\,\nu_a^2-2\,\eta_O\,G_a\left(\tilde\lambda_a+3\,c_a\,k_a\right)\right]}{2\,k_a^2\,\nu_a}\,,\quad
h^{(10)}_a=\chi_\nu\,\frac{v_a\left(\eta_O\,G_a-\eta_B\,N_a\,\nu_a\right)}{2\,k_a^2}\,,\nn
& q^{(2)}_a=-\,\frac{v_a\left[\eta_O^2\,G_a^2\left(\tilde\lambda_a^2-4\,\tilde\lambda_a\,\nu_a+11\,\nu_a^2\right)+2\,\eta_B\,\eta_O\,G_a\,N_a\,\nu_a^2\left(\nu_a-\tilde\lambda_a\right)-2\,\eta_B^2\,N_a^2\,\nu_a^4\right]}{8\,k_a^4\,\nu_a^3}\,,\nn
& h^{(21)}_a=\frac{v_a\left[2\,\eta_O^2\,G_a^2+\eta_B\,\eta_O\,G_a\,N_a\left(\nu_a-\tilde\lambda_a\right)-\eta_B^2\,N_a^2\,\nu_a^2\right]}{4\,k_a^4\,\nu_a}\,.
\label{eqingB3c}
\end{align}

\subsection{QTSM and Weyl nodes}
\label{secappres}

\subsubsection{Zero-field reduction of the kinetic equation}
\label{secappzero}

For $B=0$, we have $\xi_s=\epsilon_s$, $D_a=1$, and $\bw_a=\partial_k\epsilon_s\,\hat{\bk}$. The FS radii and the weights are those of Eq.~\eqrf{eqfsb0}, and they are independent of the direction. The source for $\bE=E\,\boldsymbol{\hat z}$ is $h_a=w_a\cos\theta$, with $w_a=\partial_k\epsilon_s$ at $k_a$. The LFO vanishes, and Eq.~\eqrf{eqboltzop} reads
\begin{align}
\frac{\Lambda_a(\hat{\bk})}{\tau_a}-\sum_{a'}\int d^2\hat{\bk}'\,\rho_{a'}\,\mathcal W_{aa'}(\hat{\bk},\hat{\bk}')\,\Lambda_{a'}(\hat{\bk}')=w_a\,\hat k_z\,.
\end{align}
All the overlaps depend on $C=\hat{\bk}\cdot\hat{\bk}'$, since $\hat{\bs n}\cdot\hat{\bs n}'=C$ for two sheets of the same chirality, and since the overlaps between $\Gamma$ and $R$ are taken as in Eq.~\eqrf{eqovgriso}. We expand $\mathcal O_{aa'}=\sum_l o^{aa'}_l\,P_l(C)$ in the Legendre polynomials. By the addition theorem and the orthogonality of the spherical harmonics,
\begin{align}
\int d^2\hat{\bk}'\,P_l(C)\,Y_{l'm}(\hat{\bk}')=\frac{4\,\pi}{2\,l+1}\,\delta_{ll'}\,Y_{l'm}(\hat{\bk})\,.
\label{eqappproj}
\end{align}
The overlaps of Eq.~\eqrf{eqoverlap} are polynomials in $C$. With $C^2=\left(1+2\,P_2(C)\right)/3$, we obtain
\begin{align}
 \frac14\left(1+C\right)^2=\frac13+\frac12\,P_1+\frac16\,P_2\,,\quad
C^2=\frac13+\frac23\,P_2\,,\quad
\frac12\left(1-C^2\right)=\frac13-\frac13\,P_2\,,\quad
 \frac12\left(1+C\right)=\frac12+\frac12\,P_1\,.
\end{align}
The coefficients $o_0$ and $o_1$ that enter below are collected in Table~\ref{tabappo}. Eq.~\eqrf{eqovgriso} gives $o_0=\bar o_1$ and $o_1=\bar p_1$ for $(\Gamma,1)(R_i,+)$, and $o_0=\bar o_0$ and $o_1=0$ for $(\Gamma,0)(R_i,+)$.

\begin{table}[t!]
\centering
\small
\renewcommand{\arraystretch}{1.6}
\begin{tabular}{|c|c|c|c|c|c|c|}
\hline
Pair & $(\Gamma,1)(\Gamma,1)$ & $(\Gamma,1)(\Gamma,0)$ & $(\Gamma,0)(\Gamma,0)$ & $R_iR_i$ & $(\Gamma,1)R_i$ & $(\Gamma,0)R_i$ \\
\hline
$o_0$ & $1/3$ & $1/3$ & $1/3$ & $1/2$ & $\bar o_1$ & $\bar o_0$ \\
\hline
$o_1$ & $1/2$ & $0$ & $0$ & $1/2$ & $\bar p_1$ & $0$ \\
\hline
\end{tabular}
\caption{\label{tabappo} The Legendre coefficients $o_0$ and $o_1$ of the squared overlaps.}
\end{table}

By Eq.~\eqrf{eqappproj}, the contraction of the kernel $\mathcal W_{aa'}=\left(\beta_{aa'}^2/\pi\right)\mathcal O_{aa'}$ with $\Lambda=1$ gives $4\,\beta_{aa'}^2\,o_0^{aa'}$ times $\rho_{a'}$, and with $\Lambda=\hat k_z$ it gives $\left(4/3\right)\beta_{aa'}^2\,o_1^{aa'}\,\rho_{a'}\,\hat k_z$. The sum over the final sheets contains both copies at $R$ for the sheets at $\Gamma$, and the sheets $(\Gamma,1)$, $(\Gamma,0)$ and both copies at $R$ for a sheet at $R$, whose amplitudes combine into $\gamma_R$. The rates are therefore
\begin{align}
 \tau_1^{-1}=4\left[\frac{\beta_{11}^2\,\rho_1}{3}+\frac{\beta_{10}^2\,\rho_0}{3}+2\,\bar o_1\,g_1^2\,\rho_R\right],\quad
\tau_0^{-1}=4\left[\frac{\beta_{10}^2\,\rho_1}{3}+\frac{\beta_{00}^2\,\rho_0}{3}+2\,\bar o_0\,g_0^2\,\rho_R\right],\quad
 \tau_R^{-1}=4\left[\bar o_1\,g_1^2\,\rho_1+\bar o_0\,g_0^2\,\rho_0+\frac{\gamma_R\,\rho_R}{2}\right],
\end{align}
which is Eq.~\eqrf{eqtauB0}. The source $\propto\hat k_z$ is closed under the in-scattering, and the ansatz $\Lambda_a=\mathcal A_a\,\hat k_z$ gives
\begin{align}
 \left(\tau_1^{-1}-\frac23\,\beta_{11}^2\,\rho_1\right)\mathcal A_1-\frac83\,\bar p_1\,g_1^2\,\rho_R\,\mathcal A_R=w_1\,,\quad
 -\frac43\,\bar p_1\,g_1^2\,\rho_1\,\mathcal A_1+\left(\tau_R^{-1}-
 \frac23\,\gamma_R\,\rho_R\right)\mathcal A_R=w_R\,,\quad
\tau_0^{-1}\,\mathcal A_0=w_0\,.
\label{eqappA}
\end{align}
The sheet $(\Gamma,0)$ has $o_1=0$ for every partner, and its equation is therefore decoupled. The diagonal entries are $N_{\Gamma\Gamma}=\tau_1^{-1}-\left(2/3\right)\beta_{11}^2\,\rho_1$ and $N_{RR}=\tau_R^{-1}-\left(2/3\right)\gamma_R\,\rho_R$, which are the entries of Eq.~\eqrf{eqNmat}. The factor $2$ in $N_{\Gamma R}$ is due to the two copies at $R$.

\subsubsection{Solution and conductivity}
\label{secappsol}

The system of Eq.~\eqrf{eqappA} is $N_{\Gamma\Gamma}\,\mathcal A_1-N_{\Gamma R}\,\mathcal A_R=w_1$ and $-N_{R\Gamma}\,\mathcal A_1+N_{RR}\,\mathcal A_R=w_R$. Cramer's rule with $\mathcal N=N_{\Gamma\Gamma}\,N_{RR}-N_{\Gamma R}\,N_{R\Gamma}$ gives Eq.~\eqrf{eqAamp}. The conductivity is obtained from Eq.~\eqrf{eqsigmw}, in which the AH term vanishes at $B=0$. We have
\begin{align}
\sigma_{zz}=\frac{e^2}{8\,\pi^3}\sum_a\int d^2\hat{\bk}\,\rho_a\,w_a\,\hat k_z\,\mathcal A_a\,\hat k_z=\frac{e^2}{8\,\pi^3}\,\frac{4\,\pi}{3}\sum_a\rho_a\,w_a\,\mathcal A_a\,.
\end{align}
With $\rho_a\,w_a=k_a^2$, and with the two copies at $R$ counted separately, this is Eq.~\eqrf{eqsig0}. The sum rule of Eq.~\eqrf{eqsolv} holds trivially, because the source has no $l=0$ component.

\subsubsection{Isotropy of the zero-field solution}
\label{secappiso}

For $B=0$ and $\omega_a=0$, the sector $m=1$ of Eq.~\eqrf{eqell} is $\tau_a^{-1}\,\ell_a=w_a\sin\theta+\sum_{a'}\int d\theta'\,\sin\theta'\,\rho_{a'}\,\beta_{aa'}^2\,\mathcal O^{(1)}_{aa'}\,\ell_{a'}$. With $\ell_a=\mathcal A_a\sin\theta$ and the harmonics of Eqs.~\eqrf{eqoverlapm} and \eqrf{eqovgriso}, the integrals follow from $\int_0^\pi d\theta'\,\sin^3\theta'=4/3$ and from the vanishing of the integrals of the terms that are odd in $\cos\theta'$. For the pairs $(\Gamma,1)(\Gamma,1)$ and $R_iR_i$, $\mathcal O^{(1)}=\frac12\,Y\left(1+X\right)$ and $\frac12\,Y$ give $\frac23\,\beta^2\,\rho'\,\mathcal A'\sin\theta$ each. For $(\Gamma,1)(\Gamma,0)$ and $(\Gamma,0)(\Gamma,0)$, $\mathcal O^{(1)}=-X\,Y$ and $2\,X\,Y$ are odd in $\cos\theta'$, and they give zero. For $(\Gamma,1)R_i$, $\mathcal O^{(1)}=\bar p_1\,Y$ gives $\frac43\,\bar p_1\,\beta^2\,\rho'\,\mathcal A'\sin\theta$, and the pair $(\Gamma,0)R_i$ gives zero. These values coincide with $\left(4/3\right)\beta^2\,\rho'\,o_1\,\mathcal A'$ of Table~\ref{tabappo}, and the rates are those of the sector $m=0$. The amplitudes $\mathcal A_a$ of Eq.~\eqref{eqappA} therefore solve the sector $m=1$, and they are real. Eq.~\eqrf{eqsigperp} gives $\sigma_H^{\rm B}=0$ and
\begin{align}
\sigma_\perp=\frac{e^2}{8\,\pi^2}\sum_a\rho_a\,w_a\,\mathcal A_a\int_0^\pi d\theta\,\sin^3\theta=\frac{e^2}{6\,\pi^2}\sum_a k_a^2\,\mathcal A_a=\sigma_0\,.
\end{align}
The AH term vanishes at $B=0$, and $\sigma_\parallel=\sigma_0$ follows from Appendix~\ref{secappsol}.

\subsubsection{Hall conductivity to the leading order in the magnetic field}
\label{secappHall}

The sector $m=1$ is
\begin{align}
\left[\tau_a^{-1}(\theta)+i\,\omega_a(\theta)\right]\ell_a=\frac{q_a}{D_a}\sin\theta+\sum_{a'}\int_0^\pi d\theta'\,\sin\theta'\,\rho_{a'}\,\beta_{aa'}^2\,\mathcal O^{(1)}_{aa'}\,\ell_{a'}\,.
\end{align}
The quantities $\tau_a^{-1}$, $\rho_a$, $q_a$, $D_a$, and the kernel are real and analytic in $B$. Only $\omega_a$ is imaginary, and it equals $e\,B\,w_a/k_a$ at the leading order, since $q_a\to w_a$, $D_a\to1$, and $k_a(\theta)\to k_a$. Writing $\ell_a=\ell_a^{\rm R}+i\,\ell_a^{\rm I}$, the imaginary part of the equation at the first order in $B$ is
\begin{align}
\tau_a^{-1}\,\ell_a^{\rm I}-\sum_{a'}\int_0^\pi d\theta'\,\sin\theta'\,\rho_{a'}\,\beta_{aa'}^2\,\mathcal O^{(1)}_{aa'}\,\ell_{a'}^{\rm I}=-\,\omega_a\,\mathcal A_a\sin\theta\,,
\end{align}
where the zeroth-order solution $\ell_a^{\rm R}=\mathcal A_a\sin\theta$ has been used. The right-hand side is proportional to $\sin\theta$. Hence $\ell^{\rm I}_a=\mathcal A'_a\sin\theta$, where the amplitudes obey the system of Eq.~\eqrf{eqappA} with the sources $w_a$ replaced by $-\,e\,B\,\mathcal Y_a$ and $\mathcal Y_a=w_a\,\mathcal A_a/k_a$. The solution is $\mathcal A'_a=-\,e\,B\,\mathcal Z_a$, with the $\mathcal Z_a$ defined above Eq.~\eqref{eqsigHB}. In Eq.~\eqrf{eqsigperp}, the imaginary part of $\ell_a/\sin\theta$ is $\mathcal A_a'$ at this order, and $\rho_a\,q_a/D_a\to k_a^2$. This gives
\begin{align}
\sigma_H^{\rm B}=-\frac{e^2}{8\,\pi^2}\cdot\frac43\sum_a k_a^2\,\mathcal A'_a=\frac{e^3\,B}{6\,\pi^2}\sum_a k_a^2\,\mathcal Z_a\,,
\end{align}
which is Eq.~\eqrf{eqsigHB}, where the sum counts the two copies at $R$ separately.

\subsubsection{Anomalous-Hall conductivity to the leading order in the magnetic field}
\label{secappAH}

For axisymmetric sheets, the second term of Eq.~\eqrf{eqsigmw} has only the component $\sigma_{xy}$. With $\bOm^a$ of Eq.~\eqrf{eqbcomm}, $\int_0^{k_a}dk\,k^2\,\Omega^a_z=-\chi_\nu\,N_a\,\hat k_z\,k_a(\theta)/2$, and $\sigma_H\equiv\sigma_{yx}=-\sigma_{xy}$ gives
\begin{align}
\sigma_H^{\rm AH}=-\frac{\eta_B\,e^2}{8\,\pi^2}\sum_a\chi_\nu\,N_a\int_{-1}^{1}du\,u\,k_a(u)\,,
\end{align}
as in Eq.~\eqrf{eqsigah2}. To the first order in $B$, the radius is $k_a(u)=k_a+\delta k_a(u)$ with $\partial_k\epsilon_s\,\delta k_a=-\,\eta_O\,\chi_\nu\,e\,v_\nu\,G_a\,B\,u/(2\,k_a)$, which follows from $\xi_s=\mu$ in Eq.~\eqrf{eqxiw}. The term $k_a$ does not contribute, because $\int_{-1}^1du\,u=0$. Using $\int_{-1}^1du\,u^2=2/3$, we obtain
\begin{align}
\sigma_H^{\rm AH}\simeq\frac{\eta_B\,\eta_O\,e^3\,B}{24\,\pi^2}\sum_a\frac{N_a\,G_a\,v_\nu}{k_a\,w_a}\,,
\end{align}
where the sum is over the four sheets and $\chi_\nu^2=1$ has been used. The numbers of Eq.~\eqrf{eqbcomm} give $N_a\,G_a\,v_\nu/(k_a\,w_a)=2/\left[k_1\left(1+2\,c\,k_1\right)\right]$ for $(\Gamma,1)$, zero for $(\Gamma,0)$, and $\tilde v_F/(\mu+\Delta)$ for each copy at $R$, which is Eq.~\eqrf{eqsigHAH}. For the two conjugate QTSM nodes, the sum over the two nodes doubles the term of $(\Gamma,1)$, which is the result of Eq.~\eqref{eqsigah3}.

\subsubsection{Decoupled sets of sheets}
\label{secappdec}

For $g_0=g_1=0$, we have $N_{\Gamma R}=N_{R\Gamma}=0$, and $\tau_1^{-1}$ and $\tau_0^{-1}$ lose the terms with $g_s$. Eq.~\eqrf{eqAamp} gives $\mathcal A_1=w_1/N_{\Gamma\Gamma}$ with $N_{\Gamma\Gamma}=\left(2/3\right)\beta_{11}^2\,\rho_1+\left(4/3\right)\beta_{10}^2\,\rho_0$, $\mathcal A_0=w_0\,\tau_0$ with $\tau_0^{-1}=\left(4/3\right)\left(\beta_{10}^2\,\rho_1+\beta_{00}^2\,\rho_0\right)$, and $\mathcal A_R=w_R/N_{RR}$ with $N_{RR}=\left(4/3\right)\gamma_R\,\rho_R$. Inserting $k_R^2=\rho_R\,\tilde v_F$ and $w_R=\tilde v_F$ in the term of the Weyl nodes of Eq.~\eqrf{eqsig0}, we obtain
\begin{align}
\sigma_R=\frac{e^2}{6\,\pi^2}\cdot2\,k_R^2\,\frac{3\,\tilde v_F}{4\,\gamma_R\,\rho_R}=\frac{e^2\,\tilde v_F^2}{4\,\pi^2\,\gamma_R}\,.
\end{align}

\subsubsection{The limit of vanishing quadratic correction}
\label{secappc0}

For $c\to0$, $k_1$ and $\rho_1$ tend to $\mu/v_F$ and $\mu^2/v_F^3$, and $\rho_R$ stays finite, whereas $k_0^2=\mu/(c\,v_F)$, $w_0=2\sqrt{c\,v_F\,\mu}$, and $\rho_0=\sqrt{\mu/(c\,v_F)}/(2\,c\,v_F)$ scale as $c^{-1}$, $c^{1/2}$, and $c^{-3/2}$. We assume $\beta_{10}\neq0$, $\beta_{00}\neq0$, and $\bar o_0\,g_0\neq0$. The entries of Eq.~\eqrf{eqNmat} then satisfy $N_{\Gamma\Gamma}=\frac43\,\beta_{10}^2\,\rho_0\left[1+\mathcal O(c^{3/2})\right]$ and $N_{RR}=4\,\bar o_0\,g_0^2\,\rho_0\left[1+\mathcal O(c^{3/2})\right]$, while $N_{\Gamma R}$ and $N_{R\Gamma}$ are finite. Hence $\mathcal N=N_{\Gamma\Gamma}\,N_{RR}\left[1+\mathcal O(c^{3})\right]$, $\mathcal A_1=\mathcal O(c^{3/2})$, and $\mathcal A_R=\mathcal O(c^{3/2})$. The terms of $(\Gamma,1)$ and $R$ in Eq.~\eqrf{eqsig0} are therefore $\mathcal O(c^{3/2})$. For the sheet $(\Gamma,0)$, $\tau_0^{-1}=\frac43\,\beta_{00}^2\,\rho_0\left[1+\mathcal O(c^{3/2})\right]$, and
\begin{align}
k_0^2\,\mathcal A_0=\frac{\mu}{c\,v_F}\cdot2\sqrt{c\,v_F\,\mu}\cdot\frac{3}{4\,\beta_{00}^2}\cdot\frac{2\,c\,v_F}{\sqrt{\mu/(c\,v_F)}}\left[1+\mathcal O(c^{3/2})\right]=\frac{3\,c\,v_F\,\mu}{\beta_{00}^2}\left[1+\mathcal O(c^{3/2})\right].
\end{align}
Inserting this in Eq.~\eqrf{eqsig0} gives Eq.~\eqrf{eqsigc0}.

\subsubsection{Kernels and rates of the second-order coefficients}
\label{secappB3}

The sheets are $a=1$, $a=0$, and $a=R$, which stand for $(\Gamma,1)$, $(\Gamma,0)$, and each of the two copies at $R$. The copies have the same weights and amplitudes, and the first-order equation is the same for both. We therefore have $\Lambda_{1,R_1}=\Lambda_{1,R_2}$, and the sums over the sheets in Eq.~\eqref{eqforms} count the sheet $R$ twice. We write $n_1=n_0=1$ and $n_R=2$ for these multiplicities.

Let $\bar{\mathcal K}^{(m)}_{aa'}$ be the kernel of Eq.~\eqref{eqsector} for the scattering into one representative sheet of $a$ from all the copies of $a'$. With the intranode overlaps of Eq.~\eqref{eqoverlapm}, the internode overlaps of Eq.~\eqref{eqovgriso}, and the sums over the copies at $R$, the kernels of the sector $m=0$ are
\begin{align}
& \bar{\mathcal K}^{(0)}_{11}=\frac{\beta_{11}^2}{4}
\left ( 3+4\cos\theta\,\cos\theta'-\cos^2\theta-\cos^2\theta'
+3\cos^2\theta\,\cos^2\theta'\right ),\quad
%%%%%%%%%%
 \bar{\mathcal K}^{(0)}_{10}=\bar{\mathcal K}^{(0)}_{01}=\frac{\beta_{10}^2}{2}\left (1+\cos^2\theta+\cos^2\theta'-3\cos^2\theta\,\cos^2\theta'\right ),
%%%%%%%%%%%%%%%%%% 
\nn & \bar{\mathcal K}^{(0)}_{00}=\beta_{00}^2\left( 1-\cos^2\theta-\cos^2\theta'+3\cos^2\theta\,\cos^2\theta'\right ),\quad
%%%%%%%%%%%%%%
\bar{\mathcal K}^{(0)}_{1R}=4\,g_1^2\left(\bar o_1+\bar p_1\cos\theta\,\cos\theta'\right),\quad
\bar{\mathcal K}^{(0)}_{R1}=2\,g_1^2\left(\bar o_1+\bar p_1\cos\theta\,\cos\theta'\right),
%%%%%%%%%%%%%%%%%%
\nn & \bar{\mathcal K}^{(0)}_{0R}=4\,g_0^2\,\bar o_0\,,\quad
 %%%%%%%%%%%%%%%%%%%%%%
\bar{\mathcal K}^{(0)}_{R0}=2\,g_0^2\,\bar o_0\,,\quad
\bar{\mathcal K}^{(0)}_{RR}=\gamma_R\left(1+\cos\theta\,\cos\theta'\right).
\label{eqkerm0}
\end{align}
We denote by $\kappa^{(0)}_{aa',pq}$ the coefficient of $\cos^p\theta\,\cos^q\theta'$ in $\bar{\mathcal K}^{(0)}_{aa'}$. The kernels of the sector $m=1$ have the form $\bar{\mathcal K}^{(1)}_{aa'}=\sin\theta\,\sin\theta'\left(\kappa^{(1)}_{aa',00}+\kappa^{(1)}_{aa',11}\cos\theta\,\cos\theta'\right)$, with
%%%%%%%%%%%%%%%%%%
\begin{align}
& \left(\kappa^{(1)}_{11,00},\kappa^{(1)}_{11,11}\right)=\frac{\beta_{11}^2}{2}\left(1,1\right),\quad
\left(\kappa^{(1)}_{10,00},\kappa^{(1)}_{10,11}\right)=\left(0,-\beta_{10}^2\right),\quad
\left(\kappa^{(1)}_{00,00},\kappa^{(1)}_{00,11}\right)=\left(0,2\,\beta_{00}^2\right),\nn
& \left(\kappa^{(1)}_{1R,00},\kappa^{(1)}_{1R,11}\right)=\left(2\,g_1^2\,\bar p_1,0\right),\quad
\left(\kappa^{(1)}_{R1,00},\kappa^{(1)}_{R1,11}\right)=\left(g_1^2\,\bar p_1,0\right),\quad
\left(\kappa^{(1)}_{RR,00},\kappa^{(1)}_{RR,11}\right)=\left(\frac{\gamma_R}{2},0\right),
\label{eqkerm1}
\end{align}
%%%%%%%%%%%%%%%%%%%%%
and all the elements for the pairs $(0,R)$ and $(R,0)$ vanish. The kernels of the pair $(0,1)$ are equal to those of the pair $(1,0)$. The rate $\tau_a^{-1}$ is that of Eq.~\eqref{eqtauB0}, and its coefficients of the first and the second order, which follow from the weights of Eq.~\eqref{eqingB3} in Eq.~\eqref{eqtau}, are $\tau_a^{-1,(1)}=t^{(1)}_a\cos\theta$ and $\tau_a^{-1,(2)}=t^{(20)}_a+t^{(22)}_a\cos^2\theta$ with
\begin{align}
& t^{(1)}_a=\frac23\sum_{a'}\kappa^{(0)}_{aa',11}\,\varrho^{(1)}_{a'}\,,\quad
 t^{(20)}_a=\sum_{a'}\varrho^{(2)}_{a'}\left(\frac23\,\kappa^{(0)}_{aa',00}+\frac25\,\kappa^{(0)}_{aa',02}\right),\quad
t^{(22)}_a=\sum_{a'}\varrho^{(2)}_{a'}\left(\frac23\,\kappa^{(0)}_{aa',20}+\frac25\,\kappa^{(0)}_{aa',22}\right).
\label{eqtcoef}
\end{align}

\subsubsection{Longitudinal second-order coefficient}

Let $M^{(i)}_a(r)$ be the moments $\int_0^\pi d\theta\,\sin\theta\,\rho^{(i)}_a(\theta)\cos^r\theta$ of the coefficients of the weights, with $\rho^{(0)}_a=\rho_a$. They are $M^{(0)}_a(r)=2\,\rho_a/\left(r+1\right)$ for even $r$, $M^{(1)}_a(r)=2\,\varrho^{(1)}_a/\left(r+2\right)$ for odd $r$, and $M^{(2)}_a(r)=2\,\varrho^{(2)}_a/\left(r+3\right)$ for even $r$, and they vanish otherwise. For $m=0$, the solution at $B=0$ is $\Lambda_{0,a}=\mathcal A_a\cos\theta$ with the amplitudes of Eq.~\eqref{eqAamp}. The first-order equation below Eq.~\eqref{eqlam1} reads $\tau_a^{-1}\Lambda_{1,a}-\sum_{a'}\int_0^\pi d\theta'\,\sin\theta'\,\rho_{a'}\,\bar{\mathcal K}^{(0)}_{aa'}\,\Lambda_{1,a'}=s_a$, where $s_a$ collects the first-order terms of the weights, the rates, and the source acting on $\Lambda_0$. It is the even polynomial $s_a=s^{(0)}_a+s^{(2)}_a\cos^2\theta$ with
\begin{align}
& s^{(0)}_a=h^{(10)}_a+\sum_{a'}\varrho^{(1)}_{a'}\,\mathcal A_{a'}\left(\frac23\,\kappa^{(0)}_{aa',00}+\frac25\,\kappa^{(0)}_{aa',02}\right),\quad
 s^{(2)}_a=q^{(1)}_a-t^{(1)}_a\,\mathcal A_a+\sum_{a'}\varrho^{(1)}_{a'}\,\mathcal A_{a'}\left(\frac23\,\kappa^{(0)}_{aa',20}+\frac25\,\kappa^{(0)}_{aa',22}\right).
\label{eqsm0}
\end{align}
The in-scattering term lies in the span of the even powers of $\cos\theta$ up to the second one [cf. Eq.~\eqref{eqspans}]. The function $\Lambda_{1,a}=c_{a,0}+c_{a,2}\cos^2\theta$ therefore has six coefficients, which obey, for $p=0,2$,
\begin{align}
\tau_a^{-1}\,c_{a,p}=s^{(p)}_a+\sum_{a'}\rho_{a'}\sum_{q=0,2}\kappa^{(0)}_{aa',pq}\left(\frac{2\,c_{a',0}}{q+1}+\frac{2\,c_{a',2}}{q+3}\right).
\label{eqcm0}
\end{align}
The matrix of this system has the null vector $c_{a,0}=\text{const}$ and $c_{a,2}=0$. It does not matter for $\tilde S_2^{(0)}$, since $$\sum_a n_a\,\rho_a\left(2\,s^{(0)}_a+2\,s^{(2)}_a/3\right)=0 $$ by Eq.~\eqref{eqsolv}. We obtain
\begin{align}
\tilde S_2^{(0)}=\mathfrak l^{(0)}+\sum_a n_a\,\rho_a\left[2\,c_{a,0}\,s^{(0)}_a+\frac23\left(c_{a,0}\,s^{(2)}_a+c_{a,2}\,s^{(0)}_a\right)+\frac25\,c_{a,2}\,s^{(2)}_a\right],
\label{eqS2m0}
\end{align}
with the local term
\begin{align}
 \mathfrak l^{(0)} & =\sum_a n_a \Big\{2\,\mathcal A_a\Big[\rho_a\left(\frac23\,h^{(21)}_a+\frac25\,q^{(2)}_a\right)+\varrho^{(1)}_a\left(\frac23\,h^{(10)}_a+\frac25\,q^{(1)}_a\right)+\frac25\,w_a\,\varrho^{(2)}_a\Big]\nn
& \qquad \quad -\mathcal A_a^2\Big[\rho_a\left(\frac23\,t^{(20)}_a+\frac25\,t^{(22)}_a\right)+\frac25\left(\varrho^{(1)}_a\,t^{(1)}_a+\varrho^{(2)}_a\,\tau_a^{-1}\right)\Big]\Big\}\nn
& \quad +\sum_{a,a'}n_a\,\mathcal A_a\,\mathcal A_{a'}\sum_{p,q=0}^{2}\kappa^{(0)}_{aa',pq}\sum_{i=0}^{2}M^{(i)}_a(p+1)\,M^{(2-i)}_{a'}(q+1)\,.
\label{eqL0}
\end{align}

\subsubsection{Transverse second-order coefficient}

For $m=1$, the solution at $B=0$ is $\Lambda_{0,a}=\mathcal A_a\sin\theta$ with the same amplitudes, and we write $\Lambda_1=\Lambda_1^{\rm R}+i\,\Lambda_1^{\rm I}$. The imaginary part obeys the equation of the sector $m=1$ with the source $-\,\omega_a^{(1)}\Lambda_{0,a}$, where $\omega_a^{(1)}=w_a/k_a$, and its solution is $\Lambda^{\rm I}_{1,a}=-\mathcal Z_a\sin\theta$ with the amplitudes $\mathcal Z_a$ of Sec.~\ref{secresph}. Its contribution to $\tilde S_2^{(1)}$ is $-\left(4/3\right)\sum_a n_a\,k_a\,\mathcal Z_a\,\mathcal A_a$. The real part is odd in $\cos\theta$ and has the form $\Lambda^{\rm R}_{1,a}=\check\phi_a\sin\theta\,\cos\theta$. The kernels of Eq.~\eqref{eqkerm1} act on this form through $\kappa^{(1)}_{aa',11}$ only, and the sheet $R$ is therefore decoupled from the sheets at $\Gamma$. With the coefficients
\begin{align}
b_a=q^{(1)}_a-t^{(1)}_a\,\mathcal A_a+\frac{4}{15}\sum_{a'}\kappa^{(1)}_{aa',11}\,\varrho^{(1)}_{a'}\,\mathcal A_{a'}\,,
\label{eqbm1}
\end{align}
the three coefficients $\check\phi_a$ obey
\begin{align}
\tau_a^{-1}\,\check\phi_a=b_a+\frac{4}{15}\sum_{a'}\kappa^{(1)}_{aa',11}\,\rho_{a'}\,\check\phi_{a'}\,.
\label{eqphi}
\end{align}
For $a=R$, the sum vanishes, and $\check\phi_R=\tau_R\,b_R$. For the sheets at $\Gamma$, Eq.~\eqref{eqphi} is a $2\times2$ system. The real part of $\tilde S_2^{(1)}$ is
\begin{align}
\Re\tilde S_2^{(1)}=\mathfrak l^{(1)}+\frac{4}{15}\sum_a n_a\,\rho_a\,\check\phi_a\,b_a-\frac43\sum_a n_a\,k_a\,\mathcal Z_a\,\mathcal A_a\,,
\label{eqS2m1}
\end{align}
with the local term
\begin{align}
& \mathfrak l^{(1)}=\sum_a n_a\Big\{\frac{8}{15}\,\mathcal A_a\left(\rho_a\,q^{(2)}_a+\varrho^{(1)}_a\,q^{(1)}_a+w_a\,\varrho^{(2)}_a\right)
-\mathcal A_a^2\Big[\rho_a\left(\frac43\,t^{(20)}_a+\frac{4}{15}\,t^{(22)}_a\right)+\frac{4}{15}\left(\varrho^{(1)}_a\,t^{(1)}_a+\varrho^{(2)}_a\,\tau_a^{-1}\right)\Big]\Big\}\nn
& \qquad+\sum_{a,a'}n_a\,\mathcal A_a\,\mathcal A_{a'}\left[\frac{16}{45}\,\kappa^{(1)}_{aa',00}\left(\rho_a\,\varrho^{(2)}_{a'}+\varrho^{(2)}_a\,\rho_{a'}\right)+\frac{16}{225}\,\kappa^{(1)}_{aa',11}\,\varrho^{(1)}_a\,\varrho^{(1)}_{a'}\right].
\label{eqL1}
\end{align}
The imaginary part of $\tilde S_2^{(1)}$ vanishes.

%%%%%%%%%%%%%%%%%%%%%
\subsection{Two untilted QTSM nodes}
\label{secappres3}

For $\bB=B\,\boldsymbol{\hat z}$, the two nodes contribute equally, and we use the notation of Eq.~\eqref{eqxiamp3}.
\subsubsection{Zero-field reduction of the two sectors}
\label{secappzero3}

For $B=0$, the weights $\rho_s$ and the velocities $w^0_s$ do not depend on $u$, and $q_s/D_s=w^0_s$. The moments of the weight are $\rho^{(0)}_s=2\,\rho_s$, $\rho^{(1)}_s=0$, and $\rho^{(2)}_s=2\,\rho_s/3$. Inserting them into Eq.~\eqref{eqkappa} with $X=\rho$ and $\gamma=\gamma^+$, we obtain
\begin{align}
 \mathfrak K_{1,0}=\frac43\left(\gamma^+_{11}\,\rho_1+\gamma^+_{10}\,\rho_0\right),\quad
\mathfrak K_{0,0}=\frac43\left(\gamma^+_{10}\,\rho_1+\gamma^+_{00}\,\rho_0\right),\quad
 \mathfrak K_{s,1}=\mathfrak K_{s,2}=0\,.
\end{align}
The coefficients $\mathfrak K_{s,0}$ are the rates $\tau_1^{-1}$ and $\tau_0^{-1}$ of Eq.~\eqref{eqlampar}, which are constants.

For $\bE\parallel\bB$, we insert $\Lambda_s=\mathcal A_s\,u$ into Eq.~\eqref{eqlampar}. The moments $M^{(0)}_s$ and $M^{(2)}_s$ vanish by parity, and $M^{(1)}_s=2\,\rho_s\,\mathcal A_s/3$. Hence $\mathfrak K_{1,1}(M,\gamma^-)=2\,\gamma^-_{11}\,\rho_1\,\mathcal A_1/3$ and $\mathfrak K_{0,1}=0$, and the equation $\mathcal A_s\,\tau_s^{-1}=w^0_s+\mathfrak K_{s,1}$ gives $\Xi^{\parallel}_s\,\mathcal A_s=w^0_s$, which defines $\mathcal A^{\parallel}_s$. Eq.~\eqref{eqsigpar3} then yields $\sigma_\parallel\vert_{B=0}=\left(e^2/3\,\pi^2\right)\sum_s k_s^2\,\mathcal A^{\parallel}_s$, with $\rho_s\,w^0_s=k_s^2$ from Eq.~\eqref{eqfsb0}.

For $\bE\perp\bB$, the integrals needed are $\int_{-1}^{1}du\left(1-u^2\right)=4/3$, $\int_{-1}^{1}du\left(1-u^2\right)u^2=4/15$, and $\int_{-1}^{1}du\left(1-u^2\right)u=0$. With $\mathcal G_s=\tau_s$ at $B=0$, the angular brackets of Eq.~\eqref{eqmomsysperp} are $\langle 1\rangle^{\perp}_s=4\,\rho_s\,\tau_s/3$, $\langle u\rangle^{\perp}_s=0$, and $\langle u^2\rangle^{\perp}_s=4\,\rho_s\,\tau_s/15$. The sources are $\langle q_s/D_s\rangle^{\perp}_s=4\,\rho_s\,\tau_s\,w^0_s/3$ and $\langle u\,q_s/D_s\rangle^{\perp}_s=0$. The system of Eq.~\eqref{eqmomsysperp} then splits into a first row, which determines $\mathcal L^{(0)}_1$, and a homogeneous $2\times2$ block for $\mathcal L^{(1)}_1$ and $\mathcal L^{(1)}_0$. The block has a nonvanishing determinant, since the operator of Eq.~\eqref{eqboltzop} is invertible in this sector, and therefore $\mathcal L^{(1)}_1=\mathcal L^{(1)}_0=0$. The first row gives
\begin{align}
\mathcal L^{(0)}_1=\frac{4\,\rho_1\,\tau_1\,w^0_1/3}{1-\tfrac23\,\gamma^+_{11}\,\rho_1\,\tau_1}=\frac43\,\rho_1\,\mathcal A^{\perp}_1\,.
\end{align}
Eq.~\eqref{eqlamperp} then gives $\check\lambda^{(0)}_1=2\,\gamma^+_{11}\,\rho_1\,\mathcal A^{\perp}_1/3$ and $\check\lambda^{(0)}_0=\check\lambda^{(1)}_s=0$, and the mean free-path is $\ell_s=\mathcal A^{\perp}_s\sin\theta$, with $\Xi^{\perp}_s\,\mathcal A^{\perp}_s=w^0_s$. Eq.~\eqref{eqsigperp3} becomes
\begin{align}
\sigma_\perp\big\vert_{B=0}=\frac{e^2}{4\,\pi^2}\sum_s\frac43\,\rho_s\,\tau_s\,w^0_s\left(w^0_s+\check\lambda^{(0)}_s\right)=\frac{e^2}{3\,\pi^2}\sum_s k_s^2\,\mathcal A^{\perp}_s\,,
\end{align}
where $w^0_s+\check\lambda^{(0)}_s=\mathcal A^{\perp}_s\,\tau_s^{-1}$ has been used. The Hall conductivity vanishes, since $\ell_s$ is real. Subtracting the two conductivities, we obtain
\begin{align}
\left(\sigma_\perp-\sigma_\parallel\right)\big\vert_{B=0}=\frac{e^2}{3\,\pi^2}\,k_1^2\,w^0_1\left(\frac{1}{\Xi^{\perp}_1}-\frac{1}{\Xi^{\parallel}_1}\right)=\frac{e^2}{3\,\pi^2}\,k_1^2\,w^0_1\,\frac{\Xi^{\parallel}_1-\Xi^{\perp}_1}{\Xi^{\parallel}_1\,\Xi^{\perp}_1}\,,
\end{align}
since the sheet $s=0$ has the same amplitude in both sectors. With $\Xi^{\parallel}_1-\Xi^{\perp}_1=\tfrac23\left(\gamma^+_{11}-\gamma^-_{11}\right)\rho_1=\tfrac43\left(\beta^{\rm inter}_{11}\right)^2\rho_1$ and $\rho_1\,w^0_1=k_1^2$, this is Eq.~\eqref{eqdsig0}.

\subsubsection{Lorentz-force-only limit of the sector $m=1$}
\label{secapplfo}

For $\eta_B=\eta_O=0$, we have $\xi_s=\epsilon_s$ and $D_a=1$. The FS radii $k_s$, the weights $\rho_s$, the rates $\tau_s^{-1}$ of Eq.~\eqref{eqlampar}, and $q_s=w^0_s$ are independent of $u$ and of $B$. The cyclotron frequency of Eq.~\eqref{eqresalg} is $\omega_s=e\,B\,w^0_s/k_s$. The sector $m=0$ does not contain $B$, since the LFO annihilates it, and $\sigma_\parallel$ equals its zero-field value.

In the sector $m=1$, we make the ansatz $\ell_s=c_s\sin\theta$ with complex constants $c_s$ that are the same for both nodes, in accordance with the even solution of Appendix~\ref{secappsym}. The moments are $\mathcal L^{(0)}_s=4\,\rho_s\,c_s/3$ and $\mathcal L^{(1)}_s=0$, and the coefficients of Eq.~\eqref{eqlamperp} are $\check\lambda^{(0)}_1=2\,\gamma^+_{11}\,\rho_1\,c_1/3$, $\check\lambda^{(0)}_0=0$, and $\check\lambda^{(1)}_s=0$. Eq.~\eqref{eqell} then reduces to
\begin{align}
\left(\tau_s^{-1}+i\,\omega_s\right)c_s=w^0_s+\delta_{s,1}\,\frac23\,\gamma^+_{11}\,\rho_1\,c_1\,,\quad
c_s=\frac{w^0_s}{\Xi^{\perp}_s+i\,\omega_s}\,.
\end{align}
The solution is unique, since $\Xi^{\perp}_1=\tfrac23\,\gamma^+_{11}\,\rho_1+\tfrac43\,\gamma^+_{10}\,\rho_0$ and $\Xi^{\perp}_0=\tau_0^{-1}$ are positive. Eq.~\eqref{eqsigperp}, summed over the two nodes, gives $\sigma_\perp-i\,\sigma_H^{\rm B}=\left(e^2/4\,\pi^2\right)\sum_s\int_{-1}^{1}du\left(1-u^2\right)\rho_s\,w^0_s\,c_s$. With $\int_{-1}^{1}du\left(1-u^2\right)=4/3$ and $\rho_s\,w^0_s=k_s^2$, this is Eq.~\eqref{eqlfoexact}. The AH term vanishes for $\eta_B=0$. The real and imaginary parts are
\begin{align}
\sigma_\perp=\frac{e^2}{3\,\pi^2}\sum_s\frac{k_s^2\,w^0_s\,\Xi^{\perp}_s}{\left(\Xi^{\perp}_s\right)^2+\omega_s^2}\,,\quad
\sigma_H=\frac{e^2}{3\,\pi^2}\sum_s\frac{k_s^2\,w^0_s\,\omega_s}{\left(\Xi^{\perp}_s\right)^2+\omega_s^2}\,.
\end{align}
For $\omega_s\ll\Xi^{\perp}_s$, the expansion $\left(\Xi+i\,\omega\right)^{-1}=\Xi^{-1}-i\,\omega\,\Xi^{-2}-\omega^2\,\Xi^{-3}+\dots$ and the relation $k_s^2\,w^0_s\,\omega_s^2=e^2\,B^2\left(w^0_s\right)^3$ give Eq.~\eqref{eqlfosmall}, where $w^0_s/\Xi^{\perp}_s=\mathcal A^{\perp}_s$ has been used. For $\omega_s\gg\Xi^{\perp}_s$, we have $\sigma_H\to\left(e^2/3\,\pi^2\right)\sum_s k_s^2\,w^0_s/\omega_s=e\sum_s k_s^3/\left(3\,\pi^2\,B\right)$, and $\sigma_\perp\to\left(e^2/3\,\pi^2\right)\sum_s k_s^2\,w^0_s\,\Xi^{\perp}_s/\omega_s^2\propto B^{-2}$.

\subsubsection{Hall conductivity to the leading order in the magnetic field}
\label{secappHall3}

For general $\eta_B$ and $\eta_O$, the sector $m=1$ is
\begin{align}
\left[\tau_s^{-1}(u)+i\,\omega_s(u)\right]\ell_s=\frac{q_s}{D_s}\sin\theta+\text{in-scattering term of Eq.~\eqref{eqell}}\,.
\end{align}
The argument of Appendix~\ref{secappHall} applies. The quantities $\tau_s^{-1}$, $\rho_s$, $q_s$, $D_s$, and the coefficients $\check\lambda^{(n)}_s$ are real and analytic in $B$, and only $\omega_s$ is imaginary. We write $\ell_s=\ell^{\rm R}_s+i\,\ell^{\rm I}_s$. At $B=0$, the real part is $\ell^{\rm R}_s=\mathcal A^{\perp}_s\sin\theta$, and the imaginary part vanishes. The imaginary part of the equation reads
\begin{align}
\tau_s^{-1}\,\ell^{\rm I}_s-\left(\text{in-scattering term of }\ell^{\rm I}\right)=-\,\omega_s\,\ell^{\rm R}_s\,,
\end{align}
The right-hand side is of the first order in $B$, and so is $\ell^{\rm I}_s$. The kernel, the rates, and the weights on the left-hand side are therefore those of $B=0$, since the first-order corrections of $\rho_s$, $q_s$, $D_s$, and $\tau_s^{-1}$ are odd in $u$ and enter $\ell^{\rm I}_s$ only at the second order. At the leading order, $\omega_s=e\,B\,w^0_s/k_s$ is independent of $u$. The equation is then that of Appendix~\ref{secappzero3} with the source $w^0_s$ replaced by $-\,e\,B\,w^0_s\,\mathcal A^{\perp}_s/k_s$, and its solution is
\begin{align}
\ell^{\rm I}_s=-\,\frac{e\,B\,w^0_s\,\mathcal A^{\perp}_s}{k_s\,\Xi^{\perp}_s}\sin\theta=-\,\frac{e\,B}{k_s}\left(\mathcal A^{\perp}_s\right)^2\sin\theta\,.
\end{align}
In Eq.~\eqref{eqsigperp}, summed over the two nodes, we have $\mathrm{Im}\!\left(\ell_s/\sin\theta\right)=-\,e\,B\left(\mathcal A^{\perp}_s\right)^2/k_s$ and $\rho_s\,q_s/D_s\to k_s^2$. Hence
\begin{align}
\sigma_H^{\rm B}=-\,\frac{e^2}{4\,\pi^2}\cdot\frac43\sum_s k_s^2\left(-\,\frac{e\,B}{k_s}\left(\mathcal A^{\perp}_s\right)^2\right)=\frac{e^3\,B}{3\,\pi^2}\sum_s k_s\left(\mathcal A^{\perp}_s\right)^2\,,
\end{align}
which is Eq.~\eqref{eqsigHB3}.

For the AH part, Eq.~\eqref{eqsigah3} involves only the sheet $(+,1)$. To the first order in $B$, the radius is $k_{(+,1)}(u)=k_1+\delta k$ with $w^0_1\,\delta k=-\,\eta_O\,e\,v_F\,B\,u/\left(2\,k_1\right)$, which follows from $\xi_s=\mu$ in Eq.~\eqref{eqxiw}. The term $k_1$ does not contribute, since $\int_{-1}^{1}du\,u=0$. Using $\int_{-1}^{1}du\,u^2=2/3$, we obtain
\begin{align}
\sigma_H^{\rm AH}\simeq\frac{\eta_B\,\eta_O\,e^3\,v_F\,B}{6\,\pi^2\,k_1\,w^0_1}=\frac{\eta_B\,\eta_O\,e^3\,B}{6\,\pi^2\,k_1\left(1+2\,c\,k_1\right)}\,.
\end{align}
The sum of the two parts is Eq.~\eqref{eqsigH3}.

%%%%%%%%%%%%%%%%%
\subsubsection{Kernels and rates of the second-order coefficients}
\label{secappB33}

The sheets are $s=1$ and $s=0$ of the node $+$, and those of the node $-$ follow from the exchange symmetry of Sec.~\ref{secresB33}. The sums over the four sheets in Eq.~\eqref{eqforms} are therefore twice the sums over $s$, and we write $n_s=2$.

The sum over the two nodes in Eq.~\eqref{eqlampar} reduces the kernels of Eq.~\eqref{eqoverlap3} to those of the node $+$. The rates involve $\gamma^+_{ss'}$, the in-scattering of the sector $m=0$ involves $\gamma^-_{ss'}$, and both involve $\gamma^+_{ss'}$ for $m=1$. For $\gamma=\gamma^{\pm}$, let $\bar{\mathcal K}^{(0),\pm}_{ss'}$ be the kernels of the sector $m=0$,
\begin{align}
& \bar{\mathcal K}^{(0),\pm}_{11}=\frac{\gamma^{\pm}_{11}}{4}(\left[3+4\cos\theta\,\cos\theta'-\cos^2\theta-\cos^2\theta'+3\cos^2\theta\,\cos^2\theta'\right),\nn
%%%%%%%%%%%%%%%%%
& \bar{\mathcal K}^{(0),\pm}_{10}=\bar{\mathcal K}^{(0),\pm}_{01}
=\frac{\gamma^{\pm}_{10}}{2}\left(1+\cos^2\theta+\cos^2\theta'
-3\cos^2\theta\,\cos^2\theta'\right),\quad
%%%%%%%%
\bar{\mathcal K}^{(0),\pm}_{00}=
\gamma^{\pm}_{00}\left(1-\cos^2\theta-\cos^2\theta'+3\cos^2\theta\,\cos^2\theta'\right).
\label{eqkermiv0}
\end{align}
We denote by $\kappa^{(0),\pm}_{ss',pq}$ the coefficient of $\cos^p\theta\,\cos^q\theta'$ in $\bar{\mathcal K}^{(0),\pm}_{ss'}$. The kernels of the sector $m=1$ have the form $\bar{\mathcal K}^{(1)}_{ss'}=\sin\theta\,\sin\theta'\left(\kappa^{(1)}_{ss',00}+\kappa^{(1)}_{ss',11}\cos\theta\,\cos\theta'\right)$, with
\begin{align}
\left(\kappa^{(1)}_{11,00},\kappa^{(1)}_{11,11}\right)=\frac{\gamma^+_{11}}{2}\left(1,1\right),\quad
\left(\kappa^{(1)}_{10,00},\kappa^{(1)}_{10,11}\right)=\left(0,-\gamma^+_{10}\right),\quad
\left(\kappa^{(1)}_{00,00},\kappa^{(1)}_{00,11}\right)=\left(0,2\,\gamma^+_{00}\right),
\label{eqkermiv1}
\end{align}
and the kernels of the pair $(0,1)$ equal those of the pair $(1,0)$. The rates $\tau_s^{-1}$ are those below Eq.~\eqref{eqmomsys3}. Their coefficients of the first and the second order, which follow from the weights of Eq.~\eqref{eqingB3} in Eq.~\eqref{eqtau}, are $\tau_s^{-1,(1)}=t^{(1)}_s\cos\theta$ and $\tau_s^{-1,(2)}=t^{(20)}_s+t^{(22)}_s\cos^2\theta$ with
\begin{align}
 t^{(1)}_s=\frac23\sum_{s'}\kappa^{(0),+}_{ss',11}\,\varrho^{(1)}_{s'}\,,\quad 
t^{(20)}_s=\sum_{s'}\varrho^{(2)}_{s'}\left(\frac23\,\kappa^{(0),+}_{ss',00}+\frac25\,\kappa^{(0),+}_{ss',02}\right),\quad
t^{(22)}_s=\sum_{s'}\varrho^{(2)}_{s'}\left(\frac23\,\kappa^{(0),+}_{ss',20}+\frac25\,\kappa^{(0),+}_{ss',22}\right).
\label{eqtcoefiv}
\end{align}

%%%%%%%%%%%%%%%%%%%%%%%%%%%%%%%%%%%%%%%%%
\subsubsection{Longitudinal second-order coefficient}

Let $M^{(i)}_s(r)$ be the moments $\int_0^\pi d\theta\,\sin\theta\,\rho^{(i)}_s(\theta)\cos^r\theta$ of the coefficients of the weights, with $\rho^{(0)}_s=\rho_s$. They are $M^{(0)}_s(r)=2\,\rho_s/\left(r+1\right)$ for even $r$, $M^{(1)}_s(r)=2\,\varrho^{(1)}_s/\left(r+2\right)$ for odd $r$, and $M^{(2)}_s(r)=2\,\varrho^{(2)}_s/\left(r+3\right)$ for even $r$, and they vanish otherwise. For $m=0$, the solution at $B=0$ is $\Lambda_{0,s}=\mathcal A^{\parallel}_s\cos\theta$ with the amplitudes of Eq.~\eqref{eqxiamp3}. The first-order equation below Eq.~\eqref{eqlam1} reads $\tau_s^{-1}\Lambda_{1,s}-\sum_{s'}\int_0^\pi d\theta'\,\sin\theta'\,\rho_{s'}\,\bar{\mathcal K}^{(0),-}_{ss'}\,\Lambda_{1,s'}=s_s$, where $s_s$ collects the first-order terms of the weights, the rates, and the source acting on $\Lambda_0$. It is the even polynomial $s_s=s^{(0)}_s+s^{(2)}_s\cos^2\theta$ with
\begin{align}
& s^{(0)}_s=h^{(10)}_s+\sum_{s'}\varrho^{(1)}_{s'}\,\mathcal A^{\parallel}_{s'}\left(\frac23\,\kappa^{(0),-}_{ss',00}+\frac25\,\kappa^{(0),-}_{ss',02}\right),\quad
 s^{(2)}_s=q^{(1)}_s-t^{(1)}_s\,\mathcal A^{\parallel}_s
 +\sum_{s'}\varrho^{(1)}_{s'}\,\mathcal A^{\parallel}_{s'}\left(\frac23\,\kappa^{(0),-}_{ss',20}+\frac25\,\kappa^{(0),-}_{ss',22}\right).
\label{eqsm0iv}
\end{align}
The in-scattering term lies in the span of the even powers of $\cos\theta$ up to the second one. The function $\Lambda_{1,s}=c_{s,0}+c_{s,2}\cos^2\theta$ therefore has four coefficients, which obey, for $p=0,2$,
\begin{align}
\tau_s^{-1}\,c_{s,p}=s^{(p)}_s+\sum_{s'}\rho_{s'}\sum_{q=0,2}\kappa^{(0),-}_{ss',pq}\left(\frac{2\,c_{s',0}}{q+1}+\frac{2\,c_{s',2}}{q+3}\right).
\label{eqcm0iv}
\end{align}
The matrix of this system is regular, since the null vector of the full problem has the opposite symmetry [cf. Sec.~\ref{secresB33}]. We obtain
\begin{align}
\tilde S_2^{(0)}=\mathfrak l^{(0)}+\sum_s n_s\,\rho_s\left[2\,c_{s,0}\,s^{(0)}_s+\frac23\left(c_{s,0}\,s^{(2)}_s+c_{s,2}\,s^{(0)}_s\right)+\frac25\,c_{s,2}\,s^{(2)}_s\right],
\label{eqS2m0iv}
\end{align}
with the local term
\begin{align}
 \mathfrak l^{(0)} & =\sum_s n_s\Big\{2\,\mathcal A^{\parallel}_s\Big[\rho_s\left(\frac23\,h^{(21)}_s+\frac25\,q^{(2)}_s\right)+\varrho^{(1)}_s\left(\frac23\,h^{(10)}_s+\frac25\,q^{(1)}_s\right)+\frac25\,w_s\,\varrho^{(2)}_s\Big]\nn
& \qquad \quad -\left(\mathcal A^{\parallel}_s\right)^2\Big[\rho_s\left(\frac23\,t^{(20)}_s+\frac25\,t^{(22)}_s\right)+\frac25\left(\varrho^{(1)}_s\,t^{(1)}_s+\varrho^{(2)}_s\,\tau_s^{-1}\right)\Big]\Big\}\nn
& \qquad+\sum_{s,s'}n_s\,\mathcal A^{\parallel}_s\,\mathcal A^{\parallel}_{s'}\sum_{p,q=0}^{2}\kappa^{(0),-}_{ss',pq}\sum_{i=0}^{2}M^{(i)}_s(p+1)\,M^{(2-i)}_{s'}(q+1)\,.
\label{eqL0iv}
\end{align}

\subsubsection{Transverse second-order coefficient}

For $m=1$, the solution at $B=0$ is $\Lambda_{0,s}=\mathcal A^{\perp}_s\sin\theta$, and we write $\Lambda_1=\Lambda_1^{\rm R}+i\,\Lambda_1^{\rm I}$. The imaginary part obeys the equation of the sector $m=1$ with the source $-\,\omega_s^{(1)}\Lambda_{0,s}$, where $\omega_s^{(1)}=w_s/k_s$. The sheet $s=1$ has the vertex correction of $\Xi^{\perp}_1$ and the sheet $s=0$ has none, and the solution is therefore $\Lambda^{\rm I}_{1,s}=-\left(\mathcal A^{\perp}_s\right)^2\sin\theta/k_s$. Its contribution to $\tilde S_2^{(1)}$ is $-\left(4/3\right)\sum_s n_s\left(\mathcal A^{\perp}_s\right)^3$. The real part is odd in $\cos\theta$ and has the form $\Lambda^{\rm R}_{1,s}=\phi_s\sin\theta\,\cos\theta$. The kernels of Eq.~\eqref{eqkermiv1} act on this form through $\kappa^{(1)}_{ss',11}$ only. With the coefficients
\begin{align}
b_s=q^{(1)}_s-t^{(1)}_s\,\mathcal A^{\perp}_s+\frac{4}{15}\sum_{s'}\kappa^{(1)}_{ss',11}\,\varrho^{(1)}_{s'}\,\mathcal A^{\perp}_{s'}\,,
\label{eqbm1iv}
\end{align}
the two coefficients $\phi_s$ obey the $2\times2$ system
\begin{align}
\tau_s^{-1}\,\phi_s=b_s+\frac{4}{15}\sum_{s'}\kappa^{(1)}_{ss',11}\,\rho_{s'}\,\phi_{s'}\,.
\label{eqphiiv}
\end{align}
The real part of $\tilde S_2^{(1)}$ is
\begin{align}
\Re\tilde S_2^{(1)}=\mathfrak l^{(1)}+\frac{4}{15}\sum_s n_s\,\rho_s\,\phi_s\,b_s-\frac43\sum_s n_s\left(\mathcal A^{\perp}_s\right)^3,
\label{eqS2m1iv}
\end{align}
with the local term
\begin{align}
& \mathfrak l^{(1)}=\sum_s n_s\Big\{\frac{8}{15}\,\mathcal A^{\perp}_s\left(\rho_s\,q^{(2)}_s+\varrho^{(1)}_s\,q^{(1)}_s+w_s\,\varrho^{(2)}_s\right)\nn
& \qquad-\left(\mathcal A^{\perp}_s\right)^2\Big[\rho_s\left(\frac43\,t^{(20)}_s+\frac{4}{15}\,t^{(22)}_s\right)+\frac{4}{15}\left(\varrho^{(1)}_s\,t^{(1)}_s+\varrho^{(2)}_s\,\tau_s^{-1}\right)\Big]\Big\}\nn
& \qquad+\sum_{s,s'}n_s\,\mathcal A^{\perp}_s\,\mathcal A^{\perp}_{s'}\left[\frac{16}{45}\,\kappa^{(1)}_{ss',00}\left(\rho_s\,\varrho^{(2)}_{s'}+\varrho^{(2)}_s\,\rho_{s'}\right)+\frac{16}{225}\,\kappa^{(1)}_{ss',11}\,\varrho^{(1)}_s\,\varrho^{(1)}_{s'}\right].
\label{eqL1iv}
\end{align}
The imaginary part of $\tilde S_2^{(1)}$ vanishes.

\subsection{Two tilted QTSM nodes}
\label{secappres4}

The two nodes contribute equally, and we set $v_F=1$. We write $\tilde\nu_s=\sqrt{s^2+4\,\bar\mu}$, $\rho_s=k_s^2/\tilde\nu_s$, and $\tau_s$ for the rates at $t=0$. The amplitudes $\mathcal A_s$ are $\mathcal A^\parallel_s$ for the longitudinal and $\mathcal A^\perp_s$ for the transverse conductivity.

\subsubsection{Expansion of the weights, the rates, and the sources at $B=0$}
\label{secappexp4}

At $B=0$, the radius of the sheet $(+,s)$ obeys $c\,k^2+\left(s+\tilde\upsilon\right)k=\mu$ with $\tilde\upsilon=t\,u$, and $\rho_{+,s}(u)=k^2/\sqrt{\left(s+\tilde\upsilon\right)^2+4\,\bar\mu}$ is a function of $\tilde\upsilon$ alone. Since $\mathrm d k/\mathrm d \tilde\upsilon=-k/\tilde\nu_s$ at $\tilde\upsilon=0$, a Taylor expansion gives
\begin{align}
\rho_{+,s}(u)=\rho_s\left[1+\mathfrak p_s\,t\,u+\mathfrak q_s\,t^2\,u^2\right]+\mathcal O\!\left(t^3\right),\quad
\mathfrak p_s=-\frac{2\,\tilde\nu_s+s}{\tilde\nu_s^2}\,,\quad
\mathfrak q_s=\frac{3\left(\tilde\nu_s+s\right)^2}{2\,\tilde\nu_s^4}\,.
\label{eqappw4}
\end{align}
The sources of the two sectors are $h_s=\tilde\nu_s\,u+t\left[1-\mathfrak d_s\,u^2\right]+\mathfrak e_s\,t^2\,u^3$ and $q_s(u)=\tilde\nu_s-\mathfrak d_s\,t\,u+\mathfrak e_s\,t^2\,u^2$, with
\begin{align}
\mathfrak d_s=1-\frac{s}{\tilde\nu_s}\,,\quad
\mathfrak e_s=\frac{\tilde\nu_s^2-s^2}{2\,\tilde\nu_s^3}\,.
\label{eqappsrc4}
\end{align}
The moments of the weight are $\rho^{(0)}_s=2\,\rho_s\left(1+\mathfrak q_s\,t^2/3\right)$, $\rho^{(1)}_s=2\,\rho_s\,\mathfrak p_s\,t/3$, and $\rho^{(2)}_s=2\,\rho_s\left(1+3\,\mathfrak q_s\,t^2/5\right)/3$. Inserting them into Eq.~\eqref{eqtau40} gives the rates
\begin{align}
\tau_1^{-1}(u)&=\tau_1^{-1}+\mathfrak a\,t\,u+\left(\mathfrak b_1+\mathfrak c_1\,u^2\right)t^2\,,\quad
\tau_0^{-1}(u)=\tau_0^{-1}+\left(\mathfrak b_0+\mathfrak c_0\,u^2\right)t^2\,,\quad
\mathfrak a =\frac{2}{3}\,\gamma^+_{11}\,\rho_1\,\mathfrak p_1\,,\quad
%%%%%%%%%%%%%%%%%
\mathfrak b_1 =\frac{2}{5}\,\gamma^+_{11}\,\rho_1\,\mathfrak q_1+\frac{8}{15}\,\gamma^+_{10}\,\rho_0\,\mathfrak q_0\,,\nn
%%%%%%%%%%%%%%%%%%%%%%%%%%
\mathfrak c_1 &= \frac{2}{15}\,\gamma^+_{11}\,\rho_1\,\mathfrak q_1-\frac{4}{15}\,\gamma^+_{10}\,\rho_0\,\mathfrak q_0\,,\quad
\mathfrak b_0 =\frac{8}{15}\,\gamma^+_{10}\,\rho_1\,\mathfrak q_1+\frac{4}{15}\,\gamma^+_{00}\,\rho_0\,\mathfrak q_0\,,\quad
\mathfrak c_0=-\frac{4}{15}\,\gamma^+_{10}\,\rho_1\,\mathfrak q_1+\frac{8}{15}\,\gamma^+_{00}\,\rho_0\,\mathfrak q_0\,.
\label{eqapptau4}
\end{align}
The times follow from $\tau_s(u)=\tau_s\left[1-\tau_s\,\delta\tau^{-1}_s+\left(\tau_s\,\delta\tau^{-1}_s\right)^2\right]$, where $\delta\tau^{-1}_s$ is the sum of the terms of the orders $t$ and $t^2$ in Eq.~\eqref{eqapptau4}. The moment $\rho^{(1)}_1$ is the only source of the term linear in $u$.

\subsubsection{Longitudinal conductivity at the order $t^2$}
\label{secappll4}

For $t=0$, the solution is $\Lambda_s=\mathcal A_s\,u$ with $M^{(1)}_1=2\,\rho_1\,\mathcal A_1/3$. We expand the moments as $M^{(n)}_s=t\,m^{(n)}_s$ for the even moments and $M^{(1)}_1=2\,\rho_1\,\mathcal A_1/3+t^2\,m^{(1)}_1$. Inserting Eqs.~\eqref{eqappw4}--\eqref{eqapptau4} into Eq.~\eqref{eqmomsys4}, the terms of the order $t$ give a linear system for the four even moments,
\begin{align}
&\left(1-\tfrac{4}{3}\,\rho_1\,\tau_1\,\gamma^-_{11}\right)m^{(0)}_1-\tfrac{4}{3}\,\rho_1\,\tau_1\,\gamma^-_{10}\,m^{(0)}_0=\mathcal S^{(0)}_1\,,\nonumber\\
&-\tfrac{2}{5}\,\rho_1\,\tau_1\,\gamma^-_{11}\,m^{(0)}_1+\left(1-\tfrac{2}{15}\,\rho_1\,\tau_1\,\gamma^-_{11}\right)m^{(2)}_1-\tfrac{8}{15}\,\rho_1\,\tau_1\,\gamma^-_{10}\,m^{(0)}_0+\tfrac{4}{15}\,\rho_1\,\tau_1\,\gamma^-_{10}\,m^{(2)}_0=\mathcal S^{(2)}_1\,,\nonumber\\
&-\tfrac{4}{3}\,\rho_0\,\tau_0\,\gamma^-_{10}\,m^{(0)}_1+\left(1-\tfrac{4}{3}\,\rho_0\,\tau_0\,\gamma^-_{00}\right)m^{(0)}_0=\mathcal S^{(0)}_0\,,\nonumber\\
&-\tfrac{8}{15}\,\rho_0\,\tau_0\,\gamma^-_{10}\,m^{(0)}_1+\tfrac{4}{15}\,\rho_0\,\tau_0\,\gamma^-_{10}\,m^{(2)}_1-\tfrac{4}{15}\,\rho_0\,\tau_0\,\gamma^-_{00}\,m^{(0)}_0+\left(1-\tfrac{8}{15}\,\rho_0\,\tau_0\,\gamma^-_{00}\right)m^{(2)}_0=\mathcal S^{(2)}_0\,,
\label{eqappm4}
\end{align}
with the sources
\begin{align}
\mathcal S^{(0)}_1&=2\,\rho_1\,\tau_1-\tfrac{2}{3}\,\rho_1\,\tau_1\,\mathfrak d_1+\tfrac{2}{3}\,\rho_1\,\mathcal A_1\,\mathfrak p_1-\tfrac{4}{9}\,\rho_1^2\,\tau_1\,\gamma^+_{11}\,\mathcal A_1\,\mathfrak p_1\,,\nonumber\\
\mathcal S^{(2)}_1&=\tfrac{2}{3}\,\rho_1\,\tau_1-\tfrac{2}{5}\,\rho_1\,\tau_1\,\mathfrak d_1+\tfrac{2}{5}\,\rho_1\,\mathcal A_1\,\mathfrak p_1-\tfrac{4}{15}\,\rho_1^2\,\tau_1\,\gamma^+_{11}\,\mathcal A_1\,\mathfrak p_1\,,\nonumber\\
\mathcal S^{(0)}_0&=2\,\rho_0\,\tau_0-\tfrac{2}{3}\,\rho_0\,\tau_0\,\mathfrak d_0+\tfrac{2}{3}\,\rho_0\,\mathcal A_0\,\mathfrak p_0\,,\quad
\mathcal S^{(2)}_0=\tfrac{2}{3}\,\rho_0\,\tau_0-\tfrac{2}{5}\,\rho_0\,\tau_0\,\mathfrak d_0+\tfrac{2}{5}\,\rho_0\,\mathcal A_0\,\mathfrak p_0\,.
\end{align}
The terms of the order $t^2$ of the equation for $M^{(1)}_1$ fix $m^{(1)}_1$ in terms of these moments. Inserting the result into Eq.~\eqref{eqsigpar4}, the terms of the order $t$ cancel, and
\begin{align}
\sigma_\parallel\big\vert_{B=0}=\frac{e^2}{3\,\pi^2}\sum_s k_s^2\,\mathcal A_s+\frac{e^2}{2\,\pi^2}\,\mathfrak T_\parallel\,t^2\,,
\label{eqappsigpar4}
\end{align}
where $\mathfrak T_\parallel=\sum_{n,s}C^{(n)}_s\,m^{(n)}_s+\mathfrak T^{\rm loc}_\parallel$ with the coefficients
\begin{align}
C^{(0)}_1&=\tfrac{2}{15}\Big[4\,\rho_0\,\mathcal A_0\,\gamma^-_{10}\,\mathfrak p_0-2\,\rho_1^2\,\tau_1\,\mathcal A_1\,\gamma^-_{11}\,\gamma^+_{11}\,\mathfrak p_1+3\,\rho_1\,\mathcal A_1\,\gamma^-_{11}\,\mathfrak p_1
-4\,\rho_0\,\tau_0\,\gamma^-_{10}\,\mathfrak d_0+10\,\rho_0\,\tau_0\,\gamma^-_{10}-3\,\rho_1\,\tau_1\,\gamma^-_{11}\,\mathfrak d_1+10\,\rho_1\,\tau_1\,\gamma^-_{11}\Big],\nonumber\\
%%%%%%%%%%%%%%%%
C^{(2)}_1&=-\tfrac{2}{45}\Big[6\,\rho_0\,\mathcal A_0\,\gamma^-_{10}\,\mathfrak p_0+2\,\rho_1^2\,\tau_1\,\mathcal A_1\,\gamma^-_{11}\,\gamma^+_{11}\,\mathfrak p_1-3\,\rho_1\,\mathcal A_1\,\gamma^-_{11}\,\mathfrak p_1-6\,\rho_0\,\tau_0\,\gamma^-_{10}\,\mathfrak d_0+3\,\rho_1\,\tau_1\,\gamma^-_{11}\,\mathfrak d_1\Big],\nonumber\\
%%%%%%%%%%%%%%%%%%%%%%%
C^{(0)}_0 &=\tfrac{4}{45}\Big[3\,\rho_0\,\mathcal A_0\,\gamma^-_{00}\,\mathfrak p_0-4\,\rho_1^2\,\tau_1\,\mathcal A_1\,\gamma^-_{10}\,\gamma^+_{11}\,\mathfrak p_1+6\,\rho_1\,\mathcal A_1\,\gamma^-_{10}\,\mathfrak p_1-3\,\rho_0\,\tau_0\,\gamma^-_{00}\,\mathfrak d_0+15\,\rho_0\,\tau_0\,\gamma^-_{00}-6\,\rho_1\,\tau_1\,\gamma^-_{10}\,\mathfrak d_1+15\,\rho_1\,\tau_1\,\gamma^-_{10}\Big],\nonumber\\
%%%%%%%%%%%%%%%%%%%%%
C^{(2)}_0&=\tfrac{4}{45}\Big[6\,\rho_0\,\mathcal A_0\,\gamma^-_{00}\,\mathfrak p_0+2\,\rho_1^2\,\tau_1\,\mathcal A_1\,\gamma^-_{10}\,\gamma^+_{11}\,\mathfrak p_1-3\,\rho_1\,\mathcal A_1\,\gamma^-_{10}\,\mathfrak p_1-6\,\rho_0\,\tau_0\,\gamma^-_{00}\,\mathfrak d_0+3\,\rho_1\,\tau_1\,\gamma^-_{10}\,\mathfrak d_1\Big].
\end{align}
The local part is
\begin{align}
\mathfrak T^{\rm loc}_\parallel&=
\tfrac{4}{5}\left(\rho_0\,\mathcal A_0\,\mathfrak e_0+\rho_1\,\mathcal A_1\,\mathfrak e_1\right)-\tfrac{2\,\rho_0\,\mathfrak q_0}{225}\left(44\,\rho_0\,\mathcal A_0^2\,\gamma^+_{00}-\tfrac{45\,\mathcal A_0^2}{\tau_0}+28\,\rho_1\,\mathcal A_1^2\,\gamma^+_{10}\right)
-\tfrac{2\,\rho_1\,\mathfrak q_1}{225}\left(28\,\rho_0\,\mathcal A_0^2\,\gamma^+_{10}+36\,\rho_1\,\mathcal A_1^2\,\gamma^+_{11}-\tfrac{45\,\mathcal A_1^2}{\tau_1}\right)\nonumber\\
%%%%%%%%%%%%
&\quad-\tfrac{2}{45}\Big[18\,\rho_0\,\mathcal A_0\,\mathfrak d_0\,\mathfrak p_0-30\,\rho_0\,\mathcal A_0\,\mathfrak p_0-4\,\rho_1^3\,\tau_1\,\mathcal A_1^2\left(\gamma^+_{11}\right)^2\mathfrak p_1^2+6\,\rho_1^2\,\mathcal A_1^2\,\gamma^+_{11}\,\mathfrak p_1^2
-12\,\rho_1^2\,\tau_1\,\mathcal A_1\,\gamma^+_{11}\,\mathfrak d_1\,\mathfrak p_1+20\,\rho_1^2\,\tau_1\,\mathcal A_1\,\gamma^+_{11}\,\mathfrak p_1
\nn & \quad \quad \quad
+18\,\rho_1\,\mathcal A_1\,\mathfrak d_1\,\mathfrak p_1-30\,\rho_1\,\mathcal A_1\,\mathfrak p_1-9\,\rho_0\,\tau_0\,\mathfrak d_0^2+30\,\rho_0\,\tau_0\,\mathfrak d_0-45\,\rho_0\,\tau_0-9\,\rho_1\,\tau_1\,\mathfrak d_1^2+30\,\rho_1\,\tau_1\,\mathfrak d_1-45\,\rho_1\,\tau_1\Big].
\end{align}
The coefficient of Sec.~\ref{secresll4} is $\mathcal C_\parallel=\mathfrak T_\parallel/\left[\tfrac{2}{3}\sum_s\rho_s\,\tilde\nu_s\,\mathcal A_s\right]$.

\subsubsection{Transverse conductivity at the order $t^2$}
\label{secappph4}

For $t=0$, the solution is $\mathcal L^{(0)}_1=4\,\rho_1\,\mathcal A_1/3$ with $\mathcal A_s=\mathcal A^\perp_s$. We expand $\mathcal L^{(1)}_s=t\,\ell_s$ and use Eqs.~\eqref{eqappw4}--\eqref{eqapptau4} and $q_s(u)$ in Eq.~\eqref{eqmomsysperp}. The terms of the order $t$ give
\begin{align}
&\left(1-\tfrac{2}{15}\,\rho_1\,\tau_1\,\gamma^+_{11}\right)\ell_1+\tfrac{4}{15}\,\rho_1\,\tau_1\,\gamma^+_{10}\,\ell_0=\tfrac{4}{15}\,\rho_1\,\mathcal A_1\,\mathfrak p_1-\tfrac{4}{15}\,\rho_1\,\tau_1\,\mathfrak d_1-\tfrac{8}{45}\,\rho_1^2\,\tau_1\,\gamma^+_{11}\,\mathcal A_1\,\mathfrak p_1\,,\nonumber\\
&\tfrac{4}{15}\,\rho_0\,\tau_0\,\gamma^+_{10}\,\ell_1+\left(1-\tfrac{8}{15}\,\rho_0\,\tau_0\,\gamma^+_{00}\right)\ell_0=\tfrac{4}{15}\,\rho_0\,\mathcal A_0\,\mathfrak p_0-\tfrac{4}{15}\,\rho_0\,\tau_0\,\mathfrak d_0\,.
\label{eqappl4}
\end{align}
The terms of the order $t^2$ fix the correction to $\mathcal L^{(0)}_1$. Eq.~\eqref{eqsigperp4} then gives
\begin{align}
\sigma_\perp\big\vert_{B=0}=\frac{e^2}{3\,\pi^2}\sum_s k_s^2\,\mathcal A_s+\frac{e^2}{4\,\pi^2}\,\mathfrak T_\perp\,t^2\,,
\end{align}
where $\mathfrak T_\perp=C^\perp_1\,\ell_1+C^\perp_0\,\ell_0+\mathfrak T^{\rm loc}_\perp$ with
\begin{align}
C^\perp_1&=-\tfrac{2}{45}\Big[6\,\rho_0\,\mathcal A_0\,\gamma^+_{10}\,\mathfrak p_0+2\,\rho_1^2\,\tau_1\,\mathcal A_1\left(\gamma^+_{11}\right)^2\mathfrak p_1-3\,\rho_1\,\mathcal A_1\,\gamma^+_{11}\,\mathfrak p_1-6\,\rho_0\,\tau_0\,\gamma^+_{10}\,\mathfrak d_0+3\,\rho_1\,\tau_1\,\gamma^+_{11}\,\mathfrak d_1\Big],\nonumber\\
C^\perp_0&=\tfrac{4}{45}\Big[6\,\rho_0\,\mathcal A_0\,\gamma^+_{00}\,\mathfrak p_0+2\,\rho_1^2\,\tau_1\,\mathcal A_1\,\gamma^+_{10}\,\gamma^+_{11}\,\mathfrak p_1-3\,\rho_1\,\mathcal A_1\,\gamma^+_{10}\,\mathfrak p_1-6\,\rho_0\,\tau_0\,\gamma^+_{00}\,\mathfrak d_0+3\,\rho_1\,\tau_1\,\gamma^+_{10}\,\mathfrak d_1\Big],\nonumber\\
%%%%%%%%%%%%%%
\mathfrak T^{\rm loc}_\perp&=\tfrac{8}{15}\left(\rho_0\,\mathcal A_0\,\mathfrak e_0+\rho_1\,\mathcal A_1\,\mathfrak e_1\right)
-\tfrac{4\,\rho_0\,\mathfrak q_0}{225}\left(28\,\rho_0\,\mathcal A_0^2\,\gamma^+_{00}-\tfrac{15\,\mathcal A_0^2}{\tau_0}+36\,\rho_1\,\mathcal A_1^2\,\gamma^+_{10}\right)
-\tfrac{4\,\rho_1\,\mathfrak q_1}{225}\left(36\,\rho_0\,\mathcal A_0^2\,\gamma^+_{10}+32\,\rho_1\,\mathcal A_1^2\,\gamma^+_{11}-\tfrac{15\,\mathcal A_1^2}{\tau_1}\right)\nonumber\\
&\quad-\tfrac{4}{135}\Big[18\,\rho_0\,\mathcal A_0\,\mathfrak d_0\,\mathfrak p_0-4\,\rho_1^3\,\tau_1\,\mathcal A_1^2\left(\gamma^+_{11}\right)^2\mathfrak p_1^2+6\,\rho_1^2\,\mathcal A_1^2\,\gamma^+_{11}\,\mathfrak p_1^2 -12\,\rho_1^2\,\tau_1\,\mathcal A_1\,\gamma^+_{11}\,\mathfrak d_1\,\mathfrak p_1+18\,\rho_1\,\mathcal A_1\,\mathfrak d_1\,\mathfrak p_1-9\,\rho_0\,\tau_0\,\mathfrak d_0^2-9\,\rho_1\,\tau_1\,\mathfrak d_1^2\Big].
\end{align}
The coefficient of Sec.~\ref{secresph4} is $\mathcal C_\perp=\mathfrak T_\perp/\left (\tfrac{4}{3}\sum_s\rho_s\,\tilde\nu_s\,\mathcal A^\perp_s\right )$.

\subsubsection{Series of the anomalous-Hall conductivity}
\label{secappah4}

For the sheet $(+,1)$, the radius at $B\neq0$ is the root of $c\,k^3+A\,k^2-\mu\,k+\eta_O\,b\,u/2=0$ with $A=1+t\,u$ and $b=e\,B$. Let $k$ be the root at $b=0$, which obeys $\mu=c\,k^2+A\,k$. The derivative of the cubic at $b=0$ is $k\,\tilde\nu$ with $\tilde\nu=A+2\,c\,k$, and the expansion in $b$ gives Eq.~\eqref{eqkhat4}. Inserting it into Eq.~\eqref{eqsigH4}, we obtain
\begin{align}
\sigma_H^{\rm AH}=-\frac{\eta_B\,e^2}{2\,\pi^2}\sum_{n=0}^{3}\eta_O^n\left(e\,B\right)^nJ_n\,,\quad
J_n=\int_{-1}^{1}du\,u^{n+1}\,\mathfrak k_n(u)\,,
\end{align}
with $\mathfrak k_0=k_{+,1}(u)$. The functions $\mathfrak k_n$ depend on $u$ only through $t\,u$. We expand them in $\tilde\upsilon=t\,u$ at fixed $k_1=k_{\Gamma,1}$ and $\tilde\nu_1=1+2\,c\,k_1$ and integrate term by term. The integrals are
\begin{align}
J_0&=-\frac{2\,k_1}{3\,\tilde\nu_1}\,t-\frac{k_1\left(\tilde\nu_1+1\right)}{5\,\tilde\nu_1^5}\,t^3+\mathcal O\!\left(t^5\right),\quad
J_1=-\frac{1}{3\,k_1\,\tilde\nu_1}+\frac{3\left(\tilde\nu_1-1\right)}{10\,k_1\,\tilde\nu_1^5}\,t^2+\mathcal O\!\left(t^4\right),\nonumber\\
J_2&=-\frac{8\,\tilde\nu_1^2-9\,\tilde\nu_1+3}{20\,k_1^3\,\tilde\nu_1^5}\,t+\frac{5\left(\tilde\nu_1-1\right)^3}{8\,k_1^3\,\tilde\nu_1^9}\,t^3+\mathcal O\!\left(t^5\right),\nonumber\\
J_3&=-\frac{8\,\tilde\nu_1^2-5\,\tilde\nu_1+1}{40\,k_1^5\,\tilde\nu_1^5}-\frac{128\,\tilde\nu_1^4-325\,\tilde\nu_1^3+345\,\tilde\nu_1^2-175\,\tilde\nu_1+35}{112\,k_1^5\,\tilde\nu_1^9}\,t^2+\mathcal O\!\left(t^4\right),
\end{align}
and they give Eq.~\eqref{eqahser4}. The integrals of the odd integrands vanish, which fixes the parity of Eq.~\eqref{eqparity4}.

\subsubsection{Terms of the order $t\,B$}
\label{secappmix4}

The terms of $\sigma_\parallel$ and $\sigma_\perp$ that are linear in both $B$ and $t$ are the leading terms odd in $B$ [cf. Eq.~\eqref{eqparity4}]. We write $b=e\,B$ and keep $v_F=1$. The radius of the sheet $(+,s)$ obeys the cubic equation of Sec.~\ref{secappah4} with $A=s+t\,u$ and the term $\eta_O\,G_s\,b\,u/2$, and all the sheet functions therefore depend on $t$ and $b$ only through the products $t\,u$ and $b\,u$, apart from the explicit factors in the sources. Expanding them to the first order in each parameter, including the mixed term of the order $t\,b$, gives
\begin{align}
\rho_{+,s}(u)&=\rho_s\left[1+\left(\mathfrak p^t_s\,t+\mathfrak p^b_s\,b\right)u+\mathfrak p^{tb}_s\,t\,b\,u^2\right],\quad
h_s(u) = \tilde\nu_s\,u+\sum_{\iota=t,b}\iota\left(\mathfrak h^\iota_{0,s}+\mathfrak h^\iota_{2,s}\,u^2\right)+t\,b\left(\mathfrak h^{tb}_{1,s}\,u+\mathfrak h^{tb}_{3,s}\,u^3\right),\nn
\frac{q_s(u)}{D_s(u)}&=\tilde\nu_s+\sum_{\iota=t,b}\iota\,\mathfrak g^\iota_s\,u+t\,b\,\mathfrak g^{tb}_s\,u^2\,,
\label{eqappmixdata}
\end{align}
where $q_s$ and $D_s$ are the quantities of Appendix~\ref{secappingr}. The coefficients for the tilt are $\mathfrak p^t_s=\mathfrak p_s$, $\mathfrak h^t_{0,s}=1$, and $\mathfrak h^t_{2,s}=\mathfrak g^t_s=-\mathfrak d_s$ of Eqs.~\eqref{eqappw4} and \eqref{eqappsrc4}. With $k_s=k_{\Gamma,s}$, the coefficients for the field are
\begin{align}
& \mathfrak p^b_1 =-\frac{2\,\eta_B\,\tilde\nu_1^2+\eta_O}{2\,\tilde\nu_1^2\,k_1^2}\,,\quad
\mathfrak p^b_0=0\,,\quad
\mathfrak p^{tb}_1=\frac{\left(2\,\eta_B-\eta_O\right)\tilde\nu_1^2+3\,\eta_O}{2\,\tilde\nu_1^4\,k_1^2}\,,\quad
\mathfrak p^{tb}_0=-\frac{\eta_O}{\tilde\nu_0^2\,k_0^2}\,,\quad
\mathfrak h^b_{0,1} =-\frac{2\,\eta_B\,\tilde\nu_1-\eta_O}{2\,k_1^2}\,,\quad
\mathfrak h^b_{0,0}=\frac{\eta_O}{k_0^2}\,,\nn &
%%%%%%%%%%%%%%%%%%%%%%
\mathfrak h^b_{2,1}=\mathfrak g^b_1
=\frac{2\,\eta_B\,\tilde\nu_1^2-3\,\eta_O\,\tilde\nu_1+\eta_O}{2\,k_1^2\,\tilde\nu_1}\,,\quad
\mathfrak h^b_{2,0}=\mathfrak g^b_0=-\frac{3\,\eta_O}{k_0^2}\,,\quad
%%%%%%%%%%%%%%%%%%%%%%%%%%%%%%5
\mathfrak h^{tb}_{1,1} =-\frac{\eta_B\left(\tilde\nu_1+1\right)-\eta_O}{k_1^2\,\tilde\nu_1}\,,\quad
\mathfrak h^{tb}_{1,0}=\frac{2\,\eta_O}{k_0^2\,\tilde\nu_0}\,,\nn &
%%%%%%%%%%%%
\mathfrak h^{tb}_{3,1} =\mathfrak g^{tb}_1=\frac{2\,\eta_B\,\tilde\nu_1^3+2\,\eta_B\,\tilde\nu_1^2-5\,\eta_O\,\tilde\nu_1^2+2\,\eta_O\,\tilde\nu_1-\eta_O}{2\,k_1^2\,\tilde\nu_1^3}\,,\quad
\mathfrak h^{tb}_{3,0}=\mathfrak g^{tb}_0=-\frac{5\,\eta_O}{k_0^2\,\tilde\nu_0}\,.
\label{eqappmixtab}
\end{align}
The coefficients $\mathfrak p^b_0$ and $\mathfrak p^{tb}_0$ reflect that the sheet $s=0$ has no BC, so that only the OMM modifies its radius.

For the sector $m=0$, we write $M^{(n)}_s=t\,m^{t(n)}_s+b\,m^{b(n)}_s$ for the even moments. For $\iota=t,b$, the moments solve Eq.~\eqref{eqappm4} with the sources
\begin{align}
& \mathcal S^{\iota(0)}_s 
=\rho_s\,\tau_s\left(2\,\mathfrak h^\iota_{0,s}+\tfrac{2}{3}\,\mathfrak h^\iota_{2,s}\right)+\tfrac{2}{3}\,\rho_s\,\mathcal A_s\,\mathfrak p^\iota_s-\delta_{s1}\,\tfrac{4}{9}\,\rho_1^2\,\tau_1\,\gamma^+_{11}\,\mathcal A_1\,\mathfrak p^\iota_1\,,\nn &
\mathcal S^{\iota(2)}_s
=\rho_s\,\tau_s\left(\tfrac{2}{3}\,\mathfrak h^\iota_{0,s}+\tfrac{2}{5}\,\mathfrak h^\iota_{2,s}\right)+\tfrac{2}{5}\,\rho_s\,\mathcal A_s\,\mathfrak p^\iota_s-\delta_{s1}\,\tfrac{4}{15}\,\rho_1^2\,\tau_1\,\gamma^+_{11}\,\mathcal A_1\,\mathfrak p^\iota_1\,.
\end{align}
For $\iota=t$, they reduce to those of Appendix~\ref{secappll4}. The coefficient of $t\,b$ in Eq.~\eqref{eqsigpar4} is $\left(e^2/2\,\pi^2\right)e\,\mathfrak T^{tb}_\parallel$, where
\begin{align}
\mathfrak T^{tb}_\parallel=\sum_{n,s}\left(C^{(n)}_s[b]\,m^{t(n)}_s+C^{(n)}_s[t]\,m^{b(n)}_s\right)+\mathfrak T^{\rm loc}_{tb}\,.
\label{eqapptbpar}
\end{align}
The coefficients are linear functions of the data of one parameter,
\begin{align}
C^{(0)}_1[\iota]&=\tfrac{8}{15}\,\rho_0\,\mathcal A_0\,\gamma^-_{10}\,\mathfrak p^\iota_0-\tfrac{4}{15}\,\rho_1^2\,\tau_1\,\mathcal A_1\,\gamma^-_{11}\,\gamma^+_{11}\,\mathfrak p^\iota_1+\tfrac{2}{5}\,\rho_1\,\mathcal A_1\,\gamma^-_{11}\,\mathfrak p^\iota_1
+\tfrac{4}{3}\,\rho_0\,\tau_0\,\gamma^-_{10}\,\mathfrak h^\iota_{0,0}+\tfrac{4}{3}\,\rho_1\,\tau_1\,\gamma^-_{11}\,\mathfrak h^\iota_{0,1}
\nn & \quad 
+\tfrac{8}{15}\,\rho_0\,\tau_0\,\gamma^-_{10}\,\mathfrak h^\iota_{2,0}+\tfrac{2}{5}\,\rho_1\,\tau_1\,\gamma^-_{11}\,\mathfrak h^\iota_{2,1}\,,\nn
%%%%%%%%%%%%%%%%%%%%%%%%
C^{(2)}_1[\iota]&=-\tfrac{4}{15}\,\rho_0\,\mathcal A_0\,\gamma^-_{10}\,\mathfrak p^\iota_0-\tfrac{4}{45}\,\rho_1^2\,\tau_1\,\mathcal A_1\,\gamma^-_{11}\,\gamma^+_{11}\,\mathfrak p^\iota_1+\tfrac{2}{15}\,\rho_1\,\mathcal A_1\,\gamma^-_{11}\,\mathfrak p^\iota_1
-\tfrac{4}{15}\,\rho_0\,\tau_0\,\gamma^-_{10}\,\mathfrak h^\iota_{2,0}+\tfrac{2}{15}\,\rho_1\,\tau_1\,\gamma^-_{11}\,\mathfrak h^\iota_{2,1}\,,\nn
%%%%%%%%%%%%%%%%%%%%%%%%%%%%%%%%%%%%
C^{(0)}_0[\iota]&=\tfrac{4}{15}\,\rho_0\,\mathcal A_0\,\gamma^-_{00}\,\mathfrak p^\iota_0-\tfrac{16}{45}\,\rho_1^2\,\tau_1\,\mathcal A_1\,\gamma^-_{10}\,\gamma^+_{11}\,\mathfrak p^\iota_1+\tfrac{8}{15}\,\rho_1\,\mathcal A_1\,\gamma^-_{10}\,\mathfrak p^\iota_1
+\tfrac{4}{3}\,\rho_0\,\tau_0\,\gamma^-_{00}\,\mathfrak h^\iota_{0,0}+\tfrac{4}{3}\,\rho_1\,\tau_1\,\gamma^-_{10}\,\mathfrak h^\iota_{0,1}
\nn & \quad
+\tfrac{4}{15}\,\rho_0\,\tau_0\,\gamma^-_{00}\,\mathfrak h^\iota_{2,0}+\tfrac{8}{15}\,\rho_1\,\tau_1\,\gamma^-_{10}\,\mathfrak h^\iota_{2,1}\,,\nn
%%%%%%%%%%%%%%%%%%%%%%%%%%%%%%%%%%%%%
C^{(2)}_0[\iota]&=\tfrac{8}{15}\,\rho_0\,\mathcal A_0\,\gamma^-_{00}\,\mathfrak p^\iota_0+\tfrac{8}{45}\,\rho_1^2\,\tau_1\,\mathcal A_1\,\gamma^-_{10}\,\gamma^+_{11}\,\mathfrak p^\iota_1-\tfrac{4}{15}\,\rho_1\,\mathcal A_1\,\gamma^-_{10}\,\mathfrak p^\iota_1
+\tfrac{8}{15}\,\rho_0\,\tau_0\,\gamma^-_{00}\,\mathfrak h^\iota_{2,0}-\tfrac{4}{15}\,\rho_1\,\tau_1\,\gamma^-_{10}\,\mathfrak h^\iota_{2,1}\,,
\end{align}
and the local part is
\begin{align}
\mathfrak T^{\rm loc}_{tb}&=\tfrac{4}{3}\sum_s\rho_s\,\mathcal A_s\,\mathfrak h^{tb}_{1,s}+\tfrac{4}{5}\sum_s\rho_s\,\mathcal A_s\,\mathfrak h^{tb}_{3,s}
-\tfrac{2\,\rho_0\,\mathfrak p^{tb}_0}{225}\left(44\,\rho_0\,\mathcal A_0^2\,\gamma^+_{00}-\tfrac{45\,\mathcal A_0^2}{\tau_0}+28\,\rho_1\,\mathcal A_1^2\,\gamma^+_{10}\right)
\nn & \quad
-\tfrac{2\,\rho_1\,\mathfrak p^{tb}_1}{225}\left(28\,\rho_0\,\mathcal A_0^2\,\gamma^+_{10}+36\,\rho_1\,\mathcal A_1^2\,\gamma^+_{11}-\tfrac{45\,\mathcal A_1^2}{\tau_1}\right)
+\tfrac{4}{45}\Big[\rho_0\,\mathcal A_0\,\mathfrak p^t_0\left(15\,\mathfrak h^b_{0,0}+9\,\mathfrak h^b_{2,0}\right)+\rho_0\,\mathcal A_0\,\mathfrak p^b_0\left(9\,\mathfrak h^t_{2,0}+15\right)\nn
&\qquad+\rho_1\,\mathcal A_1\,\mathfrak p^t_1\left(15\,\mathfrak h^b_{0,1}+9\,\mathfrak h^b_{2,1}\right)+\rho_1\,\mathcal A_1\,\mathfrak p^b_1\left(9\,\mathfrak h^t_{2,1}+15\right)
-\rho_1^2\,\tau_1\,\mathcal A_1\,\gamma^+_{11}\,\mathfrak p^t_1\left(10\,\mathfrak h^b_{0,1}+6\,\mathfrak h^b_{2,1}\right)-\rho_1^2\,\tau_1\,\mathcal A_1\,\gamma^+_{11}\,\mathfrak p^b_1\left(6\,\mathfrak h^t_{2,1}+10\right)\nn
&\qquad+4\,\rho_1^3\,\tau_1\,\mathcal A_1^2\left(\gamma^+_{11}\right)^2\mathfrak p^t_1\,\mathfrak p^b_1-6\,\rho_1^2\,\mathcal A_1^2\,\gamma^+_{11}\,\mathfrak p^t_1\,\mathfrak p^b_1
+\rho_0\,\tau_0\left(15\,\mathfrak h^b_{0,0}\,\mathfrak h^t_{2,0}+45\,\mathfrak h^b_{0,0}+9\,\mathfrak h^b_{2,0}\,\mathfrak h^t_{2,0}+15\,\mathfrak h^b_{2,0}\right)\nn
&\qquad+\rho_1\,\tau_1\left(15\,\mathfrak h^b_{0,1}\,\mathfrak h^t_{2,1}+45\,\mathfrak h^b_{0,1}+9\,\mathfrak h^b_{2,1}\,\mathfrak h^t_{2,1}+15\,\mathfrak h^b_{2,1}\right)\Big],
%%%%%%%%%%%%
\quad \mathcal A_s=\mathcal A^\parallel_s\,.
\end{align}
%%%%%%%%%%%%%%%%%%%%

For the sector $m=1$, the moments $\mathcal L^{(1)}_s=t\,\ell^t_s+b\,\ell^b_s$ solve, for $\iota=t,b$,
\begin{align}
\left(1-\tfrac{2}{15}\,\rho_1\,\tau_1\,\gamma^+_{11}\right)\ell^\iota_1+\tfrac{4}{15}\,\rho_1\,\tau_1\,\gamma^+_{10}\,\ell^\iota_0&=\tfrac{4}{15}\,\rho_1\,\mathcal A_1\,\mathfrak p^\iota_1-\tfrac{8}{45}\,\rho_1^2\,\tau_1\,\gamma^+_{11}\,\mathcal A_1\,\mathfrak p^\iota_1+\tfrac{4}{15}\,\rho_1\,\tau_1\,\mathfrak g^\iota_1\,,\nn
\tfrac{4}{15}\,\rho_0\,\tau_0\,\gamma^+_{10}\,\ell^\iota_1+\left(1-\tfrac{8}{15}\,\rho_0\,\tau_0\,\gamma^+_{00}\right)\ell^\iota_0&=\tfrac{4}{15}\,\rho_0\,\mathcal A_0\,\mathfrak p^\iota_0+\tfrac{4}{15}\,\rho_0\,\tau_0\,\mathfrak g^\iota_0\,,
\end{align}
with $\mathcal A_s=\mathcal A^\perp_s$ from here on. The frequency $\omega_s(u)$ contributes only to the Hall part at this order, since it enters $\sigma_\perp$ through even powers. The coefficient of $t\,b$ in the real part of Eq.~\eqref{eqsigperp4} is $\left(e^2/4\,\pi^2\right)e\,\mathfrak T^{tb}_\perp$ with $\mathfrak T^{tb}_\perp=\sum_s\left(C^\perp_s[b]\,\ell^t_s+C^\perp_s[t]\,\ell^b_s\right)+\mathfrak T^{\rm loc}_{\perp,tb}$. The coefficients are
\begin{align}
C^\perp_1[\iota]&=-\tfrac{4}{15}\,\rho_0\,\mathcal A_0\,\gamma^+_{10}\,\mathfrak p^\iota_0-\tfrac{4}{45}\,\rho_1^2\,\tau_1\,\mathcal A_1\left(\gamma^+_{11}\right)^2\mathfrak p^\iota_1+\tfrac{2}{15}\,\rho_1\,\mathcal A_1\,\gamma^+_{11}\,\mathfrak p^\iota_1-\tfrac{4}{15}\,\rho_0\,\tau_0\,\gamma^+_{10}\,\mathfrak g^\iota_0+\tfrac{2}{15}\,\rho_1\,\tau_1\,\gamma^+_{11}\,\mathfrak g^\iota_1\,,\nn
C^\perp_0[\iota]&=\tfrac{8}{15}\,\rho_0\,\mathcal A_0\,\gamma^+_{00}\,\mathfrak p^\iota_0+\tfrac{8}{45}\,\rho_1^2\,\tau_1\,\mathcal A_1\,\gamma^+_{10}\,\gamma^+_{11}\,\mathfrak p^\iota_1-\tfrac{4}{15}\,\rho_1\,\mathcal A_1\,\gamma^+_{10}\,\mathfrak p^\iota_1+\tfrac{8}{15}\,\rho_0\,\tau_0\,\gamma^+_{00}\,\mathfrak g^\iota_0-\tfrac{4}{15}\,\rho_1\,\tau_1\,\gamma^+_{10}\,\mathfrak g^\iota_1\,,
\end{align}
and the local part is
\begin{align}
\mathfrak T^{\rm loc}_{\perp,tb}&=-\tfrac{4}{675}\Big[\mathcal A_0^2\,\rho_0\,\mathfrak p^{tb}_0\left(84\,\rho_0\,\gamma^+_{00}-\tfrac{45}{\tau_0}\right)+108\,\mathcal A_0^2\,\rho_0\,\rho_1\,\gamma^+_{10}\,\mathfrak p^{tb}_1
-90\,\rho_0\,\mathcal A_0\left(\mathfrak p^b_0\,\mathfrak g^t_0+\mathfrak p^t_0\,\mathfrak g^b_0+\mathfrak g^{tb}_0\right)-90\,\rho_0\,\tau_0\,\mathfrak g^b_0\,\mathfrak g^t_0\nn
&\qquad \quad
+\mathcal A_1^2\left(108\,\rho_0\,\rho_1\,\gamma^+_{10}\,\mathfrak p^{tb}_0+96\,\rho_1^2\,\gamma^+_{11}\,\mathfrak p^{tb}_1-\tfrac{45\,\rho_1}{\tau_1}\,\mathfrak p^{tb}_1\right)
+\mathcal A_1^2\,\rho_1^2\,\gamma^+_{11}\,\mathfrak p^t_1\,\mathfrak p^b_1\left(60-40\,\rho_1\,\tau_1\,\gamma^+_{11}\right)\nn
&\qquad \quad
-90\,\rho_1\,\mathcal A_1\left(\mathfrak p^b_1\,\mathfrak g^t_1+\mathfrak p^t_1\,\mathfrak g^b_1+\mathfrak g^{tb}_1\right)+60\,\rho_1^2\,\tau_1\,\mathcal A_1\,\gamma^+_{11}\left(\mathfrak p^b_1\,\mathfrak g^t_1+\mathfrak p^t_1\,\mathfrak g^b_1\right)
-90\,\rho_1\,\tau_1\,\mathfrak g^b_1\,\mathfrak g^t_1\Big].
\end{align}

\subsubsection{Hall conductivity at the orders $t^2B$ and $t\,B^2$}
\label{secappmixH4}

The Hall conductivity at the orders $t^2B$ and $t\,B^2$ is linear in the frequency $\omega_s(u)$, since the terms with three powers of $\omega_s$ are of the order $B^3$. We therefore expand Eq.~\eqref{eqsigperp4} to the first order in $\omega_s$ and keep the weights, the sources, and the in-scattering term at $\omega_s=0$. The in-scattering kernel is symmetric under the exchange of the sheets with the weights $\left(1-u^2\right)\rho_s(u)$, and this gives
\begin{align}
\sigma_H^{\rm B}=\frac{e^2}{4\,\pi^2}\sum_s\int_{-1}^{1}du\left(1-u^2\right)\rho_s(u)\,\omega_s(u)\,\Lambda_s(u)^2\,,
\label{eqappHall4}
\end{align}
where $\Lambda_s(u)=\tau_s(u)\left[q_s(u)/D_s(u)+\check\lambda_s(u)\right]$ is the real solution of the sector $m=1$ at $\omega_s=0$, with $\check\lambda_s(u)$ of Eq.~\eqref{eqlam4}. For $t=0$ and $\eta_B=\eta_O=0$, Eq.~\eqref{eqappHall4} gives the Hall conductivity linear in $B$ of Eq.~\eqref{eqsigHB3}. We write $\omega_s=b\,\bar\omega_s$ with $b=e\,B$ and expand $\bar\omega_s(u)$ in powers of $t\,u$ and $b\,u$ as
\begin{align}
\bar\omega_s(u)=\mathfrak w_{0,s}+\left(\mathfrak w^t_s\,t+\mathfrak w^b_s\,b\right)u+\left(\mathfrak w^{tt}_s\,t^2+\mathfrak w^{tb}_s\,t\,b\right)u^2+\dots
\end{align}
with
\begin{align}
&\mathfrak w_{0,1}=\frac{\tilde\nu_1}{k_1}\,,\quad
\mathfrak w^t_1=\frac{1}{k_1\,\tilde\nu_1}\,,\quad
\mathfrak w^b_1=\frac{2\,\eta_B\,\tilde\nu_1^2-2\,\eta_O\,\tilde\nu_1+\eta_O}{2\,k_1^3\,\tilde\nu_1}\,,\quad
\mathfrak w^{tt}_1=\frac{\tilde\nu_1-1}{2\,k_1\,\tilde\nu_1^3}\,,
\nn &
%%%%%%%%%%%%%
\mathfrak w^{tb}_1=\frac{4\,\eta_B\,\tilde\nu_1^3
+2\,\eta_B\,\tilde\nu_1^2-6\,\eta_O\,\tilde\nu_1^2+3\,\eta_O\,\tilde\nu_1-\eta_O}{2\,k_1^3\,\tilde\nu_1^3}\,,
%%%%%
\mathfrak w_{0,0}=\frac{\tilde\nu_0}{k_0}\,,\quad
\mathfrak w^t_0=\mathfrak w^{tt}_0=0\,,\quad
\mathfrak w^b_0=-\frac{2\,\eta_O}{k_0^3}\,,\quad
\mathfrak w^{tb}_0=-\frac{6\,\eta_O}{k_0^3\,\tilde\nu_0}\,.
\end{align}
The remaining data are those of Eqs.~\eqref{eqappw4}, \eqref{eqappsrc4}, and \eqref{eqappmixtab}, with $\mathfrak p^{tt}_s=\mathfrak q_s$ and $\mathfrak g^{tt}_s=\mathfrak e_s$. We use $\mathcal A_s=\mathcal A^\perp_s$ and the moments $\ell^\iota_s$ of the previous subsection. The notation $\left(\iota,\iota'\right)\in\tilde\iota$ denotes the ordered pairs with $\iota\,\iota'=\tilde\iota$, which are $\left(t,t\right)$ for $\tilde\iota=tt$ and $\left(t,b\right)$ and $\left(b,t\right)$ for $\tilde\iota=tb$.

We expand the solution as $\Lambda_s(u)=\mathcal A_s+\sum_\iota\iota\,\Lambda^\iota_s\,u+\sum_{\tilde\iota}\tilde\iota\left(\Lambda^{\tilde\iota}_{0,s}+\Lambda^{\tilde\iota}_{2,s}\,u^2\right)$. The odd part of the rate at the first order is $\mathfrak a^\iota=\tfrac{2}{3}\,\gamma^+_{11}\,\rho_1\,\mathfrak p^\iota_1$ for $s=1$ and vanishes for $s=0$. The coefficients of the first order are
\begin{align}
\Lambda^\iota_1&=\tau_1\left[\mathfrak g^\iota_1+\tfrac{1}{2}\,\gamma^+_{11}\,\ell^\iota_1-\gamma^+_{10}\,\ell^\iota_0-\mathfrak a^\iota\,\mathcal A_1\right],\quad
\Lambda^\iota_0=\tau_0\left[\mathfrak g^\iota_0-\gamma^+_{10}\,\ell^\iota_1+2\,\gamma^+_{00}\,\ell^\iota_0\right].
\end{align}
For the second order, we need the rate coefficients $\mathfrak b^{\tilde\iota}_s$ and $\mathfrak c^{\tilde\iota}_s$, which are those of Eq.~\eqref{eqapptau4} with $\mathfrak q_s$ replaced by $\mathfrak p^{\tilde\iota}_s$, and the second-order part $\tfrac{1}{2}\,\gamma^+_{11}\,\mathcal X^{\tilde\iota}$ of $\check\lambda_1$. With $\mathcal D_1=\left(1-\tfrac{2}{3}\,\rho_1\,\tau_1\,\gamma^+_{11}\right)^{-1}$, the moment $\mathcal X^{\tilde\iota}$ of the order $\tilde\iota$ is
%%%%%%%%%%%%%%
\begin{align}
\mathcal X^{\tilde\iota}&=\tfrac{2}{15}\,\rho_1\,\tau_1\sum_{(\iota,\iota')\in\tilde\iota}\mathfrak p^\iota_1\left(\gamma^+_{11}\,\ell^{\iota'}_1-2\,\gamma^+_{10}\,\ell^{\iota'}_0\right)
+\tfrac{4\,\rho_1\,\mathcal D_1}{675}\Big[\mathcal A_1\left(20\,\rho_1^2\,\tau_1^2\left(\gamma^+_{11}\right)^2-30\,\rho_1\,\tau_1\,\gamma^+_{11}\right)\sum_{(\iota,\iota')\in\tilde\iota}\mathfrak p^\iota_1\,\mathfrak p^{\iota'}_1\nn
&\quad-108\,\mathcal A_1\,\rho_0\,\tau_1\,\gamma^+_{10}\,\mathfrak p^{\tilde\iota}_0-96\,\mathcal A_1\,\rho_1\,\tau_1\,\gamma^+_{11}\,\mathfrak p^{\tilde\iota}_1
+45\,\mathcal A_1\,\mathfrak p^{\tilde\iota}_1
+\tau_1\left(45-30\,\rho_1\,\tau_1\,\gamma^+_{11}\right)\sum_{(\iota,\iota')\in\tilde\iota}\mathfrak p^\iota_1\,\mathfrak g^{\iota'}_1+45\,\tau_1\,\mathfrak g^{\tilde\iota}_1\Big],
\end{align}
and the second-order coefficients of $\Lambda_s$ are
\begin{align}
\Lambda^{\tilde\iota}_{0,s} =\tau_s\left[\delta_{s1}\,\tfrac{1}{2}\,\gamma^+_{11}\,\mathcal X^{\tilde\iota}-\mathfrak b^{\tilde\iota}_s\,\mathcal A_s\right],\quad
\Lambda^{\tilde\iota}_{2,s} =\tau_s\left[\mathfrak g^{\tilde\iota}_s-\mathfrak c^{\tilde\iota}_s\,\mathcal A_s-\delta_{s1}\sum_{(\iota,\iota')\in\tilde\iota}\mathfrak a^\iota\,\Lambda^{\iota'}_1\right].
\end{align}
Inserting these expansions into Eq.~\eqref{eqappHall4} and integrating over $u$ with the weight $1-u^2$ gives
\begin{align}
\sigma_H^{\rm B} & \supset\frac{e^2}{4\,\pi^2}\left[e\,B\,t^2\,\mathfrak H_{tt}+\left(e\,B\right)^2t\,\mathfrak H_{tb}\right],\nn
%%%%%%%%%%%%%%%%%
\mathfrak H_{\tilde\iota}&=\frac{4}{15}\sum_s\rho_s\Big\{\mathcal A_s^2\left(\mathfrak p^{\tilde\iota}_s\,\mathfrak w_{0,s}+\mathfrak w^{\tilde\iota}_s\right)+\mathfrak w_{0,s}\,\mathcal A_s\left(10\,\Lambda^{\tilde\iota}_{0,s}+2\,\Lambda^{\tilde\iota}_{2,s}\right)
+\sum_{(\iota,\iota')\in\tilde\iota}\Big[\mathcal A_s^2\,\mathfrak p^\iota_s\,\mathfrak w^{\iota'}_s+2\,\mathcal A_s\,\Lambda^\iota_s\left(\mathfrak p^{\iota'}_s\,\mathfrak w_{0,s}+\mathfrak w^{\iota'}_s\right)+\mathfrak w_{0,s}\,\Lambda^\iota_s\,\Lambda^{\iota'}_s\Big]\Big\}.
\end{align}
The term of the order $t^2B$ is the correction to the Hall conductivity linear in $B$. It does not depend on $\eta_B$ and $\eta_O$, because $B$ enters only through $\omega_s$ at this order. The term of the order $t\,B^2$ is the leading Hall term that is even in $B$. For $\eta_B=\eta_O=0$, the latter vanishes, as stated in Sec.~\ref{secresph4}.

\section{Numerical implementation}
\label{secnum}

The numerical solution of the SBE that underlies the figures of Secs.~\ref{secsys}, \ref{secsys2}, and \ref{secsys3} uses $e=\hbar=1$. The sheets are specified by the dispersion $\epsilon_a(k)=v_a\left(\tilde\lambda_a\,k+c_a\,k^2\right)$, the chemical potential $\mu_a$ measured from the node, the chirality $\chi_\nu$, and the numbers $N_a$ and $G_a$ and the velocity $v_\nu$ of Eq.~\eqref{eqbcomm}. The offset $\Delta$ enters through $\mu_R=\mu+\Delta$.

\subsection{Fermi surfaces and kinetic ingredients}

The radius $k_a(\theta)$ of the sheet $a$ is the positive root of $\xi_a(k,\theta)=\mu_a$ [cf. Eq.~\eqref{eqxiw}] that is continued from the radius at $B=0$. It is obtained by the Newton iteration at every direction of the grid. The quantities $D_a$, $w_{a,k}$, $q_a$, $\rho_a$, $\omega_a$, and $h_a$ of Eq.~\eqref{eqwqd} follow from $k_a(\theta)$ with the analytical derivatives. Sheets that do not cross $\mu$, or that are cut by the quadratic term at a large radius, are not included, as stated in Secs.~\ref{secresnum} and \ref{secresnum2}. The grids in $\theta$ are Gauss-Legendre grids in $u=\cos\theta$, and the weights of the quadrature are denoted by $w_i$.

\subsection{Magnetic field along the polar axis}

For $\bB=B\,\boldsymbol{\hat z}$, the sector equations of Eq.~\eqref{eqsector} are discretised on the grid $u_i$ with the weights $w_i$. In the sector $m$, they read
\begin{align}
\left[\tau_a^{-1}(u_i)+i\,m\,\omega_a(u_i)\right]\Lambda^{(m)}_a(u_i)-\sum_{a'}\sum_{i'}w_{i'}\,\rho_{a'}(u_{i'})\,\mathcal K^{(m)}_{aa'}(u_i,u_{i'})\,\Lambda^{(m)}_{a'}(u_{i'})=h^{(m)}_a(u_i)\,,
\end{align}
with the out-scattering rate of Eq.~\eqref{eqtau} evaluated by the same quadrature. The in-scattering term lies in the finite spans of Eq.~\eqref{eqspans}, and the discretised equations are therefore the moment system of Eq.~\eqref{eqmomsys} on a grid. In the sector $m=0$, the null vector of Eq.~\eqref{eqboltzop} is removed by appending the number constraint of Eq.~\eqref{eqnumber} as an extra row, and the overdetermined system is solved by least squares. The conductivities $\sigma_\parallel$, $\sigma_\perp$, and $\sigma_H^{\rm B}$ follow from Eqs.~\eqref{eqsigpar} and \eqref{eqsigperp} by the same quadrature, and $\sigma_H^{\rm AH}$ follows from Eq.~\eqref{eqsigah2}.

\subsection{General orientation of the magnetic field}

For the cases (b) and (c), the sectors are coupled, and Eq.~\eqref{eqtiltsol} is solved with the moments of Eq.~\eqref{eqmomsys}. The functions $f^a_p$ are the real orthonormal spherical harmonics of the frame of the nodes with $l\leq l_a$, where $l_a$ is the highest rank of the kernel at the sheet $a$. The overlaps are polynomials in the Cartesian components of $\hat{\bk}$ and $\hat{\bk}'$ of the degrees given in Appendix~\ref{secappingr}. The matrices
\begin{align}
Q^{aa'}_{pp'}=\frac{\beta_{aa'}^2}{\pi}\int d^2\hat{\bk}\int d^2\hat{\bk}'\,f^a_p(\hat{\bk})\,\mathcal O_{aa'}(\hat{\bk},\hat{\bk}')\,f^{a'}_{p'}(\hat{\bk}')
\end{align}
are therefore evaluated exactly by a product of a Gauss-Legendre grid in $\cos\theta$ and a uniform grid in $\varphi$, and the finite-rank form of Sec.~\ref{secfinrank} is exact. The overlaps with the full dependence on both azimuths are used, namely Eq.~\eqref{eqovgr4} for the QTSM and Weyl nodes, and Eq.~\eqref{eqovgr} with $\mathcal O=\mathcal O^{(0)}+\mathcal O^{(1)}\cos\left(\varphi-\varphi'\right)+\mathcal O^{(2)}\cos\left(2\,\varphi-2\,\varphi'\right)$ for the RSW and TSM nodes. The intranode overlaps are the functions of $C=\hat{\bk}\cdot\hat{\bk}'$ of Eqs.~\eqref{eqovrsw} and \eqref{eqovtsm}.

For $\hat{\bB}$ along $\left(\sin\theta_B,0,\cos\theta_B\right)$ in the frame of the nodes, we introduce the frame $\left(\hat{\bs e}_1,\hat{\bs e}_2,\hat{\bB}\right)$ with the polar angle $\theta'$ and the azimuth $\varphi'$ about $\bB$. The FS of every sheet is axisymmetric about $\bB$, and the LFO of Eq.~\eqref{eqlfoaxi} is $\omega_a(\theta')\,\partial_{\varphi'}$ in this frame. The resolvent $\hat{\mathcal R}_a=\left(\tau_a^{-1}+\omega_a\,\partial_{\varphi'}\right)^{-1}$ is evaluated on a product grid of Gauss-Legendre points in $\cos\theta'$ and uniform points in $\varphi'$, with the Fourier spectral differentiation matrix for $\partial_{\varphi'}$. At every $\theta'$, the matrix $\tau_a^{-1}+\omega_a\,\partial_{\varphi'}$ is inverted, and the result is applied to the sources $h^j_a$ and to the functions $f^a_p$. The sources are
\begin{align}
h^j_a=\frac{1}{D_a}\left[w_{a,j}+\eta_B\,e\,B_j\left(\bOm^a\cdot\bw_a\right)\right],
\end{align}
where $\bw_a=w_{a,k}\,\hat{\bk}+w_{a,\theta}\,\hat{\bs\theta}'$ is rotated into the frame of the nodes, and $B_j$ is the component of $\bB$. The matrices $K^a_{pp'}$ and the sources $S^j_{a,p}$ of Eq.~\eqref{eqmomsys} are the moments with the weight $\rho_a$ on the same grid. The number constraint closes the moment system as in the sectors, and one matrix serves all of $j=x,y,z$. The mean free-path is then reconstructed from Eq.~\eqref{eqtiltsol}, and the tensor follows from
\begin{align}
\sigma_{ij}=\frac{e^2}{8\,\pi^3}\sum_a\int d^2\hat{\bk}\,\rho_a\,h^i_a\,\Lambda^j_a-\sigma_H^{\rm AH}\,\epsilon_{ijl}\,\hat B_l\,,
\end{align}
with $\sigma_H^{\rm AH}$ of Eq.~\eqref{eqsigah2}. The unknowns, the ranks $l_a$, and the grids are collected in Table~\ref{tabnum}.

\begin{table}[t!]
\centering
\small
\renewcommand{\arraystretch}{1.5}
\begin{tabular}{|c|c|c|}
\hline
Quantity & QTSM and Weyl nodes & RSW and TSM nodes \\
\hline
Ranks $l_a$ & \makecell{$2$ for $(\Gamma,1)$ \\ $1$ for $R_i$} & \makecell{$3$ for $(\Gamma,\tilde\psi)$ \\ $2$ for $R_i$} \\
\hline
Moments for each $j$ & $17$ & $50$ \\
\hline
Grid of the resolvent, $\theta'\times\varphi'$ & $24\times24$ & $24\times24$ \\
\hline
Grid of the matrices $Q^{aa'}$, $\theta\times\varphi$ & $10\times20$ & $16\times32$ \\
\hline
Grid in $u$ for $\bB\parallel\hat{\boldsymbol z}$ & $120$ & $120$ \\
\hline
\end{tabular}
\caption{\label{tabnum} Ranks of the harmonics, numbers of moments, and grids of the numerical solution. The sheet $(\Gamma,0)$ is absent for the parameters of Sec.~\ref{secresnum}, and the moments of the two copies at $R$ are counted separately.}
\end{table}

\subsection{Normalisation}

The amplitude $\beta_{\rm intra}$ is fixed by the out-scattering rate of the reference sheet at $B=0$, which is set to $1/\tau_0=\mu/\left(\mu\,\tau_0\right)$ with $\mu\,\tau_0=15$. The reference sheet is $(\Gamma,1)$ for the QTSM and Weyl nodes and $(\Gamma,3/2)$ for the RSW and TSM nodes. The amplitudes between the nodes are $r_\beta$ times $\beta_{\rm intra}$. For the overlaps of Eqs.~\eqref{eqovgriso} and \eqref{eqovgr}, the rate of the reference sheet at $B=0$ is independent of the direction. For the overlaps of Eq.~\eqref{eqovgr8c}, the rate depends on the direction, and its average over the grid with the quadrature weights fixes $\beta_{\rm intra}$. The field is measured by $\alpha=e\,B/\left(2\,k_{\rm ref}^2\right)$ with the radius $k_{\rm ref}$ of the smallest sheet at $B=0$, and the conductivities are given in units of the values at $B=0$ stated in the text.

\section{Symmetry of the nodes at the P point}
\label{secappP}

In this appendix, we derive the statements about the little group at the point P in the space groups 199, 214, and 220 that are used in the comparison with earlier work. The space groups are generated from the general positions of their conventional cells together with the centring translation $\left(1/2,1/2,1/2\right)$ of the body-centred lattice. The operations are enumerated, and the rotation part of each of them is tested against the condition for the little group at P. The numbers obtained in this way are collected in Table~\ref{tabP}.

\subsection{Action of the point group on P}

The rotation part of every operation of the three space groups is a signed permutation matrix $g=\operatorname{diag}\left(\mathfrak r_1,\mathfrak r_2,\mathfrak r_3\right)\mathcal B$, where $\mathcal B$ is a permutation matrix and $\mathfrak r_i=\pm1$. In units of $2\pi/a$, the point P is at $\bk_{\rm P}=\left(1/2,1/2,1/2\right)$. Since every row of $g$ contains a single nonzero entry, we have
\begin{align}
g\,\bk_{\rm P}=\frac{1}{2}\left(\mathfrak r_1,\mathfrak r_2,\mathfrak r_3\right).
\end{align}
The reciprocal lattice of the body-centred lattice consists of the integer triples with an even sum. The difference $g\,\bk_{\rm P}-\bk_{\rm P}$ has the components $\left(\mathfrak r_i-1\right)/2$, which are $0$ or $-1$, and their sum equals minus the number $\mathfrak n_-$ of minus signs among the $\mathfrak r_i$. Therefore $g\,\bk_{\rm P}$ is equivalent to $\bk_{\rm P}$ for an even $\mathfrak n_-$. For an odd $\mathfrak n_-$, the sum $g\,\bk_{\rm P}+\bk_{\rm P}$ has the components $\left(\mathfrak r_i+1\right)/2$ with the sum $3-\mathfrak n_-$, which is even, and $g\,\bk_{\rm P}$ is equivalent to $-\bk_{\rm P}$. The points P and $-$P are inequivalent, since $2\,\bk_{\rm P}=\left(1,1,1\right)$ has an odd sum. The relation $\det g=\mathfrak r_1\,\mathfrak r_2\,\mathfrak r_3\,\operatorname{sgn}\mathcal B$ turns the condition for an even $\mathfrak n_-$ into
\begin{align}
g\,\bk_{\rm P}\simeq\bk_{\rm P}\ \Longleftrightarrow\ \mathfrak r_1\,\mathfrak r_2\,\mathfrak r_3=1\ \Longleftrightarrow\ \det g=\operatorname{sgn}\mathcal B\,,
\end{align}
where $\simeq$ denotes the equality up to a reciprocal-lattice vector. A proper rotation belongs to the little group at P if its permutation is even, and an improper operation belongs to it if its permutation is odd. The twelve proper rotations with an even permutation form the tetrahedral group $\mathrm T$. They are the identity, the three twofold rotations about the coordinate axes, and the eight threefold rotations about the $\langle111\rangle$ axes. The twelve improper operations that satisfy the condition are the six mirrors, which exchange two coordinates, such as $m_{1\bar{1}0}$, and the six fourfold rotoinversions about the coordinate axes. Together with $\mathrm T$, they form the group $\mathrm T_{\rm d}$. The remaining twelve rotations of the group $\mathrm O$ have an odd permutation and map P onto $-$P. They are the six twofold rotations about the $\langle110\rangle$ axes and the six fourfold rotations about the coordinate axes.

\begin{table}[t!]
\centering
\small
\renewcommand{\arraystretch}{1.5}
\begin{tabular}{|c|c|c|c|c|c|}
\hline
Group & \makecell{Point \\ group} & \makecell{Operations in \\ the conventional cell} & \makecell{Rotation parts \\ mapping P to P} & \makecell{Rotation parts \\ mapping P to $-$P} & \makecell{Invariant \\ vectors} \\
\hline
199 & $\mathrm T$ & 24 & \makecell{12 \\ (12 proper)} & 0 & 0 \\
\hline
214 & $\mathrm O$ & 48 & \makecell{12 \\ (12 proper)} & \makecell{12 \\ (6 twofold, 6 fourfold)} & 0 \\
\hline
220 & $\mathrm T_{\rm d}$ & 48 & \makecell{24 \\ (12 proper, 12 improper)} & 0 & 0 \\
\hline
220.91 & $\overline{4}'3\,m'$ & \makecell{48 \\ (24 unitary)} & \makecell{12 \\ (all unitary)} & \makecell{12 \\ (all antiunitary)} & 0 \\
\hline
\end{tabular}
\caption{\label{tabP} Numbers of operations of the space groups 199, 214, 220, and of the magnetic group 220.91, and their action on the point P. The last column gives the dimension of the space of vectors $\mathbf t$ that are invariant under the unitary little group at P.}
\end{table}

\subsection{Absence of a tilt at P}

Let $\mathbf t$ be the tilt, such that the tilt term at the node is $\mathbf t\cdot\bk$ with $\bk$ measured from P. This term is invariant under the unitary little group if and only if $g\,\mathbf t=\mathbf t$ for every rotation part $g$ of the little group, since both $\mathbf t$ and $\bk$ are polar vectors. In the space groups 199 and 214, the little group has the twelve rotation parts of $\mathrm T$, and in the space group 220 it contains them. The group $\mathrm T$ is generated by the threefold rotation $C_3$ about $[111]$ and the twofold rotation $C_{2z}$ about $\boldsymbol{\hat z}$ \cite{cano-multifold-magnetic}. The first operation permutes the components of $\mathbf t$ cyclically, and the second acts as
\begin{align}
C_{2z}\,\mathbf t=\left(-t_x,-t_y,t_z\right).\label{eqtiltinv}
\end{align}
The invariance under $C_3$ requires $t_x=t_y=t_z$, and the invariance under $C_{2z}$ requires $t_x=t_y=0$. Hence $\mathbf t=0$ in the space groups 199, 214, and 220. The additional improper operations of the space group 220 do not change this conclusion, since they only add conditions on $\mathbf t$.

\subsection{Chirality of the nodes at P and $-$P}

The rotations of the space groups 199 and 214 are proper, and they preserve the flux of the BC through a sphere around the node. Time reversal maps the node at P onto a node at $-$P with the same flux. In the space group 199, all operations map P onto itself, and none of them relates P to $-$P. In the space group 214, the twelve rotation parts that map P onto $-$P are proper, and they again relate two nodes of the same chirality. No operation of these groups maps a node onto one of opposite chirality, and a pair of nodes with opposite chiralities at P and $-$P does not occur.

For the space group 220, we consider an improper operation $g$ of the little group at P. It maps a small sphere $\mathscr S$ around P onto itself, and it maps each band onto the band of the same energy. The BC is a pseudovector, and it transforms as
\begin{align}
\bOm_s\left(g\,\bk\right)=\det g\;g\,\bOm_s\left(\bk\right).
\end{align}
The outward normal of the sphere obeys $\hat{\mathbf n}\left(g\,\bk\right)=g\,\hat{\mathbf n}\left(\bk\right)$, and $g$ is orthogonal. The flux of the band $s$ through $\mathscr S$, $\Phi_s=\oint_{\mathscr S} dA\,\hat{\mathbf n}\cdot\bOm_s$, thus satisfies
\begin{align}
\hat{\mathbf n}\left(g\,\bk\right)\cdot\bOm_s\left(g\,\bk\right)=\det g\;\hat{\mathbf n}\left(\bk\right)\cdot\bOm_s\left(\bk\right)\ \Longrightarrow\ \Phi_s=\det g\,\Phi_s\,.
\end{align}
For $\det g=-1$, this gives $\Phi_s=0$. If a band is degenerate on the sphere, the same argument applies to the trace of the non-Abelian curvature over the degenerate multiplet. The mirror $m_{1\bar{1}0}$ is an improper operation of the little group at P, and the flux of every band through $\mathscr S$ vanishes. The space group 220 therefore hosts no chiral node at P, and the nodes at P and $-$P are related by time reversal only.

%%%%%%%%%%%%%5
\subsection{The magnetic group 220.91}

The unitary subgroup of the magnetic group 220.91 is the space group 199. The improper operations of $\mathrm T_{\rm d}$ enter only in the product with time reversal \cite{cano-multifold-magnetic}. An antiunitary operation, which is the product of time reversal and the improper operation $g$, maps $\bk$ onto $-g\,\bk$. Since all operations of $\mathrm T_{\rm d}$ map P onto itself, it maps P onto $-$P, and it is not an element of the little group at P. The little group at P thus consists of the twelve unitary rotations of the space group 199, as given in Table~\ref{tabP}, and it has no invariant vector [cf. Eq.~\eqref{eqtiltinv}]. A tilt at P is therefore excluded. The antiunitary operations relate P to $-$P. Since the improper operation reverses the flux of the BC and time reversal preserves it, they relate nodes of opposite chirality. The tilt of these nodes vanishes by the argument above, and a pair of opposite chiralities with non-vanishing tilts is not obtained.

\end{document}